\documentclass[12pt ]{book} 
\usepackage[utf8]{inputenc}
\usepackage{amssymb}
\usepackage{amsfonts}
\newcommand\Snn{\mathcal{S}_{nn}(\vec{k},\,\omega)}
\let\cleardoublepage=\clearpage
\usepackage{amsmath,amssymb, textcomp,titlesec}
\usepackage{graphicx,subfigure,color,cite,etoolbox,epsfig}
\usepackage{setspace,changepage}
\usepackage[toc,page,titletoc]{appendix}
\usepackage{enumitem,environ,tikz}
\usepackage{slashed}

\usepackage{epsfig}
\usetikzlibrary{shapes,arrows,matrix}
\tikzstyle{process} = [rectangle, minimum width=3cm, minimum height=4cm, text centered, draw=black, fill=orange!30]
\tikzstyle{arrow} = [thick,->,>=stealth]
\usepackage{tabularx,colortbl,fancyhdr}
\usepackage[margin=10pt,font=small,labelfont=bf]{caption}
\graphicspath{{./Figures/}}
\titleformat{\chapter}[display]
{\normalfont\bfseries\filcenter}
{\LARGE\MakeUppercase{\chaptertitlename} \thechapter}
{1ex}
{\titlerule[2pt]
\vspace{2ex}%
\LARGE}
[\vspace{1ex}%
{\titlerule[2pt]}]
\usepackage[top    = 2.55cm,bottom = 3.50cm,left   = 3.10cm,right  = 3.0cm]{geometry}

\def\nn{\nonumber\\}

\def\nn{\nonumber\\}

\def\[{\left[}
\def\]{\right]}  

\usepackage{bm}
\usepackage{color}
\usepackage{setspace}
\usepackage{amsmath}
\usepackage{amssymb}
\usepackage{pdfpages}
\usepackage[utf8]{inputenc}
\def \beq{\begin{equation}}
\def \eeq{\end{equation}}
\def \beqa{\begin{eqnarray}}
\def \eeqa{\end{eqnarray}}
\def \pd{\partial}
\def\nn {\nonumber}
\usepackage{graphicx}
\usepackage{epstopdf} 

\usepackage{xspace}

\usepackage{epsfig}
\usepackage{amsthm}
\usepackage{amsfonts}
\usepackage{float}
\usepackage{amsmath,amssymb}
\usepackage{color}
\usepackage{slashed}
\usepackage{relsize}
\usepackage{bbold}
\usepackage{graphicx}
\usepackage{dcolumn}
\usepackage{bm}
\usepackage[normalem]{ulem}
\usepackage{cancel}
\usepackage{tikz}

\newcommand{\ep}{\epsilon}

\newcommand{\vk}{\bold  k}

\newcommand{\rw}{\rightarrow}
\newcommand{\mn}{\mu\nu}

\newcommand{\bseq}{\begin{subequations}}
	\newcommand{\eseq}{\end{subequations}}

\newcommand{\B}{ {\mathcal B} }

\newcommand{\K}{ {\mathcal K} }

\newcommand{\Q}{ {\mathcal Q} }

\newcommand{\R}{ {\mathcal R} }

\usepackage{physics}

\def\sl{\hspace{-0.15cm}/}

\def\vec#1{\mathchoice
        {\mbox{\boldmath $#1$}}
        {\mbox{\boldmath $#1$}}
        {\mbox{\boldmath $\scriptstyle #1$}}
        {\mbox{\boldmath $\scriptscriptstyle #1$}}
}
\usepackage{color} 
\usepackage[normalem]{ulem}  

  \usepackage{pdfpages}
\usepackage[colorlinks = true,
            linkcolor = blue,
            urlcolor  = blue,
            citecolor = blue,
            anchorcolor = black]{hyperref}

\author{By\\Md Hasanujjaman}
\begin{document} 

\begin{center}
 \vskip 0.9 cm
{\Large {\bf {\huge 
{ \doublespacing
	Effects Of Critical Point On \\Quark Gluon Plasma}}}

}
\vskip 3.0cm
{\bf{ \large Thesis submitted for the degree of \\
DOCTOR OF PHILOSOPHY (Science)\\
in\\
PHYSICS (Theoretical)}}
\vskip 01.70cm
{\bf {\em By}} 
\vskip -0.2cm
{\bf {\large MD HASANUJJAMAN}}

\vskip 2.7cm
{\bf {\Large Department of Physics}}
\vskip 0.5cm
{\bf {\Large UNIVERSITY of CALCUTTA}}
\vskip 0.5cm
{\bf {\large 2022}}
\end{center}

\frontmatter
\centerline{{\bf{\large{DECLARATION} }}}
\vskip 1.2cm
I, hereby declare that this dissertation represents my works in my own words and where others' ideas have been included, I have adequately cited and referred the original sources. The work is original and has not been submitted earlier as a whole or in part of a degree/diploma at this or any other Institution/University.
\vskip 2.5cm
\leftline{Md Hasanujjaman}
\vskip 0.1cm
\leftline{Kolkata, India}
 \newpage
\centerline{{\bf {\large DEDICATIONS}}}
\vskip 5cm
\hskip 5cm
\hspace{4cm}{to \Large {{\bf{AMMA}}}} 
\newpage
~
\vskip -.50cm
\centerline{{\bf{\large ACKNOWLEDGEMENT}}}
\vskip 0.30cm

{\it{Firstly, I would like to thank my thesis advisor Prof. Jan-e Alam for his continuous encouragement, involvement, and suggestions in completing the thesis. 
His expertise was invaluable in formulating the research questions and methodology. No finite word can encompass my gratitude to his efforts toward completing the thesis. Especially, after I have joined Darjeeling Government College, he kept on pushing for the works to complete the thesis. The door was always open whenever I ran into a trouble or had a question about my research or writing. He consistently steered me in the right direction whenever he thought I needed it. Without his guidance, this thesis would not have been a reality. I also thank my Joint-Supervisor, Prof. Abhijit Bhattacharyya for his constant efforts and supports to carry out the works in completion of the thesis. I am really grateful to have them as my supervisors.

 I am also thankful to all my collaborators, Dr. Mahfuzur Rahaman, Dr. Golam Sarwar, Dr. Sandeep Chatterjee, Dr. Trambak Bhattacharyya, Sushant Kumar Singh, Dr. Snigdha Ghosh, and Dr. Sourav Sarkar for the academic discussions. Their mentality towards the academic activity always encourages me to enhance the academic skills. We have discussed numerous things beyond the academics, which was very much enjoyable and fascinating.

I do not want to miss the chance to thank my fellow mates in VECC in my Ph.D tenure. For that I thank Rajendra, Arghya, Pingal, Somnath, Prabrisa, and Vijay. I specially thank Sarwar, Ashik, Atta da, Noor, and Mahfuzur for the distinctive mental support. We have spent quality time together even beyond my Ph.D tenure at VECC.

In this occasion, I would like to thank my teacher Partha Sarkar, who always appreciated and encouraged to pursue my career in Physics. I also thank Raju da, Dr. Shibaji Banerjee, and Dr. Suparna Roychowdhury for making the subject interesting. Their guidance are very much appreciated.

I would like to thank Department of Atomic Energy (DAE) for the financial support for the period I have spent there. I also thank the Department of Higher Education (WBHED) of Govt. of West Bengal, India for the support.

 Besides them, I am thankful to Amma, Abba, and Raju for the support whenever I needed it the most. The debt can never be repaid. I also want to thank my lifelong friends Bablu, Shibu, Milton Da, Moid, Saradindu and Nayan for the character installed in me. I guess the friendship among ourselves will grow stronger with time. Your support will always intrigue me.

 Finally, I would like to thank my wife Najnin Islam for her love and constant support for tolerating me over the past few years. Thank you for being my best friend. I owe you a lot.
 
 Over all, I thank the almighty for everything.}}

\vskip 3cm
\rightline{Md Hasanujjaman \hspace{0.9cm}}

\newpage
\centerline{{\bf{\Large{List of publications arising from the thesis}}}} 
\begin{enumerate}
	\item
		{\bf ``Dispersion and suppression of sound near the QCD critical point''},
		\\{} {\bf{Md. Hasanujjaman}}, {M. Rahaman}, A.  Bhattacharyya and J. Alam,
		 \\{}\href{https://journals.aps.org/prc/abstract/10.1103/PhysRevC.102.034910} {\underline{Phys. Rev. C } {\bf 102}, (2020) 034910.}
		\href{https://arxiv.org/abs/2003.07575}{[arXiv:2003.07575[nucl-th]].} 
	\item
	{\bf ``Dynamical spectral structure of density fluctuation near the QCD critical point''},
	{} {\bf{Md. Hasanujjaman}}, Golam Sarwar, {M. Rahaman}, A.  Bhattacharyya and J. Alam,  
{}\href{https://doi.org/10.1140/epja/s10050-021-00589-3}{\underline{Eur. Phys. J. A} {\bf 57},  283 (2021).} \href{https://arxiv.org/abs/2008.03931}{	arXiv:2008.03931 [nucl-th].}	  
\item
{\bf ``Role of slow, out-of-equilibrium modes on the dynamic structure
factor near the QCD critical point''},
{}G. Sarwar,  {\bf{Md. Hasanujjaman}}, and J. Alam, 
 \href{https://arxiv.org/abs/2205.01136}{\underline{arXiv:2205.01136}.} {\it{Submitted to Phys. Rev. D}}.
\item
{\bf ``The fate of nonlinear perturbations near the QCD critical point''},
{}G. Sarwar,  {\bf{Md. Hasanujjaman}}, {M. Rahaman}, A. Bhattacharyya and J. Alam, {}\href{https://doi.org/10.1016/j.physletb.2021.136583}{\underline{Phys.  Lett.  B } {\bf 820}, (2021)  10, 136583: } \href{https://arxiv.org/abs/2012.12668}{arXiv:2012.12668 [nucl-th].}
\end{enumerate} 

 \centerline{{\bf{\Large{Other publications}}}} 
\begin{enumerate}
	\item
	{\bf ``Initial condition from the shadowed Glauber model''},
	{}S.~Chatterjee, S.~K.~Singh, S.~Ghosh, {\bf{M. Hasanujjaman}}, J.~Alam and S.~Sarkar,\\{}\href{https://doi.org/10.1016/j.physletb.2016.05.022}{\underline{Phys. Lett. B} \textbf{758} (2016), 269-273; \href{https://arxiv.org/abs/1510.01311}{[arXiv:1510.01311 [nucl-th]].}}
	
	\item
	{\bf ``Nonlinear waves in a hot, viscous and nonextensive quark gluon plasma''},
	{}G.~Sarwar, {\bf{M. Hasanujjaman}}, T.~Bhattacharyya, M.~Rahaman, A.~Bhattacharyya and J.~Alam, {}\href{https://link.springer.com/article/10.1140/epjc/s10052-022-10122-5}{\underline{Eur. Phys. J. C} {\bf 82},  189 (2022). } 

\end{enumerate} 

\vskip 2cm
\rightline{Md Hasanujjaman \hspace{0.9cm}}
%

\cleardoublepage
 \addcontentsline{toc}{chapter}{Contents}
\tableofcontents
\chapter*{Notation}
$\mathcal{L}_{QCD}$: The QCD Lagrangian\\
 $N_c$: number of colors \\ 
$	D_{\mu} =\partial_\mu-i\,g_sT_a A^a_\mu  $: covariant derivative\\ 
$D=u^{\mu}\partial_{\mu}$: Comoving derivative in local rest frame.\\
$g_s$: strong coupling constant.\\
$A_\mu ^a$: non-Abelian gauge fields with $a=1, 2, \cdots, N_c^2-1$ being the color representation  \\
$T_a$: generators of the $SU(N_c)$ group satisfying the group algebra $[T_a, T_b]=if_{abc}T_c$ \\
 $\psi_{i,f}$: Dirac spinor field for quark \\ 
$	G_{\mu\nu}^a$: gluon field strength tensor\\
$\sqrt{s}$: centre of mass energy of nucleus-nucleus collision\\
 $\Lambda _{QCD}$: QCD scale parameter \\
$T$: temperature\\
$T_{c}$: temperature at critical end point\\
$\mu_{B}$: baryon chemical potential \\
$\xi$: correlation length\\
$\mu_{c}$: baryon chemical potential at critical end point\\
$P$: thermodynamic pressure\\
$V$: volume of a system\\
$R$: molar gas constant.\\
$M$: magnetization\\
$t=(T-T_{c})/T_{c}$: reduced temperature\\
$\alpha, \beta, \nu, \gamma, \delta$: critical exponents\\
$h$: magnetic field strength\\
$\chi, \chi_{B}$: magnetic susceptibility, baryon susceptibility\\
$C_{i}$: cumulants of $i^{th}$ order\\
$K_{n}$: Knudsen number\\
$\lambda$: mean free path\\
$R$: size of a system\\
$\eta $: shear viscosity\\
$\zeta $:  bulk viscosity
\\$\kappa $: thermal conductivity  \\
$s$: entropy density \\
$ u^\mu$: fluid four velocity\\
$T^{\mu\nu }$: energy  momentum tensor (EMT)\\   
$N^\mu $: particle four-current \\ 
 $S^\mu $: entropy four-current\\ 
$\Delta^{\mu\nu}=g^{\mu\nu}-u^\mu u^\nu$: projection 
operator   \\
$n$ : net baryon number density\\
 $\Pi $ : bulk pressure \\ 
  $\epsilon$:   energy density\\   
 $h (=\epsilon+P)$: enthalpy density  \\
$\tau^{\mu\nu}$ and $n^\mu$:   dissipative currents to $T^{\mu\nu}$ and $N^{\mu}$ respectively  \\
$\Pi$: bulk viscous pressure\\ 
$\pi^{\mu\nu}$: shear-stress tensor. \\
$h^\mu$: energy-diffusion four-current\\
$\beta_\nu=u_\nu /T$\\
 $\beta  =\frac{1}{T}$: inverse of temperature\\
 $\alpha=\mu/T$: thermal potential\\
$\mathcal{Q}^\mu$: deviations from local equilibrium\\ 
$\delta N^\mu\equiv N^\mu-N^\mu_{(0)}$, $\delta T^{\mu\nu}\equiv 
T^{\mu\nu}-T^{\mu\nu}_{(0)}$, where subscript $(0)$ indicates equilibrium \\  
$\beta _0,\beta_1,\beta_2$ : relaxation coefficients\\
$\alpha_0$ and $\alpha_1$ : coupling coefficients\\ 
 $\tau_{\Pi}$: relaxation times for bulk pressure 
 \\$\tau_q$:   relaxation times for heat flux  \\ $\tau_{\pi}$:  relaxation times for  tensor\\
$l_{\Pi q}, l_{q\Pi} $:   relaxation lengths  couple  the  heat flux and bulk  pressure \\
 $ l_{q\pi}, l_{\pi q} $: relaxation lengths heat flux and  the shear tensor  \\
 $\mathcal{B}$: Bag constant\\
 $s_{c}$: entropy density at the critical end point\\
 $s_{Q}$: entropy density for qgp gas\\
 $s_{H}$: entropy density for hadron gas\\
 $g_{q}, g_{g}$: degeneracy factor of quarks and gluons\\
$F, G$: Helmholtz free energy and Gibbs free energy\\
$\omega$: frequency\\
 $P_1,\epsilon_1, n_1, T_1$ and $ u^\alpha_1$:  tiny perturbations in   \\ 
 $\vk$:  wave vector \\
 $\omega_{\Re}$: real part of frequency.\\
  $\omega_{im}$: imaginary part of frequency\\
 $c_s$: speed of sound  in  fluid \\
   $k=3hc_s/(2\eta) =k_v$\\
    $L_\eta=\eta/(hc_s)$  
    $k_{th}$: threshold wave-vector\\
  $\lambda_{th}$: threshold wavelength\\ 
  $\rho $: particle number density\\
  $\mathcal{F}$: fluidity of system\\
   $R_v\sim k_v^{-1}$, scale of viscous horizon\\
  $n_{v}$: Order of harmonics\\
  $\mathcal{S}_{nn}(k, \omega)$: structure factor\\
  $I_{R}, I_{B}$: intensity of Rayleigh and Brillouin peak\\
  $C_{P}, C_{V}$: isobaric and isochoric specific heat\\
  $\kappa_{T}, \kappa_{s}$: isothermal and adiabatic compressibility\\
  $\alpha_{P}$: volume expansivity coefficients\\
  $\mathcal{K}$: opacity factor\\
  $R_{AA}$: nuclear suppression factor\\
  $R_{HBT}$: Hanburry-Brown-Twiss radius\\
  $\phi$: out-of-equilibrium slow mode\\
  $\pi$: chemical potential corresponds to $\phi$\\
  $C_{A\pi}=(\frac{\partial \pi}{\partial A})$\\
  $\hat{\epsilon_{i}}$: perturbation in energy density of $i^{th}$ order
 \chapter*{Effects Of Critical Point On Quark Gluon Plasma}
\hspace*{6cm}\underline{\Huge{{\bf{Abstract}}}}
\vspace*{1cm}

 We have used the relativistic second-order causal viscous hydrodynamics equipped with an equation of state which includes the QCD critical point to study the propagation of linear and nonlinear perturbations in quark gluon plasma created in a relativistic heavy-ion collision. We have also studied the behaviour of dynamic structure factor, which is calculated by taking the correlation of dynamical density fluctuation. It is found that the critical point has significant effects on the propagation of linear as well as on nonlinear perturbations. It is also found that the dynamic structure factors are substantially modified with the presence of the QCD critical point.

{\doublespacing

}

{\onehalfspacing


}
{\doublespacing

\cleardoublepage
\addcontentsline{toc}{chapter}{List of Figures}
\listoffigures

 \addcontentsline{toc}{chapter}{List of Tables}
\listoftables

\mainmatter
\pagestyle{fancy}
\fancyhead{}
\fancyfoot{}
\fancyfoot{}
\lhead{\leftmark}
\setlength{\headheight}{16.5pt}
\cfoot{\thepage}
 
\chapter[Introduction]{Introduction}
\label{chapter1}  
\section{The elementary particles and interactions in Standard Model}
\label{sec0101}
Human beings are always fascinated to think what's next? The hunger to know even beyond our limitations is what drives us towards discovering new things. By looking up the stars, we could discover astronomy. Starting from our own Milky way galaxy (size, $d\sim 10^{16}$m) to stars in galaxy ($d\sim 10^{9}$m) to planets (earth with $d\sim 10^{6}$m) to human ($d\sim 2$m) to atoms ($d\sim 10^{-10}$m) to nucleus ($d\sim 10^{-14}$m) to nucleons ($d\sim 10^{-15}$m) to partons (constituents of proton, $d\sim 10^{-16}$m), we have come a long way to find out the basic ingredients of matter. Since 1930s, physicists have put efforts in discovering an exceptional insight into the basic structure of matter: everything in the universe is consisting of a few basic building blocks called fundamental particles, which are governed by four fundamental forces which are known to be the strong force, electromagnetic force, weak force and the gravitational force. So far with our best understanding, we know how the fundamental particles interact with three known forces (except the gravity) belongs to the Standard Model~\cite{Oerter2006} of particle physics. The theory of the Standard Model is described by the Quantum Field Theory (QFT). The Standard Model has potential to explain vast amount of experimental results and could precisely predict wide variety of phenomenon. It has become established as a well-tested physics theory over time. The Standard Model, as presently drawn up, has sixty one ($61$) elementary particles (including anti-particles), are able to form composite particles, attributing for the hundreds of different species of particles, have been discovered since the 1960s.
\begin{table}[h]
\centering
\includegraphics[width=15.cm,height=8.5cm]{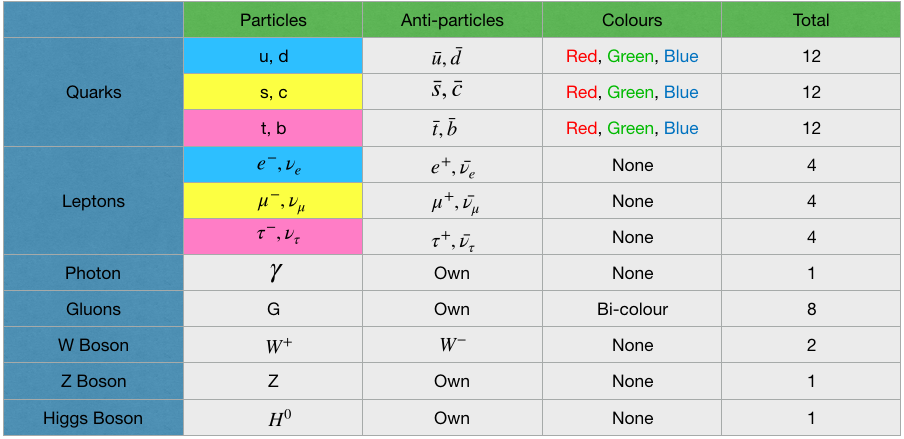}	
\caption{Elementary particles, in total turns out to be 61.}
\label{table1}
\end{table}

Elementary particles~\cite{Griffiths2008} are arranged in two groups, called quarks and leptons. Each group are subdivided into three `generations'. The most stable and lightest particles are kept in the first generation, whereas the less stable and heavier particles belong to both the second and third generations. Therefore, the universe with stable elements are consisting of particles that belong to the first generation because the heavier ones rapidly decays into the lightest stable ones. Amongst six quarks (flavours), the `up quark' and the `down quark' make up the first generation. The `charm quark' and `strange quark, and then the `top quark' and `bottom (or beauty) quark are kept in the second and the third generations respectively. A quark may also come in three different `colours' and a colourless object may be created by mixing the quarks or a quark and an anti-quark. Three leptons are similarly organized in three generations: the `electron' in the first generation, the `muon' in the second generation, and the `tauon' in the third generation. Each leptons have their corresponding neutrinos, which are electrically neutral with very little (or zero) mass. 


Among the four basic interactions, we are concerned with the strong interaction here. The theory of strong interaction is formulated in Quantum Chromodynamics (QCD). It is kind of a quantum field theory (QFT) which belongs to non-abelian gauge theory, with symmetry group SU(3). It characterizes the interaction between the quarks with the help of the gauge boson: gluons. The dynamics of the quarks and gluons (together called as partons) are dictated by the QCD Lagrangian is written as
\beqa
\mathcal{L_{QCD}}= \sum_{f}\, \bar{\psi_{i,f}}\big(i\gamma^{\mu}D_{\mu}-m\big)\,\psi_{i,f}-\frac{1}{4}G^{\mu \nu}_{a}G^{a}_{\mu\nu}
\eeqa
where, $D_{\mu}=\partial_{\mu}-i\,g_{s}\,T_{a}\,A^{a}_{\mu}$ is the gauge covariant derivative and $g_{s}$ is the strong coupling constant. $A^{a}_{\mu}$ 's are the non-Abelian gauge fields with $a=1,\,2,\,...., \,N_{c}^{2}-1$ being the color representation. $T_{a}$'s are the generators of the SU($N_{c}$) group satisfying the group algebra $[T_{a},\,T_{b}]=i\,f_{abc}T_{c}$. $\psi_{i,f}$'s are the Dirac spinor for the quark fields. The index $i$ represents the quark color for $N_{c}=3$ (red, green, blue) and the index $f$ refer to the quark/anti-quark flavor, $N_{f}=6$ (up, down, strange, charm, bottom, top). For $N_{c}=3$, we have $N_{c}^{2}-1=8$ gluon fields. The $\gamma^{\mu}$ are the Dirac matrices connecting the spinor representation to the vector representation of the Lorentz group. The symbol $G^{\mu\nu}_{a}$ is gauge invariant gluon field tensor, like the term $F^{\mu\nu}$ (photon field tensor) in Quantum Electrodynamics (QED), is given by
\beqa
G^{a}_{\mu\nu}=\pd_{\mu}A^{a}_{\nu}-\pd_{\nu}A^{a}_{\mu}+g_{s}\,f_{abc}A^{b}_{\mu}A^{c}_{\nu}
\eeqa
The third term in the right hand side of the above equation represents the self interaction between gluons.

 The QCD exhibits two salient features:
 
 a) {\bf{Color confinement}}~\cite{PhysRevD.10.2445,Mandelstam,tHooft:1975krp}: It is the phenomenon that color-charged particles {\it{i.e.}} quarks and gluons cannot be isolated at low energies, therefore in normal condition direct observation of those partons are not possible. The partons stay together inside the hadrons, and can not be separated from their parent hadron without producing new hadrons.
 
 b) {\bf{Asymptotic freedom}}~\cite{PhysRev.96.191,Gross1973,Politzer1973,CollinsandPerry1975,Cabibbo1975}: This phenomena signifies that the interaction between quarks gets weaker as the distance between them gets smaller, and quarks are asymptotically free. This is called asymptotic freedom. This indicates that at high energies, the partons interact weakly.

 

After the discovery of asymptotic freedom in QCD, the first successful applications were to deep inelastic scattering (DIS)~\cite{DIS} to probe the structure of hadrons (particularly the baryons, such as protons and neutrons), using beams of electrons, muons and neutrinos. It provided the first convincing evidence of the existence of quarks.


From the idea of confinement and asymptotic freedom, it can be concluded that at very high energies, when the quarks gets very closer to each other, the interaction between them gets weaker. Just after the discovery of the asymptotic freedom, authors in Ref.\cite{CollinsandPerry1975} had shown that at very high density, the nuclear matter will dissolve to its basic constituents, {\it{i.e.}} quarks and gluons. Subsequently, studies~\cite{Shuryak1978,Shuryak1980} confirmed that at very high temperature and density the properties of nuclear matter is governed by quarks and gluons, not by the hadrons. A thermalized state of quarks and gluons is called quark gluon plasma (QGP), which may have existed after a few microsecond of the Big Bang~\cite{Yagi:2005yb}. But on further expansion of the universe, the partons get confined within the hadrons and till then no free quarks and gluons are observed in the nature. However, such a deconfined state as QGP is believed to exist in the core of the neutron stars~\cite{Shuryak1980,Nature2020}.

 In laboratory, the QGP can be created by colliding nuclei at relativistic energies. The magnitude of density and temperature required to form a deconfined state of quarks and gluons can be achieved by colliding heavy ions at ultra-relativistic high energies. The typical value of temperature for transition from hadronic matter to QGP is $\sim 10^{12}\,K$, and density $\gtrsim 3-4$ times the normal nuclear matter density. At the high temperature and density, the hadrons overlap substantially, and as a result, the quarks and gluons confined within a hadronic volume get deconfined and roam throughout the nuclear volume in a system formed in heavy-ion collisions. The QGP state produced in nuclear collision is very transient in nature and having a small volume. Therefore, the detection of the QGP is extremely challenging. However, there are many studies which confirms that QGP can be created by colliding heavy ions at ultra-relativistic high energies~\cite{PhysRevD.31.545,SHURYAK1978150,SINHA198391,PhysRevLett.58.101,ALAM1996243,ALAM2000159,Niida:2021wut,SINGH1993147,KOCH1986167,Gazdzicki:1996pk,PhysRevD.51.3408,PhysRevC.51.1444,TIWARI1997225,Muller:1994rb,Bjorken:1982tu,Gyulassy1990,Gyulassy1992,Matsui:1986dk,Kluberg:2009wc,NA50:2004sgj,Wong:1997rm}.

\section{Signatures of QGP in Heavy-ion-collision}
\label{sec0103}
The official announcement of CERN on Feb 10, 2010, stated-``compelling evidence now exists for the formation of a new state of matter at energy densities about 20 times larger than that the centre of atomic nuclei and temperature about 100000 times higher than in the centre of the sun. This state exhibits characteristic properties which can not be understood with conventional hadronic dynamics but which are qualitatively consistent with expectations from the formation of a state of matter in which quarks and gluons no longer feel the constraint of color confinement.''

There are numerous signatures which confirms that a medium of QGP is created by a collision. In this dissertation, we are not going to discuss the details of the signatures confirming the formation of QGP. However, one may look into the details of such signatures, {\it{e.g}} 

i) Direct photons and dileptons~~\cite{PhysRevD.31.545,SHURYAK1978150,SINHA198391,PhysRevLett.58.101,ALAM1996243,ALAM2000159,Niida:2021wut,SINGH1993147}~,

ii) Strangeness enhancement~\cite{KOCH1986167,Gazdzicki:1996pk,PhysRevD.51.3408,PhysRevC.51.1444,TIWARI1997225,Muller:1994rb}~,

iii) Jet quenching~\cite{Bjorken:1982tu,Gyulassy1990,Gyulassy1992}~,

iv) $J/\psi$ suppression~\cite{Matsui:1986dk,Kluberg:2009wc,NA50:2004sgj,Wong:1997rm,SINGH1993147}~,

v) and many more.

\section{Experimental station}
\label{sec0104}
To recreate the quark matter in the laboratory, physicists need the help of particle colliders which can provide huge centre of mass (CM) energy. The CM energy ranges from $\sqrt{s}=2.4 \,GeV-5.04 \,TeV$ are available at the moment to conduct experiments with beam of atomic nuclei, with mass numbers ranging from proton $(A = 1)$ to Uranium $(A = 238)$. The Relativistic Heavy-Ion Collider (RHIC) at Brookhaven National Laboratory (BNL), and the Large Hadron Collider (LHC) at CERN, both of which collide either protons or heavy ions, or protons with ions. The ALICE (A Large Ion Collider Experiment) Collaboration, is one of eight detector experiments at the LHC, is optimized for studying heavy-ion collisions, and at RHIC there are two principal experiments in operation, STAR (for Solenoidal Tracker at RHIC) and PHENIX (for Pioneering High Energy Nuclear Interaction eXperiment) to conduct. There are other experiments, such as the Compressed Baryonic Matter Experiment (CBM) at the Facility for Antiproton and Ion Research (FAIR) and the Multi Purpose Detector (MPD) at NICA (Nuclotron-based Ion Collider fAcility) at the Joint Institute for Nuclear Research (Dubna, Russia) which are both in advanced stages of construction. A typical heavy-ion collision is depicted in Fig.\ref{fig0103}.

\begin{figure}[h]
\centering
\includegraphics[width=14.2cm,height=9.cm]{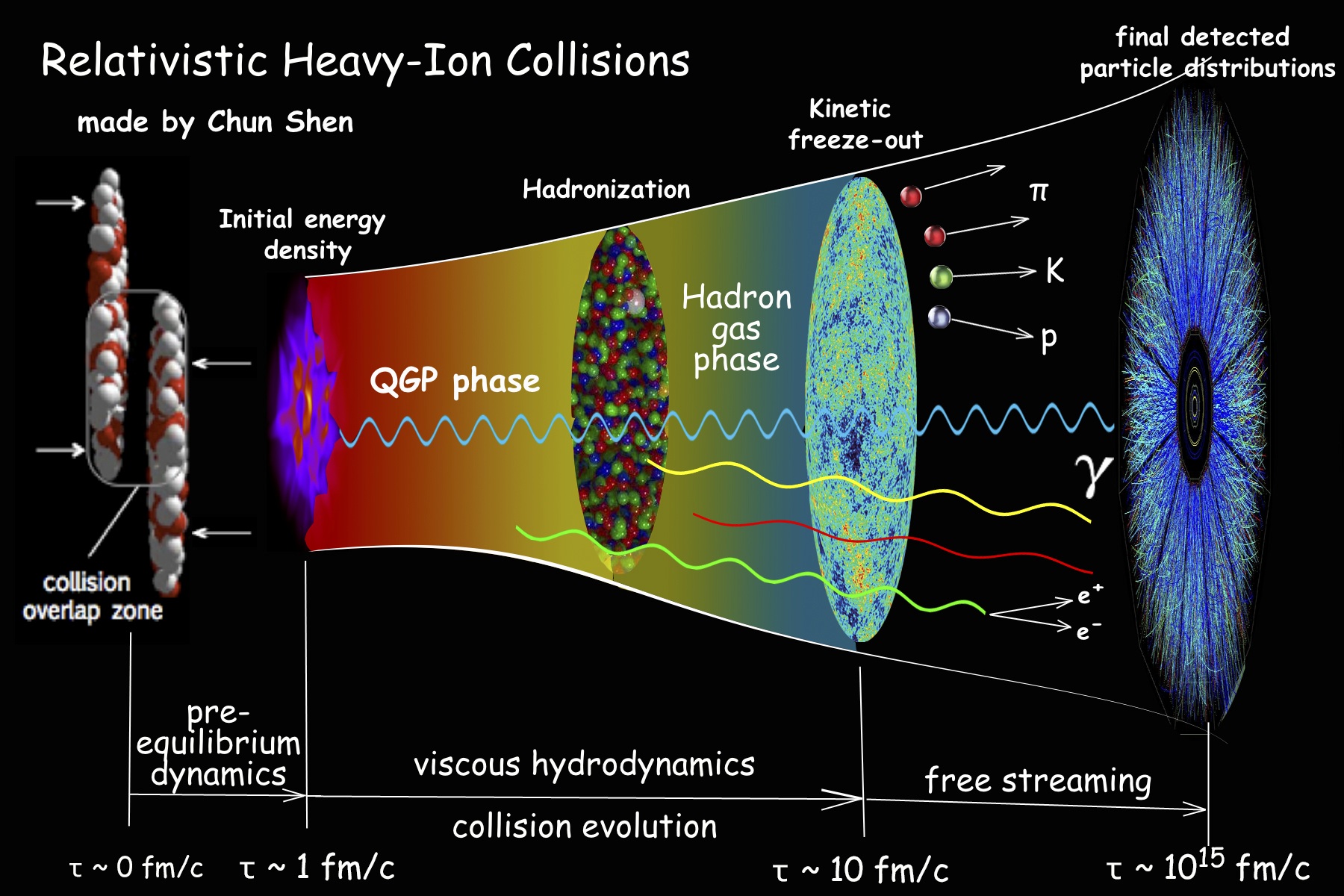}	
\caption{Typical diagram of heavy-ion collision~\cite{Shen:2014vra}.}
\label{fig0103}
\end{figure}

\section{QCD phase diagram and its prospects}
\label{sec0105}
\label{phase_diagram}
QCD is a remarkably successful non-abelian quantum field theory to describe the strong interaction. Asymptotic freedom of QCD allows to use perturbative technique to calculate physical quantities with quarks and gluons as fundamental degrees of freedom. Full analytical treatment of QCD is not trivial because if we neglect the masses of the quarks, the theory has no numerically rudimentary parameters, and the confinement scale ($\Lambda_{QCD}\sim 200 MeV \sim 1 fm^{-1}$) becomes solely the independent intrinsic scale of this theory. But for large values of the thermodynamic parameters such as $T$ (temperature) and baryon chemical potential ($\mu_{B}$), the laws of thermodynamics is dominated by short-distance QCD dynamics, and due to the asymptotic freedom the theory can be studied purturbatively.

Thermodynamic properties of a system are generally expressed in thermodynamic parameter space, simply known as a phase diagram. For the case of the QCD phase diagram, it is represented in $T-\mu_{B}$ parameter space. Any point on the phase diagram corresponds to an equilibrium state, generally characterized by different thermodynamic parameters, such as, $P$ (pressure), baryon density ($n_{B}$) as well as kinetic coefficients, like diffusion or viscous coefficients, or other characteristics of different correlation functions.

Lattice simulations has been applied to describe the phase diagram very accurately at $\mu_{B}=0.$ But the study of thermodynamics of the QCD at $\mu_{B}\ne 0$ is very difficult. The major hindrance in lattice simulations is the so called `sign problem'~\cite{Gavai:2014ela}, and so far no method has been devised to overcome this problem. However, the interesting part (especially the QCD critical point) in the QCD phase diagram lie at the region of non-zero $\mu_{B}$. But the study of critical point is not convergent due to lack of first-principle calculations, therefore, work in this direction will be very much needed to understand the QCD critical point at $\mu_{B} \ne 0$.

The lattice calculations predict that the transition governed by the temperature at $\mu_{B}=0$ is not a associated with a thermodynamic singularity, rather, it is a smooth crossover~\cite{Fodor_2002,MASAYUKI1989668,PhysRevD.58.096007,DEFORCRAND2002290,Aoki:2006we}  from the phase of  hadrons to the phase of the quarks and gluons. Whereas, the $\mu_{B}$ driven transition at $T=0$ is a first-order phase transition~\cite{DEFORCRAND2002290,Endrodi:2011gv}. This conclusion about the first-order phase transition is less robust due to the sign problem in this regime of $\mu_{B} \ne 0$. At $T=0$, the first-order phase transition line does not end at the vertical axis $\mu_{B}=0$, but stops somewhere in the middle of the phase diagram. This point is known as the critical end point (CEP) of the first-order phase transition line, as shown in Fig.\ref{fig0107}. In condensed matter physics, critical end points are observed where most liquids possess such a singularity, including water.

The existence of the CEP of QCD phase diagram was suggested theoretically in Refs.~\cite{Halasz:1998qr,PhysRevD.49.426,PhysRevD.42.1757,BERGES1999215,PhysRevD.62.105008} and predicted later in lattice simulation ~\cite{FODOR200287,Fodor:2004nz}. Due to the scarcity of first principle calculations, the exact location of the CEP is still unknown. Some of the QCD based effective models such as NJL and PNJL predict the location of the CEP~\cite{Stephanov:2004wx} with uncertainties of 266-504 MeV in $\mu_{c}$ and 115-162 MeV in $T_{c}$. Therefore, locating the CEP in the phase diagram is still a big, and challenging task. It is one of the main aims of the Beam Energy Scan (BES) programme at RHIC~\cite{STAR:2010vob} to find the CEP by creating the QGP with different $\mu_{B}$ and $T$ by tuning the colliding energy ($\sqrt{s}$ ) of the heavy nuclei. So, the lack of conclusiveness on the position of the CEP pushes for more theoretical as well as experimental study. 

\begin{figure}[h]
\centering
\includegraphics[height=9.cm,width=10.9cm]{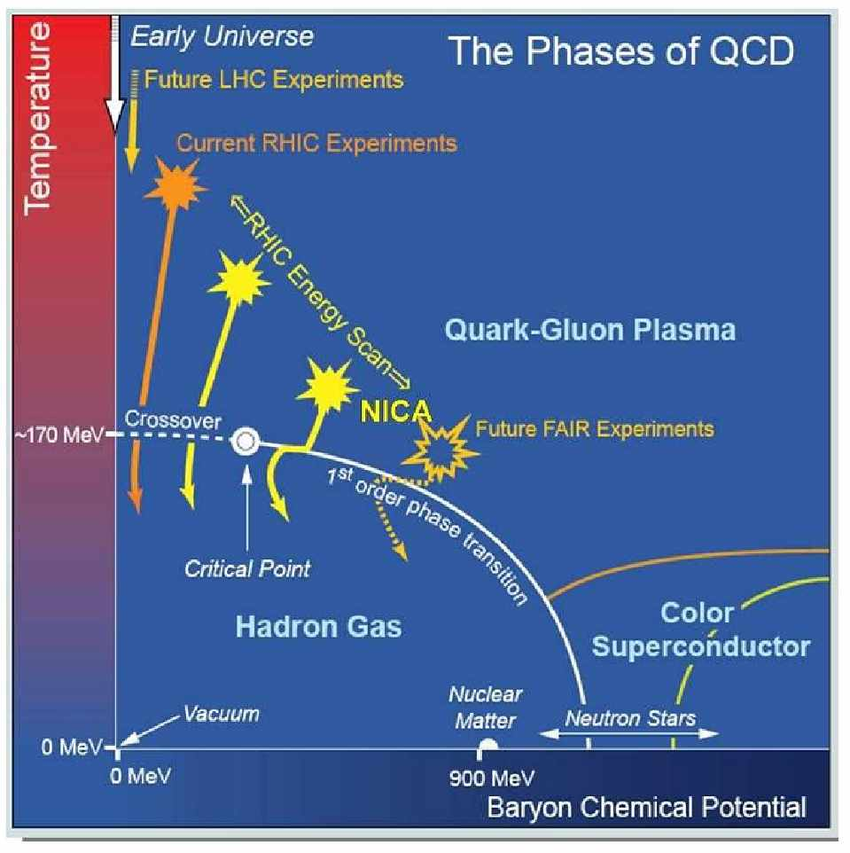}
\caption{ A conjectured phase diagram of QCD~\cite{NICA}}
\label{fig0107}
\end{figure}

A well known feature of a critical point is that the correlation length diverges, and the system is characterised by large fluctuations~\cite{Stanley,SKMa,Minami,PhysRevLett.81.4816,PhysRevD.60.114028}. Therefore, if the system created in Relativistic Heavy-Ion Collider Experiments (RHIC-E) passes through (near) the CEP, the effects will be reflected on the particle spectra. One of the most promising signatures of the CEP is a non-monotonic behaviour of beam energy dependence of higher-order cumulants of the baryon fluctuations reflected through the net proton production. The net proton yield has been  calculated by using both QCD based models~\cite{Stephanov:1998dy,Stephanov:1999zu,Stephanov:2008qz,Stephanov:2011pb} and  gauge/gravity correspondence~\cite{Critelli}. When the CEP is approached, the fluctuations become very large, which is related to the diverging nature of the correlation length ($\xi$). However, due to the critical slowing down, correlation length does not grow as much and limits within 2-3 fm at most~\cite{Berdnikov:1999ph}. The effect of critical slowing down has been taken into account with the slow hydrodynamic modes, recently developed in Ref.~\cite{Stephanov:2017ghc}. In order to find the location of the CEP, we need to better understand the effect of the CEP on the evolution of the QGP, such that the effects could be translated into the particle spectra, and from the analysis of the spectra of the particles, we can have ideas about the QGP which was created with particular $\sqrt{s}$. Therefore, the BES program can help finding a range of $\sqrt{s}$ where we can see significant effects of the QCD critical point. In such scenario, we need more studies in this direction to understand the effects of the critical point.

\section{Organization of the thesis}
In the present dissertation, we are not searching for the location of the CEP, but we want to observe its effects on hydrodynamic evolution of the QGP, by assuming its existence in the phase diagram at some point $(T_{c}, \,\mu_{c})=(154,\,367) MeV$~\cite{Nonaka,Hasan1,Hasan2}, which certainly satisfies the constraints of $\mu_{c}> 2T_{c}$~\cite{BazavovCEP}. However, the choice of the position of the CEP is kept as parameter in the EoS. So, even if we change the position of the CEP in the phase diagram, the obtained result will not change significantly.

The theories as well as experimental results of critical point is well established in the condensed matter physics. However, I want to discuss the phenomenological consequences of the critical point in the context of condensed matter physics and also in the context of the domain of RHIC-E in the next chapter, {\textit{i.e.}} in Ch.~\ref{chapter2}. The evolution dynamics (space-time evolution) of the QGP is modelled by the relativistic viscous hydrodynamics. Therefore, the relevant equations of the relativistic viscous hydrodynamics will be discussed in Ch.~\ref{chapter3}. Hydrodynamic equations are closed by an important equation, which is known as the equation of state (EoS), and to observe the effects of the CEP, we have numerically constructed the EoS which contains the critical point. Thus, the construction of the EoS will be discussed in Ch.~\ref{chapter4}. The next chapters, {\textit{i.e.}} in Ch.~\ref{chapter5} and Ch.~\ref{chapter6}, we will discuss about the propagation of the perturbations (sound waves) based on the linear analysis and compare the propagation of sound near and away from the CEP. In Ch.~\ref{chapter5}, the propagation is studied in the context of dispersion relations, whereas in Ch.~\ref{chapter6}, the propagation of sound wave is studied in the context of the dynamic structure factor. Critical point is accompanied by an effect, called the critical slowing down, where the non-equilibrium modes will resist the system to come to equilibrium state. Therefore, inclusion of those out-of-equilibrium slow modes plays an important role in critical phenomena, and will be discussed in Ch.~\ref{chapter8} in the context of the dynamic structure factor. Now as the perturbations in QGP may not be always small in magnitudes, thus linear analyses are not enough to study their propagation. For that we have studied those perturbations in terms of nonlinear waves, will be discussed in Ch.~\ref{chapter7}.  We finally conclude the studies and provide some outlook in the context of heavy-ion collisions in Ch.~\ref{chapter9}.

\chapter[Phenomenology of critical point]{Phenomenology of critical point}
\label{chapter2}  
\section{Historical Overview}
\label{sec0201}
The study of critical phenomena was started with a paper by Thomas Andrews in 1869~\cite{Andrews2}. He plotted the variation of pressure ($P$) with volume ($V$) for different temperature ($T$) for carbon-di-oxide ($CO_{2}$). Such diagrams are commonly known as $P-V$ diagram. He found that gases could only be liquified when the gas were below a specific temperature, which are different for each gas. He labelled this characteristic specific temperature as the critical temperature ($T_{c}$) of the gas. 
\begin{figure}[h]
\centering
\includegraphics[width=14.7cm,height=6cm]{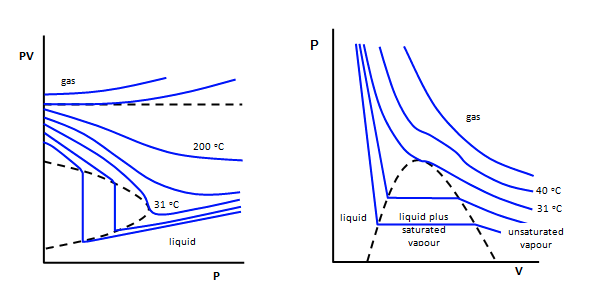}
\caption{$PV$ diagram for $CO_{2}$ gas.}
\label{fig0201}
\end{figure}
Fig. \ref{fig0201} depicts the plot of $P$ on the vertical axis and $V$ on the horizontal axis, where individual curves are plotted different constant temperature ($T$). The curves are mentioned as the isotherms of $P$ against $V$. The state of the $CO_2$ depends on the values of $P, V$ and $T$. The liquid state of the $CO_{2}$ only exists at or below the critical temperature. This specific behaviour observed by Andrews was finally explained in 1873 by J. D. Van der Waals with his equation of state (EoS)~\cite{Vanderwaals} for gases as: 
\begin{equation}
(P+\frac{a}{V^2})(V-b)=RT~,
\end{equation}
where, $a$ and $b$ are some constants, which may vary numerically form one gas to another, and $R$ is known as molar gas constant. The above equation describes the behaviour of gases above the critical temperature as well as the two coexisting phases below the $T_{c}$. Critical phenomena are universal and was precisely described by the laws of corresponding state, and it expresses the behaviour of a gas solely in terms of ratios of thermodynamic variables, $P,\, V,\, T$ to their values at the critical point (a point where first-order phase transition line stops) {\it{i.e.}} $P_{c},\, V_{c},\, T_{c}$. As Van der Waals EoS is not independent on the type of gases due to the variations of $a$ and $b$, thus the corrected equation was suggested as law of corresponding states. According to mathematical description a critical point is said to be a point of inflection. Thus to calculate the critical values, $P_{c}, V_{c}, T_{c}$ from the Van der Waals EoS, we need to put
\begin{eqnarray}
\Big(\frac{\partial P}{\partial V}\Big) = 0,\,\,\,\, \Big(\frac{\partial^2 P}{\partial V^2}\Big)= 0,
\end{eqnarray}
and calculated the critical coefficient as 
\begin{eqnarray}
T_c = \frac{8a}{27Rb}; \hspace{1cm} P_c=\frac{a}{27b^2}; \hspace{1cm} V_c=3b~.
\end{eqnarray}
Now the reduced temperature, pressure, volume was defined as
\beqa
P_r=\frac{P}{P_c}; \hspace{1cm} T_r=\frac{T}{T_c}; \hspace{1cm} V_r=\frac{V}{V_c}~.
\eeqa
Substituting these reduced variables into the Van der Waals EoS we landed into the equation of corresponding states
\beq
(P_r+\frac{3}{V_r^2})(V_r-\frac{1}{3})=\frac{8}{3}T_r~.
\eeq
This approach works remarkably well at temperatures between the normal boiling point and the critical point for many compounds but tends to break down near and below the triple-point temperature. At these temperatures the liquid is influenced more by the behaviour of the solid, which has not been successfully correlated by corresponding states methods.
\section{Water-vapour and ferromagnetic-paramagnetic critical point}
\label{sec0202}

  \begin{figure}[h]
  	\vspace*{+0cm}
  	\minipage{0.49\textwidth}
  	\includegraphics[width=8.cm,height=5cm]{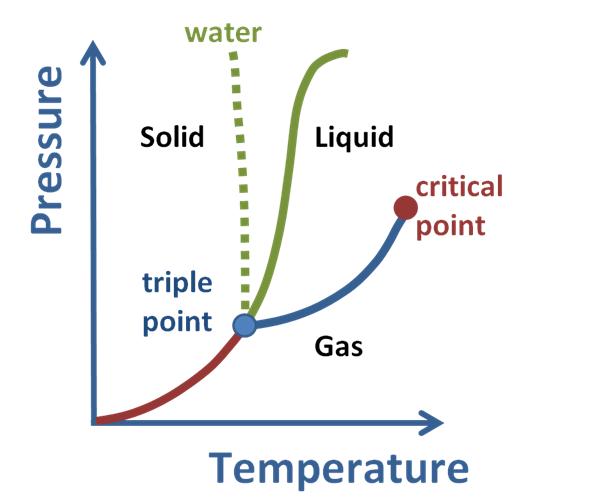}
  	\caption{Phase diagram of ice-water-vapour phase transition. The critical end point is observed at the end of water-vapour transition line~\cite{doi:10.1063/1.1461829}.}
  	\label{fig0202}
  	\endminipage\hfill
  	\minipage{0.49\textwidth}
  	\vspace*{0.8cm}
  	\hspace*{0.2cm}
  	\includegraphics[width=7.cm,height=5cm]{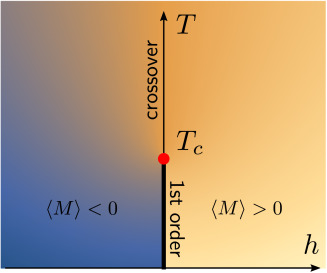}
  	\caption {The phase diagram of the Ising ferromagnet. The transition between the phases with $<M>\, > \,0$ and $<M>\,<\,0$ is a first-order transition for $T<T_{c}$ and a smooth crossover at $T>T_{c}$. The transition changes its character at the critical point~\cite{Bzdak:2019pkr}.}
  	\label{fig0203}
   	\endminipage\hfill 
  \end{figure}
For a simple and clear view of a critical point, the common specific examples are the vapour-liquid critical point and the Curie point of ferromagnetic-paramagnetic transition. These are still the best-known and most studied critical points. Fig. \ref{fig0202} shows the schematic $P-T$ diagram of a pure substance. The commonly known phases (solid, liquid, and vapour) are separated by the explicit phase boundaries (along which two phases can coexist). At the triple point, all the phase boundaries intersect thus all three phases can coexist in equilibrium. However, along the vapour-liquid phase boundary, it cease to terminate at the critical point. This termination point where the latent heat vanishes and first-order phase transition line stops, is identified as the critical end point or simply the critical point. For water-vapour phase transition, the critical point is observed at $(T_{c}, P_c) \sim (647 \,K, 218 \,atm)$. Up to now, no such endpoint is observed for ice-water transition.

A magnetic material behaves as a ferromagnet at low temperatures, whereas, it shows paramagnetic nature at higher temperatures. Even without application of any external magnetic field, a ferromagnet possesses non-vanishing magnetization ($M$), but a paramagnetic material only possesses magnetization under a magnetic field. Therefore, if there is no external field, a magnetic material at high temperature can not have any magnetization. But when it cools down below the Curie temperature, $T_{c}$, the material suddenly possesses the magnetization. This phenomena is an example of a phase transition, which is explained by the Ising model.
\section{Phase transition and characterization of its order}
\label{sec0203}
Phase transitions are usually studied by recording the response of the system to external perturbations (disturbances). For example, the liquid-gas phase transition may be understood by the response of the change in volume with a change in pressure, which is represented as the isothermal compressibility. A first-order phase transition is a transition form one phase to another at which the first-order derivative of the free energy to a thermodynamic variable shows discontinuity at $T = T_c$. An example of such kind of phase transition is the water-ice or water-vapour transition at normal pressure. The water entirely changes to ice (vapour) at the freezing point (boiling point). A second-order phase transition exhibits a discontinuity in the second-order derivatives of the free energy, but it is continuous in the first order derivative of the free energy. More clearly, first-order or second-order phase transitions are accompanied by discontinuity in the derivative of free energy. Transition from one phase to other can be accompanied by another type of transition, known as crossover, where it does not possess any discontinuity in the free energy or in the derivatives of the free energy. 


In liquid-gas phase transition (first-order), the equilibrium pressure and temperature in $P-T$ diagram is governed by the Clausius-Clapeyron’s equation, where latent heat is associated with the phase transition. Whereas, in the second-order transition due to the fact of unchanged volume and entropy, the Clausius-Clapeyron's equation is not applicable, and this leads to Ehrenfest’s equation. Therefore, association of latent heat is missing in second-order phase transition.
\section{Order parameters and critical behaviour}
\label{sec0204}
The phase transition for any thermodynamic system is usually accompanied by a change in the symmetry of the system. For a magnetic material, the two phases (the paramagnetic and the ferromagnetic) have different symmetries. Beyond the critical temperature, there is vanishing magnetization, which implies that the system is rotationally invariant. whereas, below the critical temperature, the non-vanishing magnetization infers that there is a preferred direction of the spin of the electrons. Therefore, the rotational symmetry for the system is explicitly broken, and since the symmetry is absent in one of the phases, we need a parameter in order to make distinction between the two phases. Such a parameter is called the order parameter.


Thus, order parameter plays a crucial role in phase transition. By looking into the behaviour of the order parameter, we can infer about the system. For paramagnetic-ferromagnetic system, the order parameter is the magnetization. For water-vapour transition, the order parameter is the density of the system, and for the QCD deconfinement phase transition, it is the Polyakov loop~\cite{Fukushima:2017csk}. 

\subsection{Critical Opalescence}
Critical opalescence~\cite{Stanley} is a beautiful light scattering phenomenon, which was elegantly explained by Einstein in 1910. The phenomena was first observed by Thomas Andrews for $CO_2$ gas, and since then this phenomena has been observed in many other systems. When $CO_2$ passes through the critical point, the order parameter (here, density of $CO_2$) fluctuates over the whole dimension of the gas, which is comparable to the wavelength of light. Thus light can not pass through the system, causing the transparent liquid to appear cloudy as shown in Fig.\ref{fig0204}. 
\begin{figure}[h]
\centering
\includegraphics[width=14.4cm,height=5.3cm]{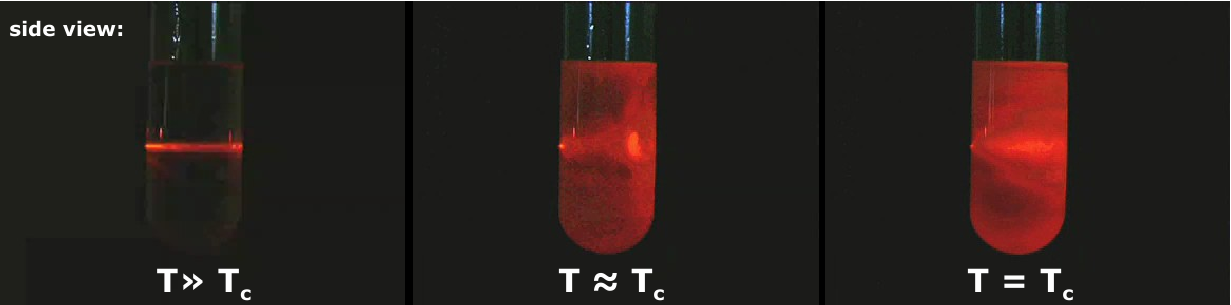}
\caption{The liquid appears clear below ($T < T_c$) or above $T > T_c$, but looking cloudy near or at critical point $T=T_c$}
\label{fig0204}
\end{figure}

Now if we consider the magnetic system where the order parameter is the magnetization of the system. A ferromagnetic material possesses magnetization due to the alignment of the spins in some preferred direction. When the ferromagnetic material is being heated up and reaches the critical temperature, the magnetization fluctuates over the whole volume of the ferromagnet. This situation can be thought of as when $T=T_c$, a spin flips from its alignment and due to the fluctuation this flipping will cause the other spins to de-align and the magnetization disappears.

The critical opalescence is also studied in case of the QCD critical point~\cite{Csorgo:2009wc,Antoniou:2006zb,Kunihiro:2009hv}, will be discussed in preceding chapters.

\section{Scaling laws, critical exponents and universality hypothesis}
\label{sec0205}
Renormalization group for scalar field theory is capable to predict the behaviour of correlations near the critical point of a thermodynamic system. It is predicted that the correlation length ($\xi$) will diverge as the critical point is approached. This follows a relation
\beqa
\xi \sim \big|T-T_{c}\big|^{-\nu},
\label{scaling_law}
\eeqa
where, $T_{c}$ is the critical temperature and $\nu$ is some critical exponent. This is known as scaling law. Statistical mechanics interprets that it is directly measurable in the realistic case of three dimensions. The value of $\nu$ is closed to $0.5$, suggested in Landau approximation~\cite{Landau_statistical_mechanics}, but may differ by some systematic corrections.

It is familiar that the thermodynamic properties at the critical point can be portrayed by critical exponents. Thus, critical exponents describe the behaviour of the singularity of the thermodynamic quantities at the critical point. To illustrate this, firstly we need to introduce the quantity known as reduced temperature
\beq
t=\frac{T-T_c}{T_c},
\eeq
such that the scaling law in Eq.\eqref{scaling_law}, can be written as
\beqa
\xi \sim |t|^{-\nu},
\eeqa

Near the critical point, the behaviour of the thermodynamic quantities is defined by a number of additional exponents. The specific heat at fixed external magnetic field ($h$) is defined as
\beqa
C_{h}\sim |t|^{-\alpha}\,.
\eeqa
Since, the ordering of spin sets in at $t=0$, the magnetization $M(t, h)$ at $h=0$ goes to zero at $t \to 0$ from below
\beqa
M\sim |t|^{\beta}\,.
\eeqa
Now even at $t=0$, finite magnetization remains at $h= 0$. Thus,
\beqa
M \sim h^{1/\delta}\,.
\eeqa
And as the magnetic susceptibility diverges at the critical point, it can be represented as
\beqa
\chi \propto |t|^{-\gamma} \,.
\eeqa
Here, $\alpha, \beta, \delta$ and $\gamma$ are the critical exponents, and are proficiently measured experimentally for a variety of thermodynamic systems.

One of the astounding attributes of critical behaviour is that it is universal, meaning that the properties of the thermodynamic variables can be understood without considering the microscopic details of the system. All the systems that behave in similar fashion belong to the same universality class. This is because of the divergence of the correlation lengths near the critical point, where the large wavelength feature dictates the dynamics of the system. It is experimentally discovered that a large variety of systems share the same critical exponents. The variety of systems that share the same sets of critical exponents are categorized into the same universality class. The property of any universality class is a rigorous prediction of the renormalization group theory of phase transitions. It is evidently found that the thermodynamic properties near the critical point depend on few features such as dimensionality and symmetry of the system.

Calculation shows that 3D Ising model and QCD belongs to the $\mathcal{O}(4)$ universality class~\cite{Halasz:1998qr,BERGES1999215}. Thus, the calculations of the 3D Ising model can be mapped onto the QCD phase diagram. This feature of universality will be used to construct the EoS which contains the effects of the critical point.
\section{Exploring the QCD critical point and its signatures}
\label{sec0206}
As discussed earlier in Ch.\ref{chapter1}, the first principle calculations {\textit{i.e.}} lattice simulations are not possible at finite $\mu_{B}$, the exact location of the critical point is not known so far. But the model calculations predicts the existence the CEP at finite $\mu_{B}$ region of the phase diagram, and thus raises the probability of discovering the CEP in such experiments~\cite{PhysRevLett.81.4816}. In Ref. \cite{PhysRevLett.81.4816}, the signatures were proposed on the basis of the fact that the CEP is an authentic thermodynamic singularity where different susceptibilities must diverge and the order parameter fluctuates over long wavelengths. All the resulting signatures due to the effect of the critical point share a usual property that they possess non-monotonic behaviour when passes through (near) the CEP.  The signatures get strengthen and then weaken again when the CEP is approached and then passed. After a collision, the evolution of the produced thermalized QGP follows a path of an isentropic trajectory that passes through (or near) the CEP is infrequent. Thus the imprint on the QGP due to the effects of the CEP through the critical fluctuations is less probable. Fortunately, long ago, it was shown that it is surprisingly easy to pass through (near) the critical region from a broad range of initial parameters~\cite{CSERNAI1986223}. In addition, let the QGP fireball pass through the critical region, the dynamics of critical fluctuations will be affected by an effect, known as `critical slowing-down'~\cite{Berdnikov:1999ph}, which allows the system to stay longer time near the CEP. This has both advantages and disadvantages to detect the CEP. If critical fluctuations induced by the CEP relax very quickly to thermodynamic equilibrium, the induced critical dynamics might have been washed away by the time it reaches the chemical freeze-out. However, the evolution of fluctuations is slowed down in the vicinity of the critical point, some signals of critical fluctuations may survive until freeze-out.

The existence of the CEP in the way of evolution process leads to a phenomena which is referred to as the `focusing' effect of trajectories towards the CEP~\cite{Nonaka}. This effect leniently permits the analysis of critical phenomena in the vicinity of the CEP that may not require a fine-tuned collision energy. Another feature of the focusing effect arises through the diverging nature of the isochoric specific heat capacity $\big(C_{V}=(\frac{\pd E}{\pd T})\big)$ at the CEP. As a result, the trajectories which pass near (through) the CEP will linger for a longer time near the vicinity of the CEP. This makes the system to freeze-out at a temperature fairly close to the critical temperature. So, while scanning in collision energies, we expect to find certain deviations in particle spectra if the isentropic trajectory passes by the vicinity of the critical point.

From the above discussion, it is inferred that the final state of the heavy-ion collisions (HICs) will be affected by the critical phenomena. Therefore, isentropic trajectories of events passing through (away) the CEP can be distinguished from the analysis of the particle spectra. For that event-by-event analysis of suitable observables could be a standard practice to get the effects of the CEP~\cite{PhysRevD.60.114028}. Another important point to note is that the critical signatures which is directly reflected in thermodynamic properties of the system near the CEP are not very subtle to the finer details of the evolution. It is argued that the event-by-event fluctuations in both $T$ and $\mu_{B}$ should be anomalously small as the system passes near the critical point because event-by-event fluctuations of $T$ can be related to $C_{V}$ at freeze-out~\cite{Stodolsky_PhysRevLett.75.1044,Shuryak_1998} as
\beqa
\Big(\frac{\Delta T}{T}\Big)^{2}=\frac{1}{C_{V}}\,.
\eeqa
Now as the CEP is approached, $C_{V}$ diverges, thus the fluctuation in $T$ is suppressed. Also the fluctuation in $\mu_{B}$ is suppressed via the divergence of the other susceptibilities.

The event-by-event analysis is applied to the fluctuations of pion's multiplicity and momentum distributions. There are two promising reasons why the pions are the most sensitive to critical fluctuations. First, the pions are the lightest hadrons observed in a relativistic heavy-ion collisions, are the most abundant hadron. The second and very crucial reason is that pions couple strongly to the fluctuations of the sigma field (proportional to the magnitude of the chiral condensate). If the freeze-out occurs near the CEP, then it was predicted for the soft pions to follow a large non-thermal multiplicity as well as an enhanced event-by-event fluctuations~\cite{PhysRevD.60.114028}. Also due to the straggling of the system near the CEP, the lifetime of the QGP may get enhanced, causing the suppression of $J/\psi$~\cite{1998}.

So far the discussion on the signatures of detecting the CEP is based on the assumption that fluctuations are enhanced near the CEP. Unfortunately, there are several reasons to doubt the arguments~\cite{PhysRevLett.101.122302,NPA2009,Luo2009}. In Ref.~\cite{PhysRevLett.101.122302}, transverse velocity dependence $(\beta_{T})$ of the anti-proton to proton ratio ($\overline{p}/p$) is considered as a prominent signature of the CEP. The arguments are established on the fact that the CEP acts as an attractor to the hydrodynamic isentropic trajectories in the phase diagram~\cite{Nonaka}. It is shown that the evolution of the $\overline{p}/{p}$ ratio following a isentropic trajectory between the phase boundary in the QCD phase diagram to the point of chemical freeze-out is highly dependent on the existence of the CEP. With the presence of the CEP, the isentropic trajectory is deformed, and the ratio $\overline{p}/{p}$ is enhanced when approaching from the CEP to the chemical freeze-out point.

Concerning the study of the critical point, the fluctuations and correlations have been considered to be sensitive observables to explore the phase structure of the QCD matter~\cite{AsakawaPhysRevLett.85.2072,Jeon2003eventbyevent,Koch2008hadronic}. They have an explicit physical interpretation for a system in thermal equilibrium and can provide essential information about the effective degrees of freedom. The crucial feature of a critical point driven by the divergence (in the ideal thermalized limit) of $\xi$ and the large scale magnitude of the fluctuations. The simplest and easiest measure of the fluctuations are the variances ($\sigma^{2}$) of the event-by-event observables namely the multiplicities or mean transverse momenta of particles. For that, a measurable and directly related observable, the net proton number cumulants can be obtained from protons and anti-protons multiplicities on an event-by-event basis, which is believed to be sensitive to the QCD critical point. The baryon susceptibilities, $\chi_{i}^{B}$, can be calculated from the EoS can be related to the cumulants, $C_{i}$ as~\cite{Sourendu2011,Nahrgang2016,Luo2017,Bzdak2020}:
\beqa
\chi_{i}^{B}=\frac{\pd^{i}(P/T^{4})}{\pd (\mu/T)^{i}}=\frac{1}{VT^{3}}C_{i}\,
\label{barsusc}
\eeqa
As the pressure, volume or temperature can fluctuate, one must cancel them out by taking ratios between the susceptibilities or the cumulants to obtain
\beqa
\frac{\sigma^{2}}{M}=\frac{\chi_{2}^{B}}{\chi_{1}^{B}}=\frac{C_{2}}{C_{1}}\,\,,\,\,\,\,\,\, S\sigma=\frac{\chi_{3}^{B}}{\chi_{2}^{B}}=\frac{C_{2}}{C_{2}}\,\,,\,\,\,\,\, \kappa\sigma^{2}=\frac{\chi_{4}^{B}}{\chi_{2}^{B}}=\frac{C_{4}}{C_{2}}\,,
\eeqa
where, $M, \sigma^{2}, S, \kappa$ are the moments of net baryon including mean, variance, skewness and kurtosis respectively, and can be defined in terms of baryon number $N$ as:
\beqa
M=<N>, \sigma^{2}={<N^{2}>-<N>^{2}}, S=<(\delta N)^{2}>,  \kappa=\frac{<(\delta N)^{4}>}{<(\delta N)^{2}>^{2}}-3\,.
\eeqa
The skewness and kurtosis describe how the shape of a probability distribution deviates from the Gaussian distribution, where the first one gives the asymmetry and the later
the `tailednes' of the distribution. They are also called the non-Gaussian fluctuations.

Now, the critical contribution to these variances diverges as $\xi^{2}$ and would demonstrate in a non-monotonic behaviour when the CEP is approached and passed in the system created in the beam energy scan program~\cite{PhysRevLett.81.4816,PhysRevD.60.114028}. But in a real situation of HICs, due to finite size effect and finite time effect (critical slowing down), the correlation length can grow up to 2-3 fm~\cite{Berdnikov:1999ph}, whereas, in absence of the CEP, the correlation length is of the order of $0.5-1$ fm, when the system is away from the CEP. Therefore, the contribution to the variance may not be a significant effect to detect the CEP. However, non-Gaussian, higher moments of the fluctuations are more sensitive on $\xi$~\cite{Stephanov:2008qz,Stephanov:2011pb}. For example, the third moment (skewness) depends as $\xi^{9/2}$, and the fourth moment (kurtosis) grows as $\xi^{7}$ near the CEP, making it a more reliable quantity to be reflected on the multiplicity fluctuations of pions and protons. Since the protons are much heavier in mass than the pions, the critical effects on their multiplicity fluctuations are expected to be more stronger in protons than for pions~\cite{Stephanov:2008qz}, and thus physicists are more interested in proton multiplicity fluctuations when searching for the CEP experimentally~\cite{Bzdak2020,Adam2021,Abdallah2021}.

It is also pointed out that the sign of the fourth moment may become negative as the critical point is approached from the side of the crossover region of the QCD phase transition. As a measurable and directly related observable, the net proton number cumulants can be obtained from protons and anti-protons multiplicities on an event-by-event basis.

\chapter[Relativistic Hydrodynamics]{Relativistic Hydrodynamics}
\label{chapter3}  
\section{Introduction}
\label{sec0301}
A fluid (liquid or gas) is defined as a substance that does not support a shear stress, and the motion of a fluid is a great concern of fluid dynamics~\cite{Landau_fluid_mechanics}. The fluid dynamical description is simple because the information of the system is encoded in its thermodynamic properties {\it{i.e.}} it provides the macroscopic description of the system. It deals with the collective behaviour, assuming the system attains the local thermal equilibrium {\it{i.e.}} the system has to be described as a complete entity like a liquid or gas rather than a collection of individual particles. If the mean free path of the fluid is defined by $\lambda$ and the characteristic system size is $R$, then a quantity, Knudsen number is defined by taking their ratio as $K_{n}=\lambda/R$. If the Knudsen number is very small ($K_{n}<<1$) then the local equilibrium is assumed to be established and the hydrodynamic description is allowed. There is no other assumption made concerning the nature of the particles and fields, and their interactions. Therefore, a fluid is a continuous medium, implies that even a small volume element of the fluid is an ensemble of a large number of particles. So, whenever we talk about a point of the fluid, it actually is not a mathematical point but it contains large number of particles moving together. A relativistic fluid is a classical fluid which follows the laws of special relativity and/or general relativity. The practical applications of hydrodynamics are extremely diverse. Hydrodynamics is used in designing ships, aircraft, pipelines, pumps, hydraulic turbines, and spillway dams and in studying sea currents, river drifts, and many more.


A fluid can be also categorized into two types: inviscid fluid (the ideal fluid) and viscid fluid (the viscous fluid). Relativistic ideal hydrodynamics deals with the inviscid fluid, considers local thermodynamic equilibrium, where each fluid element is homogeneous, infer vanishing spatial gradients (zeroth order in gradient expansion). Thus, the independent variables used here are $\epsilon$ (energy density), $P$ (pressure), and the fluid four-velocity ($u^{\mu}$). On the other hand, the dissipative hydrodynamical description does not depend on the assumption of local thermodynamic equilibrium. However, the fluid should not be far away from an equilibrium state. The theory of dissipative fluid was firstly formulated by Carl Eckart in 1940~\cite{Eckart:1940te} and then by Landau and Lifshitz in 1987 in a covariant manner~\cite{Landau_fluid_mechanics}. The dissipative hydrodynamics are established from an expansion of entropy four-current ($S^{\mu}$), concerning of dissipative fluxes, and $S^{\mu}$ contains terms linear in dissipative quantities ({\it{i.e.}} first-order gradient in hydrodynamic fields). This is why these are called a first-order theories and corresponding equation is known as Naiver-Stokes (NS) equation. Differential equations in first-order theories are parabolic in nature, thus violating causality: a signal can propagate with an arbitrary high speed.

The speed of the propagation of signals can be finite even in a parabolic theory. Therefore, first-order theories cannot be excluded by referring to causality only~\cite{Van:2007pw}. However, recently there has been a lot of progress on overcoming the problem of causality and unstable solution. In Ref.~\cite{Bemfica_firstorder}, it has been shown that if hydrodynamic variables are chosen differently, distinct from the thermodynamic variables ($T, \mu, u^{\mu}$) adopted by Eckart and Landau-Lifshitz, the energy-momentum tensor (EMT) may provide a causal theory. The authors have also shown that linear perturbations of equilibrium states are stable. The first-order stable theory with non-vanishing baryon-chemical potential is found in Refs.~\cite{Kovtun:2019hdm,Bemfica:2020zjp}. Also, in Ref.~\cite{Bemfica:2020zjp}, authors have claimed that when the system coupled to Einstein's equations ($R^{\mn}-\frac{1}{2}Rg_{\mn}+\Lambda g_{\mn}=8\pi G T_{\mn}$, where, $R^{\mn}$ is the Ricci tensor, $R=g_{\mn}R^{\mn}$, $\Lambda$ is cosmological constant, and $G$ is the gravitational constant) it became causal, hyperbolic, and the solutions are well-posed with all dissipative contributions (shear viscosity, bulk viscosity, and heat flow) are included. There are also frame stabilized hydrodynamics in first-order theory and correspondence with the second-order theory can be found in Refs.~\cite{ArpanDas:2020fnr,ArpanDas:2020gtq,ArpanDas:2020grz}.

The second-order theory was developed by M\"{u}ller~\cite{Muller:1967zza}, and later by Israel and Stewart~\cite{Israel:1976tn,Stewart,Israel:1979wp}, generally known as M\"{u}ller-Israel-Stewart (MIS) theory. Differential equations concerning the first-order theories are in general parabolic, therefore they are considered as acausal. Whereas, the differential equations concerning the second-order theories, are generally hyperbolic, therefore they provide causal theories. The problem of acausality was remedied by introducing a time delay in the response of the dissipative currents. In second-order theories, the dissipative fluxes are treated as independent variables, which follow equations of motion, describing their relaxation towards their respective NS values. In Refs.~\cite{Hiscock:1983zz,Hiscock:1985zz}, the authors have studied the stability of the MIS theory by examining the behaviour of small perturbations about equilibrium and found that MIS theory is causal, stable, and well-posed. Recently~\cite{Bemfica:2019cop,Bemfica:2020xym}, significant progress was made to understand causality in the nonlinear regime of such theories. It is established in Ref.~\cite{Bemfica:2020xym} that a set of conditions is needed to hold the causality in the nonlinear regime of MIS-like theories at zero chemical potential. The new formalism of second-order theories was proposed, where covariance and causality are satisfied by incorporating the memory effect in dissipative currents~\cite{Denicol:2008hb,Denicol:2008ha}. Further reviews on second-order theory can be found in Refs.~\cite{Muronga:2006zw,Denicol:2008rj,Koide:2006ef}.


In this dissertation, without going into an argument on the choice of order of the theory, we are only considering the second-order theory, specifically, the MIS hydrodynamics to serve our purpose. Another issue of dissipative hydrodynamics is that of choice of frame. In general, there are number of ways of choosing the frames of reference, but the mostly used choices are Landau-Lifshitz (LL)~\cite{Landau_fluid_mechanics} and Eckart's~\cite{Eckart:1940te} frame of reference. As the physics in all the possible frames must be independent of the choice of the frame, in this dissertation, we have used both the frames of reference as mentioned specifically below. Definition of dissipative fluxes are also modified on the choice of the metric tensor, $g^{\mn}= (-1,\,1,\,1,\,1)$ or $g^{\mn}=(1,\,-1,\,-1,\,-1)$. However, in this dissertation, we are going to use the former definition of the metric. In this dissertation, we have also used units, $\hbar=K_{B}=c=1$.

\section{Thermodynamics in covariant form}
\label{sec0302}
The covariant form of the thermodynamics was started with the Fourier's law of heat conduction where heat and temperature gradients are related in an un-relaxed manner. This infers that if two bodies at different temperatures are placed together, heat spontaneously flows from the warmer to the colder body without any time delay, depicted by an equation as:
\beqa
\vec{q}=-\kappa\vec{ \nabla} T\,,
\label{eq0301}
\eeqa
where, $q, \kappa, T$ are heat flux, thermal conductivity, and temperature respectively. Cattaneo figured a way out of this kind of situation by modifying the Fourier's law in which he introduced some characteristic time in the response for the material takes to react~\cite{Cattaneo2011,Lopez}, and finally got
\beqa
\vec{q}=-\kappa \vec{\nabla} T \,\,\,\,\rw \,\,\,\,\vec{q}+\tau \dot{\vec{q}}=-\kappa \vec{\nabla} T\,.
\label{eq0302}
\eeqa
Here, $\tau$ is the relaxation time of the medium. Such an adaptation leads to the famous telegrapher equation for the propagation of heat signals. The very first attempt of a relativistic extension of the heat equation was put forward by Charles Eckart in 1940~\cite{Eckart:1940te}. His proposed model has become a typecast of theories known as first-order relativistic viscous theories. In this type of theories, the entropy current, characterized by a vector field, $S^{\mu}$, depends on terms that are linear in deviations from equilibrium, specifically the heat flow or the shear viscosity. The modest way in such type of theories is to impose the second law, which takes the form as $\pd_{\mu} S^{\mu} \ge 0$, leads to the relativistic edition of the Fourier's law
\beqa
q^{\mu}=-\kappa \Delta^{\mn}\big[ \pd_{\nu}T+ T \dot{u}_{\nu} \big]\,,
\label{eq0303}
\eeqa
where, $q^{\mu}$ is the heat flux, $u^{\mu}=\gamma(1,\vec{v})$ is the fluid four velocity, and it follows
\beqa
u_{\mu}u^{\mu}=-1\,. 
\label{eq0118}
\eeqa
Here, $\gamma=1/(1-v^{2})$, is the Lorentz factor. 

The operator $ \Delta^{\mn}$ is a projection operator to $u^{\mu}$, and is defined as
\beqa
\Delta^{\mn}=g^{\mn}+u^{\mu}u^{\nu}\,.
\label{eq0119}
\eeqa
 It has the properties 
 \beqa
 \Delta^{\mu\nu\rho\lambda}=\frac{1}{2}\Big [\Delta^{\mu\rho}\Delta^{\nu\lambda}+\Delta^{\nu\rho}\Delta^{\mu\lambda}-\frac{2}{3}g^{\mu\nu}\Delta^{\rho\lambda}\Big ]
 \eeqa
 \beqa
 \Delta^{\mu\nu}u_{\mu}=\Delta^{\mu\nu}u_{\nu}=0, \,\,\,\,\,  \,\,\,\, \Delta^{\mu\nu}\Delta^{\alpha}_{\nu}=\Delta^{\mu\alpha}, \,\,\,\, \Delta^{\mn}\pd_{\nu}=\nabla^{\mu},\,\,\,\,\Delta^{\mu}_{\mu}=3\,,
 \label{eq0120}
 \eeqa

Before discussing the covariant thermodynamics, we recall the useful thermodynamic relations derived from first, second, and third law of thermodynamics as follows:
\beqa
Ts=\ep+P-\mu n\,.
\label{eq0304}
\eeqa
This relation is called the Euler's equation. Here, $T, \ep, P, \mu, n,$ and $s$ are the temperature, energy density, pressure, chemical potential, number density, and entropy density respectively.  
\beqa
dP=sdT+nd\mu\,.
\label{eq0305}
\eeqa
This is known as Gibbs-Duhem relation. Using Eqs.\eqref{eq0304} and \eqref{eq0305}, we find
\beqa
Tds=d\ep -\mu dn\,.
\label{eq0306}
\eeqa
To write the above three equation in covariant form, we define
\beqa
\beta=\frac{1}{T},\,\,\,\,\, \beta_{\mu}=\frac{u_{\mu}}{T},\,\,\,\,\, \text{and}\,\,\,\,\,\, \alpha=\frac{\mu}{T}\,.
\label{eq0307}
\eeqa
With this definition, Eqs.\eqref{eq0304}, \eqref{eq0305}, and \eqref{eq0306} are postulated in the following covariant form~\cite{Israel:1979wp}:
\beqa
S^{\mu}_{(0)}&=&P\beta^{\mu} +\beta_{\nu}T^{\mn}_{(0)} -\alpha N^{\mu}_{(0)}\,,
\label{eq0308}
\eeqa
\beqa
dS^{\mu}&=& -\alpha dN^{\mu}-\beta_{\nu} dT^{\mn}\,,
\label{eq0309}
\eeqa
\beqa
d(P\beta^{\mu})&=& N^{\mu}_{(0)} d\alpha + T^{\mn}_{(0)} d\beta_{\nu}\,.
\label{eq0310}
\eeqa
where, $T^{\mn}, N^{\mu}$ are the energy-momentum tensor and particle flow vector respectively. Also, $T^{\mn}_{(0)}$ and  $N^{\mu}_{(0)}$ are respectively the equilibrium states of energy-momentum tensor and the particle four current in equilibrium. The transition from the equilibrium to non-equilibrium state is affected by the assumption that Eq.\eqref{eq0310} holds for arbitrary virtual displacement (not just the neighbourhood of equilibrium state). Therefore, adding Eqs.\eqref{eq0308} and \eqref{eq0309}, we get an arbitrary non-equilibrium state
\beqa
S^{\mu}_{(0)}+dS^{\mu}&=& P\beta^{\mu} - \alpha \big[N^{\mu}_{(0)}+dN^{\mu}\big]-\beta_{\nu}\big[T^{\mn}_{(0)}+dT^{\mn}\big]\,.
\label{eq0311}
\eeqa
If the contribution of the higher order terms are present, the definition of the entropy density of any arbitrary non-equilibrium state can be written using Eq.\eqref{eq0311}
\beqa
S^{\mu}=P\beta^{\mu}-\alpha N^{\mu} -\beta_{\nu}T^{\mn}-\mathcal{Q^{\mu}}\,,
\label{eq0312}
\eeqa
Here, $\mathcal{Q}^{\mu}$ is function of $\delta T^{\mn}$ and $\delta N^{\mu}$, where $\delta T^{\mn}=T^{\mn}_{(0)}-dT^{\mn}$ and $\delta N^{\mu}=N^{\mu}_{(0)}-dN^{\mu}$ are just deviations from local equilibrium. Thus, $\mathcal{Q}^{\mu}$ can be expanded in Taylor's series in terms of dissipative currents up to any arbitrary order, and keeping up to second order gives rise to the second-order theory. Hence, the choice of $\mathcal{Q}^{\mu}$ fixes the structures of the dissipative fluxes.
\section{M\"{u}ller-Israel-Stewart Hydrodynamics}
\label{sec0303}
The set of hydrodynamic equations are partial differential equations which require a well defined set of initial conditions. However, existence of some hydrodynamic modes in the equations can travel backwards in time, thus the initial conditions cannot be put arbitrarily~\cite{Mario2000}. As a consequence solving the relativistic NS equation numerically becomes a challenging task. One feasible way to manage the theory was worked out in Maxwell-Cattaneo law. This law was thought to be a successful phenomenological extension of the NS equation. But it is unsatisfactory since it is not deduced from a first-principle framework, but rather introduced `by hand'. It is also found out that this law can not even correctly explain the causality of larger wave number of relativistic viscous hydrodynamic theory~\cite{Romatschke:2009im}. Therefore, it can not be generalized in order to formulate the theory of relativistic viscous hydrodynamics.

The covariant formulation of relativistic dissipative hydrodynamics was formulated by Eckart~\cite{Eckart:1940te} and Landau-Lifshitz~\cite{Landau_fluid_mechanics}. These are first-order theory, suffers from the problem of causality and stability~\cite{Hiscock:1983zz,Hiscock:1985zz}. Later second-order theory was established by firstly M\"{u}ller in 1967 and then Eckart's theory was generalized extensively by Israel and Stewart~\cite{Israel:1976tn,Stewart,Israel:1979wp} in 1976. The causality and stability of the Israel-Stewart (IS) hydrodynamics was tested by Hiscock and Lindblom~\cite{Hiscock:1983zz} in 1983 and found to be causal and stable in a far wider range of circumstances. 
\subsection{Equations of motion in general form}
Before formulating the equations in relativistic viscous fluid, one need to go through the equations of the ideal fluid. Therefore, to a relativistic ideal fluid, the general form of the EMT, $T^{\mn}_{(0)}$, (net) particle four-current, $N^{\mu}_{(0)}$, and the entropy four-current, $S^{\mu}_{(0)}$, have to be formulated from the fluid four velocity, $u^{\mu}$, and the metric tensor, $g^{\mu\nu}$. However, since, $T^{\mn}_{(0)}$ must be a symmetric tensor and are transformed by tensorial set of rules, and $N^{\mu}_{(0)}$ and $S^{\mu}_{(0)}$ transform by vector rules under the Lorentz transformations, the general form of these quantities can be obtained as
\beqa
T^{\mn}_{(0)}=\ep u^{\mu}u^{\nu}+P\Delta^{\mn}~,\,\,\,\,\, N^{\mu}_{(0)}=nu^{\mu}~,\,\,\,\,\,S^{\mu}_{(0)}=su^{\mu}~.
\label{eq0113}
\eeqa
The effective representation of an ideal fluid is derived by using the conservation laws of energy, momentum, and net particle number, usually by taking the four-divergences to those quantities as:
\beqa
\pd_{\mu}T^{\mn}_{(0)}=0~,\,\,\,\, \pd_{\mu}N^{\mu}_{(0)}=0
\label{eq0114}
\eeqa
The different dissipative effects lead to the dissipative currents such as $\tau^{\mn}$ and $\nu^{\mu}$ are to be added with the ideal currents $T^{\mn}_{(0)}$ and $N^{\mu}_{(0)}$ respectively. Here, $\tau^{\mn}$ must be symmetric tensor ($\tau^{\mn}=\tau^{\nu\mu}$) in order to satisfy the conservation of angular momentum. Now, the main concern is to find the suitable equations, which need to be satisfied by all of these dissipative currents. Therefore, the EMT appears as
\beqa
T^{\mn}&=&T^{\mn}_{(0)}+\tau^{\mn}\nn\\
&=& \ep u^{\mu}u^{\nu}+P\Delta^{\mn}+\tau^{\mn}\nn\\
&=& \epsilon u^{\mu}u^{\nu}+(P+\Pi)\Delta^{\mu \nu}+h^{\mu}u^{\nu}+h^{\nu}u^{\mu}+\pi^{\mu \nu} \,,
\label{eq0115}
\eeqa

where, $\Pi, \,\pi^{\mn}$ and $h^{\mu}$ are the bulk viscous pressure, shear stress tensor and energy dissipation respectively. 

Now, the particle four current with dissipation will appear as
\beqa
N^{\mu}&=&N^{\mu}_{(0)}+\nu^{\mu}\nn\\
&=&nu^{\mu}+\nu^{\mu} \,,
\label{eq0116}
\eeqa
where, $n$ is number density (baryon, charge, strangeness), and $\nu^{\mu}$ is the particle diffusion current, is related to heat flow vector ($q^{\mu}$) by the following relation
\beqa
q^{\mu}=h^{\mu}- \nu^{\mu}\,\,\frac{\ep+P}{n}\,,
\label{eq0117}
\eeqa
and it follows
\beqa
u_{\mu}h^{\mu}=0, \,\,\,\, u_{\mu}\nu^{\mu}=0\,.
\label{eq0121}
\eeqa
 The viscous fluxes follow the following relations:
\beqa
u_{\mu}q^{\mu}&=& q^{\mu}u_{\mu}=0,\nn\\
\pi^{\mn}&=&\pi^{\nu\mu},\,\,\,\,
u_{\mu}\pi^{\mu\nu}=0,\,\,\,\, \pi^{\mu}_{\mu}=0~.
\label{eq0122}
\eeqa
The relations between energy density, pressure and the dissipative fluxes and EMT are given by the following relations:
\beqa
q_{\alpha}&=&u_{\mu}\tau^{\mu\nu}\Delta _{\nu\alpha}, \,\,\,\, u_{\mu}\tau^{\mu\nu}=q^{\nu} \nn\\
P+\Pi&=&-\frac{1}{3}\Delta_{\mu\nu}T^{\mu\nu}, \,\,\,\, \Pi=-\frac{1}{3}\Delta_{\mu\nu}\tau^{\mu\nu}~.
\label{eq0123}
\eeqa
The equations of motion will be found out by the conservation equations as:
\beqa
\pd_{\mu}T^{\mn}=0~,\,\,\,\, \pd_{\mu}N^{\mu}=0~.
\label{eq0124}
\eeqa
In Eq.\eqref{eq0115}, $T^{\mn}$ is a second rank symmetric tensor and has $10$ independent components, and Eq.\eqref{eq0116}, $N^{\mu}$ is a four vector, has $4$
 independent components, {\it{i.e.}} a total of $14$ independent components. In tensor decomposition of $T^{\mn}$ and $N^{\mu}$, if we use Eq.\eqref{eq0121}, we have $3+3=6$ independent variables. The shear stress, $\pi^{\mn}$ is a symmetric traceless second rank tensor, and has $5$ independent variables. Taking into account the quantities, $\Pi, \ep, n, u^{\mu},$ and $P$, we have total 17 independent coordinates, which is extra in $3$ than expected, can cause ambiguity. But fluid four-velocity is defined as an arbitrary, normalized, time-like 4-vector, may need proper definition to reduce the number of independent coordinates.
\subsection{Fitting conditions, definition of fluid velocity, and the choice of a frame}
The flow in ideal hydrodynamics is uniquely calculated since the local energy fluxes and the charge densities are in the same direction, {\it{i.e.}}, the directions of the eigenvector of the energy-momentum tensor and the conserved current match as
\beqa
\ep=u_{\mu}u_{\nu}T^{\mn}_{(0)},\,\,\,\,\text{and}\,\,\,\,\,n=u_{\mu}N^{\mu}_{(0)}~.
\label{eq0124}
\eeqa
But the presence of the dissipative fluxes lead to the separation of the two local fluxes in the systems. Therefore, it is necessary to define a local rest frame (LRF) for the fluid. In such scenario, the fitting conditions also demands
\beqa
u_{\mu}u_{\nu}\tau^{\mn}=0,\,\,\,\, \text{and}\,\,\,\, u_{\mu}\nu^{\mu}=0~.
\label{eq0126}
\eeqa

So far $u^{\mu}$ is arbitrary. It attains a physical meaning by relating it to $T^{\mn}$ and $N^{\mu}$ only, and therefore it needs to be defined in some LRF. In general, there could be many choices for the frame of references (a frame can be attached to each conserved charge of the system). However, the most widely used choices are Eckart~\cite{Eckart:1940te} and Landau-Lifshitz (LL)~\cite{Landau_fluid_mechanics} frames of references. 
\section{Eckart frame of reference}
\label{sec0304}
Eckart frame represents a local rest frame where the net-charge dissipation is vanishing but the energy dissipation is non-vanishing {\it{i.e.}}
\beqa
\nu^{\mu}=0,\,\,\,\,\text{and}\,\,\,\,h^{\mu}\ne 0~,
\label{eq0127}
\eeqa
And, the velocity of fluid is defined as
\beqa
u^{\mu}_{E}=\frac{N^{\mu}}{\sqrt{-N^{\nu}N^{\nu}}},\,\,\,\,n=-u^{E}_{\mu}N^{\mu}~.
\label{eq0128}
\eeqa
 The EMT and the particle current will become
 \beqa
 T^{\mn}_{E}&=&\epsilon u^{\mu}u^{\nu}+(P+\Pi)\Delta^{\mu \nu}+q^{\mu}u^{\nu}+q^{\nu}u^{\mu}+\pi^{\mu \nu}~,
 \label{eq0129}
 \eeqa
and
\beqa
 N^{\mu}_{E}&=&nu^{\mu}~.
 \label{eq0130}
 \eeqa
 The $14$ unknowns are $\ep (1), n (1), P (1), h^{\mu} (3), \pi^{\mn} (5), u^{\mu}_{E} (3)$.
 \section{Landau-Lifshitz frame of reference}
 \label{sec0305}
 The Landau-Lifshitz (LL) frame represents a local rest frame where the energy dissipation is zero but the net-number dissipation (diffusion) is non-zero implies
 \beqa
 h^{\mu}=0\,\,\,\,\text{and}\,\,\,\,\nu^{\mu}\ne0~.
 \label{eq0131}
 \eeqa
 The velocity of the fluid is defined as
 \beqa
 u^{\mu}_{L}=-\frac{u_{\nu}T^{\mn}}{\sqrt{u_{\alpha}T^{\alpha \beta}T_{\beta\gamma}u^{\gamma}}},\,\,\,\, u_{\mu}u_{\nu}T^{\mu\nu}=\ep~.
 \label{eq0132}
 \eeqa
 The EMT and the particle current will become
 \beqa
 T^{\mn}_{L}&=&\epsilon u^{\mu}u^{\nu}+(P+\Pi)\Delta^{\mu \nu}+\pi^{\mu \nu}~,
 \label{eq0133}
 \eeqa
 and
\beqa
 N^{\mu}_{L}&=&nu^{\mu}+\nu^{\mu}=nu^{\mu}-\frac{nq^{\mu}}{\ep+P}~.
 \label{eq0134}
 \eeqa
 The $14$ unknowns are $\ep (1), n (1), P (1), \nu^{\mu} (3), \pi^{\mn} (5), u^{\mu}_{L} (3)$.

 The above definitions of $u^{\mu}$ on Eckart and LL frame impose some constraints on the dissipative currents. In Eckart frame of reference, the particle diffusion ($\nu^{\mu}$) is set to zero, whereas in LL frame of reference, the energy diffusion ($h^{\mu}$) is set to be zero. In simple words, the Eckart definition eliminates any diffusion of particles, whereas, the Landau-Lifshitz definition of the velocity field eliminates any diffusion of energy. 
\section{Choice of $\mathcal{Q^{\mu}}$ and the dissipative fluxes}
\label{sec0306}
 The equilibrium state of a thermodynamic system is interpreted as a stationary state, where the intensive and extensive thermodynamic variables of the system do not change with time. The second law of thermodynamics says that the entropy of an isolated thermodynamic system should either increase or remain constant. Hence, for a thermodynamic system in equilibrium, the entropy, being an extensive variable, settles to remain constant. On contrary, for an out of equilibrium system, the entropy must invariably increase. This is an exceptionally powerful concept that will be mainly used in this section to derive the equations of motion of a fluid with dissipation. Therefore, for our study we take space-time derivative of Eq.\eqref{eq0311} which satisfies the relation:
\beqa
\pd_{\mu}S^{\mu}\ge 0\,.
\label{eq0135}
\eeqa
 But before deriving the EOM and evaluating the dissipative fluxes, we need to fix the $\mathcal{Q^{\mu}}$ (entropy current due to heat flow). The choice of $\Q^{\mu}$ fixes the order of the theory. In Eckart's definition, a simple form of $\mathcal{Q^{\mu}}$ was considered as
 \beqa
 \mathcal{Q^{\mu}}=\frac{q^{\mu}}{T}\,\,\,\, \Rightarrow \,\,\,\,\,\,S^{\mu}=su^{\mu}+\frac{q^{\mu}}{T}\,.
 \label{eq0136}
 \eeqa 
 The first term in $S^{\mu}$ is the representation of the entropy carried along the motion of the fluid, whereas, the second term, represents the entropy flux due to the heat flow inside the fluid. Therefore using the second law {\textit{i.e.}}, Eq.\eqref{eq0135} we can derive
 \beqa
 T\pd_{\mu}S^{\mu}=-\Pi \pd_{\mu}u^{\mu}-q^{\mu}\Big[\frac{1}{T}\pd_{\mu}T+u^{\nu}\pd_{\nu}u_{\mu}\Big]-\pi^{\mu\nu}\Big<\pd_{\mu}u_{\nu}\Big>\,,
 \label{eq0137}
 \eeqa
 where, $\Big<\pd_{\mu}u_{\nu}\Big>$ is defined for any second rank tensor as
 \beqa
 \Big<\pd_{\mu}u_{\nu} \Big>=\frac{1}{2}\Delta^{\alpha}_{\mu}\Delta^{\beta}_{\nu}\Big[\pd_{\alpha}u_{\beta}+\pd_{\beta}u_{\alpha}-\frac{2}{3}\Delta_{\alpha\beta}\Delta^{\gamma\delta}\pd_{\gamma}u_{\delta}\Big]
 \label{eq0138}
 \eeqa
The simplest way to ensure that the Eq.\eqref{eq0135} holds, is to write
\beqa
\Pi&=&- \zeta \pd_{\mu}u^{\mu}\,,\nn\\
\pi^{\mu\nu}&=&-2\eta \Big<\pd_{\mu}u_{\nu}\Big>\,,\nn\\
q^{\mu}&=&-\kappa \Big[\frac{1}{T}\pd_{\mu}T+u^{\nu}\pd_{\nu}u_{\mu}\Big]\,.
\label{eq0139}
\eeqa
The three quantities $\zeta, \eta$, and  $\kappa$ are known as bulk viscosity, shear viscosity, and thermal conductivity respectively, and they must be positive. These three equations (in Eq.\eqref{eq0139}) are the dissipative fluxes correspond to three transport coefficients. With these, the second law becomes
\beqa
\frac{\Pi^{2}}{\zeta T}+\frac{q^{\mu}q_{\mu}}{\kappa T^{2}}+\frac{\pi^{\mu\nu}\pi_{\mu\nu}}{2\eta T}\ge 0\,.
\label{eq0140}
\eeqa
Therefore, the dissipative fluxes are decided with the choice of $\Q^{\mu}$, and this is coming from the NS theory, which is acausal. Now motivated by the above approach the theory, M\"{u}ller\cite{Muller:1967zza}, then Israel \& Stewart~\cite{Israel:1976tn,Stewart,Israel:1979wp} generalised the theory to overcome the difficulty of causality by proposing the following expression for the entropy current as
\beqa
S^{\mu}=~ su^{\mu}+\frac{q^{\mu}}{T} &&- \Big[\beta_0\Pi^2 - \beta_1 q_\nu q^\nu 
	+ \beta_2\pi_{\rho\sigma} \pi^{\rho\sigma}\Big] \frac{u^\mu}{2T} \nonumber\\
	&&+ \Big[\alpha_0\Pi\Delta^{\mu\nu} + \alpha_1\pi^{\mu\nu}\Big]\frac{q_\nu}{T}~,
	\label{eq0141}
\eeqa
The coefficients $\beta_{0}, \beta_{1}, \beta_{2}$ are called the relaxation coefficients, are causing the deviation from the physical entropy density, and $\alpha_{0}, \alpha_{1}$ are known as the coupling coefficients, and causes due to the thermo-viscous coupling. To hold the second law of thermodynamics, we must have~\cite{Israel:1979wp} in Eckart's frame of reference
\beqa
\label{eq0142}
\Pi&=& -\frac{1}{3}\zeta\Big[\pd_{\mu}u^{\mu}+\beta_{0}D\Pi-\tilde{\alpha_{0}}\pd_{\mu}q^{\mu}\Big]~, \\
\label{eq0143}
\pi^{\mu\nu}&=&-2\eta \Delta^{\mu\nu\alpha\beta}\Big[\partial_{\alpha}u_{\beta}+\beta_{2}D\pi_{\alpha\beta}-\tilde{\alpha_{1}}\partial_{\alpha}q_{\beta}\Big]~,\\
\label{eq0144}
q^{\mu}&=&-\kappa T\Delta^{\mu\nu} \Big[\frac{1}{T}\partial_\nu T +Du_{\nu}+\tilde{\beta_1} D{q_\nu}-\tilde{\alpha_0}\partial_\mu \Pi -\tilde{\alpha_1}\pd_{\lambda}\pi ^{\lambda}_{\nu}~ \Big] . 
\eeqa
whereas, the dissipative fluxes are modified in Landau-Lifshitz frame as
\begin{eqnarray}
\Pi &=&-\frac{1}{3}\zeta\Big[\pd_{\mu}u^\mu +\beta_0 D \Pi-\alpha_0  \pd_{\mu}q^\mu \Big]~,\nonumber\\   
\pi^{\mu\nu}&=&-2\eta \Delta^{\mu\nu\alpha\beta}\Big[\partial_{\alpha}u_{\beta}+\beta_{2}D\pi_{\alpha\beta}-\alpha_{1}\partial_{\alpha}q_{\beta}\Big]~,\nonumber\\
q^{\mu}&=&\kappa T\Delta^{\mu\nu} \Big[\frac{nT}{\epsilon+P}(\partial_\nu \alpha )-\beta_1 D{q_\nu}+\alpha_0\partial_\nu \Pi +\alpha_1\pd_{\lambda}\pi ^{\lambda}_{\nu} \Big]~. 
\label{eq0145}
\end{eqnarray}
where,  $D\equiv u^\mu\partial_\mu$, is known as co-moving derivative and in LRF, $D\Pi =\dot{\Pi }$ represents the time derivative. Also to note that the coupling and relaxation coefficients are different in two different frames and these coefficients in  the Eckart frame ($\tilde{\alpha_{0}}, \tilde{\alpha_{1}}, \tilde{\beta_{1}}$) are connected to the 
corresponding coupling and relaxation coefficients in the Landau frame ($\alpha_{0}$, $\alpha_{1}$, $\beta_{1}$) by the following relation~\cite{Israel:1979wp}
\begin{equation}
\tilde{\alpha_{0}}-\alpha_{0}=\tilde{\alpha_{1}}-\alpha_{1}=-(\tilde{\beta_{1}}-\beta_{1})=-[(\epsilon+p)]^{-1}~.
\label{eq0146}
\end{equation}
The $\beta _0, {\beta_1}$ and $\beta_2$ are also related to the relaxation time scale as~\cite{Muronga:2003ta,Muronga:2001zk}:
\begin{equation}
\tau_{\Pi}=\zeta \beta_0, \,\,\,\,\tau_q=\kappa T\beta_1,\,\,\,\, \tau _{\pi}=2\eta \beta_2~.
\label{eq0147}
\end{equation}
The coupling coefficients are connected to the scale of relaxation lengths, which again couple to heat flux, and bulk pressure $(l_{q\Pi}, l_{\Pi q})$, the heat flux, and shear tensor $(l_{q\pi}, l_{\pi q})$ by the following relation
\begin{equation}
l_{\Pi q}=\zeta \alpha_0,\,\,\,\, l_{q\Pi}=\kappa T \alpha _0, \,\,\,\,l_{q\pi}=\kappa T\alpha_1,\,\,\,\, 
l_{\pi q}=2\eta \alpha_1~.
\label{eq0148}
\end{equation}  
The  expressions for relaxation and coupling coefficients are given in Appendix-\ref{appendix01_A}, which are evaluated through the thermodynamic integral from kinetic theory prescription.~\cite{Israel:1979wp}. 

Once the dissipative fluxes are chosen by the choice of frame, the structure of $T^{\mn}$ and $N^{\mu}$ are determined. Therefore, the equation of motion of the fluid will be dictated by the following relations as:
\beqa
\pd_{\mu}T^{\mn}=0\,;\,\,\,\,\,\pd_{\mu}N^{\mu=0}\,.
\label{eq0349}
\eeqa
The hydrodynamic equations are not in general closed. Therefore, to close the equations we need an extra equation, known to be the equation of state (EoS). Also, the above equations are partial differential equations, which must be solved by the proper initial condition. However, in this dissertation, we will not discuss about the initial conditions (one can see for the references~\cite{Glauber1955,IPGlasma2012,Sandeep2016}). But we shall construct an EoS which will contain the critical point, shall be discussed in the next chapter to study the effects of critical point on the evolution of the QGP.

\chapter[Equation of State (EoS) with critical point]{Equation of State with critical point}
\label{chapter4}  
\section{Introduction}
\label{sec0401}
After the formation of the QGP in a nuclear collision at relativistic energies, the bulk evolution of the thermalized system is controlled by the hydrodynamic equations, equation of state (EoS) and initial condition. The EoS of the QGP characterizes the transition from the hadronic phase to the QGP phase, is understood by the rise in the number of degrees of freedom, in the  deconfined phase. One of the first EoS of the QGP was the one derived from the MIT bag model~\cite{MITBag}. Because of its simplicity, it has been widely used in the analysis in data from Relativistic Heavy-Ion Collider Experiments (RHIC-E) and in astrophysics and cosmology too.

The bag model was first suggested by the theoretical physicist Nikolay Bogoliubov in 1967. His idea was to consider the quarks with enormous mass making them unable to move. However, this seems to contradict the asymptotic freedom observed at a very close range. He solved this by confining the quarks within a spherical cavity of radius $R$ in which they feel an attractive field. This resulted in Bogoliubov’s bag model, where the quarks could move freely inside the bag but were completely confined within it. Although Bogoliubov’s model was a very simple one, it still gave some reasonable predictions~\cite{Thomas}.

The improvement in the MIT bag-model has an additional term in its Lagrangian density, called the bag constant $\B$. It corresponds to the outwards pressure exerted on the bag. This simple correction has shown to greatly improve agreement between predictions and tabulated values of the masses of different quarks and hadrons. In the Bag model, the energy density and the pressure are given by~\cite{MITBag,Fogaca1,FogacaEOS}:
\beqa
\epsilon=\frac{37\pi^{2}}{30}T^{4}+\B \,\,\,\,\,\,\,\text{and}\,\,\,\,\,\, p=\frac{37\pi^{2}}{90}T^{4}-\B~.
\label{eq0401}
\eeqa
The numerical value of $\B^{1/4}$ is taken as $220 MeV$.

The lattice QCD provides the EoS in the strongly-coupled regime from first principles, but it is unable to serve at finite $\mu_{B}$ regime. The lattice EoS are till now the best fit to the data~\cite{Borsanyi:2016ksw,Borsanyi:2013bia,Borsanyi:2011sw}. Recently, the continuum extrapolated results for the EoS of QCD with the physical masses of light and strange quark have been incorporated~\cite{Borsanyi:2013bia,HotQCD:2014kol}. The bulk thermodynamic observables such as $P$ (pressure), $\epsilon$ (energy density) and $s$ (entropy density) have now been evaluated at $\mu_{B}=0$ for the three quark flavors $u, d$ and $s$. The thermodynamic observables behave smoothly in the transition region. At reasonably low temperature, the observables are found to be in quite good accordance with hadron resonance gas (HRG) model calculations. Although there are few systematic deviations observed, which may be attributed to the presence of additional resonances which are not accounted in the HRG model calculations in the standard particle data tables~\cite{HRG1,HRG2}.

Due to the well-known sign problem for lattice QCD formulations at $\mu_{B}\ne 0$, an immediate evaluation of the EoS at non-zero $\mu_{B}$ is unfortunately not possible. Nevertheless, it is made possible for small values of the chemical potentials, by using a Taylor expansion of the thermodynamic potential~\cite{GavaiEoS1,AltonEoS1} and have been obtained on coarse lattices~\cite{AltonEoS1,GavaiEoS2,AltonEoS2}. Such type of expansion scheme have even been extended up to sixth order in the baryon chemical potential~\cite{AltonEoS3,EjiriEoS}.

But as discussed earlier in Ch.\ref{chapter2}, Sec.\ref{sec0206}, that the CEP appears at $\mu_{B} \ne 0$ and $T\ne 0$ domain of the QCD phase diagram. Therefore, the lattice QCD can not be extended to provide the ultimate EoS near the CEP. In this chapter, we construct an EoS, which contains the CEP, based on Refs.~\cite{Nonaka,Parotto} (see Ref.~\cite{MITWathid} for pedagogical approach). 
\section{Construction of the EoS}
\label{sec0402}
As far as the study of the critical point is concerned, the two most common (for better understanding) examples of such critical points are the liquid-gas coexistence curve and
the Curie point in ferromagnetic-paramagnetic transition. Although these two systems are way out on a fundamental, microscopic level, the physical behaviour near the critical point is remarkably similar on a qualitative and even on quantitative level. This is because the diverging nature of the correlation length happens for all the systems near the CEP, and this observation is based on the concept of universality of the second-order phase transitions. As per the previous discussion on the universality hypothesis, the CEP of the QCD belongs to the same universality class as that of the 3D Ising model. Therefore, calculations carried out in the Ising model can be mapped onto the phase diagram of the QCD.

Within a small critical region around the critical point, the behaviour of thermodynamic quantities is quantitatively governed by the critical phenomena based on the universality class considerations. The order parameter, $M$ (magnetization of the system) in the Ising model is a function of the reduced temperature ($t=(T-T_{c})/T_{c}$) and the magnetic field strength ($h$), and the critical point is positioned at the origin $(t, h) = (0, 0)$ by the construction. For the Ising model, the EoS in the parametric representation is given by~\cite{Guida,Rajagopal:1992qz}
\beqa
M &=& M_0 R^{\beta} \theta~, \nonumber\\
h &=& h_0 R^{\beta \delta} \tilde{h}(\theta) ;\hspace{0.3cm} \tilde{h}(\theta)= (\theta - a\theta^3 + b\theta^5) ~,\nonumber\\
t &=& R(1-\theta^2) ~,
\label{eq0402}
\eeqa
where, $a, b$ are some numerical values given as $a=0.7620103, b=0.008040$, and $R \geq 0, -1.154 \leq \theta \leq +1.154$. The critical exponents are identified as $\beta = 0.326$ and $\delta =4.80$. The $M_0$ and $h_0$ are some normalization constant, can be found out by setting 
\beqa
\label{eq0403}
M(t=-1,h=0+) &=& 1~, \\
\label{eq0404}
M(t=0,h=1) &=&1~,\\
\label{eq0405}
M(t=0, h)&=& sgn(h) |h|^{1/\delta}~.
\eeqa
From Eq.\eqref{eq0403}
\beqa
 R(1-\theta^2)=-1 &\Longrightarrow& \theta^2 = 1+\frac{1}{R}~, \nonumber\\
\theta &=& \pm \sqrt{\frac{R+1}{R}}~.
\label{eq0406}
\eeqa
Thus, substituting in $M$, we have
\beqa
M &=& M_0R^{\beta}\theta~, \nonumber\\
1&=& M_0R^{\beta}\sqrt{\frac{R+1}{R}}~.
\label{0407}
\eeqa
\beq
M_0 =\frac{1}{R^{\beta} \sqrt{\frac{R+1}{R}}}~.
\label{0408}
\eeq
To evaluate $h_0$, we take Eq.\eqref{eq0404} and we calculate
\beqa
r &=& 0 = R(1-\theta^2) \nonumber\\
\theta &=& \pm 1 ~.\nonumber\\
\label{eq0409}
\eeqa
Considering $\theta =-1$, the magnetization defined in Eq.\eqref{eq0402}, $M$ becomes negative, which is unphysical because $M$ is always along the direction of the applied field $h$. Thus, $\theta=-1$ value is excluded and the physical value for $\theta$ will be $\theta=+1$. So
\beqa
h &=& h_0 R^{\beta \delta} \tilde{h}(\theta) \nonumber\\
&=& h_0 R^{\beta \delta} (\theta - a\theta^3 + b\theta^5) \nonumber\\
1 &=& h_0 R^{\beta \delta} (1-a+b)  
\label{eq0410}
\eeqa
Thus,
\beq
h_0 = \frac{1}{R^{\beta \delta}(1-a+b)}~.
\label{0411}
\eeq
Now to construct the EoS for Ising model, we will start with the deduction of the entropy density near the critical point, which is related to the Gibb's energy density 
\beq
G(h,t)=F(M,t)-Mh=E-Ts_c-Mh~.
\label{eq0412}
\eeq
where, $F (M, t)$ is the Helmholtz's free energy and $s_{c}$ is the entropy density at the CEP. Generally from thermodynamic relations, we have
\beqa
G =\epsilon-Ts_{c}+P+n\mu_B &\Longrightarrow& dG = dP-sdT+nd\mu_B~, \nn\\
s_{c} &=&- \Big(\frac{\partial G}{\partial T}\Big)_{\mu_B}~.
\label{eq0413}
\eeqa
But as $G \equiv G(h,t)$, we can have
\beqa
s_c =-\Big[\big(\frac{\partial G}{\partial h}\big)_t\big(\frac{\partial h}{\partial T}\big)_{\mu_B}+\big(\frac{\partial G}{\partial t}\big)_h\big(\frac{\partial t}{\partial T}\big)_{\mu_B}\Big]~.
\label{eq0414}
\eeqa
The four terms $(\frac{\partial G}{\partial h})_t, (\frac{\partial G}{\partial t})_h, \big(\frac{\partial h}{\partial T}\big)_{\mu_B}, \big(\frac{\partial t}{\partial T}\big)_{\mu_B}$ need to be evaluated as: 
\beqa
\Big(\frac{\partial G}{\partial h}\Big)_t &=& \Big(\frac{\partial F}{\partial h}\Big)_t-h\Big(\frac{\partial M}{\partial h}\Big)_t-M~.
\label{eq0415}
\eeqa
Decomposing the $\Big(\frac{\partial F}{\partial h}\Big)_t$ and substituting in the above equation, we get
\beqa
\Big(\frac{\partial G}{\partial h}\Big)_t &=& \Big(\frac{\partial F}{\partial M}\Big)_t \Big(\frac{\partial M}{\partial h}\Big)_t-h\Big(\frac{\partial M}{\partial h}\Big)_t-M~.
\label{eq0416}
\eeqa
From free energy, F, we have $h= \big(\frac{\partial F}{\partial M}\big)_t$ to impose into the previous equation to give 
\beqa
\Big(\frac{\partial G}{\partial h}\Big)_t=-M~.
\label{eq0417}
\eeqa
Now,
\beqa
\Big(\frac{\partial G}{\partial t}\Big)_h &=& \Big(\frac{\partial F}{\partial M}\Big)_t \Big(\frac{\partial M}{\partial t}\Big)_h+\Big(\frac{\partial F}{\partial t}\Big)_M-h\Big(\frac{\partial M}{\partial t}_{}\Big)_h-M\Big(\frac{\partial h}{\partial t}\Big)_{h}\nn\\
&=& \Big(\frac{\partial F}{\partial t}\Big)_M~.
\label{eq0418}
\eeqa
Therefore, the critical entropy density from Eq.\eqref{eq0413} becomes
 \beqa
 s_c(h,\,t)&=&-\Big[\big(-M\big)\Big(\frac{\partial h}{\partial T}\Big)_{\mu_B}+\Big(\frac{\partial F}{\partial t}\Big)_M\Big(\frac{\partial t}{\partial T}\Big)_{\mu_B}\Big]~,\nn\\
 &=&\Big[M\Big(\frac{\partial h}{\partial T}\Big)_{\mu_B}-\Big(\frac{\partial F}{\partial t}\Big)_M\Big(\frac{\partial t}{\partial T}\Big)_{\mu_B}\Big]~.
 \label{aaa}
 \eeqa

In the above equation, three unknown, $\big(\frac{\partial h}{\partial T}\big)_{\mu_B}, \big(\frac{\partial F}{\partial T}\big)_{M},\big(\frac{\partial t}{\partial T}\big)_{\mu_B}$ are to be evaluated.  And to evaluate the quantities, $\big(\frac{\partial h}{\partial T}\big)_{\mu_B}$, and $\big(\frac{\partial t}{\partial T}\big)_{\mu_B}$,  we need the mapping from $(t,h)\to (\mu_{B},T)$ in the QCD phase diagram.
 
\subsection{Mapping onto the QCD phase diagram}
The mapping between the two planes is essential for determining  $(\frac{\partial h}{\partial T})_{\mu_B}$ and $(\frac{\partial t}{\partial T})_{\mu_B}$. To map we consider the $t$-axis, tangential to the first-order phase transition line at the CEP of the QCD phase diagram. However, we can choose arbitrary axes for $h$-component. But for the simplicity of our calculations, we take $h$-axis to be perpendicular to the $t$-axis as shown in Fig.\ref{fig0401}.
\begin{figure}[h]
\centering
\includegraphics[width=10cm,height=7.5cm]{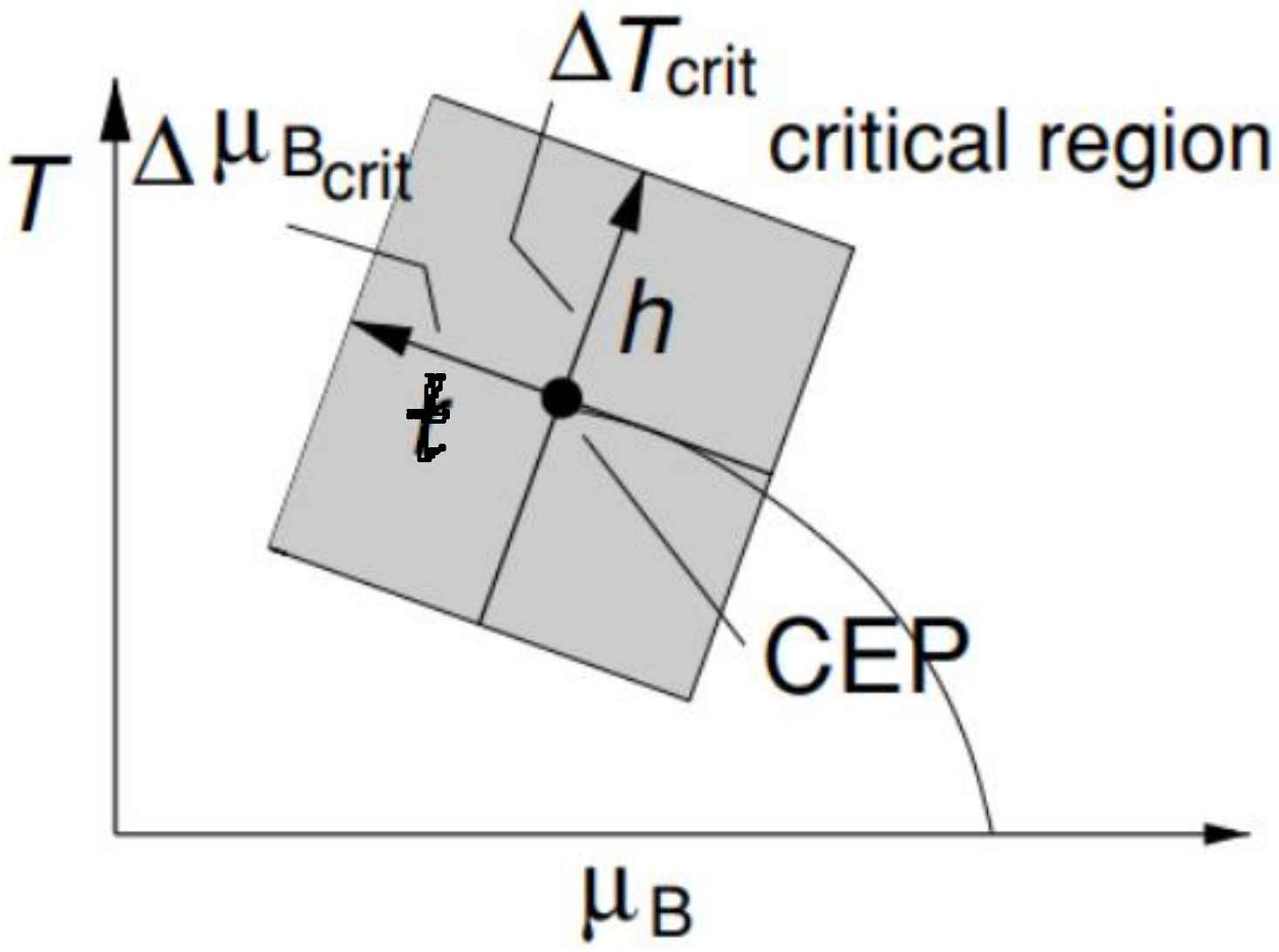}
\caption{Schematic diagram of the $t-h$ axes mapped onto the $T-\mu_{B}$ plane. The $t$-axis is tangential to the first-order phase transition line at the CEP. The $h$-axis is considered perpendicular to the $r$-axis~\cite{Nonaka} for mathematical simplicity.}
\label{fig0401}
\end{figure}
\beq
(t,h)=(0,0) \,\,\,\,\,\,\,\text{is the position of the critical point}
\label{eq0419}
\eeq
Thus on QCD phase diagram ($T-\mu_B$) plane,
\beqa
t & > & 0~, \hspace{1cm} \text{is a crossover and} \nonumber\\
t & < & 0~, \hspace{1cm} \text{is the first-order phase transition}
\label{eq0420}
\eeqa
If we consider that the CEP is located very close to $T$-axis ($\mu_B \rightarrow 0$) in the QCD phase diagram, the $t$-axis is now parallel to $\mu_B$-axis and $h$-axis is parallel to $T$-axis. The most natural thing to do is to assume a finite (but small) critical region, where the relations between the scale of the $(\mu_{B}, T)$ and the scale of $(t, h)$ is linear. Therefore, it is necessary to choose the domain of the critical region, and for that the location of the CEP must be mentioned.
\subsection{Position of the CEP and the choice of critical region} 
The exact location of the CEP in the QCD phase diagram has still remained questionable. Some of the QCD based models could also predict the location of the CEP, but the position of the CEP varies with the models of the parameters used. The location of the CEP having uncertainties ranging from 266-504 MeV in $\mu_{c}$ and 115-162 MeV in $T_c$~\cite{Luo2017}. The holographic model for five-dimensional black holes even indicated the critical coordinate as $(\mu_c,T_c)=(783$ MeV,143 MeV$)$~\cite{Holography1}. It is clear that the position of the CEP is sensitive to the parameters of the models used. It is argued in recent studies that it is unlikely for the CEP to be located at $\mu_{c}<2T_{c}$~\cite{BazavovCEP,Andronic:2017pug,Luo2017} or disfavours the position at $\mu_{c}/T_{c}\sim2-3$~\cite{VovchenkoCEP,Fodor2021PRL}. 


In this present dissertation, we chose ($\mu_{c},T_{c})= (367, 154)$MeV~\cite{Nonaka}, which certainly satisfies the criterion of $\mu_{c}/T_{c}>2$. However, the results and conclusion drawn from such a selection of $\mu_c$ and $T_c$  will remain valid for other values of $\mu_c$ and $T_c$ too.

The universality hypothesis tells that the critical exponents around second-order phase transitions are characterized by the dimensionality and symmetry of the system. The singular
part near the CEP (a function of two variables) to be mapped onto the variables characterizing the phase diagram of the 3D Ising model. In the
QCD phase diagram ($\mu_{B}-T$ plane), $t-$axis directs towards the direction of $T-axis$ but the direction of $h$ is not known~\cite{Berdnikov:1999ph,Hatta2003}. However, it is evident that the critical region is more elongated along the $t$ direction compared to the $h$ direction.  This is because of the fact that the critical exponent associated with $t$ is larger than that associated with $h$~\cite{Berdnikov:1999ph,PhysRevLett.101.122302}. Thermalized QGP follows a path of an isentropic trajectory and can be bent near the critical point (region), known as focussing effect~\cite{Nonaka,PhysRevLett.101.122302} is one of the fundamental properties of the CEP, can affect the final state of particle spectra.~\cite{PhysRevLett.101.122302,NPA2009}. Therefore, the choice of the elongation of the critical region plays an important role in the evolution of the QGP. Even model studies suggest that the size of the critical region is sensitive to affect the hydrodynamic evolution~\cite{PRD2007}. Here we simply assume that the elongation of the critical region is sufficiently large to induce a significant focusing effect and used the numerical values suggested in Ref.~\cite{Nonaka}. Finally, in the critical region, the linear mapping between $(t, h) \to (\mu_{B}, T)$ is depicted as,
\beqa
\label{eq0421}
t&=& \frac{\mu_B-\mu_{c}}{\Delta \mu_{c}}=\frac{\mu_B-0.367}{-0.2}~,\\
\label{eq0422}
h&=& \frac{T-T_{c}}{\Delta T_{c}}=\frac{T-0.154}{0.1}~.
\eeqa
It is clearly seen that the elongation of the critical region along $t$-axis than the $h$-axis. As $T$ and $\mu_{B}$ are two independent quantities in QCD phase diagram, we can calculate (using Eqs.\eqref{eq0421} and \eqref{eq0422})
\beqa
\Big(\frac{\partial h}{\partial T}\Big)_{\mu_B} = \frac{1}{\Delta T_{c}}~\,\,\,,\,\,\Big(\frac{\partial t}{\partial T}\Big)_{\mu_B} = 0~.
\label{eq0424}
\eeqa
At the end we plug into Eq.\eqref{aaa} to get
\beqa
s_c(T,\,\mu_{B}) =\frac{M(t, h)}{\Delta T_{c}}=M\Big(\frac{T-T_{c}}{\Delta T_{c}},\frac{\mu_B-\mu_{c}}{\Delta \mu_{c}}\Big)\frac{1}{\Delta T_{c}}~.
\label{eq0425}
\eeqa
\begin{figure}[h]
\centering
\includegraphics[width=11cm,height=6.5cm]{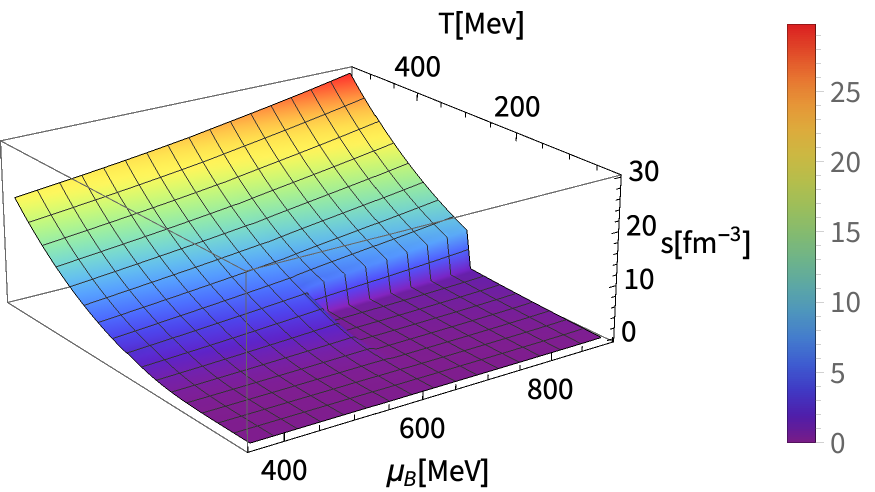}
\caption{  The constructed entropy density as a function of $T$ and $\mu_{B}$. We take the CEP at $(\mu_c,T_c)$=(367 MeV,154 MeV).}
\label{fig0402}
\end{figure}
In order to construct the EoS, we first construct a  dimensionless entropy density  as
\begin{eqnarray}
{S_{c}}= A (\Delta T_c, \Delta\mu_{c})~s_c (T,\mu_{B})~.
\label{eq0426}
\end{eqnarray}
here, $A$ is defined as 
\begin{equation}
A(\Delta T_{c},\Delta \mu_{c}) = D \sqrt{\Delta T^2_{c}+\Delta \mu_{c}^2)}~.
\label{eq0427}
\end{equation}
\begin{figure}[h]
\centering
\includegraphics[width=10cm,height=6.5cm]{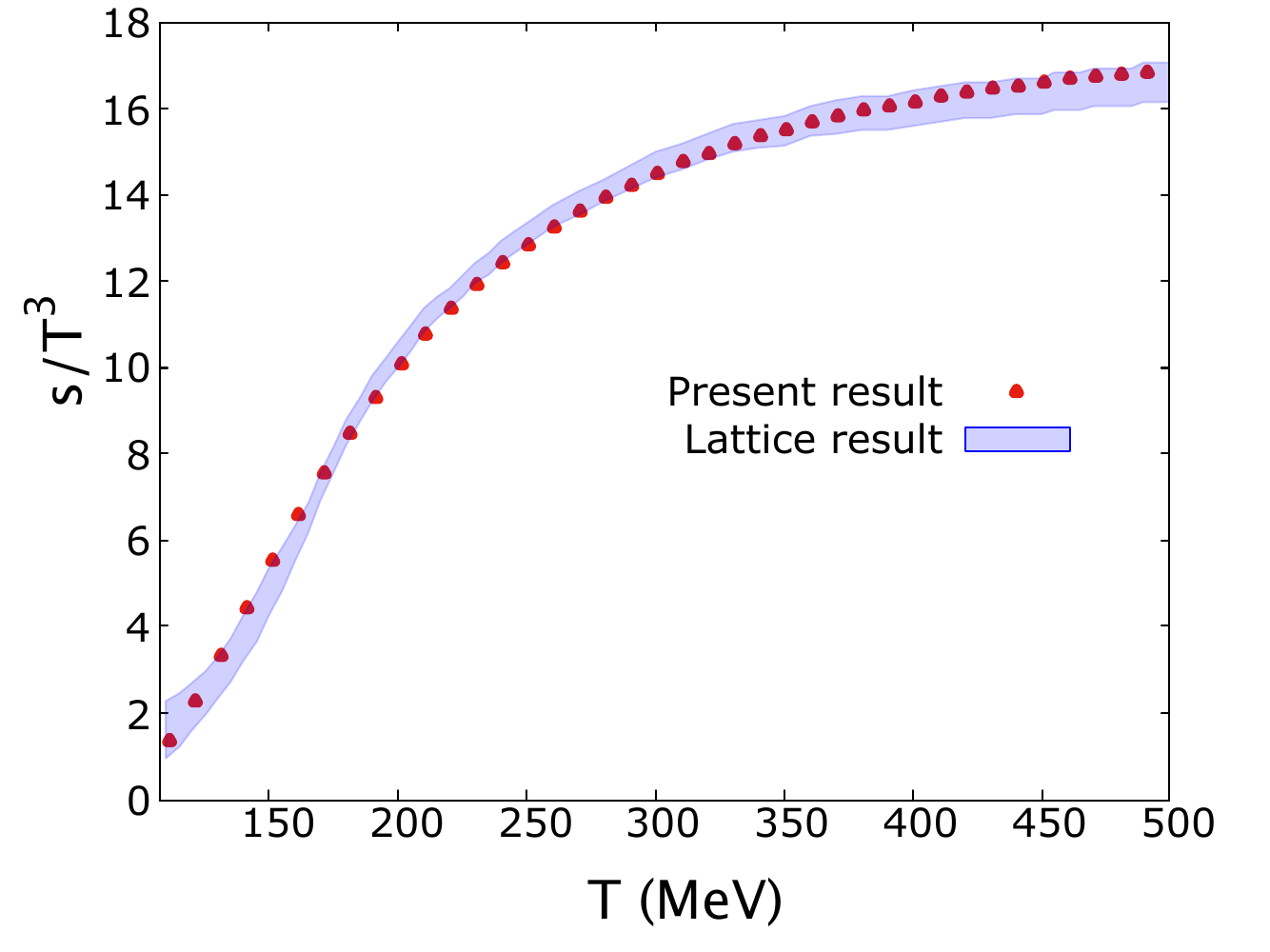}
\caption{ The variation of $s/T^{3}$ with $T$ is plotted and compared our result with Lattice result ~\cite{Borsanyi:2013bia}.}
\label{fig0403}
\end{figure}
Here $D$ is also a dimensionless quantity which represents the extension of the critical region. In this dissertation, we use $(T_c, \mu_c)= (154 \text{MeV}, 367 \text{MeV})$ with $(\Delta T_{c}, \Delta \mu_{c}, D)=(0.1\,\text{GeV}, -0.2\, \text{GeV}, 2)$. The construction of the entropy density is done by using $S_{c}$ as a switching function and by appropriately connecting entropy densities of the QGP ($s_Q$) and the hadronic ($s_H$) gases. The final result reads as:  
	\begin{eqnarray}
	s(T, \mu_{B})&=&\frac{1}{2}\Big[1-{\rm tanh}{S_{c}}(T, \mu_{B})\Big]~ s_{Q}(T, \mu_{B}) \nn\\
	&+& \frac{1}{2}\Big[1+{\rm tanh}{S_{c}}(T, \mu_{B})\Big]~s_H(T, \mu_{B})~.
	\label{eq0428}
	\end{eqnarray}
A 3D plot of $s(T,\,\mu_{B})$ is shown in Fig.\ref{fig0402}, where we clearly see a discontinuity in the entropy density at $(T_{c},\, \mu_{c})$. We calculate $s_{Q}$ by using the following expression~\cite{Satarov2009}
	\begin{eqnarray}
	s_{Q}(T, \mu_{B})=\frac{32+21N_{f}}{45}\,\pi^{2}T^{3}+\frac{N_{f}}{9}\,\mu^{2}_{B}T ~.
	\label{eq0429}
	\end{eqnarray}
with $N_{f}$ to be the number of quark flavors. In Eq.~\eqref{eq0429}, massless quarks and gluons are considered. However, the effects of thermal masses can be taken into calculation through the effective degeneracy of quarks and gluons ~\cite{Jane1996} (see Ref.~\cite{Kolb1990}). To estimate the effective degeneracy factor with thermal quark mass we consider the energy density of quarks as,
\begin{figure}[h]
\centering
\includegraphics[width=10cm,height=6.5cm]{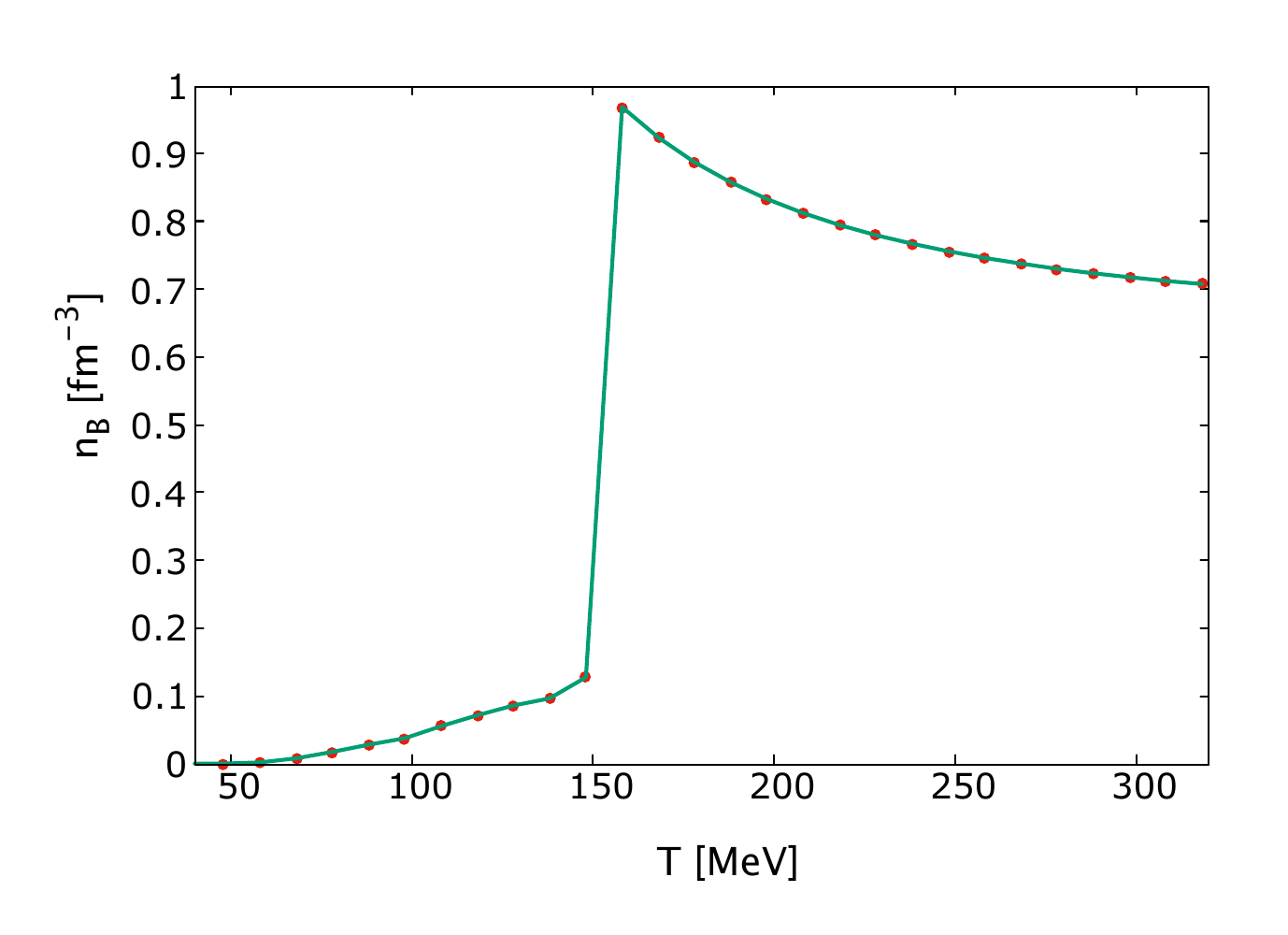}
\caption{ The variation of $n_{B}/T^{3}$ with $T$ is shown at $\mu_{B}=400 MeV$.}
\label{fig0404}
\end{figure}
\begin{equation}
\epsilon_q(T,\mu_{B},m_q)=\frac{g_q^{eff}}{(2\pi)^3}\int\,d^3p \sqrt{p^2+m_q^{2}}\,f_{FD}(E_i)~.
\label{eq0430}
\end{equation}
Here, $m_{q}$ is the thermal mass of quark, $p$ is the momentum of the quarks and $f_{FD}$ is the Fermi-Dirac distribution function. The effective degeneracy of quark is given by
\begin{equation}
g_q^{eff}=\frac{\epsilon_q(T,\mu_{B},m_q)}{g_q}~,
\label{eq0431}
\end{equation}
where, $g_{q}$ is the quark's degeneracy factor. Similarly for gluons the effective degeneracy can be estimated by using the relation
\begin{equation}
g_g^{eff}=\frac{\epsilon_g(T,m_g)}{g_g}~.
\label{eq0432}
\end{equation}
where, $m_{g}$ is the thermal mass of the gluon and $g_{g}$ is the gluon's degeneracy factor.

We have calculated $g_q^{eff}=5.76$ with thermal mass at $(\mu_{B},T)=(367,158)$ MeV, whereas for massless quark the degeneracy is found to be $g_{q}=6$. In case of gluon, with thermal mass $g_g^{eff}=14.2$ which is $g_{g}=16$ for massless gluon. For $(\mu_{B},\,T)=(367,\,158)$ MeV the thermal momenta of quarks and gluons are $1064$ MeV and $497$ MeV respectively where as the respective thermal masses are 102 MeV and 208 MeV respectively. This indicates that the effect of thermal mass at the critical point will not be
significant for the present study.
\begin{figure}[h]
\centering
\includegraphics[width=10cm,height=6.5cm]{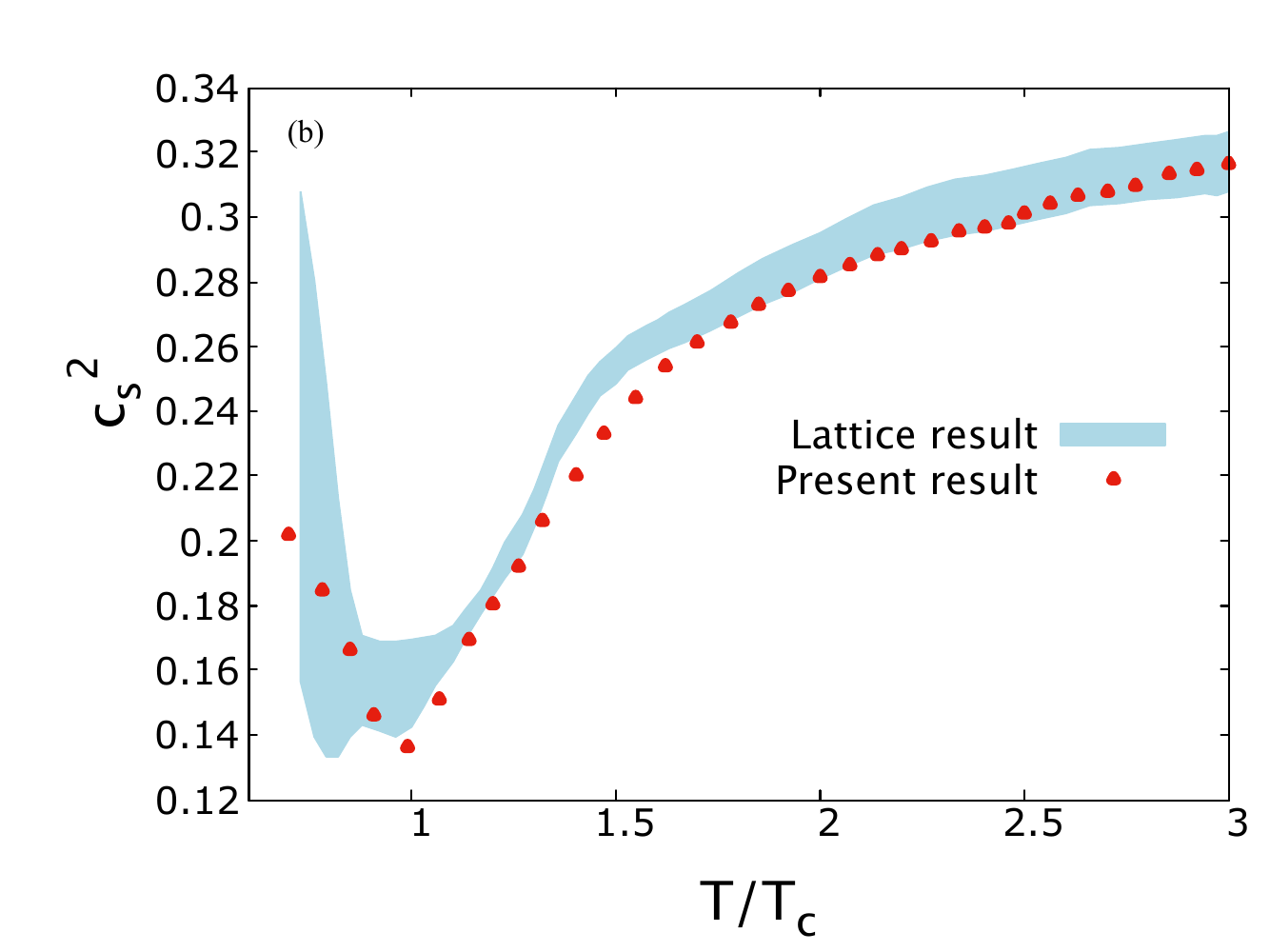}
\caption{ The variation of $c_s^2$ with $T/T_c$ from lattice QCD result~\cite{Borsanyi:2013bia} and present work have been depicted.}
\label{fig0405}
\end{figure}
\begin{figure}[h]
\centering
\includegraphics[width=10cm,height=6.5cm]{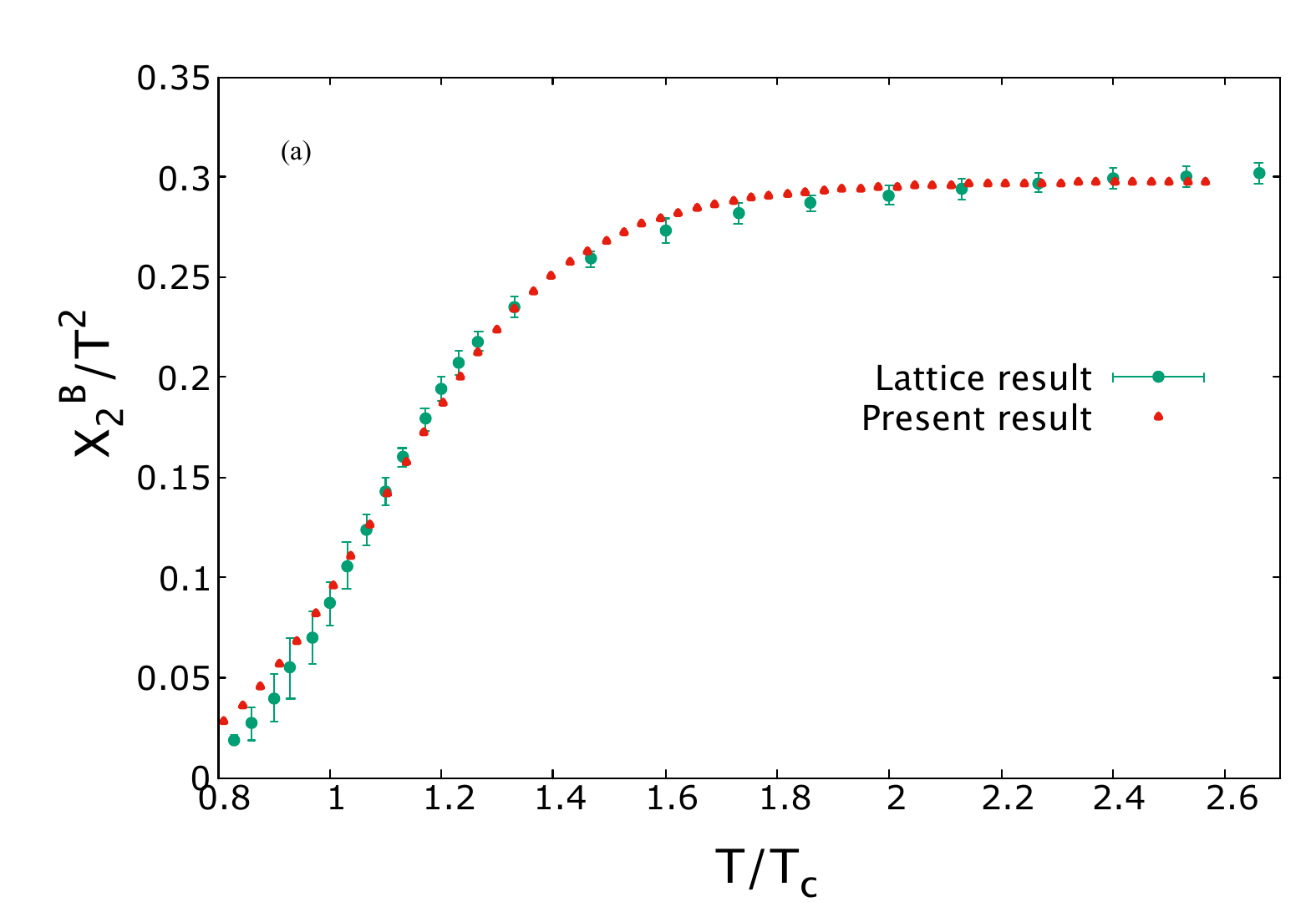}
\caption{ The $\chi^{B}_2/T^2$ obtained from the parametrization of the
present work has been displayed as a function of $T/T_c$. The result is
compared with lattice QCD results~\cite{Borsanyi:2011sw}.}
\label{fig0406}
\end{figure}

The hadronic entropy density
($s_H$) can be estimated from the following expression ~\cite{Andronic2012,Sarwar:2015irq}, 
\begin{eqnarray}
s_H(T, \mu_{B})=\pm \sum_{i}\frac{g_{i}}{2\pi^{2}}
\int^{\infty}_{0}&&dp^{\prime}{p^\prime}^{2}\Big[ln\Big(1\pm \{exp(E_{i}-\mu_{i})/T\}\Big)\nn\\
&&\pm \frac{E_{i}-\mu_{i}}{T\{exp(E_{i}-\mu_{i})/T\pm 1\}}\Big]~,
\label{eq0433}
\end{eqnarray}
where, the sum extends over all hadrons with mass up to 2.5 GeV~\cite{Sarwar:2015irq}, $g_i$  represents the statistical degeneracy factor of the $i^{th}$ hadron, and $E_{i}=\sqrt{{p^\prime}^{2}_{i}+m^{2}_{i}}$ is  the energy of the $i^{\text{th}}$ hadron of mass, $m_i$ and momentum, $p_i$. Once we know the entropy density, the thermodynamic quantities {\it e.g.} baryon number density ($n_{B}$), pressure ($P$) and energy density ($\epsilon$) can be evaluated as follows. The net baryon number density is given by 
	\begin{eqnarray}
	n_{B}(T, \mu_{B})= \int_{0}^{T} \frac{\partial s(T^{'}, \mu_{B})}{\partial \mu_{B}} dT^{'}
	+n_{B}(0, \mu_{B})~.
	\label{eq0434}
	\end{eqnarray}
	A plot of baryon number density is shown in Fig.\ref{fig0404}. From $n_{B}$, the pressure can be estimated as 
	\begin{eqnarray}
	P(T, \mu_{B})=  \int_{0}^{T} s (T^{'}, \mu_{B}) dT^{'} +P (0, \mu_{B})~,
	\label{eq0435}
	\end{eqnarray}
	where,  $n_{B}(0,T)$ and $p (0, \mu_{B})$ are the net baryon density and pressure respectively at  $T=0$. Finally, the energy density is given by, 
	\begin{eqnarray}
	\epsilon(T, \mu_{B})= Ts(T,\mu_{B})-P(T,\mu_{B})+\mu_{B} n_{B}~.
	\label{eq0436}
	\end{eqnarray}
	To get the first order phase boundary,  the discontinuity in the entropy density along the transition line also needs to be considered. We add the following term to Eq.\eqref{eq0434} to  take into account  this possibility (for $T>T_{c}$) 
	\begin{eqnarray}
	\Big|\frac{\partial T_{c}(\mu_{B})}{\partial \mu_{B}}
        \Big|\Big[s(T_c(\mu_{B})+\Delta,\mu_{B})-s(T_c(\mu_{B})-\Delta,\mu_{B})\Big]~,
	\label{eq0437}
	\end{eqnarray}
	where, $\Big|\frac{\partial T_{c}}{\partial \mu_{B}}\Big|=tan\theta_{c}$, is the tangent 
	at the $T_c$ and $\Delta (\rightarrow 0)$ is a small deviation in $T$   
        from $T_{c}$.
	The value of $T_{c}$  for the first order transition depends on $\mu_{B}$ 
	as indicated by Eq.~\eqref{eq0437}. 

In this context we contrast the entropy density, speed of sound $c_{s}^{2}=(\pd P/\pd \epsilon)_{s}$ and the baryon susceptibility, $\chi_2^B$ (Eq.\eqref{barsusc}) obtained in the present work with the corresponding lattice QCD results~\cite{Borsanyi:2013bia,Borsanyi:2011sw} in the $\mu\rightarrow 0$ limit, as shown in Figs.\ref{fig0405} and \ref{fig0406} respectively.

Recently, a variety of sophisticated EoS has been constructed~\cite{Parotto}, where authors have presented a procedure to construct a family of model EoS for QCD, each of which
contains a critical point in accordance of the universality class consideration of the 3D Ising model. They have reconstructed the EoS in a way such that it precisely features the critical behavior in the correct universality class, and as well as it matches lattice QCD results up to $\mathcal{O} (\mu_{B}^{4})$ exactly. They have also argued that the family of constructed model EoS can also be used directly to feature the divergence of quantities in the the critical region, in particular the baryon number cumulants: the second cumulant, which is related to the variance of the net-proton number distribution. For other variety of EoS at finite $\mu_{B}$, one can dig into the Refs.~\cite{Karthein2021,Fodor2021PRL,Monnai2019,Soloveva2021}.
\section{Entropy density calculation for arbitrary chosen axis}
The transport coefficients show diverging nature, or at least enhance near the CEP~\cite{Monnai2017,KapustaChi,Antoniou2017,Minami_thesis,Hohenberg1977,Son2004,Moore2008,Onuki1997}. In Ref.\cite{Martinez2019}, Martinez {\textit{et al.}} showed that the critical bulk viscosity is very sensitive to the inclination of the Ising axes in the QCD phase diagram. In most of the construction of the critical EoS, it is assumed that the $t$-axis is aligned with the $\mu_{B}$-axis of the QCD phase diagram and that the $h$-axis is perpendicular to that. In that case, the bulk viscosity is suppressed by a factor of $(n/s)^{2}$ compared to when we include a possible tilt of the $t$-axis with the $\mu_{B}$-axis. This is also true in the sense that the CEP may be located at any arbitrary point, which may cause the $t$-axis to be tilted with respect to the $\mu_{B}$-axis. Therefore, the mapping procedure can be improved to evaluate $s_{c}$ (in Eq.\eqref{eq0425}) in a more general sense when the $t$-axis makes an angle $\theta_{c}$ with the $\mu_B$-axis. The mapping procedure have to include $cos(\theta_{c})$ and $sin(\theta_{c})$ terms. So, in Eq.\eqref{aaa}, the term $(\pd t/\pd T)_{\mu_{B}}$ must not vanish, and thus contribute in the form of $s_{c}(T, \mu_{B})$ as:
\beqa
s_c (T, \mu_{B})&=& M\Big(\frac{(T-T_{c})cos(\theta_{c})}{\Delta T_{c}},\frac{(\mu_B-\mu_{c})cos(\theta_{c})}{\Delta \mu_{c}}\Big)\frac{cos(\theta_{c})}{\Delta T_{c}} \nonumber\\
&-& \Big(\frac{\partial G}{\partial t}\Big)_h\Big(\frac{(T-T_{c})cos(\theta_{c})}{\Delta T_{c}},\frac{(\mu_B-\mu_{c})cos(\theta_{c})}{\Delta \mu_{c}}\Big)\frac{sin(\theta_{c})}{\Delta T_{c}}~.
\eeqa

\section{Hydrodynamic framework in space-time evolution of the QGP}
Fixing the initial condition and after construction of the EoS, we can now solve the hydrodynamic equations uniquely. In a collision of heavy ions moving relativistically, the QGP is likely to form, and it expands very rapidly after the formation. The relativistic hydrodynamics can be a powerful tool to describe the collective flow of QCD matter, created in heavy-ion collisions (HICs). Describing elliptic flow ($v_{2}$)~\cite{Snellings:2011sz} and other flow observables ($v_{n}$)~\cite{Ollitrault:2007du} on a quantitative level is one of the greatest successes of the fluid dynamical description~\cite{Rischke:1998fq,Shuryak:2003xe,Stoecker:1986ci}. Ideal fluid dynamics quantitatively can explain only in central collisions between large $(A \sim 200)$ nuclei at mid-rapidity at top RHIC energies, but gradually break down in a smaller systems, or in a system produced from peripheral collisions, {\it{i.e.}} away from the mid-rapidity region, and at lower collision energies~\cite{Heinz:2004ar}. Even with the highest achievable centre of mass energy $(\sqrt{s})$, the lowest limit of the shear viscosity to entropy ratio is found to be $\eta/s=1/4\pi$, has been proposed based on a correspondence with black-hole physics, known as KSS bound~\cite{Kovtun:2004de}. The viscous hydrodynamics was applied extensively ever since the estimation of surprisingly small value of $\eta/s$ from the analysis of the elliptic flow data~\cite{Romatschke:2007mq}. A study of elliptic flow suggests that the magnitude of viscous corrections is at least $30 \%$~\cite{Drescher:2007cd}. Therefore, the description of strongly coupled QGP produced in HICs will be in better agreement with the viscous effect. Furthermore, if QGP fluid is formed in heavy-ion collisions, it needs to be characterized by its relevant transport coefficients, {\it{e.g.}} bulk viscosity, shear viscosity and the thermal conductivity.

\chapter[Propagation of perturbation near the QCD critical point]{Propagation of perturbation near the QCD critical point}
\label{chapter5}  
\section{Introduction}
\label{sec0501}
This chapter is based on the paper of Ref.\cite{Hasan1}. Here we study the propagation of small disturbances if the evolution of the created QGP is passing through (near) the CEP (critical end point). As discussed earlier, fixing the location of the CEP is a challenging job. However, we are not searching for the location of CEP by direct means. Rather, we want to examine its effects on the possible fate of the sound wave (disturbance) propagating through the fluid in presence of the CEP. Here, the location of the CEP is taken at: $(T_c,\, \mu_{c})= (154 \text{MeV},\, 367 \text{MeV})$ (as mentioned in Ch.\ref{chapter4}). It is expected that a system with such $(T, \mu_{B})$ may be produced through nuclear collisions at GSI-FAIR, NICA, BES-RHIC. The QGP produced in such collisions will expand very rapidly along a trajectory with constant $s/n$ ($s$ and $n$ stand for the entropy density and baryon number density respectively) and cools down consequently. We have assumed that an isentropic trajectory in the $(T, \mu_{B})$ plane will pass through the critical region of the CEP.

The space-time evolution of the thermalized QGP can be modelled by the relativistic viscous hydrodynamics (Ch.\ref{chapter3}). The first-order theory of relativistic viscous hydrodynamics is governed by Navier-Stokes (NS) equations, is known to violate causality, and gives unstable numerical solutions \cite{Hiscock:1983zz,Hiscock:1985zz}. Therefore, it makes the theory unsuitable to study the space-time evolution of QGP. These problems were rectified by M\"{u}ller~\cite{Muller:1967zza} and Grad~\cite{Grad} after including quantities in second-order dissipative flux and therefore, these theories are called the `second-order hydrodynamics'. The relativistic generalization of this theory was formulated by Israel and Stewart \cite{Israel:1979wp}, and we have used it to describe the space-time evolution of QGP. The hydrodynamic response of the fluid to the perturbation is dictated by the relevant transport coefficients (shear and bulk viscosities, and thermal conductivity) of the fluid. The effects of thermal conductivity $(\kappa)$ and the shear viscosity $(\eta)$ have been considered in this work to investigate the propagation of acoustic waves when the system passes through the CEP. The effects of the CEP in the hydrodynamic evolution enter through the EoS, is constructed on basis of the universality hypothesis (discussed in Ch.\ref{chapter4}). The behaviour of the thermodynamic quantities near the CEP is governed by the critical exponents. Dispersion relations {\it i.e.} the functional dependence of the frequency ($\omega$) on the wave vector ($k$) will be set up to study the effects of the CEP on the propagation of the sound waves in the fluid.
\section{Propagation of the perturbations in viscous fluid} 
\label{sec0502}
M\"{u}ller-Israel-Stewart (MIS) hydrodynamics is appropriate to study the evolution of the QGP as the NS theory violates causality and introduces instability in the numerical solution. Therefore, in this section, we study the propagation of perturbations through the viscous fluid by using second-order causal hydrodynamics. One of the major differences between a relativistic and a non-relativistic fluid originates from the definition of chemical potential (coming through the fugacity factor). In the non-relativistic case, the chemical potential puts a constraint on the total number of particles in the system. But in the relativistic system, the total number of particles does not remain constant due creation and annihilation of particles within the fluid. However, through the creation and annihilation processes, the conservation of certain quantum numbers remains untouched. For example, in the strong interaction, net (baryon minus antibaryon) baryon number, net electric charge, or net strangeness remain conserved (although strangeness is not conserved in weak interaction). Accordingly, the net baryon number will remain conserved throughout the evolution process of the QGP. Therefore, in the discussion below the net charge density stands for net baryon number density. Also to note that from here onwards, we will drop the subscript `$B$' from $\mu_{B}$ and $n_{B}$ to define baryon chemical potential and baryon number density.

In this chapter, we use the Landau-Lifshitz frame (LL-frame) of reference as discussed in Sec.\ref{sec0305} in Ch.\ref{chapter3}, where it is considered that the heat flux is zero but the particle current is non-zero in the local rest frame (LRF). Therefore,
\beqa
 h^\mu=0,\,\,\, n^\mu =-n q^\mu /(\epsilon+P)\,,
 \label{eq0501}
 \eeqa
  and the different viscous fluxes are given by~\cite{Israel:1979wp}-
\begin{eqnarray}
\Pi &=&-\frac{1}{3}\zeta\Big[\pd_{\mu}u^\mu +\beta_0 D \Pi-\alpha_0  \pd_{\mu}q^\mu \Big]~,\nonumber\\   
\pi^{\mu\nu}&=&-2\eta \Delta^{\mu\nu\alpha\beta}\Big[\partial_{\alpha}u_{\beta}+\beta_{2}D\pi_{\alpha\beta}-\alpha_{1}\partial_{\alpha}q_{\beta}\Big]~,\nonumber\\
q^{\mu}&=&\kappa T\Delta^{\mu\nu} \Big[\frac{nT}{\epsilon+P}(\partial_\nu \alpha )-\beta_1 D{q_\nu}+\alpha_0\partial_\nu \Pi +\alpha_1\pd_{\lambda}\pi ^{\lambda}_{\nu} \Big]~. 
\label{eq0502}
\end{eqnarray}
where  $D\equiv u^\mu\partial_\mu$, is known as co-moving derivative and in local rest frame (LRF), $D\Pi =\dot{\Pi }$ represents the time derivative. The coefficients $\alpha_{0}, \alpha_{1}, \beta_{0}, \beta_{1}$ and $\beta_{2}$ can be calculated from thermodynamics integrals (see Appendix.\ref{appendix01_A}). But we have used the ultra-relativistic limit, $\beta=m/T\rightarrow 0$ where $m$ is the mass of the particle and we have the following relations ~\cite{Israel:1979wp},
\begin{eqnarray}
\alpha_0 \approx  6\beta^{-2}P^{-1}, \alpha_1 \approx -\frac{1}{4}P^{-1},
\beta_0 \approx 216 \beta^{-4} P^{-1},\beta_1 \approx \frac{5}{4}P^{-1}, 
\beta_2 \approx \frac{3}{4}P^{-1} \,.
\label{eq0503}
\end{eqnarray}

Since in energy frame,  $h^\mu=0$, then the energy-momentum tensor (EMT) reduces to 
\begin{equation}
T^{\mu\nu}=\epsilon u^\mu u^\nu+P\Delta^{\mu\nu} +\Pi\Delta^{\mu\nu}+\pi ^{\mu\nu} \,.
\label{eq0504}
\end{equation}
Putting the explicit forms of $\Pi, q^\mu$ and $\pi ^{\mu\nu}$ given by Eq.\eqref{eq0502} into Eq.\eqref{eq0504} and keeping only the terms up to second order in space time derivatives, the EMT becomes  \cite{Rahaman:2017ezf}
\begin{eqnarray}
 T^{\mu\nu}&= & \epsilon u^\mu u^\mu+P\Delta^{\mu \nu}- \frac{1}{3}\zeta \Delta^{\mu\nu}\partial_\alpha u  ^\alpha + \frac{1}{9} \zeta \beta_0\Delta^{\mu\nu} D(\zeta\partial_\alpha u  ^\alpha)   \nn\\
 &+&\frac{\zeta \alpha _0  }{3}\Delta^{\mu\nu}\partial_\alpha\Big [\frac{n\kappa T^2}{\epsilon+P}\nabla^\alpha(\alpha) \Big]
 -2\eta \Delta^{\lambda\mu\alpha\beta} \partial_\alpha u_\beta  
  + 4\eta \beta_2\Delta^{\lambda\mu\alpha\beta} D(\eta\Delta_{\alpha\beta}^{\rho\sigma} \partial_\rho u_\sigma)  \nn\\
  &+&2\alpha _1\eta\Delta^{\mu\nu\alpha\beta}\partial_{\alpha}\Big[\frac{ n\kappa T^2}{\epsilon+P}\nabla_\beta(\alpha)\Big]\,.
\label{eq0505}
\end{eqnarray}

The solution of the MIS hydrodynamic equations grant causality and stability, which is achieved by promoting the dissipative currents (Eq.~\eqref{eq0502}) as the independent dynamical variables and introducing relaxation time scales (time delay) for these currents. In NS theory, the dissipative currents promptly respond to the hydrodynamical gradients but in MIS theory, the response of the dissipative currents is controlled by the relaxation time scales (see Eq.~\eqref{eq0503}). The EMT given in Eq.~\eqref{eq0505} represents second-order dissipative hydrodynamics which is equivalent to MIS theory for small gradients. The general form of the EMT constrained by the conformal invariance can be found in ~\cite{Baier2008}.

The charge current (up to second-order in velocity gradient) can be written as, 
\begin{eqnarray}
 N^\mu &=&nu^\mu  -\frac{n\chi T\triangle^{\mu\nu }}{(\epsilon+P)}\Big[ \frac{nT}{(\epsilon+P)} (\partial _\nu \alpha )  -\beta_1D\ \Big\{ \frac{n\chi T^2}{(\epsilon+P)}   \triangle_{\nu \sigma }(\partial^\sigma \alpha) \Big\}\nn\\
 &&-\frac{\zeta\alpha_0}{3}\partial_\nu \partial _\sigma u^\sigma-2\eta \alpha_1 \partial^\rho u_{<\rho |\nu> }\Big]\,.
\label{eq0506}
\end{eqnarray}
\section{Perturbations in the relativistic fluid}
\label{sec0503}
The motion of perturbations in the relativistic viscous fluid with one conserved current (baryonic current for the present case) is governed by Eqs.\eqref {eq0505} and \eqref{eq0506}. Now we impart small perturbations $P_1,\epsilon_1, n_1, T_1, \mu_1$ and $ u^\alpha_1$  to $P, \epsilon, n, T,\mu$ and $u^\alpha $  respectively to study the propagation of acoustic wave in the fluid with $u^\alpha=(1,0,0,0)$  as outlined in Ref.\cite{Weinberg:1971mx}. We set $u^0_1=0$ to preserve the normalization condition $u^\alpha u_\alpha =-1$.

A  space-time dependent perturbation, $\sim exp[-i(k  x -\omega t)]$, is placed into the fluid and its fate is being studied. The equation of motions that dictate the evolution of different components of the perturbations can be obtained from the conservations of the energy-momentum tensor ($T^{\mu\nu}$)  and net-baryon number flux ($N^\mu$) of the fluid:
\begin{equation}
\partial_\mu T^{\mu\nu }=0, \,\,\,\,\, \partial_\mu {N}^\mu=0\,.
\label{eq0507}
\end{equation} 
The equations of motion (EoMs) have been linearized to find the dispersion relation for small perturbation. After that the various components of energy momentum tensor are written in $\omega-k$ (frequency-wave vector) space as: 
\begin{eqnarray}
0=\omega T_1^{00}-k_iT_1^{i0}=\omega\epsilon_1-(\epsilon+P)(\vec{k}\cdot \vec{u_1})\,.
\label{eq0508}
\end{eqnarray}
and the other components of the EMT satisfies,
\begin{eqnarray}
0&=&\omega T_1^{i0}-k_jT_1^{ij}\nonumber\\
&=&\omega (\epsilon+P)u_1^i-k^iP_1+\frac{1}{3}\zeta k^i\Big[ i(\vec{k}\cdot \vec{u_1})+\frac{1}{3}\zeta\beta _0\omega(\vec{k}\cdot \vec{u_1})\Big] \nn\\
&&+i\eta\Big[ k^2u^i_1+\frac{1}{3}k^i (\vec{k}\cdot \vec{u_1})\Big]-2\eta^2 \beta_2 \omega\Big[ k^2u^i_1+\frac{1}{3}k^i (\vec{k}\cdot \vec{u_1})\Big]\nn\\
&&+\frac{nT\kappa}{(\epsilon+P)}(\mu _1-\alpha T_1)\Big[ \frac{\alpha _0\zeta k^2}{3}k^i+\frac{4}{3}\alpha_1\eta k^2 k^i\Big]\,.
\label{eq0509}
\end{eqnarray}
Using the linearization technique, the number conservation equation in the $\omega-k$ space becomes,
\begin{eqnarray}
 0&=&\omega n_1-n(\vec{k}\cdot\vec{u_1})-i \frac{n^2\kappa T k^2}{(\epsilon+P)^2}\Big[(\mu_1-\alpha T_1)(1+i\omega\kappa \beta_1)\Big]\nn\\
 &&+\frac{1}{3}\frac{n \kappa Tk^2}{(\epsilon+P)}\Big[\zeta \alpha_0+4 \eta \alpha_1\Big](\vec{k}\cdot\vec{u_1})\,.
\label{eq0510}
\end{eqnarray}
We have considered terms up to the first order in perturbation as
\beqa
n\,u^{\mu}\to~(n+n_{1})(u^{\mu}+u_{1}^{\mu})\approx~nu^{\mu}+n_{1}u^{\mu}+nu_{1}^{\mu}\,.
\label{eq0511}
\eeqa
In deriving the Eqs.\eqref{eq0508}, \eqref{eq0509}, \eqref{eq0510}, we have used
\beqa
\pd_{\mu}\alpha&=&\pd_{\mu}\Big[\frac{\mu+\mu_{1}}{T+T_{1}}\Big]~\approx \pd_{\mu}\Big[(\frac{\mu+\mu_{1}}{T})(1-\frac{T_{1}}{T})\Big]\nn\\
&=&\pd_{\mu}\Big[\frac{\mu+\mu_{1}}{T}-\frac{\mu}{T^{2}}T_{1}\Big]=~\frac{1}{T}\big[\pd_{\mu}\mu_{1}-\alpha \pd_{\mu}T_{1}\big]~,
\label{eq0512}
\eeqa
and,
\beqa
\nabla_{\mu}\alpha=\frac{1}{T}\big[\nabla_{\mu}\mu_{1}-\alpha \nabla_{\mu}T_{1}\big]\,.
\label{eq0513}
\eeqa
In Eqs.~\eqref{eq0508}, \eqref{eq0509} and \eqref{eq0510}, we have considered terms up to first order in perturbations and neglected the higher order terms. Also, we have not perturbed the different transport coefficients, as they are not hydrodynamical variables. In LRF, we take them as constant in space and time, hence their comoving derivative are zero.  As mentioned, we have considered, $\eta\neq 0$,  $\kappa\neq 0$ and $\zeta$=0. We decompose the fluid velocity into directions perpendicular and parallel to the direction of wave vector, $\vec{k}$ as:
\begin{equation}
\vec{u_1}=\vec{u_1}_\bot +\vec{k} (\vec{k}\cdot\vec{u_1})/k^2\,.
\label{eq0514}
\end{equation}
The modes propagating along the direction of $\vec{k}$ are called longitudinal 
and those perpendicular to $\vec{k}$ are called transverse modes. 

The quantities, $\epsilon_1$, $P_1$ and $\mu_1$ defined above can be 
expressed in terms of the derivatives of the thermodynamic quantities (which
are connected to response functions like specific heat, etc.) 
as follows:
\begin{eqnarray}
&&\epsilon_1=\Big(\frac{\partial \epsilon}{\partial T}\Big)_n T_1+\Big( \frac{\partial \epsilon}{\partial n}\Big)_T n_1\,,
\nn\\&&P_1=\Big( \frac{\partial P}{\partial T}\Big)_n T_1+\Big( \frac{\partial P}{\partial n}\Big)_T n_1\,,
\nn\\&&\mu_1=\Big[-\Big( \frac{\partial n}{\partial T}\Big)_n T_1+n_{1}\Big]\Big( \frac{\partial \mu}{\partial n}\Big)_T \,.
\label{eq0515}
\end{eqnarray}
\section{Dispersion Relations}
\label{sec0503}
The dispersion relation ($\omega$ (frequency) as a function of $k$ (wave vector)) plays an important role in determining the fate of the perturbation in a fluid. The $\omega$ can be a real as well as an imaginary quantity. The real part of the frequency, $\omega_{\R e}$, dictates the possibility of the propagation of the wave in the medium. Whereas, the imaginary part of the frequency, $\omega_{Im}$, decides the decay of the waves along its propagation in the medium. 

To evaluate the dispersion relation, we would like to write down a equation that dictates the $k$ dependence of $\omega$. For this purpose we use Eqs.~\eqref{eq0508}, \eqref{eq0509} and \eqref{eq0510}, and collected the coefficients $(\vec{k}\cdot\vec{u_1})$, $T_1$ and $n_1$ to put row wise to form a determinant as
\beqa
\begin{vmatrix}
&&a_{11} && a_{12} && a_{13} ~~~~ \\ 
&&a_{21} && a_{22} && a_{23} ~~~~ \\
&&a_{31} && a_{32} && a_{33} ~~~~
\end{vmatrix}
\label{eq0516}
\eeqa
 and upon the expansion of the determinant above, yield
\begin{eqnarray}
a\omega^3+b\omega^2+c\omega=0\hspace{0.5cm}\rightarrow\hspace{0.5cm}\omega(a\omega^2+b\omega+c)&=&0 \,.
\label{eq0517}
\end{eqnarray}
The coefficients $a, b$ and $c$ are determined by solving 
Eqs.~\eqref{eq0508}, \eqref{eq0509} and \eqref{eq0510}, simultaneously. 
The solutions of this equation which provide a relation between $\omega$ and $k$
which is called the dispersion relation.
The equation, \eqref{eq0517} has  
three roots, one real which is 
$\omega=0$ and two complex roots with real ($\omega_{\R e}$) and imaginary ($\omega_{im}$) 
parts given below by Eqs.~\eqref{eq0518} and \eqref{eq0521} respectively.
The real part of $\omega$ can be expressed as:
\begin{equation}
\omega_{\R e}=\sqrt{\frac{a_0k^2-a_1k^3+a_2k^4}{b_0-b_1k^2}}\,.
\label{eq0518}
\end{equation}
where,
\beqa
&&a_0=9h\Big[\Big(\frac{\partial P}{\partial T}\Big)_n
+\alpha_1n
\Big\lbrace\Big(\frac{\partial\epsilon}{\partial n}\Big)_T
\Big(\frac{\partial P}{\partial T}\Big)_n
-\Big(\frac{\partial\epsilon}{\partial T}\Big)_n
\Big(\frac{\partial P}{\partial n}\Big)_T
\Big\rbrace\Big]\,,\nn\\
&&a_1=\frac{9\alpha\beta_1n^2T^2\kappa^2}{h}+
12\alpha_1\eta\kappa nT\Big[\alpha+\frac{\alpha n}{h}\Big(\frac{\partial\epsilon}{\partial n}\Big)_T
-\frac{T}{h}\Big(\frac{\partial P}{\partial T}\Big)_n
+\frac{T}{h}\Big(\frac{\partial\epsilon}{\partial T}\Big)_n\Big]\,,
\nn\\
&&a_2=\frac{9\beta_1\kappa^2n^2T}{h}\Big[
\Big(\frac{\partial n}{\partial T}\Big)_\mu\Big(\frac{\partial\mu }{\partial n}\Big)_T
+\Big(\frac{\partial P}{\partial T}\Big)_n\Big(\frac{\partial P}{\partial n}\Big)_T\Big]\nn\\
&&\,\,\,\,\,\,\,\,\,\,\,\,\,\,+\frac{12\alpha_1\eta\kappa nT}{h}\Big[\Big(\frac{\partial P}{\partial T}\Big)_n 
+n\Big(\frac{\partial n}{\partial T}\Big)_\mu\Big(\frac{\partial\mu}{\partial n}\Big)_T
+\frac{n}{h}\Big(\frac{\partial P}{\partial T}\Big)_n\Big(\frac{\partial\epsilon}{\partial n}\Big)_T\Big]\,,\nn\\
&&b_0=9h\Big(\frac{\partial\epsilon}{\partial T}\Big)_n\,,\nn\\
&&b_1=24\beta_2\eta^2\Big(\frac{\partial\epsilon}{\partial T}\Big)_n+\frac{9\beta_1\kappa^2 n^2}{h}\Big[
T\Big(\frac{\partial\mu}{\partial n}\Big)_T\Big(\frac{\partial n}{\partial T}\Big)_\mu \nn\\
&&\,\,\,\,\,\,\,\,\,\,\,\,\,\,-T^2\Big(\frac{\partial\epsilon}{\partial n}\Big)_T\Big(\frac{\partial\mu}{\partial n}\Big)_n
+\alpha\Big(\frac{\partial\epsilon}{\partial n}\Big)_T\Big]\,.
\label{eq0519}
\eeqa
and, $h=\epsilon+P$ is the enthalpy density.   
We have kept terms up to quadratic power of
transport coefficients in Eq.~\eqref{eq0518}.  
We have also neglected the higher order terms in $\alpha_{0},\alpha_{1},\beta_{0},\beta_{1},\beta_{2}$.
Expanding $\omega_{\R e}$ in powers of $k$ and keeping terms up to $\cal{O}$$(k^4)$ we obtain,
\begin{equation}
\omega_{\R e}=\sqrt{\frac{a_0}{b_0}}\left[k-\frac{1}{2}\frac{a_1}{a_0}k^2+(\frac{1}{2}\frac{a_2}{a_0}-\frac{1}{8}{a_1}{a_0^2}+\frac{b_1}{b_0})k^3
+(\frac{1}{4}\frac{a_1a_2}{a_0^2}+\frac{1}{16}{a_1^2}{a_0^3}-\frac{1}{2}\frac{a_1b_1}{a_0b_0})k^4\right]\,.
\label{eq0520}
\end{equation}

Similarly, the expression for the imaginary part of $\omega$ reads as:
\begin{equation}
\omega_{Im}=\frac{-c_0k^2+c_1k^3+c_2k^4}{d_0+d_1k^2}\,.
\label{eq0521}
\end{equation}
where,
\begin{eqnarray}
&& c_0=2\eta h^2\Big(\frac{\partial\epsilon}{\partial T}\Big)_n
-3hn^2\kappa T\beta_1\Big[\alpha\kappa\Big(\frac{\partial\epsilon}{\partial n}\Big)_T
+h\Big(\frac{\partial n}{\partial T}\Big)_\mu
\Big(\frac{\partial\mu }{\partial n}\Big)_T\nn\\
&&\,\,\,\,\,\,\,\,\,\,\,\,+\frac{\alpha_1}{\beta_1}\Big(\frac{\partial P}{\partial T}\Big)_n\Big(\frac{\partial\epsilon}{\partial n}\Big)_T\Big]\,,\nn\\
&&c_1=2\alpha\beta_1\eta n^2 T\Big(T\kappa^2+4 h\eta\frac{\beta-2}{\beta_1}\Big)\,,\nn\\
\label{eq06}
&&c_2=8\beta_2\eta\kappa n^2T\Big[\Big(\frac{\partial\epsilon}{\partial T}\Big)_n-\Big(\frac{\partial P}{\partial n}\Big)_{T}^2
\Big(\frac{\partial\epsilon}{\partial n}\Big)_T\Big]\,,\nn\\
&&d_0=3h^3\Big(\frac{\partial \epsilon}{\partial T}\Big)_n\,,\nn\\
&&d_1=3h\beta_1n^2\kappa\Big[\alpha\kappa\Big(\frac{\partial\epsilon}{\partial n}\Big)_T+4T^2\kappa\Big(\frac{\partial\epsilon}
{\partial T}\Big)_n\Big(\frac{\partial\mu}{\partial n}\Big)_T^2-
T\Big(\frac{\partial n}{\partial T}\Big)_n\Big(\frac{\partial\epsilon}{\partial n}\Big)_T\Big]\,.
\label{eq0522}
\end{eqnarray}
The imaginary part of $\omega$ up to $\cal{O}$$(k^4)$  is given by,
\begin{equation}
\omega_{Im}=-\frac{c_0}{d_0}\left[k^2-\frac{c_1}{c_0}k^3-(\frac{d_1}{d_0}+\frac{c_2}{c_0})k^4\right]\,.
\label{eq0523}
\end{equation}
The dispersion relation for first order hydrodynamics can be obtained by setting the relaxation coefficients ($\beta _0,\beta_1,\beta_2$) and  the coupling coefficients ($\alpha_0$, $\alpha_1$) to zero which allow only $a_0$, $b_0$, $c_0$ and $d_0$ to be non-zero. Therefore, keeping terms
up to $\cal{O}$$(k^2)$ in Eqs.~\eqref{eq0520} and \eqref{eq0523} we get
(see also~\cite{Grozdanov}),
\begin{equation}
\omega (k)=c_s k -\frac{i}{2}k^{2}\frac{\eta}{s}s\frac{4/3}{h}\,,
\label{eq0524}
\end{equation}
where, $c_{s}=\sqrt{\big(\frac{\pd P}{\pd \epsilon}\big)_{s/n}}$ is the speed of sound and $\eta/s$ is 
shear viscosity ($\eta$) to the entropy density ($s$) ratio. The Eq.\eqref{eq0524} is the dispersion relation for
NS hydrodynamics, which is obtained from the limiting case of MIS theory.

\subsection{Fluidity near the critical region}
The imaginary and real parts of $\omega (k)$ provide the information respectively on the attenuation and the propagation of the sound wave in the dissipative fluid. Thus, if magnitude of the imaginary part  is greater than the real part, the wave will dissipate quickly. Therefore, the dispersion relations in Eqs.\eqref{eq0518} and \eqref{eq0521} can be used to determine the upper limit of $k$ of the sound wave for its survivability in the medium. The threshold value of $k$, {\textit{i.e.}} $k_{th}$ can be evaluated by using the 
following condition~\cite{liao}
\begin{eqnarray}
\Big|\frac{\omega_{Im}(k)}{\omega_{\R e}(k)} \Big|_{k=k_{th}}=1\,.
\label{eq0525}
\end{eqnarray}
{\it i.e.} any wave with wave vector higher than $k_{th}$ will get
dissipated in the fluid.
Solving the above equation, we get, 
\begin{equation}
k_{th}=\sqrt{\frac{\mathcal{P}}{\mathcal{Q}}}~.
\label{eq0526}
\end{equation}
where,
\begin{eqnarray}
\label{eq0527}
\mathcal{P}&=&\frac{a_{0}}{c_{0}^{2}}-\frac{a_{0}d_{1}c^{2}_{2}}{b_{0}c_{0}^{2}d_{0}^{2}}-\frac{a_{1}d_{0}}{c_{0}^{3}}+\frac{a_{1}c_{2}^{2}d_{1}}{b_{0}c_{0}^{3}d_{0}}-\frac{b_{1}^{2}c_{2}^{3}d_{1}}{b_{0}c_{0}^{2}d_{0}^{3}}+\frac{a_{1}d_{0}}{a_{0}b_{0}c_{0}^{2}}~,\\
\label{eq0528}
\mathcal{Q}&=&\frac{b_{0}}{d_{0}^{2}}-\frac{b_{1}}{c_{0}^{2}}-\frac{b_{0}c_{2}}{c_{0}^{3}}+\frac{a_{1}}{a_{0}b_{0}d_{0}}-\frac{a_{1}b_{1}d_{0}}{a_{0}b_{0}^{2}c_{0}^{2}}+\frac{a_{1}c_{2}d_{0}}{b_{0}^{2}c_{0}^{3}}~.
\end{eqnarray}
Expanding $\mathcal{P}$ and $\mathcal{Q}$ and keeping the leading terms of the series we get, 
\begin{eqnarray}
k_{th}&=&\sqrt{\frac{a_{0}d_{0}^{2}}{b_{0}c_{0}^{2}}}\Bigg[1-\frac{1}{2}\Bigg(\frac{d_{1}c_{2}^{2}}{b_{0}d_{0}^{2}}+\frac{a_{1}d_{0}}{a_{0}c_{0}}-\frac{a_{1}c_{2}^{2}d_{1}}{a_{0}b_{0}c_{0}d_{0}}+\frac{b_{1}^{2}c_{2}^{3}d_{1}}{a_{0}b_{0}d_{0}^{3}}-\frac{a_{1}d_{0}}{a_{0}^{2}b_{0}}\Bigg)\Bigg]\nonumber\\
&&\Bigg[1+\frac{1}{2}\Bigg(\frac{b_{1}d_{0}^{2}}{b_{0}c_{0}^{2}}+\frac{c_{2}d_{0}^{2}}{c_{0}^{3}} -\frac{a_{1}d_{0}}{a_{0}b_{0}^{2}}+\frac{a_{1}b_{1}d_{0}^{3}}{a_{0}b_{0}^{3}c_{0}^{2}}-\frac{a_{1}c_{2}d_{0}^{3}}{b_{0}^{3}c_{0}^{3}}\Bigg)\Bigg]
\label{eq0529}
\end{eqnarray}
The first term of the above expression gives rise to the value of $k_{th}$ in the NS limit,  $k_{th}=\sqrt{\frac{a_{0}d_{0}^{2}}{b_{0}c_{0}^{2}}}=\frac{3}{2}c_s\frac{h}{s}\frac{1}{\eta/s}$~\cite{liao}. The subsequent terms are arising from the second-order hydrodynamical effects as indicated by the presence of coupling and relaxation coefficients appearing through the quantities defined in Eqs.~\eqref{eq0503}.

The threshold wavelength ($\lambda_{th}$) corresponding to $k_{th}$ is given by $\lambda_{th}= 2\pi/k_{th}$. Sound waves with  wavelength, $\lambda < \lambda_{th}$ will dissipate in the medium. However, sound waves with $\lambda \ge \lambda_{th}$ will survive to propagate in the fluid without much dissipative effects. The quantity $\lambda_{th}$ can also be used to define the fluidity ($\mathcal{F}$) of the system with widely varying particle density and temperature by selecting a length scale (inter-particle separation), $l\sim \rho^{-1/3}$~\cite{liao} of the system as: 
\begin{equation}
\mathcal{F}\sim \frac{\lambda_{th}}{l}.
\label{eq0530}
\end{equation}
where, $\rho$ is the particle number density of the fluid (for a relativistic fluid $l$ can be chosen as $l\sim s^{-1/3}$). The length scale, $R_v\sim 1/k_{th}$, called viscous horizon \cite{staig}, which fixes the limit for the sound wave with $\lambda$ smaller than $R_v$ will be dissipated due to highly viscous and thermal conduction effects. $R_v$ can also be used to estimate
the value of order of the highest harmonics $n_v=2\pi R/R_v$ ($n=2, 3, 4,...$)  which will survive the dissipation {\it i.e.} any harmonics of order higher than $n_v$ will not survive against dissipation. We find that $k_{th}$ is directly proportional to the $c_{s}$, which approaches  zero at the critical point, that is the sound wave loses its strength and get damped strongly near critical point. Therefore, we can argue that $k_{th}$ also vanishes or in other words $\lambda_{th}$ diverges at the critical point.
\section{Results and discussion}
\label{sec0504}
\begin{figure}[h]
\centering
\includegraphics[width=7.cm]{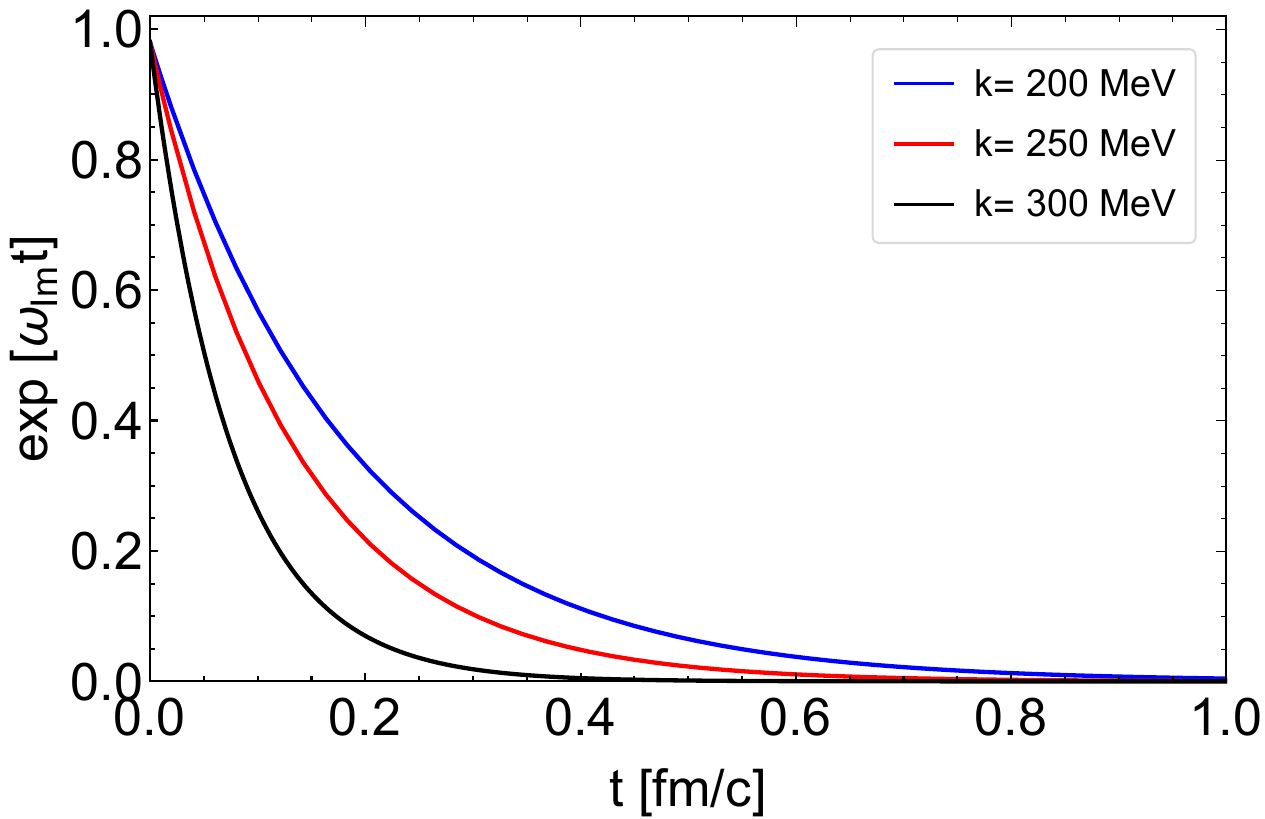}
\includegraphics[width=7.cm]{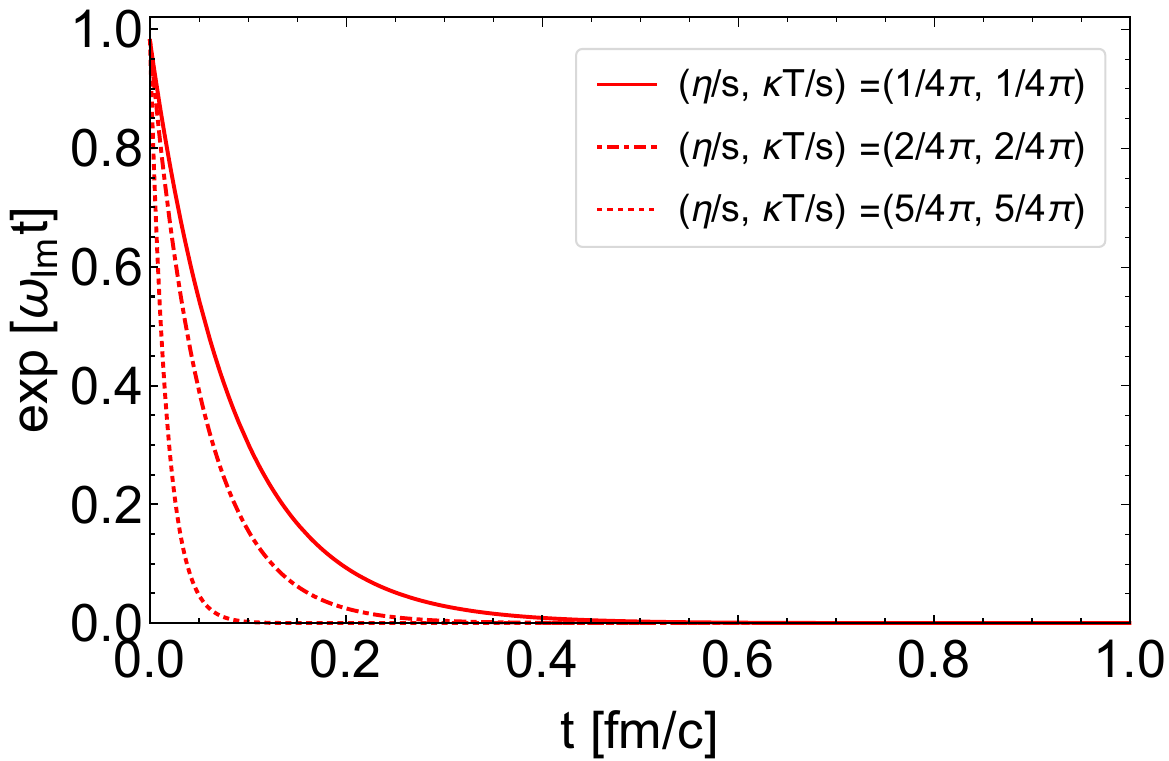} 
\caption{ a) Dissipation of sound modes with time for different k values  at $(\mu_c,T_c)=$(367 MeV,154 MeV)
and ($\eta/s=\kappa T/s=1/4\pi).$  b) Damping of the sound waves with different sets of transport 
coefficients for $k= 250 MeV$.}
\label{fig0501}
\end{figure}
Now we discuss the dissipation of the perturbation in the fluid when the system hits the CEP in the QCD phase diagram. The damping caused by the imaginary part of the frequency of hydrodynamic modes of perturbation at the critical point $(T_c,\,\mu_{c})=(154,\,367)\,MeV$ is depicted in Fig.\ref{fig0501}. It is seen that the waves with larger (smaller) values of wavenumber ($k$) damp faster (slower). The waves in fluid damp highly for larger values of transport coefficients (right panel). The results displayed in Fig.\ref{fig0502} (left panel) show that when the system is away from the critical point then the waves damp slowly for fluid with higher temperatures and densities. The waves in a medium with higher viscosity and thermal conductivity damp faster (Fig.~\ref{fig0502}, right panel) for obvious reasons.
\begin{figure}[h]
	\centering
		\includegraphics[width=7.cm]{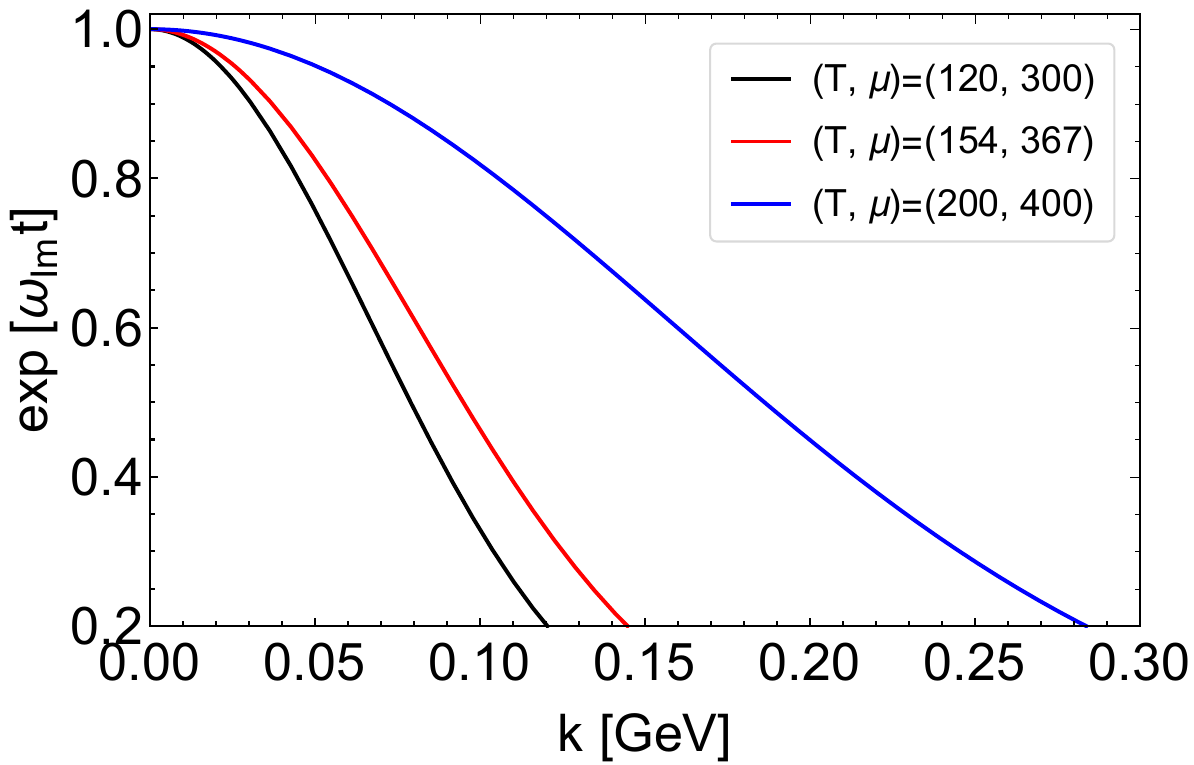}
		\includegraphics[width=7.cm]{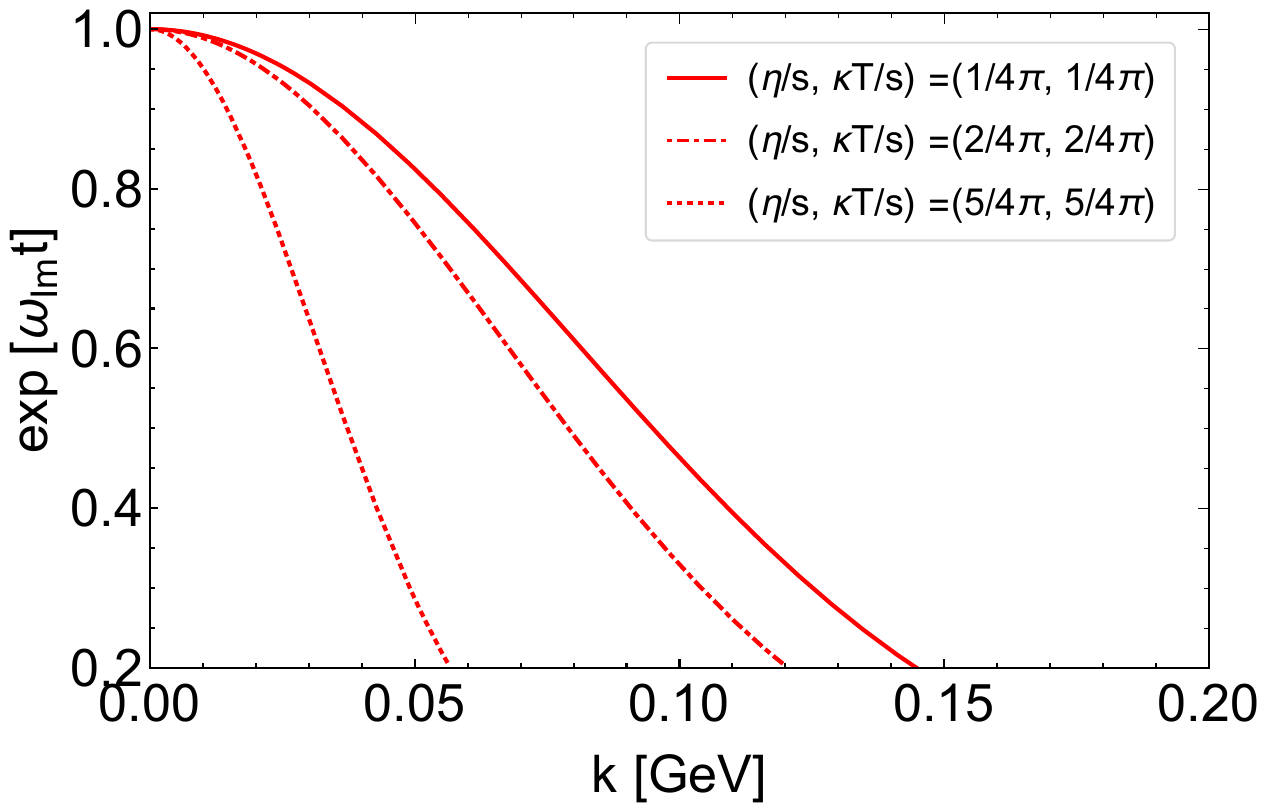} 
		\caption{ a) Dissipation of sound waves with k for different sets of values of $(\mu,T)$ at $t=0.6$ fm/c and $\eta/s=\kappa T/s=1/4\pi$. b) Damping of the sound waves with different sets of value of transport coefficients with $(\mu_c,T_c)=$ (367 MeV,154 MeV).}
\label{fig0502}
\end{figure}
\begin{figure}[h]
	\centering
		\includegraphics[width=7cm]{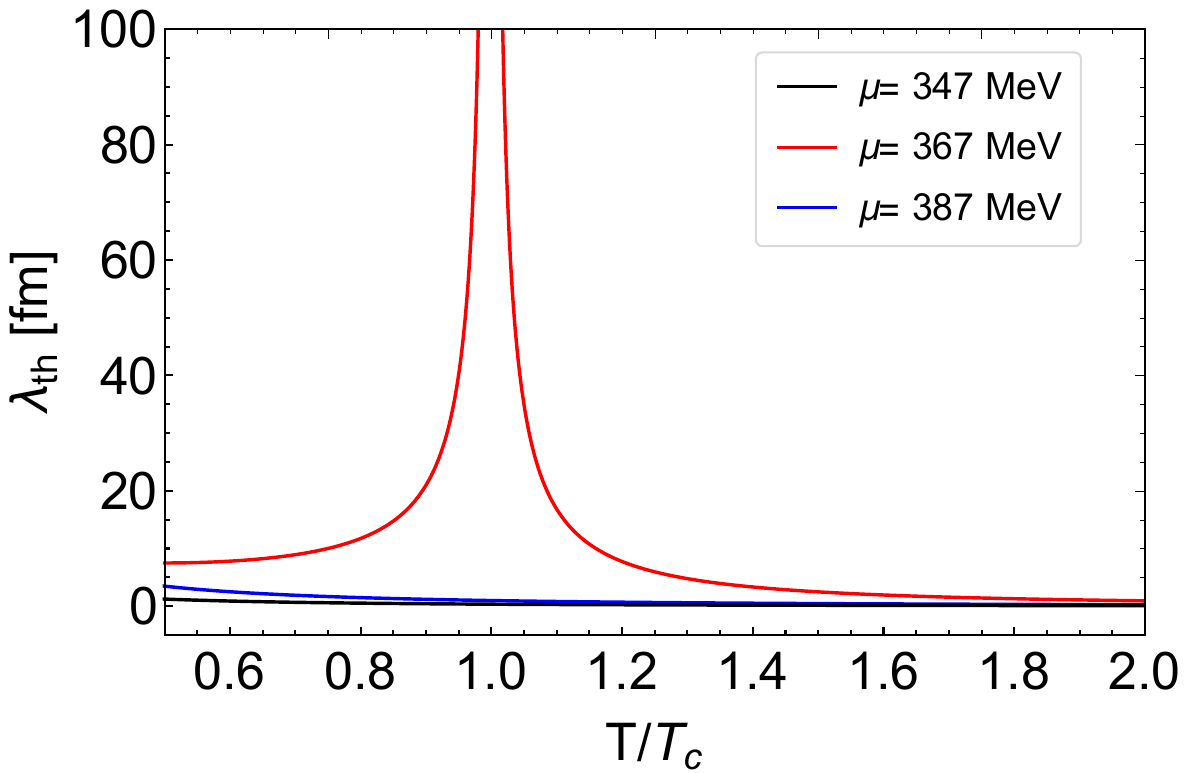}
			\includegraphics[width=7cm]{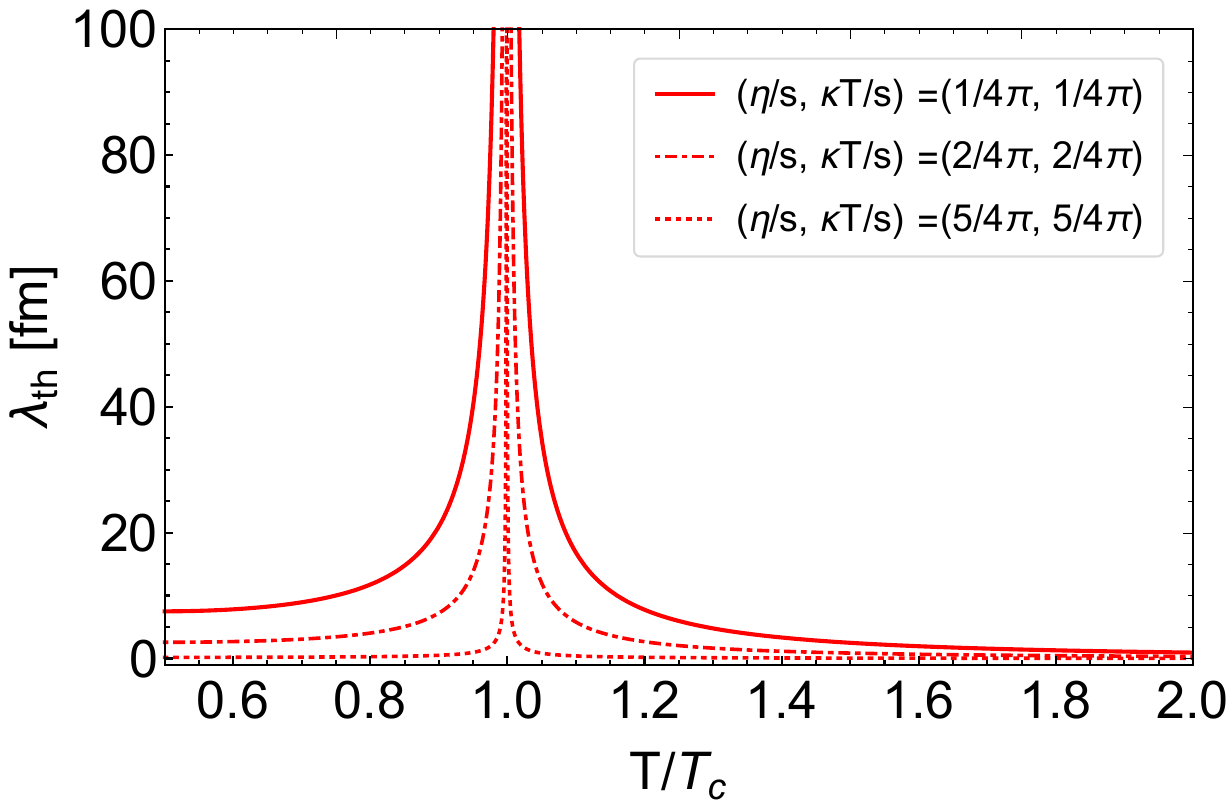}
\caption{ a) Left panel: The variation of  $\lambda_{th}$ (fm) with $T/T_c$ for different values
of $\mu$. The red line shows the diverging nature of $\lambda_{th}$ at $T/T_c=1$ for $\mu_{c}=367$ MeV. 
b)Right panel displays  $\lambda_{th}$ (fm) as a function of  $T/T_c$ at $\mu_c=367$ MeV for different set of values of the 
transport coefficients.}
\label{fig0503}
\end{figure}

The variation of $\lambda_{th}$ with temperature ($T/T_c$) is shown in Fig.\ref{fig0503}. It is evident that $\lambda_{th}(=2\pi/{k_{th}})$ depends on the transport coefficients ($\eta, \kappa$) of the system, as well as on the various response functions appearing through the thermodynamic derivatives, 
$\Big(\frac{\partial \epsilon}{\partial T}\Big)_n,\Big (\frac{\partial \epsilon}{\partial n}\Big)_T,$ $\Big(\frac{\partial P}{\partial T}\Big)_n, \Big(\frac{\partial P}{\partial n}\Big)_T, \Big(\frac{\partial n}{\partial \mu}\Big)_T, \Big(\frac{\partial n}{\partial T}\Big)_\mu$, coupling constant ($\alpha_1$), and the relaxation coefficients ($\beta_{1}, \beta_2$). To observe the behaviour of $k_{th}$ through the EoS only, we have not considered the scaling behaviour (near the CEP) of the transport coefficients but we have taken as $\eta/s=\kappa T/s=1/4\pi$. The values of $\Big(\frac{\partial \epsilon}{\partial T}\Big)_n, \Big (\frac{\partial \epsilon}{\partial n}\Big)_T, \Big(\frac{\partial P}{\partial T}\Big)_n, \Big(\frac{\partial P}{\partial n}\Big)_T, \Big(\frac{\partial n}{\partial \mu}\Big)_T, \Big(\frac{\partial n}{\partial T}\Big)_\mu$ are calculated in terms of different response function by using relevant thermodynamic relations (Appendix~\ref{appendix05_A}).  In the left panel of Fig.\ref{fig0503}, the variation of $\lambda_{th}$ with $T/T_c$ is presented. It is observed that at the CEP ($T_c= 154 \text{MeV}$, $\mu_{c}= 367 \text{MeV}$) the $\lambda_{th}$ diverges. As mentioned above, the $\lambda_{th}$ is defined as the wavelength such that waves with wavelengths, $\lambda\geq \lambda_{th}$ are only allowed to propagate and others get dissipated.  It is observed that when we consider $\mu= 347$ MeV and 387 MeV (away from the critical point) a finite value of $\lambda_{th}$ is obtained, that is wave with $\lambda\ge\lambda_{th}$ will propagate in the medium without substantial damping in such cases.  At the critical point, however, $\lambda_{th}$ diverges which infer that waves with all wavelength will dissipate in the fluid.  We also observe that for the lower value of $T$ the value of $\lambda_{th}$ is smaller. At lower temperature region, the magnitude of $\lambda_{th}$ for $\mu=387$ MeV is larger compared to $\mu = 347$ MeV.  This indicates that for higher values of temperature and chemical potential $\lambda_{th}$ is larger. The fluidity defined in Eq.\eqref{eq0530} is directly proportional to $\lambda_{th}$ which diverges at CEP, implies that fluidity also diverges at the CEP. Away from CEP,  the fluidity decreases. The fluidity is larger for $\mu= 387$ MeV compared to $\mu= 347$ MeV.

The viscous damping of a perturbation can be understood
from the relation: 
\beqa
T^{\mn}_{1}(t)= T^{\mu\nu}_{1}(0) exp({-\omega_{Im}t})~,
\label{eq0531}
\eeqa
 where, $T^{\mn}_{1}(0)$ is the perturbation in EMT at $t=0$ and $T^{\mu\nu}_{1}(t)$ is at some later time $t$ which gets dissipated as indicated by the exponential term. The spectrum of the initial $(t=0)$ perturbations can be associated with the harmonics of the shape deformations and density fluctuations \cite{Lacey:2011ug}. The dispersion relation for $\omega$ provides the value, $k_{th}$, which can
be used to define a length scale, $R_v\sim 1/k_{th}$. For system of size $R$, $R_v$ can be used to define $n_{v}= \frac{2\pi R}{R_{v}}$ which is linked to the value of the highest harmonic $n_v$ (eccentricity-driven) that will effectively survive damping. 
We have seen that the nature of the plot, $\lambda_{th}$ vs $T$ 
does not change considerably with the change in shear viscosity ($\eta/s$) but changes significantly with the variation of thermal conductivity ($\kappa T/s$). The right panel of Fig.\ref{fig0503} displays the variation of $\lambda_{th}$ with temperature for higher values of $\kappa T/s$ and $\eta/s$. As the magnitude of $\kappa T/s$ increases the gap between the divergences in the two phases gets narrower. It is well-known~\cite{KapustaChi} that the thermal conductivity diverges at the critical point. Therefore, the nature of the variation
of $\lambda_{th}$ with $T/T_c$ near the CEP will be essentially governed by the thermal conductivity. We have found that with increasing thermal conductivity the width of the divergence gets narrower.

\chapter[Dynamic structure factor near the QCD critical point]{Dynamic structure factor near the QCD critical point}
\label{chapter6}  
\section{Introduction}
\label{sec0601}
This chapter is based on the publication as shown in Ref.\cite{Hasan2}. Here, we calculate the dynamic structure factor, which is the time-dependent correlator of density fluctuations. This is evaluated from the linear analysis of the hydrodynamic equations, and we will see its behaviour near the QCD critical point. The importance of the spectral structure, more specifically the structure factor, denoted by, $\Snn$, is an accessible quantity in laboratory, in condensed matter physics by light scattering, X-ray diffraction, and neutron scattering. The spectrum of light scattered by a fluid is proportional to $\Snn$, where $k=|\vec{k}|$ is to be identified with the so-called scattering wave vector, which is related to the geometry of the scattering experiment and the wavelength of the incident light. The spectrum of scattered light contains separately identifiable Lorentzians, called Rayleigh-line (peaks) named after Lord Rayleigh~\cite{Rayleigh1881} and Brillouin-line (peaks), measured by Fleury and Boon~\cite{FlerryandBoon1969} in 1969. Rayleigh line (R-line), arises from entropy or, equivalently, temperature fluctuations at constant pressure, whereas B-lines arise from thermally excited propagating sound waves (modes) associated with adiabatic pressure fluctuations~\cite{Stanley,Linda}. The Rayleigh line and the Brillouin lines are usually investigated separately, both in theory and in experiments.

It is well-known that the correlation length ($\xi$) diverges at the critical point. The effects of the divergence on the particle correlations, baryon number fluctuations, and correlations of density  fluctuations will have a better chance to get detected provided the fluctuations to survives the evolution of the hadronic phase. It is important to understand the correlations of those fluctuations theoretically by identifying signatures of the CEP in hadronic spectra. The effects of CEP on the evolution of QGP go as input to the hydrodynamic equations through the equation of state (EoS) and via the criticality of the transport coefficients and the thermodynamic response functions of the medium~\cite{KapustaChi,Guida,Rajagopal:1992qz}. 

The dynamical structure factor, $\Snn$ has been estimated earlier in Ref.~\cite{Minami,Minami_thesis} without taking into account the effect of EoS. It is shown in this work that the EoS plays a vital role in determining the behaviour of $\Snn$, especially its strength at the R-peak, which is modified by several orders of magnitude when the effects of the CEP in EoS is incorporated. The EoS has a strong effects on B-peaks too because it determines $c_s$ (speed of sound) and hence the location of the B-peaks. It is also very crucial to understand whether all the hydrodynamic modes ($k$-modes) travel at the same speed or not. The R-peak and the B-peaks will be closer for slower modes even at points away from the CEP. Therefore, the structure of $\Snn$ will shed light on the speed of the perturbation propagating as a sound wave.
\section{Linearized Hydrodynamic equations}
\label{sec0602}
In this chapter, we use the Eckart's frame of reference as discussed in Sec.\ref{sec0303} in Ch.\ref{chapter3}, where it is considered that the heat flux is non-zero but the particle current is zero. Therefore, energy-momentum tensor (EMT) and the the particle current ($N^\mu$) are given by Eqs.\eqref{eq0129} and \eqref{eq0130}. The conservations of EMT and the net baryon number follows the Eq.\eqref{eq0349}.

The hydrodynamic Eqs.~\eqref{eq0349} are partial differential equations, which are non-linear, could not be solved analytically in general. Presently we aim to obtain the $\Snn$ of the dynamical density fluctuations in $\omega-k$ space. The hydrodynamical equations presented above can be linearized to express small perturbations (in magnitude) in the thermodynamical variables (small deviations from the equilibrium values of the variable). These linearized equations can be solved to obtain the dynamical density fluctuations. Let $Q_0$ ($Q$) represent a thermodynamic quantity in (away from) equilibrium. A small perturbation $\delta Q$ can be written as: $Q=Q_{0}+\delta Q$ where $Q$ can be any of the thermodynamic quantities such as $n, \epsilon, u^{\alpha}, q^{\alpha}, s, \Pi, \pi^{\alpha \beta}$ (baryon number density, energy density, fluid four-velocity, heat flow vector, entropy density, bulk pressure, shear stress tensor respectively), etc. The linearized hydrodynamic equations around the equilibrium~\cite{Minami,Romatschke2010,Hasan1,Sayantani2019,Grozdanov} thus become:
\beqa
0&=&\frac{\pd \delta n}{\pd t}+n_{0} \vec{\nabla}.\delta \vec{v}~,  \nn\\
0&=&h_{0}\frac{\pd \delta v}{\pd t}+\nabla(\delta P+ \delta \Pi)+\frac{\pd \delta q}{\pd t}+\vec{\nabla}.\delta\vec{\pi}~, \nn\\
0&=& \delta \Pi +\zeta[\vec{\nabla}.\delta \vec{v}+\beta_{0}\frac{\pd \delta \Pi}{\pd t}-\tilde{\alpha_{0}}\vec{\nabla}.\delta \vec{q}] ~,\nn\\
0&=& \delta \pi^{ij}+\eta[\pd^{i}\delta v^{j}+\pd^{j}\delta v^{i}-\frac{2}{3}g^{ij}\vec{\nabla}.\delta \vec{v}+2\beta_{2}\frac{\pd \delta \pi^{ij}}{\pd t} \nn\\
&&-\tilde{\alpha_{1}}(\pd^{i}\delta q^{j}+\pd^{j}\delta q^{i}-\frac{2}{3}g^{ij}\vec{\nabla}.\delta \vec{q})~,\nn\\
\eeqa
\beqa
0&=& \delta q+\kappa T_{0}[\frac{\nabla \delta T}{T_{0}}+ \frac{\pd \delta v}{\pd t}+\tilde{\beta_{1}}\frac{\pd \delta q}{\pd t}-\alpha_{0}\nabla \delta \Pi-\tilde{\alpha_{1}}\vec{\nabla}.\delta\vec{\pi}] ~,\nn\\
0&=& n_{0}\frac{\pd \delta s}{\pd t}+\frac{1}{T_{0}}\vec{\nabla}.\delta \vec{q}~,
\label{eq0605}
\eeqa
where, $h_{0}=\epsilon_{0}+P_{0}$ is the enthalpy density in equilibrium and $\vec{v}$ is the space dependent velocity of fluid four-velocity. Now we decompose the fluid velocity along the directions parallel and perpendicular to the direction of wavevector, $\vec{k}$ and denote them as $\delta \vec{v_{||}}$ and $\delta\vec{v_{\perp}}$ respectively. Here, we have considered the longitudinal components only by setting $\vec{k}.\delta \vec{v_{\perp}}=0$. The hydrodynamic equations 
can be solved for a given set of initial condition, $n(0),\, v_{||}(0),\, T(0),\, q(0),\, \Pi(0)$ and  $\pi(0)$, by using the Fourier-Laplace transformation as the following:
\beqa
\delta Q(\vec{k}, \omega)= \int^{\infty}_{-\infty} d^{3}{r}\,\, \int^{\infty}_{0}dt e^{-i(\vec{k}.\vec{r}-\omega t)} \delta \tilde{Q}(\vec{r}, t)~.
\label{eq0606}
\eeqa
And,
\beqa
\delta Q(\vec{k},\,0)=\delta Q(\vec{k}, t=0)= \int^{\infty}_{-\infty} d^{3}{r} \,\, e^{-i(\vec{k}.\vec{r})} \delta \tilde{Q}(\vec{r}, t=0)~.
\label{eq0606p}
\eeqa 
The $\delta P$ and $\delta s$ can be expressed in terms of the independent variables $n$ and $T$ as follows by using the thermodynamic relations:
\beqa
\delta P&=&\Big(\frac{\pd P}{\pd n}\Big)_{T}\delta n+ \Big(\frac{\pd P}{\pd T}\Big)_{n}\delta T~,\nn\\
\delta s&=&\Big(\frac{\pd s}{\pd n}\Big)_{T}\delta n+ \Big(\frac{\pd s}{\pd T}\Big)_{n}\delta T~.
\label{eq0607}
\eeqa
We use Eqs. \eqref{eq0606}, \eqref{eq0606p} and \eqref{eq0607} to write down the longitudinal linearized hydrodynamic equation as: 
\beqa
\delta Q(\vec{k}, \omega)=\mathcal{M}^{-1} \delta Q(\vec{k}, 0)~,
\label{eq0608}
\eeqa
where, 
\beqa
\mathcal{M} =
\begin{bmatrix}
i\omega & ikn_{0} & 0 & 0 & 0  & 0   \\
\frac{ik}{h_{0}} \Big(\frac{\pd P}{\pd n}\Big)_{T}& i\omega & \frac{ik}{h_{0}} \Big(\frac{\pd P}{\pd T}\Big)_{n}& \frac{i\omega}{h_{0}} &\frac{ik}{h_{0}} & \frac{ik}{h_{0}}  \\
0 & ik\zeta & 0 & -ik\tilde{\alpha_{0}}\zeta & 1+i\omega \beta_{0} \zeta & 0 \\
0 & -i\frac{4}{3}k\eta & 0 & i\frac{4}{3}\tilde{\alpha_{1}}k \eta & 0 & 1+2i\omega \beta_{2}\eta\\
0 & i\omega \kappa T_{0} & ik\kappa & 1+i\omega \tilde{\beta_{1}} \kappa T_{0} & ik\alpha_{0} \kappa T_{0} & ik \tilde{\alpha_{1}} \kappa T_{0} \\
-i\omega n_{0}\Big(\frac{\pd s}{\pd n}\Big)_{T} & 0 &   i\omega n_{0}\Big(\frac{\pd s}{\pd T}\Big)_{n}&\frac{ik}{T_{0}}  & 0  & 0  \\
\end{bmatrix}
\nn\\
,
\label{eq0609}
\eeqa
and 
\beqa
\delta Q(\vec{k},\omega)=
\begin{bmatrix}
\delta n(\vec{k},\omega) \\
\delta v_{||}(\vec{k},\omega)\\
\delta \Pi(\vec{k},\omega)\\
\delta \pi_{||}(\vec{k},\omega)\\
\delta q_{||}(\vec{k},\omega)\\
\delta T(\vec{k},\omega)
\end{bmatrix}
;~
\delta Q(\vec{k},\,0)=
\begin{bmatrix}
\delta n(\vec{k},0) \\
\delta v_{||}(\vec{k},0)+\frac{1}{h_{0}}\delta q_{||}(\vec{k},0)\\
i\omega \beta_{0}\zeta \delta \Pi(\vec{k},0)\\
-2\beta_{2} \eta \delta \pi_{||}(\vec{k},0)\\
-\kappa T_{0}\delta v_{||}(\vec{k},0)+\kappa T_{0}\tilde{\beta_{1}}\delta q_{||}(\vec{k},0)\\
-n_{0}\Big(\frac{\pd s}{\pd n}\Big)_{T}\delta n(\vec{k},0)+n_{0}\Big(\frac{\pd s}{\pd T}\Big)_{n}\delta T(\vec{k},0)\\
\end{bmatrix}
\label{eq0610}
\eeqa
We are concerned here 
about the density fluctuation which is given by, 
\beqa
\delta n(\vec{k},\omega)&=&\Big[ \mathcal{M}^{-1}_{11}-n_{0}\Big(\frac{\pd s}{\pd n}\Big)_{T}\mathcal{M}^{-1}_{16}\Big ] \delta n(\vec{k},0)
+\Big[\mathcal{M}^{-1}_{12} -\kappa T_{0}\mathcal{M}^{-1}_{15}\Big]\vec{\delta v_{||}}(\vec{k},0) \nn\\
&&+\mathcal{M}^{-1}_{13}\Big[ i\omega \beta_{0} \zeta\Big] \delta \Pi(\vec{k},0) 
 - \mathcal{M}^{-1}_{14}\Big[ 2\beta_{2}\eta\Big]\delta\pi_{||}(\vec{k},0) \nn\\
&&\Big[\frac{1}{h_{0}}\mathcal{M}^{-1}_{12}+\kappa T_{0}\tilde{\beta_{1}}\mathcal{M}^{-1}_{15}\Big] \delta q_{||} (\vec{k},0)
+  \mathcal{M}^{-1}_{16} \Big[ n_{0} \Big(\frac{\pd s}{\pd T}\Big)_{n}\Big]\delta T(\vec{k},0)~.
\label{eq0611}
\eeqa
We define $\mathcal{S^\prime}_{nn}(\vec{k},\omega)$ by the 
following expression: 
\beqa
\mathcal{S^\prime}_{nn}(\vec{k},\omega)=\Big< \delta n(\vec{k},\omega)\delta n(\vec{k},0)\Big>~.
\label{eq0612}
\eeqa
The correlation between two independent thermodynamic variables, say, $Q_i$ and $Q_j$  vanishes {\it{i.e.}}
\beqa
\Big< \delta Q_{i}(\vec{k},\omega)\delta Q_{j}(\vec{k},0)\Big>=0, \,\,\,\, i\neq j~.
\label{eq0613}
\eeqa
Using Eq.\eqref{eq0613} into Eq.\eqref{eq0611}, the required correlator, $\mathcal{S^\prime}_{nn}(\vec{k},\omega)$ is obtained as:
\beqa
\mathcal{S^\prime}_{nn}(\vec{k},\omega)&=&\Big[ \mathcal{M}^{-1}_{11}-n_{0}\Big(\frac{\pd s}{\pd n}\Big)_{T}\mathcal{M}^{-1}_{16}\Big ] \Big< \delta n(\vec{k},0)\delta n(\vec{k},0)\Big>~.
\label{eq0614}
\eeqa
Finally, $\Snn$ is defined as: 
\begin{equation}
\mathcal{S}_{nn}(\vec{k},\omega)=\frac{\mathcal{S^\prime}_{nn}(\vec{k},\omega)}
   {\Big< \delta n(\vec{k},0)\delta n(\vec{k},0)\Big>}~.
\label{eq0615}
\end{equation}
The variation of $\Snn$ with $k$ and $\omega$ for given values of transport coefficients, thermodynamic variables and the response functions are studied below. In the small $k$ limit with $\tilde{\alpha_0}\rightarrow 0,\tilde{\alpha_1}\rightarrow 0,\beta_0\rightarrow 0,\tilde{\beta_1}\rightarrow 0, \beta_2\rightarrow 0$ the results for Navier-Stokes hydrodynamics can be retrieved from the full expression given in the Appendix~\ref{appendix06_A}. We will see below that the variation of $\Snn$ with $\omega$ possesses with three peaks positioned at $\omega =0$ and  $\omega=\pm\omega_B$. The $\omega_B$ is a function of $k$, $c_{s}$ and other thermodynamic variables. The peak at $\omega=0$ is called the Rayleigh-peak and the doublet symmetrically situated at $\pm\omega_{B}$ are called Brillouin-peaks. The quantities $\Big(\frac{\pd P}{\pd n}\Big)_{T} ,\Big(\frac{\pd P}{\pd T}\Big)_{n},
\Big(\frac{\pd s}{\pd n}\Big)_{T}, \Big(\frac{\pd s}{\pd T}\Big)_{n}$ appearing in $\Snn$ can be evaluated in terms of relevant thermodynamic variables 
(see Appendix~\ref{appendix06_A}). 

The width of this R-line is proportional to $\kappa/nC_{P}$~\cite{Stanley}, where $\kappa$ is the thermal conductivity, $C_{P}$ is the isobaric specific heat. Therefore, if the order of divergence for $\kappa$ is less compared to $C_{P}$, then we see a narrow R-line. Two Brillouin lines are positioned at $\omega_{B} \pm c_{s}k$ with respect to the frequency of the incident light source, which may give rise to the speed of sound. The finite width of the B-line can provide information about the finite value of transport coefficients such as shear, and bulk viscosities. Also, the ratio of intensities of R-line and B-line is expressed as $I_{R}/2I_{B}=C_{P}/C_{V}-1=\kappa_{T}/\kappa_{S}-1$ (where, $\kappa_{T}$ and $\kappa_{S}$ are the isothermal and adiabatic compressibilities respectively), called the Landau-Placzek ratio. Therefore, studying the structure factor could be a useful theoretical as well as experimental aspects to investigate the behaviour of the medium near the CEP.


In  condensed matter physics, the  $\Snn$ is measured by light scattering, utilizing the relation between intensity of light with density-density correlation~\cite{Stanley}. However,
any such direct measurement of the corresponding critical opalescence in QCD is still not achieved. The possibility of such measurement of QCD opalescence has been proposed by measuring jet quenching~\cite{Csorgo:2009wc}.
 
\section{Behaviour of $\Snn$  near the CEP}
\label{sec0603}
The  dynamical structure factor, $\Snn$ defined in Eq.~\eqref{eq0615} directly depends on the transport coefficients, such as $\eta, \zeta$ and  $\kappa$ as well as on the four thermodynamic partial derivatives, $\Big(\frac{\pd P}{\pd n}\Big)_{T} ,\Big(\frac{\pd P}{\pd T}\Big)_{n}$, $\Big(\frac{\pd s}{\pd n}\Big)_{T}$ and $\Big(\frac{\pd s}{\pd T}\Big)_{n}$. 
It depends on the coupling coefficients $(\tilde{\alpha_{0}}, \tilde{\alpha_{1}})$ and also on the relaxation coefficients $(\beta_{0}, \tilde{\beta_{1}}, \beta_{2})$ which pervade into $\Snn$ through the M\"{u}ller-Israel-Stewart (MIS) hydrodynamics. In $\Snn$, the effects of the CEP (through the EoS) is carried by the thermodynamic derivatives. It is well known that the behaviour of the transport coefficients and the thermodynamic response functions near the CEP are characterized by the critical exponents. As the CEP is approached, some of those quantities start to diverge. A thermodynamic variable, $f(t)$ near the CEP can be written as~\cite{Linda}:
\beqa
f(t)=Ae^{\lambda}(1+Be^{y}+...)~,
\label{eq0616}
\eeqa
where, $y> 0$, $t=(T-T_{c})/T_{c}$ is the reduced temperature. The critical exponent $\lambda$ can be defined as:
\beqa
\lambda=\lim_{t \to 0} \frac{ln f(t)}{ln (t)}~.
\label{eq0617}
\eeqa
$\lambda$ can be either positive or negative correspondingly $f(t)$ will vanish or diverge at the CEP. 

Now the partial derivatives appearing in the expression for $\Snn$ can be expressed 
as (see Appendix~\ref{appendix06_A}): 
\beqa
\Big(\frac{\pd P}{\pd n}\Big)_{T}&=&\frac{1}{n_{0}\kappa_{T}}, \,\,\,\,\Big(\frac{\pd P}{\pd T}\Big)_{n}= \mu n_{0}c^{2}_{s}\alpha_{p}
\frac{C_{V}}{C_{P}}~, \nn\\
\Big(\frac{\pd s}{\pd n}\Big)_{T}&=&\frac{h_{0}c^{2}_{s}\alpha_{p}}{n_{0}\gamma},\,\,\,\, \Big(\frac{\pd s}{\pd T}\Big)_{n}=\frac{C_{V}}{T}~,
\label{eq0618}
\eeqa
where, $\kappa_{T}=\frac{1}{n_{0}}\Big(\frac{\pd n}{\pd P}\Big)_{T}$, is the isothermal compressibility, $\alpha_{P}=-\frac{1}{n_{0}}\Big(\frac{\pd n}{\pd T}\Big)_{P}$ is the volume  expansivity coefficient. The symbol $C_P$ and $C_V$ are the specific heats at constant pressure and volume respectively. The critical behaviour of various transport coefficients and thermodynamic response functions plays important roles to determine the strength of the signal of the presence of the CEP. We take the following $t$ dependence near CEP for the present purpose~\cite{KapustaChi,Guida,Rajagopal:1992qz}.
\beqa
&&\kappa_{T}=\kappa_{T}^{0}|t|^{-\gamma^\prime}, C_{V}=C_{0}|t|^{-\alpha}, C_{P}=\frac{\kappa_{0}T_{0}}{n_{0}}\Big(\frac{\pd P}{\pd T}\Big)^{2}_{n}|t|^{-\gamma^\prime},\nn\\
&& c^{2}_{s}=\frac{T_{0}}{n_{0}h_{0}C_{0}}\Big(\frac{\pd P}{\pd T}\Big)^{2}_{n}|t|^{\alpha}, 
\alpha_{P}=\kappa_{0}\Big(\frac{\pd P}{\pd T}\Big)_{n}|t|^{-\gamma^\prime}~, \nn\\
&& \eta=\eta_{0}|t|^{1+a_{\kappa}/2-\gamma^\prime}, \zeta= \zeta_{0}|t|^{-\alpha_{\zeta}}, 
\kappa=\kappa_{0}|t|^{-a_{\kappa}}~,
\label{eq0619}
\eeqa 
where, $\alpha, \gamma^\prime, a_{\zeta}, a_{\kappa}$ are the critical exponents. Here $\alpha=0.11$ and $\gamma^\prime= 1.2$. For a liquid-gas critical point, $a_{\zeta}=\nu d-\alpha=1.78$ (here, $d=3, \nu=0.63$), and $a_{\kappa}=0.63$. For consistency the partial derivative, $\left(\frac{\pd P}{\pd T}\right)_n$ has been calculated by using the
the EoS. The critical behaviours of the second-order coupling as well as the relaxation coefficients enter into the calculation through the thermodynamic variables which contain the effects of the CEP (see Ch.\ref{chapter4}).
\section{Results and discussion}
\label{sec0603}
Now we discuss the effects of the CEP through the EoS and subsequently on the $\Snn$. We have evaluated the $\Snn$ by including the effects of the CEP through: (i) the EoS and
(ii) the critical behaviour of various transport coefficients as discussed above. As mentioned in Ch.\ref{chapter4}, we assume that the CEP is located at $(T_{c}, \,\mu_{c})=(154,\, 367)\, MeV$ in the QCD phase diagram.

\begin{figure}[h]
\centering
\includegraphics[width=7.1cm]{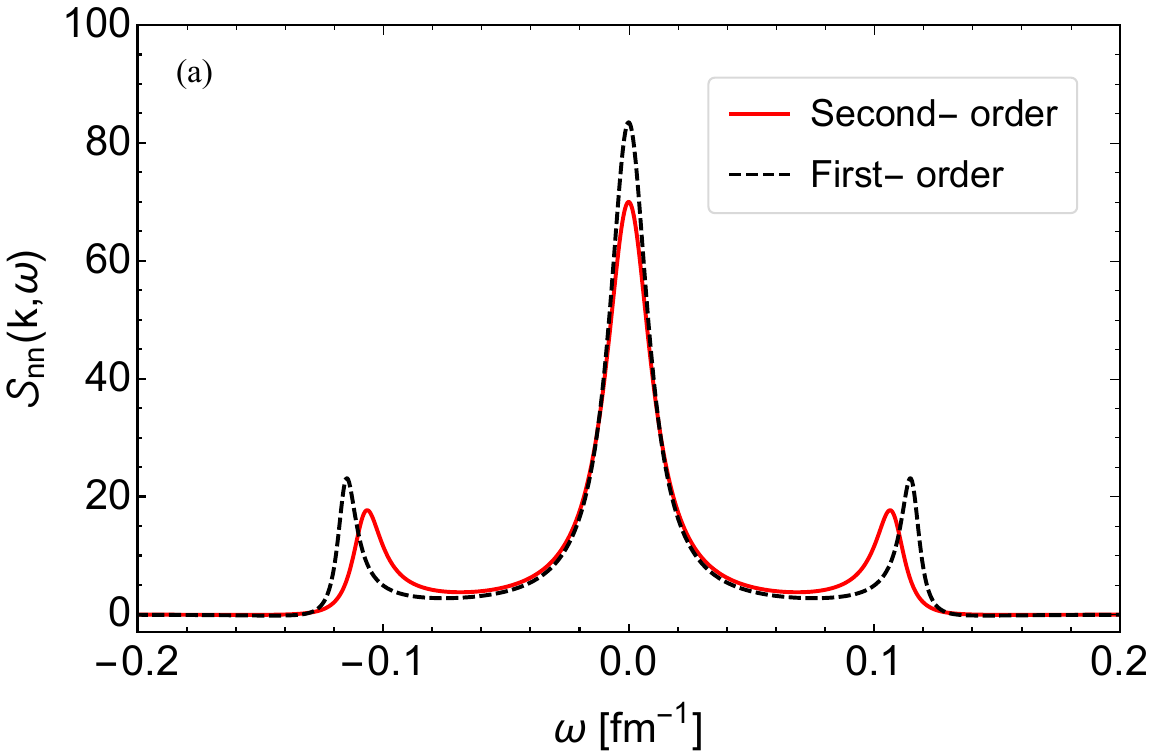}	
\includegraphics[width=7.1cm]{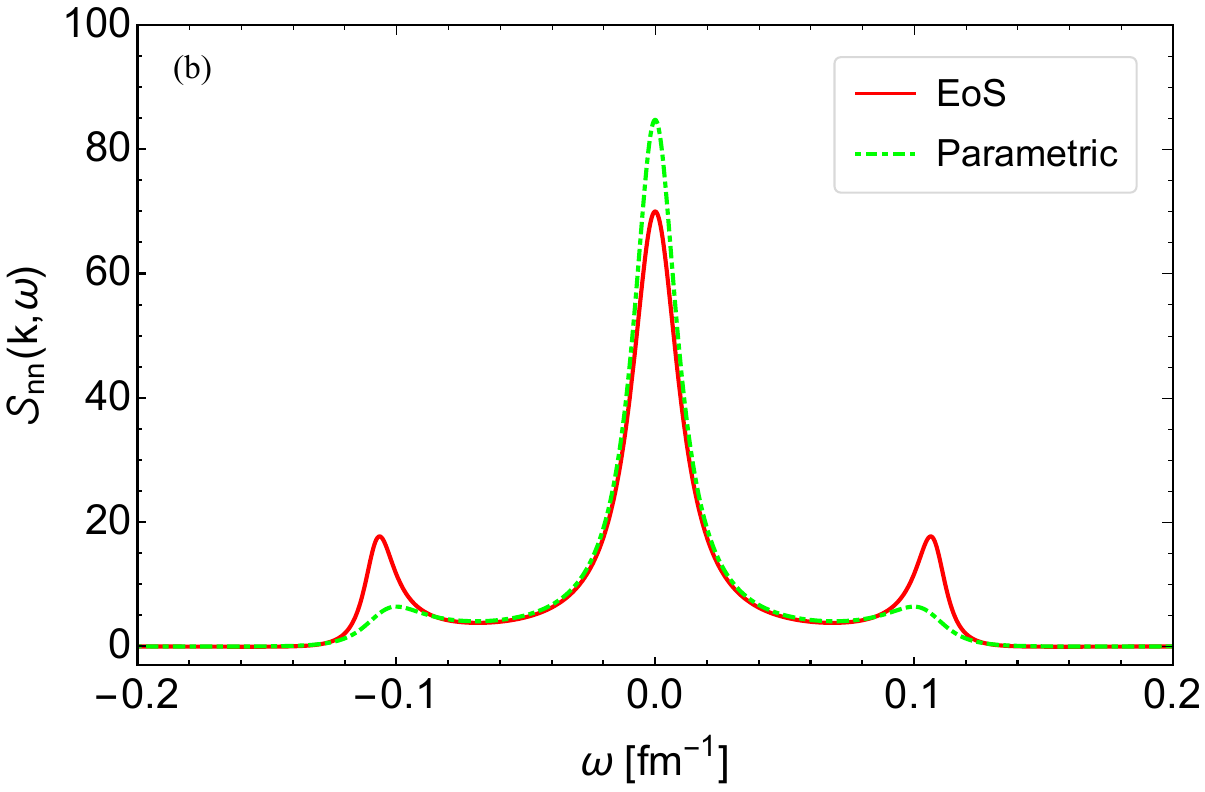}	
\caption{ a) Variation of $\Snn$ with $\omega$ for ($r=0.2$) and $k=0.1$ fm$^{-1}$ are plotted when the system is
away from CEP for first-order (NS)  and second-order (MIS) hydrodynamics.
b) Variation of $\Snn$ with $\omega$ for ($r=0.2$) and $k=0.1$ fm$^{-1}$ in second-order 
hydrodynamics. 
The results obtained with $\eta/s= \zeta/s =\kappa T/s=1/4\pi$ is shown by red line.  
The green line is obtained when parametric form of transport coefficients and response functions
are used. 
}
\label{fig0601}
\end{figure}
Fig.\ref{fig0601}(a) shows a comparison of $\Snn$ for first-order and second-order hydrodynamics. The results show a difference in the $\Snn$ estimated by using the Naiver-Stokes (NS) and the M\"{u}ller-Israel-Stewart (MIS) hydrodynamics. Fig.~\ref{fig0601}(b) depicts the variation of $\Snn$ with $\omega$ when the system is away from CEP, which is represented by $r=0.2$. To examine the effects of the EoS, we have kept the transport coefficients at their lower bounds, $\eta/s, \zeta/s, \kappa T/s=1/4\pi$. The coupling and relaxation coefficients are evaluated from Appendix~\ref{appendix01_A}. We notice that there are three distinct peaks (red line). The central one is larger in magnitude and is called the R-peak which originates from the thermal fluctuations. The symmetric doublet about the R-peak is identified as the B-peaks. The B-peaks originate from pressure fluctuation at constant entropy, which is related to the propagation of the sound waves.  We find that the B-peaks are smaller in magnitude than the R-peaks. We exclude the effects of the EoS to investigate the effects of transport coefficients and the thermodynamic response functions only near the CEP via their critical exponents in evaluating $\Snn$ (depicted by the green line).
\begin{figure}[h]
\centering
\includegraphics[width=8.5cm]{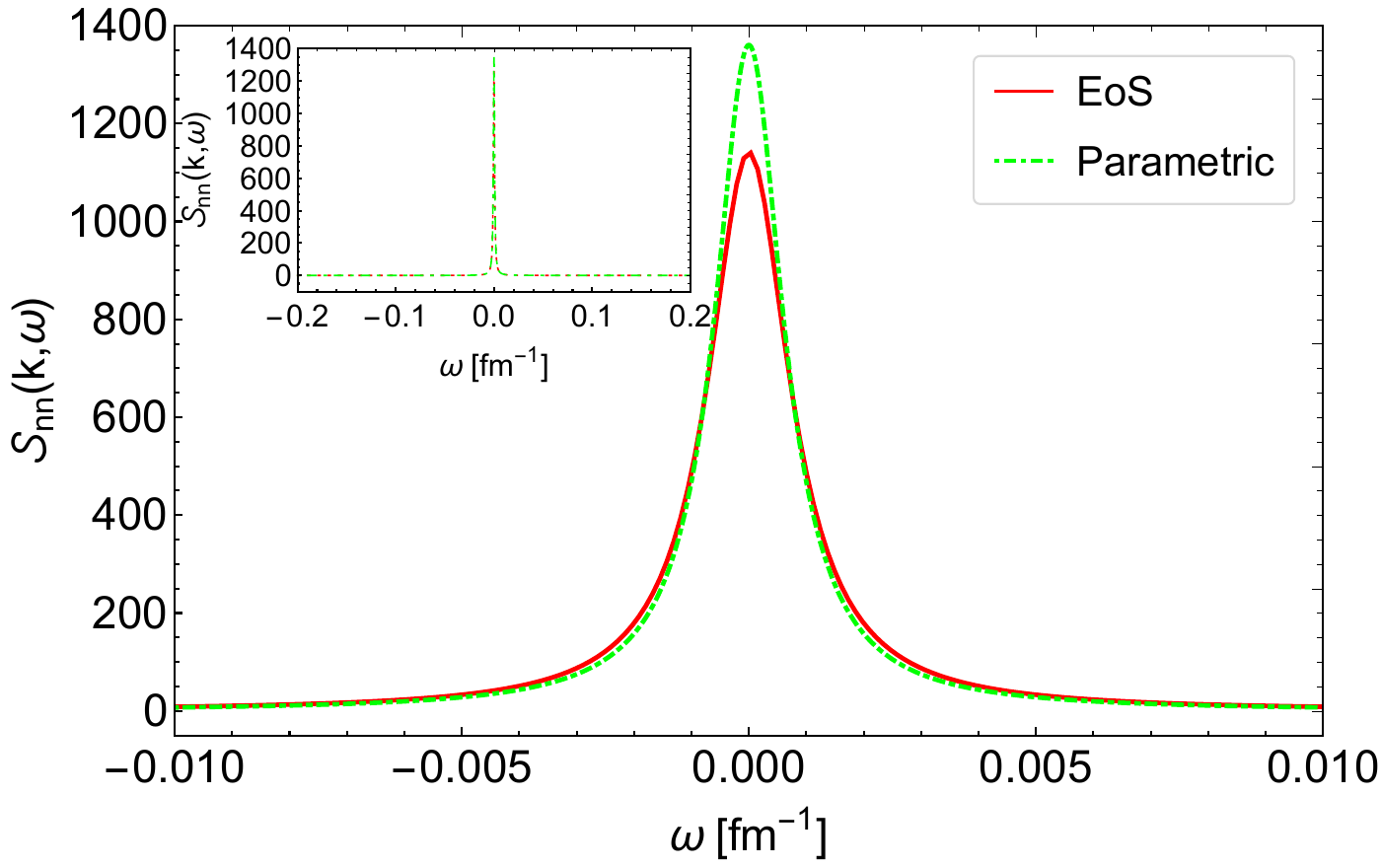}	
\caption{ $\Snn$ near the critical point for ($r=0.01$) and $k=0.1$ fm$^{-1}$. 
The red curve represents the effects of EoS and  $\eta/s= \zeta/s =\kappa T/s=1/4\pi$.
The green line is  obtained
by using the scaling hypothesis of thermodynamic variables near CEP and both effects are shown in blue curve. 
Inset plot is for the broader range in $\omega$ ($-0.2\le\omega\le 0.2$).}
\label{fig0602}
\end{figure}

Closer to the CEP, however, the scenario changes drastically. The $\Snn$ near the CEP ($r=0.01$) reveals only the R-peak with the magnitude enhanced by more than an order of magnitude. The B-peaks do not appear due to the absorption of sound waves near the CEP (Fig.\ref{fig0602}). The vanishing of the B-peaks can be understood from the fact that in the leading order of Brillouin frequencies, it depends as $\omega_B\sim \pm c_s k$. The effects of EoS are seen to reduce the magnitude of the R-peak (red line). The speed of sound is determined by EoS, therefore, the position, as well as the height of the B-peaks will depend on the EoS.
\begin{figure}[h]
\centering
\includegraphics[width=8.5cm]{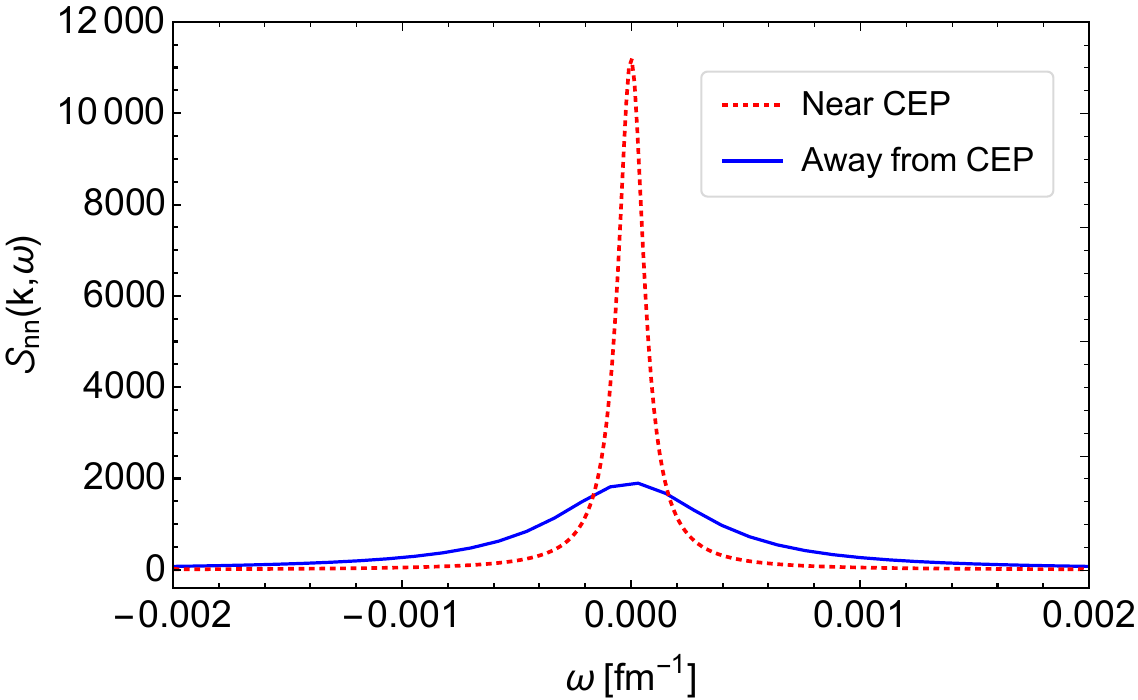}	
\caption{ Variation of $\Snn$ with $\omega$ for $r=0.01$ (red-line)
and 0.2 (blue-line) corresponding to the situation
of the system near and away from  the CEP
respectively. The value of $k$ is $0.02$ fm$^{-1}$ and $\eta/s= \zeta/s =\kappa T/s=1/4\pi$.}
\label{fig0603}
\end{figure}

The $\Snn$ is plotted as a function of $\omega$ for a smaller $k$ ($k=0.02$ fm$^{-1}$) in Fig.\ref{fig0603} when the effects of the EoS only is considered. The red (blue) line corresponds to results closer to the CEP with $r=0.01$ (away from CEP with $r=0.2$). We observe that for smaller $k$ value, the R-peak gets bigger as well as sharper (in frequency axis).  A comparison with results shown in Fig.\ref{fig0604} reveals that near the CEP, the B-peaks vanish and the R-peak gets sharper and bigger. 
\begin{figure}[h]
\centering
\includegraphics[width=8.5cm]{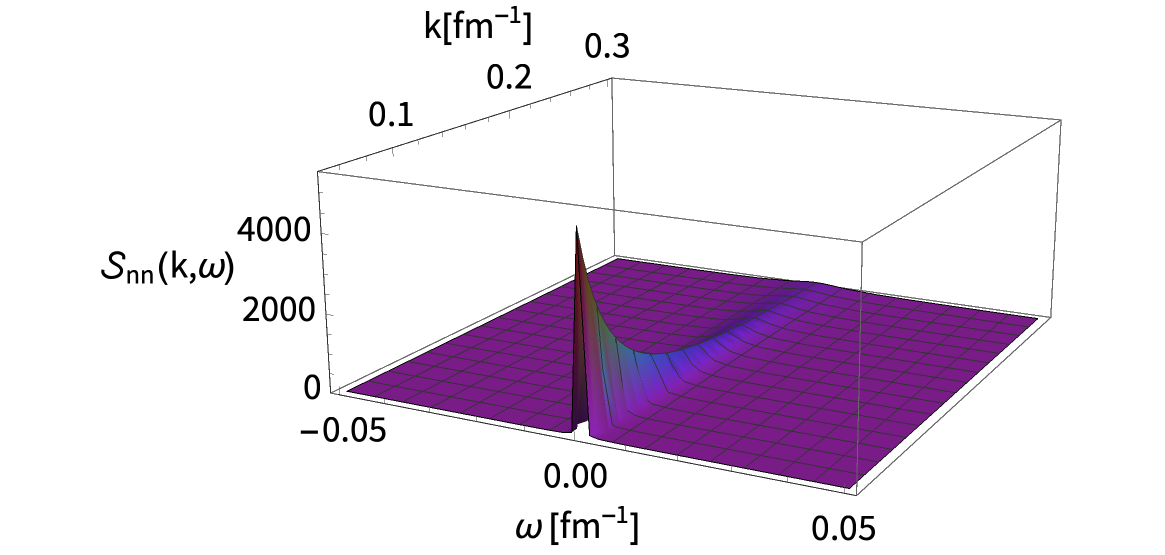}	
\caption{ Variation of $\Snn$ with $\omega$ and $k$ near the CEP ($r=0.01$)
for  $\eta/s= \zeta/s =\kappa T/s=1/4\pi$}
\label{fig0604}
\end{figure}
\begin{figure}[h]
\centering
\includegraphics[width=8.5cm]{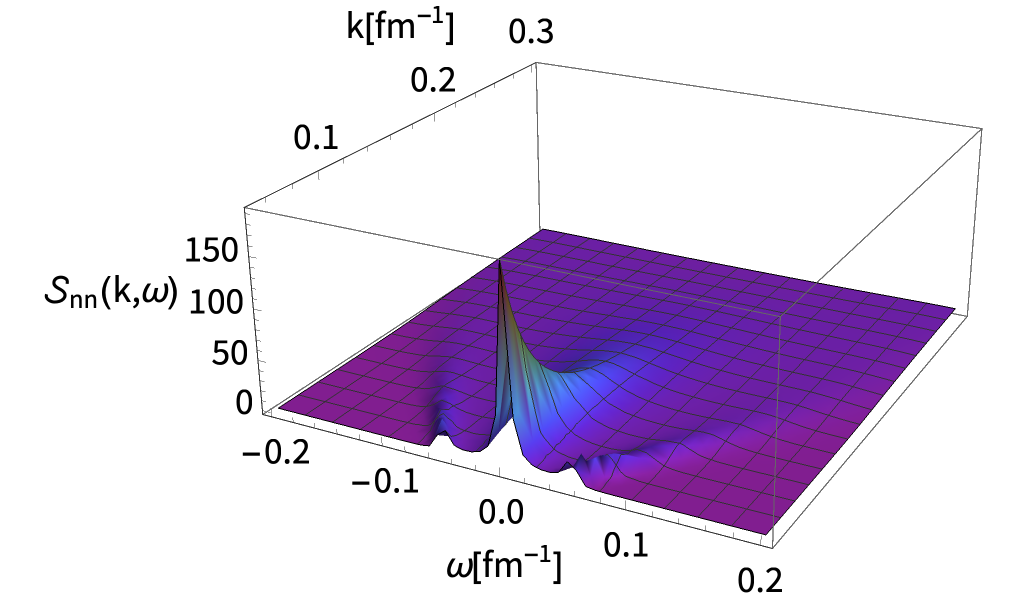}	
\caption{ Same as Fig.~\ref{fig0604} for $r=0.2$ {\it i.e.} the system is
away from CEP.}
\label{fig0605}
\end{figure}

Motivated by the results shown in Figs.~\ref{fig0602} and \ref{fig0603} {\it i.e.} by looking into the sensitivity of the results on the $k$ values we investigate the behaviour of $\Snn$ for different hydrodynamic modes ($k-$modes). The variation of $\Snn$ with $k$ and $\omega$ near the CEP is displayed in Fig.\ref{fig0604}. We see that for all values of $k$, the B-peaks vanish and the R-peak remains. The magnitude of the R-peak is maximum in the neighbourhood of $k\rightarrow 0$. The $\Snn$ is plotted with $\omega$ and $k$ in Fig.\ref{fig0605} when the system is away from the CEP. We observe that the B-peaks shift away from R-peak with the increase in $k$. This is better manifested in Fig.\ref{fig0606}, where the Brillouin frequency $\omega_B$ is plotted against $k$. A linear variation is obtained for given values of $T$ and $\mu$. The slope ($c_s$) of the line is found to be $c_s=0.532$ ({\it i.e.} $c^{2}_{s}=0.283$) close to the speed of sound $c^{2}_s=0.25$ obtained from Eq.~\ref{cs2} at the same value of temperature and chemical potential. 
\begin{figure}[h]
\centering
\includegraphics[width=8.5cm]{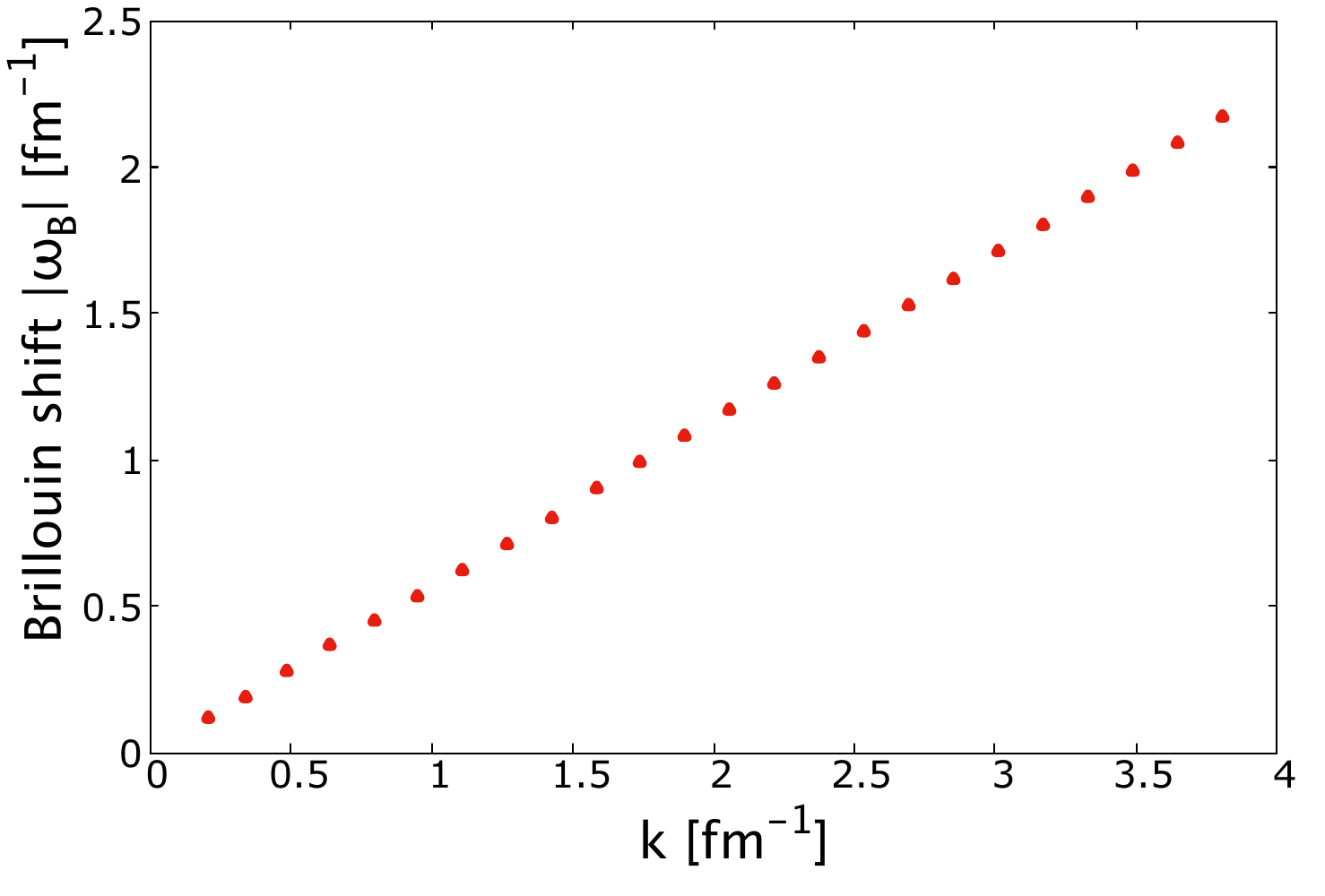}	
\caption{ Variation of the Brillouin frequency, $\omega_B$ with $k$ 
when the system is away from CEP ($r=0.2$) with $\eta/s= \zeta/s =\kappa T/s=1/4\pi$.}
\label{fig0606}
\end{figure}

Therefore, we observe that the B-peaks move toward the R-peak when the system approaches the CEP and ultimately merges with R-peak at the CEP. This indicates that the speed of propagation of all hydrodynamic modes disappears at the CEP. This is quite consistent with the finding that the speed of sound reaches its minimum at the QCD critical point. The location of the B-peaks  depend on $k$ when the system is away from the CEP, indicating that the different k-modes travel with different speeds in the fluid.

In non-relativistic limit, it is evident that the width of the R-peak is determined by $\kappa$, the widths of the $B$-peaks can be calculated in terms of $\kappa$, $\eta$, $\zeta$ and the ratio $C_P/C_V$~\cite{Stanley}. The integrated intensity of the R and B-peaks are determined by the ratio $C_V/C_P$, is connected with the ratio of isothermal to adiabatic 
compressibilities. Therefore, the spectral function contains useful information about the thermodynamic state of the system. The spectral function in QGP is calculated as a function of $\omega$ and $k$ which is not directly measurable through external probe in experiments due to transient nature of the extremely small system. In such a situation, one should depend on the indirect measurement of $\Snn$. Since the width and the integrated intensities of the R and B peaks are related to the various transport coefficients and thermodynamic response functions, thus, the determination of these coefficients by some other experimental measurements will help in constructing the spectral functions.

Some kind of mapping between $k$ and some observables is required to gain insight on $\Snn$. The spectral function can be measured by using the bin by bin density fluctuations. Then $k$'s can be associated to the inverse angular separation of the bins as $\delta\phi\sim\frac{1}{kR}$, where $R$ is the radius of the freeze-out surface (which analogous to the analysis of temperature fluctuation of CMB radiation~\cite{Durrer2008}). Therefore, the analysis of the bin by bin density correlation in $\delta\phi \sim (kR)^{-1}$ in the transverse plane for different beam energies will carry the effects of the CEP.  However, for $k \sim 0.1fm^{-1}$, $\delta\phi$ is large for typical freeze-out radius in relativistic heavy-ion collider experiments (RHIC-E), ($\sim 5$ fm). This implies that the correlation can be observed for large angular separation. That is the pattern corresponding to these $k$ values will show up over the angular size $\delta\phi$.


In the condensed matter physics, the effects of the critical point are usually investigated by measuring the intensity of the light scattering from the system. In contrast to this, such external probes are unavailable for detecting the CEP in QCD matter. However, numerous possibilities for the detection of the CEP in QCD have been discussed in the literature. The phenomenon of critical opalescence is considered as a signal of large density fluctuations in a condensed matter system~\cite{Stanley}. The possibility of detecting such phenomenon of QCD opalescence can in principle be done by measuring the suppression of hadronic spectra in heavy-ion collisions ($R_{AA}$) as demonstrated in Ref.~\cite{Csorgo:2009wc}. The $R_{AA}$ can guide to estimate the opacity factor, $\K$ as: 
\begin{equation}
\K=-\frac{ln(R_{AA})}{R_{HBT}}~,
\end{equation}
where, $R_{HBT}$ is the Hanbury-Brown-Twiss(HBT) radius of the system.

The Fourier coefficients of the azimuthal distributions of hadrons can be used to understand various properties of the matter produced in RHIC-E. The coefficient of $cos3\phi$ (triangular flow, $v_{3}$) sheds light on the quantum fluctuations of the initial state. Similarly, the coefficients of $cos2\phi$ (elliptic flow, $v_{2}$) is concerned with the EoS of the system. Near the CEP, the highest order of the harmonics to survive vary as $\sim 1/\lambda_{th}$, where $\lambda_{th}$ is some wavelength of the pressure perturbation (sound) which blows up at the CEP and hence all the flow harmonics will vanish~\cite{Hasan1} (discussed in Ch.\ref{chapter5}). However, the hadronic spectra contain contribution from all the space-time points 
{\it i.e.} from creation of the fireball to the stage of freeze out with all possible values of $T$ and $\mu$. Therefore, instead of vanishing, the suppression of flow harmonics may be observed.

The physics of the hadronic matter under extreme conditions of temperatures or densities and the QCD phase transition has been dubbed as the condensed matter physics of elementary particles~\cite{Rajagopal:2000wf} where the microscopic interaction is controlled by the non-abelian gauge theory whereas, the condensed matter physics is governed 
by the abelian gauge theory. Identification of some appropriate probes analogous to light in the abelian system will go a long way to unfold the physics of QCD matter near the CEP. It is shown~\cite{Son2004} that the dynamical universality class of the QCD critical point belong to the model H ~\cite{Hohenberg1977} with a combination (linear) of the chiral condensate 
and the baryon density as hydrodynamic mode. The evaluation of baryon number fluctuation by using the correlation length prevailed in ~\cite{Son2004} will be an interesting problem to address, however, this is beyond the scope of the present work. 
\chapter[Understanding the parametric slow mode through the dynamic structure factor]{Understanding the parametric slow mode through the dynamic structure factor}
\label{chapter8}  
\section{Introduction}
\label{sec0801}
This chapter is based on the arXiv preprint~\cite{Sarwar:2022iem}. The background fluid QGP is described by the hydrodynamic equations of second-order. Without considering the effect of critical point, hydrodynamic modelling of such collisions have been developed to a certain degree of sophistication~\cite{Lupeidu2021,Bleicher:2012qve,Aguiar2007,Nonaka}. The hydrodynamics is used to describe the slowly evolving
modes of the macroscopic system while the faster     
non-hydrodynamic modes are driven by the collision dynamics at the 
microscopic level. The time required to achieve local equilibrium
is much shorter than the time required to attain global 
equilibrium, and applicability of hydrodynamics depends on the separation of this time scale. 
The hydrodynamic modelling has been used to study the evolution 
of the fireball created in relativistic nuclear collisions with the 
inclusion of the CEP~\cite{Lupeidu2021,Bleicher:2012qve,Aguiar2007,Nonaka}, 
although strictly speaking it breaks down  near the  CEP
~\cite{Stephanov:2017ghc}

The underlying reason of the failure of validity of hydrodynamics near the CEP~\cite{Stephanov:2017ghc,Minami} happens due to long range correlations and enhanced fluctuations~\cite{AsakawaPhysRevLett.85.2072,Jeon2003eventbyevent,Koch2008hadronic}. In other words, if the system experience the CEP, the correlation 
length ($\xi$) diverges and the relaxation time
which evolves as $\sim \xi^3$, also diverges leading to critical slowing down
~\cite{Berdnikov:1999ph,Stanley,SKMa}. 
Consequently, for the system stays away from local thermal equilibrium, the  first-order or the second-order hydrodynamical model may become inapplicable.  
However, hydrodynamics can still be applied for systems far away
from equilibrium by including the higher order gradients
of hydrodynamic fields. This is a very active field of contemporary research
but beyond the scope of present work (we refer to ~\cite{Romatschke:2017ejr} 
and references therein for details).

It has been shown in Ref.~\cite{Stephanov:2017ghc} that the validity of 
the hydrodynamics can be extended near CEP by introducing a slowly evolving scalar 
non-hydrodynamic field or the slow out-of-equilibrium mode (OEM),  $\phi$, which is
subsequently incorporated in the definition
of the entropy along with other hydrodynamical variables.
The slow OEM are not treated 
separately~\cite{Stephanov:2017ghc,Rajagopal2020}
as they are coupled with hydrodynamic modes.

It is important to point out here that the description of 
far-from-equilibrium conformal systems 
may converge to hydrodynamic evolution~\cite{Heller:2015dha, Romatschke:2017vte, Strickland:2017kux, Denicol:2020eij, Soloviev:2021lhs}.
However, this may not possible for  general  non-conformal systems ~\cite{Chattopadhyay:2021ive, Jaiswal:2021uvv} 
as the gradient expansion is not always convergent~\cite{ Heller:2021oxl}. 
The above mentioned gradient expansion 
of the hydrodynamic field may not be suitable for systems with large fluctuations  near the CEP  where the OEM
relax slowly.
Instead, a separate treatment of these modes will allow to avoid the problem of convergence of gradient 
expansion of hydrodynamic fields.
The presence of slow modes and its coupling with different hydrodynamic fields 
allow the study of the fluctuations of large magnitudes near the CEP. 
In the present work, the role of $\phi$ on $\Snn$ will be studied 
using relativistic causal dissipative hydrodynamics proposed by M\"uller-Israel-Stewart (MIS)~\cite{Israel:1976tn}. 
The dispersion relations for the hydrodynamic
modes are governed by the set of hydrodynamic equations and  
the extensivity of thermodynamics restricts the coupling of 
slow modes with gradients of other hydrodynamic fields.

 In condensed matter system, the fluctuations has been extensively studied, specifically, the  spectral structure of the correlation of static as well as dynamical density fluctuations, also called the structure factor ($\Snn$). It has been shown both theoretically and experimentally that $\Snn$ depends on the transport coefficients of the system~\cite{Stanley, SKMa,Linda}. The thermal fluctuations are controlled by the the transport properties of the medium following the Onsager's hypothesis~\cite{OnsagarPhysRev.37.405}. The properties of the fluctuation are studied by the light scattering and neutron scatterings experiments. The scattered light spectrum contains separately identifiable Lorentzian peaks, called Rayleigh (R) peak~\cite{Rayleigh1881} situated at angular frequency, $\omega=0$, and two Brillouin (B) peaks symmetrically located about the R-peak, was first experimentally detected by Fleury and Boon~\cite{FlerryandBoon1969}. 
The R-line is originates from the temperature fluctuations at constant pressure, whereas the B-lines originate from the pressure fluctuations at constant entropy. The width and the integrated intensities of the R-peak and B-peaks provide information about various thermodynamic susceptibilities and transport coefficients as discussed in Ch.\ref{chapter6}. As some of the transport coefficients and response functions
change drastically, the study of the structure factor is very useful
to understand the behaviour of the fluid near the CEP where 
the B-peaks tend to vanish, giving rise to severe modification in $\Snn$. Since the width of a distribution represents the decay rate of the fluctuation a distribution of narrow width indicate the slower decay of the fluctuation.

 The CEP in QCD belongs to the same universality class~\cite{Halasz:1998qr,BERGES1999215}, $\mathcal{O}(4)$, as that of 3D Ising model and the liquid-gas critical point. The universality hypothesis is understood from the long ranged behaviour of the system and is independent on the microscopic characteristic. Therefore, understanding the slow modes in one system can be a useful guidance for the theoretical modelling relevant to the CEP of the QCD phase diagram. Near the CEP, the dynamic structure factor without any slow mode has been quite adequately studied earlier~\cite{Minami,Hasan2}. The static and dynamic  structure factors have also been 
studied for QCD matter near CEP with stochastic diffusion dynamics of net 
baryon number~\cite{Nahrgang:2018afz,Nahrgang:2020yxm,Pihan:2022xcl}. 
In this chapter, specifically, we will study the role of the slow modes on the Rayleigh and Brillouin peaks of $\Snn$.

By definition of the critical point the quantity,
$\left(\partial P/\partial V\right)_{\text{critical point}}$
approaches zero, leading to the divergence
of the isothermal compressibility ($\kappa_T\sim
\left(\partial V/\partial P\right)_T$)
and hence reulting in large fluctuation in density,
as $\overline{\left(\delta n\right)^2}\propto
\kappa_T$). This large fluctuation may reflect on
experimental observables.
Thermal fluctuations can cause entropy production  leading  to 
fluctuation in multiplicity, however, their ensemble average is 
unaffected~\cite{Nagai2016}. These fluctuations may not influence 
the lower flow harmonics but affect their correlation. The CMS 
collaborations have measured higher flow harmonics which again carry 
a strong signature of thermal fluctuations~\cite{CMS:2013jlh}, indicating 
that the measurement of fluctuations in multiplicity and higher harmonics 
as a function of collision energy at various rapidity bins may provide 
signal of the CEP (see Ref.~\cite{Yin:2018ejt} for a recent review). 
An enhancement of event-by-event fluctuation in net proton due
to the CEP has been observed in Ref.~\cite{Herold:2014zoa}.
The explicit dependence of the correlation of the multiplicity fluctuation 
on the hydrodynamic properties (speed of sound and         
shear viscosity) of the fluid in rapidity space has been shown~\cite{Kapusta:2011gt} 
within the scope of longitudinally expanding
system with boost invariance~\cite{Bjorken}. It will be
interesting to study the same correlation with the inclusion of the CEP
(as both the viscosity and speed of sound get affected by the CEP)
by using full (3+1) dimensional expansion within the ambit of
relativistic second-order viscous hydrodynamics.
The effects of out of equilibrium dynamics and
space-time inhomogeneity on the kurtosis of net baryon and $\sigma$ field
containing the signal of the CEP have been studied in Refs.~\cite{Herold:2016uvv}.

\section{Hydrodynamic framework}
\label{sec0802}
 In this chapter, we use the Eckart's frame of reference as discussed in Sec.\ref{sec0303} in Ch.\ref{chapter3}, where it is considered that the heat flux is non-zero but the particle current is zero. Therefore, energy-momentum tensor (EMT) and the the particle current ($N^\mu$) are given by Eqs.\eqref{eq0129} and \eqref{eq0130}. The conservations of EMT and the net baryon number follows the Eq.\eqref{eq0349}.

 The equation for relaxation including the additional scalar soft mode $\phi$ is introduced as~\cite{Stephanov:2017ghc}:
\begin{equation}
D\phi=-F_{\phi}+A_{\phi}\theta\,,
\end{equation}
where, $\theta=\partial_{\mu}u^{\mu}$ and $D=u^{\mu}\partial_{\mu}$. Forms of $F_{\phi}$ and $A_{\phi}$ are obtained by the second law of thermodynamics: 
\beqa
\partial_{\mu}s^{\mu}\geq 0 \,,
\eeqa
where, $s^{\mu}$ is identified as the entropy four current and is defined as
\beqa
s^{\mu}=s u^{\mu}+\Delta s^{\mu}\,.
\eeqa
In the hydro+ formalism~\cite{Stephanov:2017ghc},  the partially equilibrated entropy gets an extra contribution from the soft mode $\phi$, and the change in entropy density can be written as
\begin{equation}
ds_{+}=\beta_{+}d\epsilon-\alpha_{+} dn-\pi d\phi\,,
\label{eq0805}
\end{equation}

where, $\epsilon$ is the energy density, $n$ is the baryon number density, and $\pi$ is the cost in energy for including $\phi$ modes in the system, called the corresponding chemical potential. Here,  $\beta_{+}=1/T$ and $\alpha_{+}=\mu_{+}/T$, where, $T$ is the temperature 
$\mu_{+}$ is the chemical potential.

In the present work it is found that, imposition of the extensivity condition plays an important role to determine the form of $A_{\phi}$. From here onward, for simplicity we drop the subscript ``$+$" and write the following relations as obtained from the extensivity condition discussed in Appendix~\ref{extc}, in the presence of $\phi$ as:
\beqa
\beta\, dP&=&-(\epsilon+P) d\beta+n d\alpha+\phi d \pi\,,\\
s&=&\beta (\epsilon+P-\mu n)-\pi \phi\,.
\label{eq0811}
\eeqa
The equation for the slow mode is a form of relaxation type, therefore, keeping up to first order of slowly evolving soft mode is adequate to maintain causality. Therefore,
\begin{equation}
\partial_\mu s^{\mu}=(F_{\phi}-b  \partial_{\mu} q^{\mu})\pi -q^{\mu}\partial_{\mu}(\beta+b \pi)-\beta (\partial_{\mu} u_{\nu})\Delta T^{\mu\nu}+\partial_{\mu}(\Delta s^{\mu}+\beta q^{\mu}+b \pi q^{\mu})+S_n \theta \,,
\end{equation}
where, $b$ represents the coupling strength and 
\begin{equation}
S_n=s-\beta(\epsilon+P)+\mu \beta n+\pi A_{\phi}\,.
\label{eq0813}
\end{equation}
The soft mode couples with the heat flux satisfying $\partial_{\mu}s^{\mu}\geq 0$. Now, with $S_n=0$ and redefining $\pi$, $q^{\mu}$, $\Delta T^{\mu\nu}$ and $s^{\mu}$, we have
\beqa
\Delta s^{\mu}=-\beta q^{\mu} -b \pi q^{\mu}\,,
\eeqa 
and
\beqa
q^{\mu}&=&-\kappa T[D u^{\mu}-\frac{1}{\beta}\Delta^{\mu\nu}\partial_{\nu}(\beta+b \pi)]\,,\\
F_{\phi}&=&\gamma\pi -b \partial_{\mu}[ \kappa T D u^{\mu}-\frac{\kappa}{\beta}\Delta^{\mu\nu}\partial_{\nu}(\beta+b \pi)]\,,
\eeqa
where, the proportionality constants $\kappa\ge 0$, $\gamma\ge 0$ are respectively 
thermal conductivity and relaxation rate of slow modes. 
Along with the above expression of fluxes one also needs,  
$S_n=c \, \theta$ with $c \, \ge 0$ being a constant to
satisfy the relation $\partial_{\mu}s^{\mu}\geq 0$.  
However, since, $S_n$ contains local thermodynamic quantities at zeroth 
order in derivative, it can not be a function of the four divergence of 
fluid velocity, $\theta$. In other words, the relation 
$S_n=c\, \theta$ with non-zero $c$  will make the local equilibrium  
entropy density in fluid rest frame 
a function of the dissipative gradient of  the fluid velocity which is frame dependent. 
In that case, the local equilibrium entropy will be frame dependent which is contradictory 
to its definition as a frame independent local thermodynamic quantity, so $c$ should be zero, 
implies  $S_n=0$. Therefore, the dissipative fluxes (in Eqs.~\eqref{eq0142}-\eqref{eq0144}) get modified with the coupling to soft modes in the MIS theory as:

\begin{subequations}
\begin{eqnarray}
\label{eq0812a}
&&\Pi=-\frac{1}{3}\zeta\Big[\partial_{\mu}u^{\mu}+\beta_0 D\Pi-\tilde{\alpha}_0\partial_{\mu}q^{\mu}\Big]\,,\\
\label{eq0812b}
&& q^{\mu}=-\kappa T \Delta^{\mu\nu}\Big[\beta\partial_{\nu} (T+b \pi)+Du_{\nu}+\beta_1 D q_{\nu}-\tilde{\alpha}_0\partial_{\nu}\Pi -\tilde{\alpha}_1\partial_{\lambda} \pi^{\lambda}_{\nu}\Big]\,,\\
\label{eq0812c}
&&\pi^{\mu\nu}=-2\eta\Big[\Delta^{\mu\nu\rho\lambda}\partial_{\rho}u_{\lambda}+\beta_2 D \pi^{\mu\nu}-\tilde{\alpha}_1\Delta^{\mu\nu\rho\lambda}\partial_{\rho}q_{\lambda}\Big]\,,
\end{eqnarray}
\end{subequations}
with proportionality constants $\eta\ge0$, $\zeta\ge0$, where $\eta$ and $\zeta$ are respectively the shear and bulk viscous coefficients. The coefficients $\tilde{\alpha_{0}}, \tilde{\alpha_{1}}, \tilde{\beta_{1}}$ are the relaxation and coupling coefficients, and the expressions for those coefficients are given in Appendix-\ref{appendix01_A}, which are evaluated through the thermodynamic integral from kinetic theory prescription.~\cite{Israel:1979wp}. 

\subsection{Linearized equations and the dynamic structure factor}
To linearize the hydrodynamic equations for small deviations from equilibrium field quantities, we assume $\mathcal{Q}=\mathcal{Q}_{0}+\mathcal{\delta Q}$, 
where $\mathcal{Q}$, $\mathcal{Q}_{0}$ and $\delta\mathcal{Q}$ 
represent general hydrodynamic variables, their average values and fluctuations respectively.  
$\mathcal{Q}_{0}=0$ is its average value for dissipative degrees 
of freedom and $u_0^{\mu}=(-1,\, 0,\, 0,\, 0)$ and $\delta u^{\mu}=(0,\,\delta \vec{u})$. 
The fluxes given in Eq.~\eqref{eq0813} are 
substited in Eq.~\eqref{eq0129} and solved in the linearized
domain in frequency-wave vector space to obtain $\Snn$. 
In the linearized domain, the equations of motion for different
space-time components of the energy-momentum tensor and baryonic 
charge current obtained from Eqs.~\eqref{eq0349} read as:  
\begin{subequations}
\begin{eqnarray}
0&=&-\frac{\partial \delta \epsilon}{\partial t}-(\epsilon_0+P_0)+\vec{\nabla}\cdot \delta \vec{u}-\vec{\nabla} \cdot \delta \vec{ q}\,, \\
 0&=&  -(\epsilon_0+P_0)\frac{\partial}{\partial t} \delta u^{i}-\partial^{i}(\delta P+\delta \Pi)+\frac{\partial}{\partial t} \delta q^{i}-\partial_{j}\Pi^{ij}\,,\\
 0&=&  -\frac{\partial}{\partial t} \delta n-n_0\vec{\nabla}\cdot \delta \vec{u}\,,\\
 0&=&  \delta \Pi+\frac{1}{3}\zeta [\vec{\nabla}\cdot \delta \vec{u}+\beta_0\frac{\partial}{\partial t}\delta \Pi-\tilde{\alpha_0}\vec{\nabla}\cdot \delta \vec{q}]\,,\\
0&=&   \delta q^{i}+\kappa T_0\nabla^{i}\delta T+\kappa T_0 \frac{\partial}{\partial t} \delta u^{i}+\kappa T_0 \beta_1 
   \frac{\partial}{\partial t}\delta q^{i}-\kappa T_0 \tilde{\alpha}_0\nabla^{i}\delta \Pi -\kappa T_0\tilde{\alpha}_1\nabla_{j}\pi^{ij}\,,\\
0&=&   \delta \pi^{ij}+2\eta\delta^{ijlm}(\partial_{l}\delta u_{m}-\tilde{\alpha}_1\partial_{l}\delta q_{m})+2\eta \beta_2\frac{\partial}{\partial t} \delta \pi^{ij}\,,\\
 0&=& - \frac{\partial}{\partial t}\delta \phi-(\gamma+T_0^2\frac{K_{q\pi}^2}{\kappa}\nabla^2)C_{\phi\pi}\delta \phi -
   \big[\gamma+(T_0^2\frac{K_{q\pi}^2}{\kappa}-\frac{K_{q\pi}}{C_{T \pi}})\nabla^2\big] C_{T \pi}\delta T\nn\\
   &&-(\gamma+T_0^2\frac{K_{q\pi}}{\kappa}\nabla^2)C_{\pi n}\delta n-T_0K_{q\pi}\frac{\partial}{\partial t}(\nabla_{i}\delta u^{i})+\tilde{\phi}\nabla_{i}\delta u^{i},
\end{eqnarray}
\end{subequations}
where $K_{q\pi}=b \kappa$ and
\beqa
C_{ A\pi}=\frac{\partial \pi}{\partial A}, \,\,\,\, \text{with}\,\,\, A\equiv(T,n,\phi)\,.
\eeqa
We get a set of linear equations  
in $\omega-k$ space (Appendix~\ref{appendixA})
by taking the Fourier-Laplace transformation
of the above equations. 
The set of equations given in Eqs. \eqref{eq23}-\eqref{eq33} 
can be written in matrix form as:
\beqa
\mathcal{M}\delta\mathcal{Q}=\mathcal{A}\,,
\eeqa
where $\mathcal{M}$, is an $11 \times 11$ matrix representing the
coefficients of the column vector comprinsing the
quantities, $\delta n,\delta T,\delta u_{||}, \delta u_{\perp},\delta \Pi, 
\delta q_{||}$, $\delta q_{\perp},\delta \pi_{||\,||},\delta \pi_{||\,\perp},
\delta \pi_{\perp\,\perp}$, and $\delta \phi$.
The solution of the set of linear equations can be written  as:
\begin{equation}
\delta\mathcal{Q}=\mathcal{M}^{-1}\mathcal{A}\,,
\label{delQ}
\end{equation}
In the present work we are interested in evaluating the two point 
correlation of density fluctuation. 
The solution of the set of equations represented by Eq.~\eqref{delQ} leads to the 
following expression for density fluctuation ($\delta n$),
\beqa
\delta n (\vec{k},\omega)&=& \big[\epsilon_n \mathcal{M}^{-1}_{12}+\mathcal{M}^{-1}_{11}\big]\delta {n}(\vec{k},0)-\epsilon_T \big[\mathcal{M}^{-1}_{12}\big]\delta {T}(\vec{k},0)\nn\\
&+&\big[\epsilon_{\phi } \mathcal{M}^{-1}_{12}+\mathcal{M}^{-1}_{13}\big]\delta {\phi }(\vec{k},0)\nonumber\\
&+&\big[\epsilon_0 \mathcal{M}^{-1}_{14}-i k T_0 \mathcal{M}^{-1}_{13} \kappa _{\text{q$\pi $}}+\mathcal{M}^{-1}_{14} P_0-T_0 \chi \mathcal{M}^{-1}_{15}\big] \delta {u_{||}}(\vec{k},0)\nonumber\\
&-&\big[\beta _1 T_0 \chi  \mathcal{M}^{-1}_{15}+\mathcal{M}^{-1}_{14}\big]\delta {q_{||}}(\vec{k},0) +\frac{1}{3} \zeta \mathcal{M}^{-1}_{16}\beta _0 \delta {\pi }(\vec{k},0)\nn\\
&-&2 \eta \mathcal{M}^{-1}_{17}\beta _2 \delta { \pi }_{||||}(\vec{k},0)-2\eta \mathcal{M}^{-1}_{18}\beta _2{\pi }_{\perp \perp}(\vec{k},0)\nn\\
&-&\big[\beta _1 T_0 \chi \mathcal{M}^{-1}_{110}+\mathcal{M}^{-1}_{19}\big]\delta {q}_{\perp}(\vec{k},0)-2\eta \mathcal{M}^{-1}_{111}\beta _2 \delta {\pi }_{||\perp}(\vec{k},0),\nn\\
&+&\big[\epsilon_0 \mathcal{M}^{-1}_{19}+\mathcal{M}^{-1}_{19} P_0-T_0 \chi \mathcal{M}^{-1}_{110}\big]\delta {u}_{\perp}(\vec{k},0)\,,
\label{spfn}
\eeqa
where, $\epsilon_{x}=\frac{\partial \epsilon}{\partial x}$ and $x=n$,$T$, $\phi$. 


Now we define the correlation of density fluctuations, 
as $\mathcal{S^\prime}_{nn}(\vec{k},\omega)$ by the following expression~\cite{Stanley}: 
\beqa
\mathcal{S^\prime}_{nn}(\vec{k},\omega)=\Big< \delta n(\vec{k},\omega)\delta {n}(\vec{k},\,0)\Big>\,.
\label{eq38}
\eeqa
Since the correlation between two independent thermodynamic variables, say, $\mathcal{Q}_i$ and $\mathcal{Q}_j$  vanishes, i.e. we have,
\beqa
\Big< \delta \mathcal{Q}_{i}(\vec{k},\omega)\delta {\mathcal{Q}}_{j}(\vec{k},\,0)\Big>=0, \,\,\,\, i\neq j \,.
\label{eq39}
\eeqa
Imposing Eq.\eqref{eq39} into Eq.\eqref{spfn}, the $\mathcal{S}^{'}_{nn}$ can only contain two elements of the matrix $\mathcal{M}^{-1}$ as: 
\beqa
\mathcal{S}^{'}_{nn}(\vec{k}, \omega)=-\Big[\big(\frac{\pd{\epsilon}}{\pd n}\big) \mathcal{M}^{-1}_{12}-\mathcal{M}^{-1}_{11}\Big] \Big<{ \delta {n}}(\vec{k},\, 0){\delta {n}}(\vec{k},\,0)\Big>\,,
\eeqa
The final expression for the $\Snn$ can be obtained as:
\beqa
\mathcal{S}_{nn}(\vec{k}, \omega)=\frac{\mathcal{S}^{'}_{nn}(\vec{k}, \omega)}{\Big<{ \delta {n}}(\vec{k},\,0){\delta {n}}(\vec{k},\,0)\Big>}=-\Big[\big(\frac{\pd{\epsilon}}{\pd n}\big) \mathcal{M}^{-1}_{12}-\mathcal{M}^{-1}_{11}\Big]\,.
\eeqa
This is the correlation  in density fluctuation 
or the power spectrum with the inclusion 
of the extra degree of freedom, $\phi$.
\section{Results and discussion}
\label{sec0803}
\begin{figure}[h]
\centering
\includegraphics[width=7.cm]{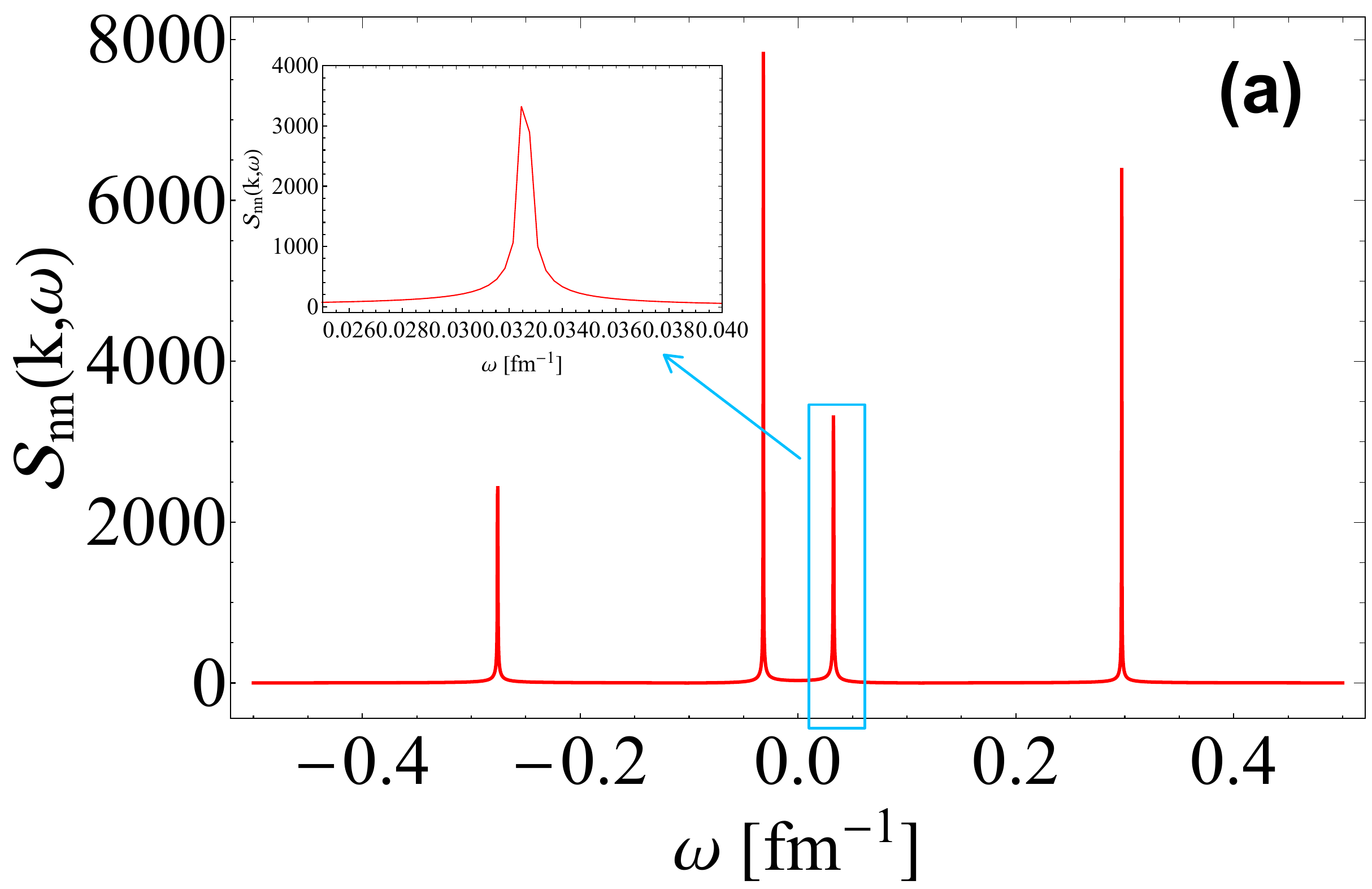}	
\includegraphics[width=7.3cm]{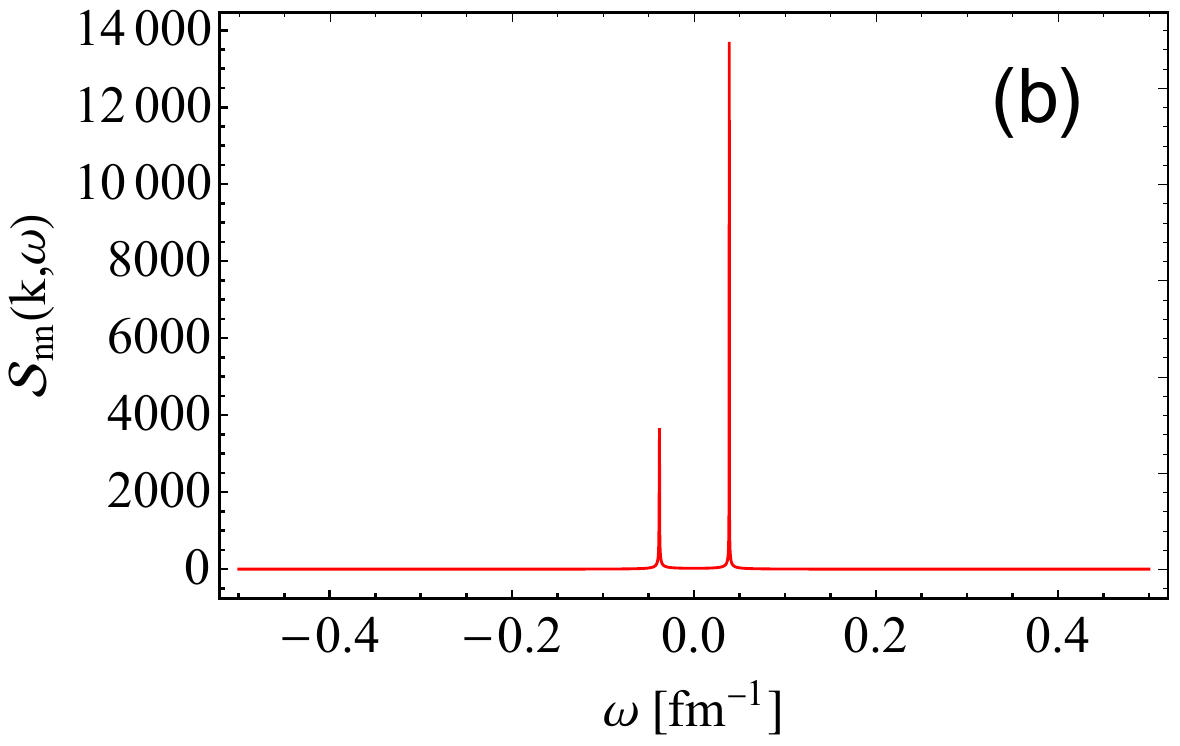}	
\caption{  (a) Variation of $\Snn$ with 
$\omega$ for  $k=0.1\, fm^{-1}$ and $\eta/s= \zeta/s =\kappa T/s=1/4\pi$, 
when the system is away from CEP ($r=0.2$). The peaks are appearing 
as very narrow due to the  scale chosen along x-axis, change in x-axis scale
makes the width visible (see the inset for the R-peak). 
(b) The system is close to the CEP ($r=0.01$). 
The results are obtained with the EoS containing the CEP and the transport coefficients,
thermodynamic response function and the relaxation coefficients 
are estimated from the scaling behaviour.}
\label{fig0801}
\end{figure}

We aim to study the dynamic structure factor near the CEP by introducing a slow scalar field. The effect of the CEP is taken through an EoS, containing the critical point. The various transport coefficients and the thermodynamic response functions near the CEP play important roles determining the structure factor are taken from the Eq.\eqref{eq0619}.
\begin{figure}[h]
\centering
\includegraphics[width=10.1cm]{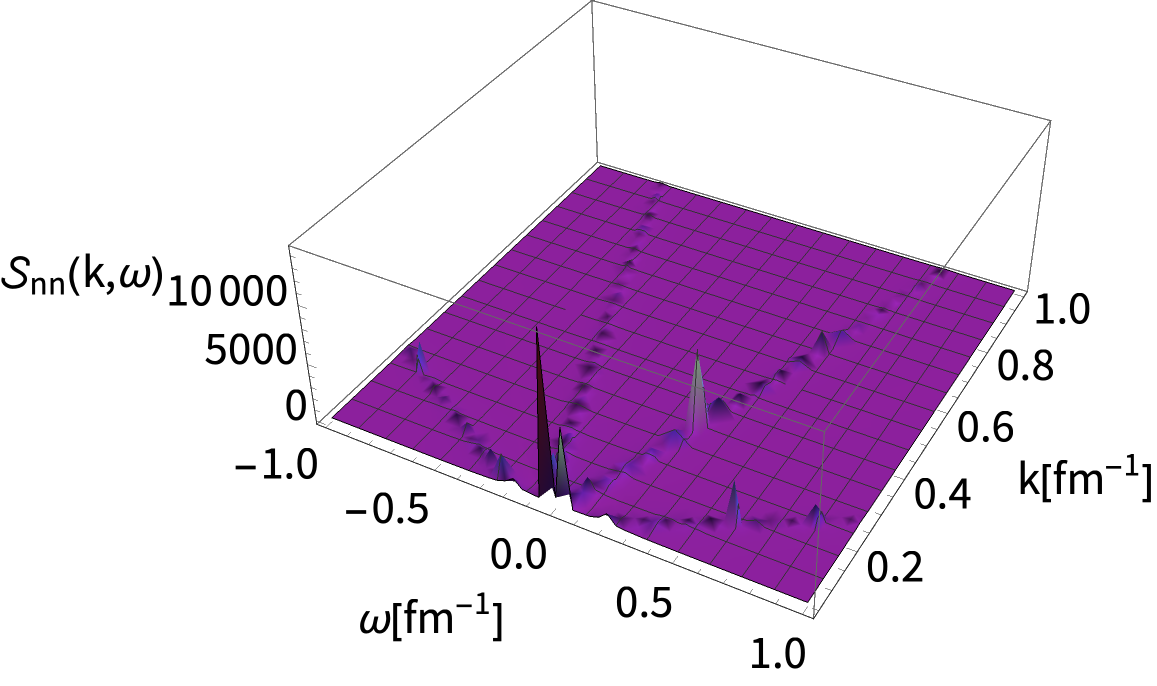}	
\caption{ Variation of $\Snn$ with $\omega$ and $k$ for $r=0.2$, 
i.e. when the system is away from CEP.}
\label{fig0802}
\end{figure}
\begin{figure}[h]
\centering
\includegraphics[width=10.1cm]{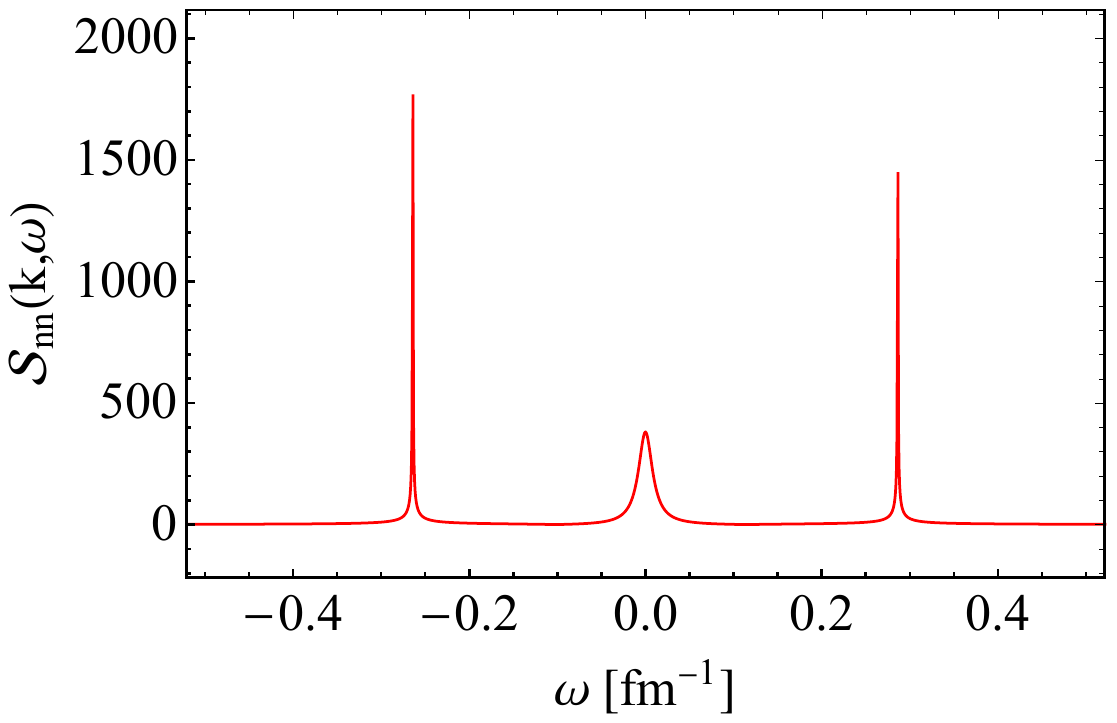}	
\caption{ Variation of $\Snn$ with $\omega$ for $k=0.1\, fm^{-1}$ and  
$r=0.2$, i.e. the system is away from CEP for $K_{q\pi}=0.$}
\label{fig0803}
\end{figure}
\begin{figure}[h]
\centering
\includegraphics[width=10.1cm]{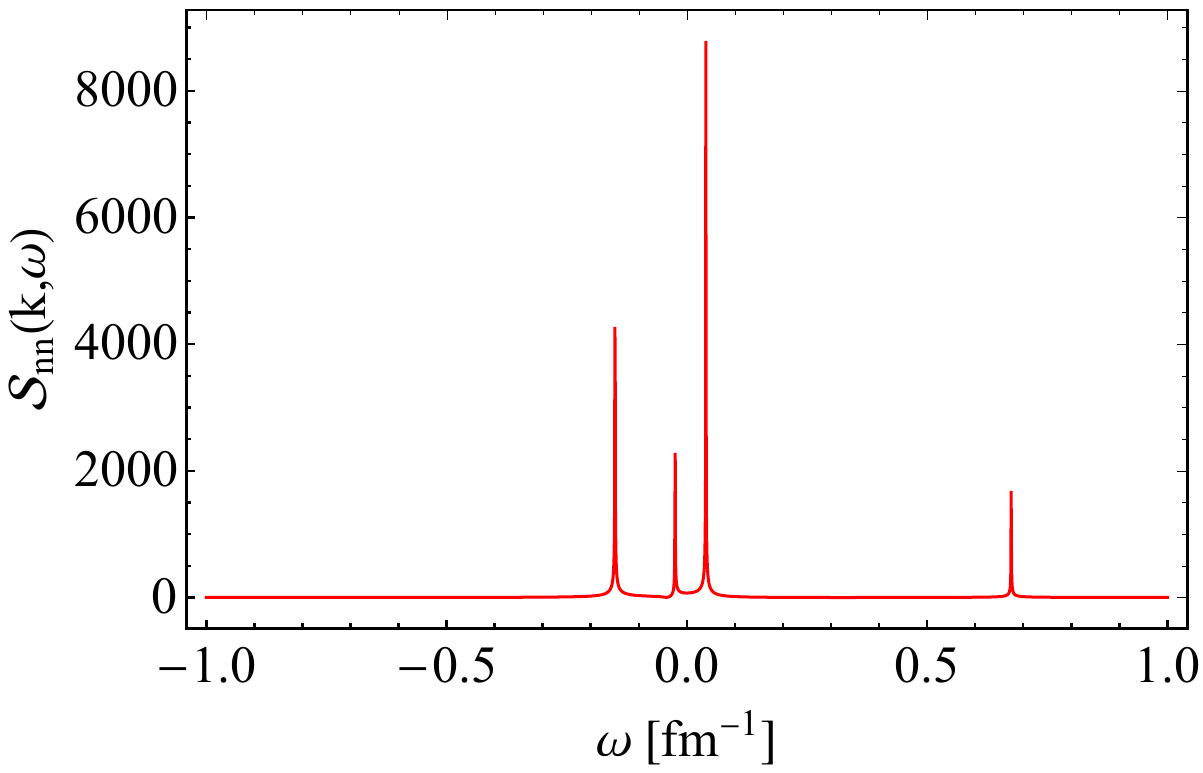}	
\caption{ Variation of $\Snn$ with $\omega$ for $k=0.1\, fm^{-1}$ 
and $r=0.2$, i.e. the system is away from CEP  but with higher value of 
$(\frac{\pd P}{\pd \phi})$. The asymmetry in the B-peaks is clearly visible.}
\label{fig0804}
\end{figure}
 
\begin{figure}[h]
\centering
\includegraphics[width=10.1cm]{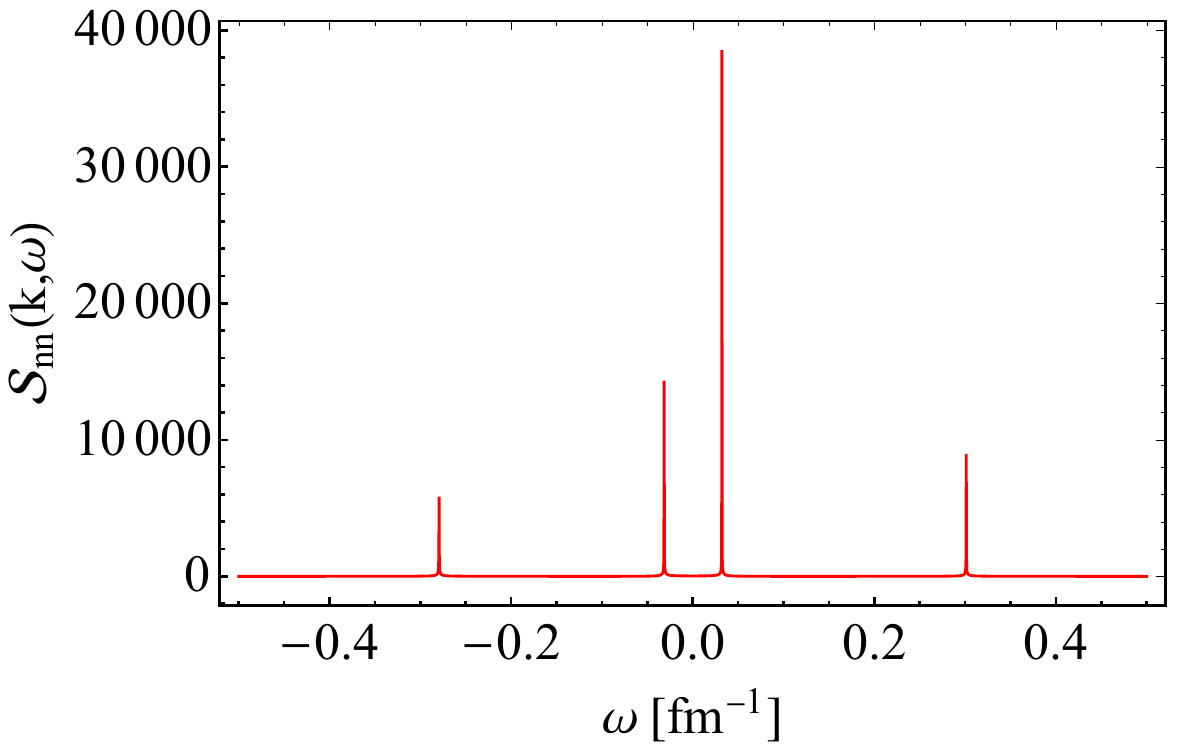}	
\caption{ Variation of $\Snn$ with $\omega$ for $k=0.1\, fm^{-1}$ 
and $r=0.2$, i.e. the system is away from CEP with increased $\phi$ modes 
in the system which increases the thermal fluctuation.}
\label{fig0805}
\end{figure}
The variation of $\Snn$ with $\omega$ is shown in Fig.\ref{fig0801}(a) for $k=0.1$ fm$^{-1}$, when the system is away from the CEP. Without the presence of $\phi$, the $\Snn$ has three peaks (see Fig.\ref{fig0601}(b)). These are identified as the  R-peak located at $\omega=0$ and the B-peaks symmetrically situated on either sides of the R-peak. While with the presence of the slow mode $\phi$, the $\Snn$  admits four peaks. The locations of the four peaks of $\Snn$ for different values of $\omega$ are obtained from the dispersion relation provided in the Appendix~\ref{appendix08_A}. The Brillouin peaks (B-peaks) are identified as the away side peaks of $\Snn$. The Stokes (left side) and the anti-Stokes (right side) components are located asymmetrically (unlike the figure shown in Fig.\ref{fig0601}) on either side of the origin with uneven magnitudes. The other two peaks (closer to $\omega=0$) 
are present even with vanishing speed of sound in the neighbourhood of CEP (as shown in Fig.\ref{fig0801}(b)), arising
due to the coupling of $\phi$ with $q^\mu$. The coupling of $\phi$ with the hydrodynamical fields 
results in one extra peak, and shifted the $R$-peak from $\omega=0$. The peak appears at  $\omega=0$ represents static thermal fluctuations 
(without time variation) on the magnitude of the fluctuations. But their appearance at non-zero $\omega$ 
represents that there is variation in the magnitude of the 
fluctuation with time. So the appearance of these peaks 
near $\omega\ne0$ is due to the 
time varying thermal fluctuations which is resulted from the coupling 
of the heat flux to slow OEM. The OEM have  much slower dynamic evolution rate, 
so that thermal fluctuations become time dependent instead of being 
time independent.
A non-relativistic fluid system 
placed in a stationary temperature gradient
showed  asymmetry 
between Stokes and anti-Stokes components,
the asymmetry  is  maximum when $k||\nabla T$ 
and vanishes with $k\perp\nabla T$. 
From Fig.\ref{fig0801}(a), it is seen that the asymmetry in Rayleigh component is exactly opposite to the Brillouin components. 
In the vicinity of the CEP (Fig.\ref{fig0801}(b)), the $\Snn$ 
shows two peaks closer to $\omega=0$ as the B-peaks disappear due to the absorption
of sound at the CEP.



In Fig.~\ref{fig0802} the $\Snn$ is plotted in $\omega-k$ plane when the system is away from CEP. The $\Snn$ with non-zero $\phi$ is quite similar (as shown in Fig.\ref{fig0604}) in a sense that the B-peaks are moving away from the R-peak for larger $k$-values, and their magnitudes gets smaller and smaller. However, in presence of $\phi$, these peaks may reappear at non-zero $k$ values too due to the coupling coupling of $\phi$ with the hydrodynamic fields. The symmetrically positioned B-peaks of $\Snn$ is reattained when the coupling $K_{q\pi}$ is set to zero (shown in Fig.~\ref{fig0803}). 

\begin{figure}[h]
\centering
\includegraphics[width=10.1cm]{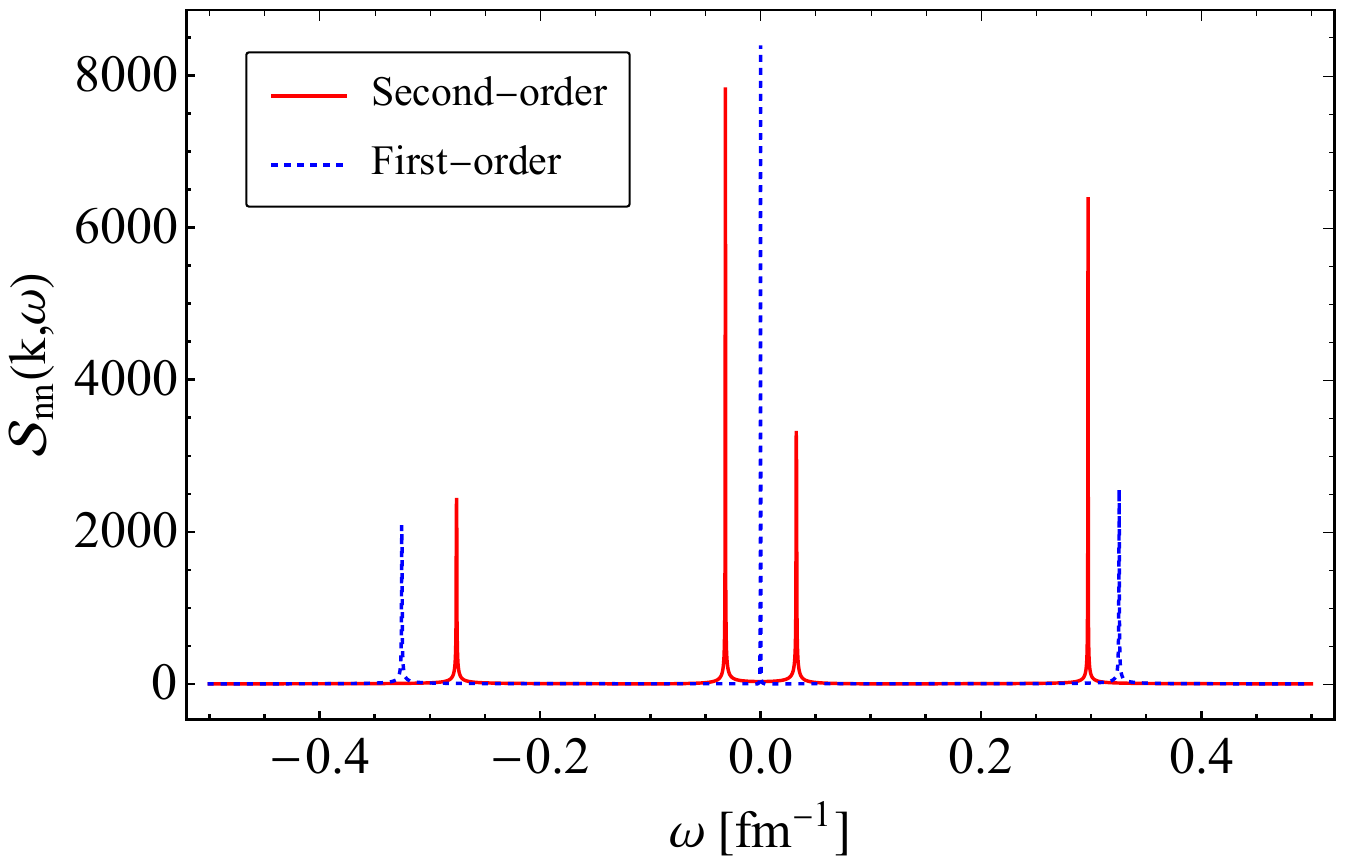}	
\caption{ Variation of $\Snn$ with $\omega$ for
$k=0.1\, fm^{-1}$ and $r=0.2$. The results for second-order
and first-order hydrodynamics are  compared here.}
\label{fig0806}
\end{figure}
\begin{figure}[h]
\centering
\includegraphics[width=10.1cm]{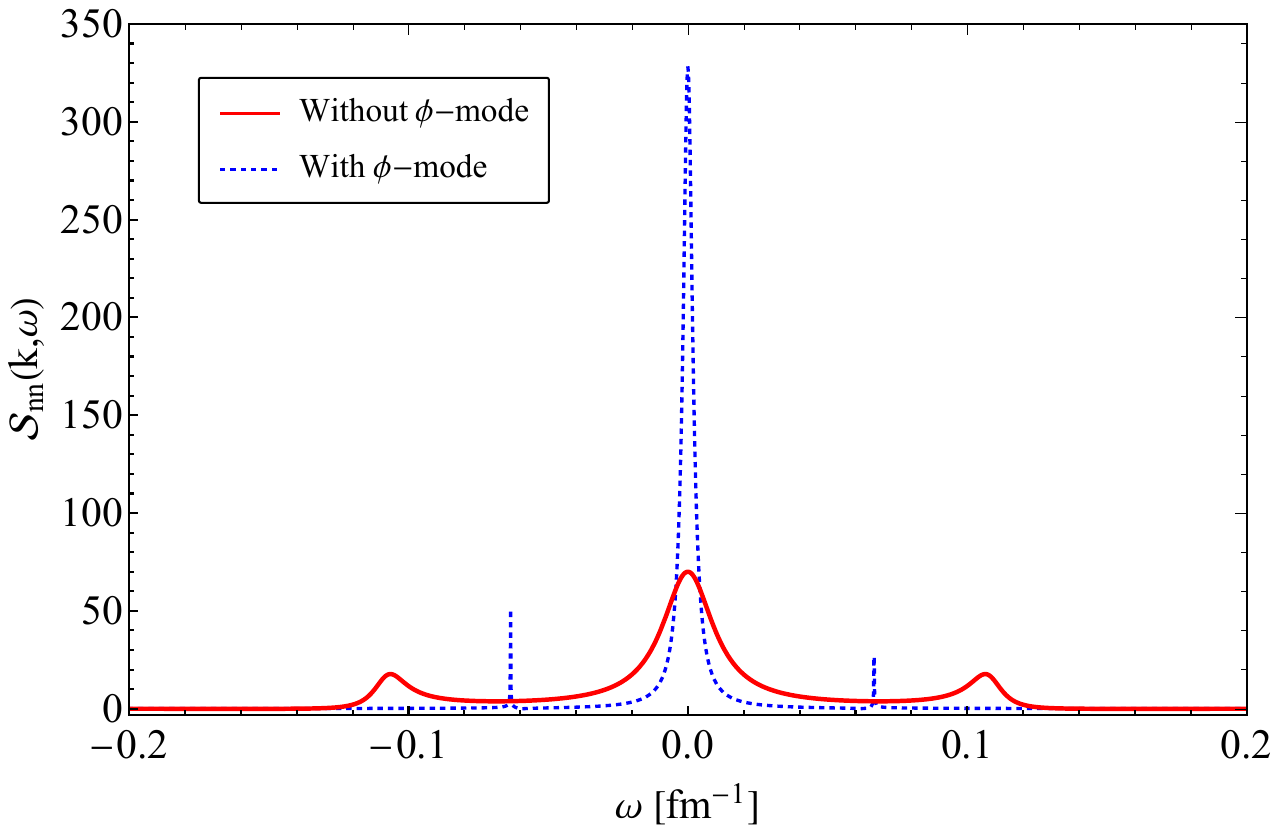}	
\caption{ Variation of $\Snn$ with $\omega$ for
$k=0.1\, fm^{-1}$ and $r=0.2$ with (dotted line) and without (solid)
$\phi$ mode, when the longitudinal modes are only considered.
}
\label{fig0807}
\end{figure}
The asymmetry in the position in $\Snn$ of the B-peaks increases with  increase in $\big(\partial P/\partial\phi\big)$, represents the effect of slow modes 
on the local pressure, which accounts for the effect of slow modes in the speed of sound 
in the system (Appendix~\ref{appendix08_A}). The asymmetry in $\Snn$ with respect to $\omega$ 
increases with  increase in $\big(\partial P/\partial\phi\big)$. 
This is distinctly visible in Fig.~\ref{fig0804} in comparison
to results displayed in Fig.~\ref{fig0801}(a).


Increasing $\phi$ modes in a system induces fluctuations in the system. In Fig.\ref{fig0805}, it is observed that the magnitudes of the peaks get enhanced due to
the enhancement in $\phi$ value. The introduction of slow modes is intended to account for the higher 
order gradients (required for system far-away from equilibrium)
which become relevant at hydrodynamic scales.

In Fig.~\ref{fig0806}, the structure factor for second-order hydrodynamics has been compared with first-order hydrodynamics. Interestingly, the structure factor for first-order hydrodynamics admits a R-peak at the origin and two symmetric B-peaks located on the opposite sides of the R-peak. The effects of $\phi$ seems to be inconsequential in-terms of the arising of extra peaks in the first-order theory because of the vanishing of various coupling and relaxation coefficients in first order approximation. However, the effect of slow modes in first-order theory is there in the asymmetry of B-peak magnitudes.

In Eq.\eqref{eq0812b} shows that the coupling of $\phi$ enters the system through $q^\mu$.
We know that first-order theory can be obtained
from second-order theory by setting the coupling coefficients ($\tilde{\alpha_0}$ and $\tilde{\alpha_1}$)
and relaxation coefficients ($\beta _0,\tilde{\beta_1}, \beta_2$) to zero in Eqs.\eqref{eq0812a}-\eqref{eq0812c}.
Therefore, the only term through which the coupling of $\phi$ enters
the first-order hydrodynamics is given by the first  term of $q^\mu$ in Eq.\eqref{eq0812c} that is,
$q^{\mu}=-\kappa T \Delta^{\mu\nu}\Big[\beta\partial_{\nu} (T+b \pi)\Big]$.
It is interesting to note that the incorporation of $\phi$ has just
shifted the temperature from $T$ to $T+b\pi$ through its chemical
potential $\pi$.  This term  does not introduce any  $\omega$ dependence,
therefore, the presence of $\phi$ in first-order hydrodynamics 
will not introduce any extra peak. 

Till now, it is evident that the extra-peaks are originating from coupling of heat flux with 
the scalar field in second-order theory. However, it is not clear whether, the coupling with 
the longitudinal or transverse modes or both  are responsible for
these peaks in second order theory. To verify, the $\Snn$ derived from the longitudinal (by neglecting the transverse modes) dispersion relation with and without $\phi$ has been depicted in Fig.~\ref{fig0807}. We find that the width of the peaks are narrowed down due to the inclusion of the $\phi$ field (blue dashed line). The decay rate of the thermal fluctuation is reflected through the width of the R-peak, which becomes smaller due to the introduction of slow mode $\phi$. Therefore, the decay of the fluctuation becomes relaxed in the presence of the $\phi$. In comparison of Figs.\ref{fig0801}(a), \ref{fig0803}, \ref{fig0806} and \ref{fig0807}, we can say that the extra peak appeared in Fig.~\ref{fig0801}(a) is due to the coupling of the transverse modes with $\phi$. It is also iimportant to mention that the presence of $\phi$ modes can result in the extra peak in the dynamic structure factor when the transverse modes in the causal theory of hydrodynamics is concerned.
 
 However, it may create confusion whether the absence of extra peaks at  $\phi\neq 0$ is due to the longitudinal modes only or because of first order hydrodynamics.
This can be resolved from the observation that the height asymmetry is different for these two scenarios.

In summary, we can argue that the dynamic structure factor with presence of a slow out-of-equilibrium mode changes significantly. We found four peaks of Lorentzian distribution, whereas the dynamic structure factor without any out-of-equilibrium mode shows three peaks. The peaks in $\Snn$ with $\phi$ are narrower in width compared to the peaks appeared in the $\Snn$ without any $\phi$. This narrower width of the distribution indicates the slow relaxation of the fluctuation in the system, causing critical slowing down. The extra peak appears in the $\Snn$ with $\phi$ is due to the coupling of the $\phi$ mode with the transverse modes when the second-order theory of hydrodynamics is concerned.

\chapter[Propagation of Nonlinear waves near the QCD critical point]{Propagation of Nonlinear waves near the QCD critical point}
\label{chapter7}  
\section{Introduction}
\label{sec0701}
In this chapter, we discuss about our publication, presented in Ref.~\cite{Sarwar:2020oux}. So far in Ch.\ref{chapter5}, Ch.\ref{chapter6} and Ch.\ref{chapter8}, we have discussed about the possible fate of small perturbations (under linear approximation) near the QCD critical point. But in this chapter, we discuss about the perturbations which are comparatively larger in magnitudes. In relativistic heavy-ion collider experiments (RHIC-E), the partons are produced with a wide range of transverse momentum ($p_T$). The partons with relatively smaller $p_T$, on subsequent scattering, contribute to produce a locally thermalized hot medium QGP, whereas the partons with high $p_T$ do not contribute in the medium formation, but they traverse the medium as jets with associated radiated partons, and produce disturbances inside the medium by depositing energy into it through interactions. The supersonic partons, can produce larger perturbations in the medium leading to a nonlinear wave. The quantum fluctuations may also lead to inhomogeneity in the medium, which may serve as perturbations on the hydrodynamically evolving medium. The hydrodynamic response of the medium to such perturbations, will be imprinted into the medium and those imprint are translated into the hadrons spectra at the freeze-out hyper-surface. This imprint will also have effects on other penetrating particles like photons and lepton pairs emitted throughout the evolution history. Specifically, the emergence of two maximas at $\Delta \phi=\pi \pm 1.2$ radian in the quenched away side jet~\cite{STAR:2005gfr,Wang2006}. This double-hump structure in the correlation function of jet is elucidated as the effect of the produced Mach cone in hydrodynamic response to the perturbation created by jets~\cite{Solana:2004fdk}. The flow harmonics (created due to the momentum anisotropy) of produced particle is attributed to the hydrodynamic response of the QGP to the initial geometry. During expansion and cooling of the QGP, on transition to hadronic phase, these anisotropies gets transmitted into hadronic momentum spectra for the conservation of momentum.

The damping of the Mach cone is related to the hydrodynamic response of the medium. Any change in the behaviour of QGP and in the nature of transition from QGP to hadron gas {\textit{e.g.}} presence of the CEP will affect the response. In general, the hydrodynamic response can also be treated as linear as well as nonlinear, depending on the magnitude of the perturbations. A small disturbance is treated as a linear perturbation~\cite{Shuryak2009,Staig:2010pn,Staig2011,Rafiei:2016zxk,Hasan1,Hasan2,Minami} 
whereas a relatively large perturbations (comparable to unperturbed value) should be treated as nonlinear perturbation~\cite{Raha1,Raha2,Raha3,Raha4,Fogaca1,Fogaca:2014gwa}. The momentum anisotropy can be accounted mostly as a linear response to the initial geometry (eccentricity) with a small contribution from nonlinear effects~\cite{Niemi2013,Giacalone2017}. Moreover, the produced Mach front has been found to travel as a shock front~\cite{AKC2006}. The propagation of this shock front is controlled by the hydrodynamic response which may be linear or nonlinear. Due to the large amplitude of this shock front one expects the perturbation to be nonlinear in nature~\cite{Osborne1946}. In Refs.~\cite{Betzthesis,BetzPRL,LiPRL} the presence of Mach cone effect on the away side jet in two particle and three particle correlations have been explained by the non-dissipative property of the nonlinear waves.

The fate of nonlinear waves in QGP has been studied by Foga\c{c}a et al.~\cite{Fogaca1,Fogaca:2014gwa} with the ambit of Naiver-Stokes (NS) as well as M\"uller-Israel-Stewart (MIS) hydrodynamics with the incorporation of the shear viscosity only. They have argued that the nonlinear waves survive, despite the presence of the shear viscosity. The propagation of nonlinear perturbation due to energy deposition by the jets into the medium have been argued to be responsible for broadening of away the side jet~\cite{Fogaca1}. The nonlinear radial spreading of localized waves (solitons) of large magnitude, which are created from the thermalization of the energy deposited by jets considered to be the cause of broadening of the peak.

The hydrodynamic response will rely on the state of the fluid phase and also on the nature of the perturbations (linear or nonlinear). The effects of linear response has already been analysed and found to be suppressed near the CEP~\cite{Minami,Hasan1,Hasan2}. In this regard, however, it will be relevant to investigate the possible fate of the nonlinear perturbations in presence of the CEP. This is particularly crucial because physical process like the energy deposition by jets can be large enough to ignite nonlinear effects. In such cases, the nonlinear wave should play dominant role in the formation of Mach-cone. Therefore, the study of the  effects of nonlinear perturbations in QGP under the influence 
of the CEP is crucially important. 

Near the CEP where fluctuations are expected to grow, the study of the nonlinear waves is important for understanding the broadening of the away side jet and the fate of double-hump structure in the correlation function. This could be a very good guide for the critical point search in QCD phase diagram. As a matter of fact to the best of the knowledge of the author of this dissertation, the fate of the nonlinear disturbances due to the presence of the CEP has not been reported till date. In this work, we address the propagation of nonlinear waves with the presence of the CEP, where we have considered the relevant dissipative coefficients. The inclusion of all those coefficients are crucial because some of these are known to diverge near the CEP~\cite{Kharzeev:2007wb,Karsch2008,Ryu2015,KapustaChi,Martinez2019}. However, the equations which govern the evolution of the nonlinear waves in the presence of all the relevant transport coefficients {\it i.e.} shear viscosity ($\eta$), bulk viscosity ($\zeta$) and thermal conductivity ($\kappa$) are not readily available within the scope of the MIS hydrodynamics, and hence we deduce those equations for the present work. 
\section{1-D flow equations}
\label{sec0702}
In this section, the relevant equations are derived to study the propagation of nonlinear perturbation in a fluid in $(1+1)$ dimension. We have used the MIS theory for the evolution of the QGP and adopted the Eckart's frame of reference (see Ch.\ref{chapter3}, Sec.\ref{sec0303}) to define fluid velocity, where it is considered that the heat flux is non-zero but the particle current is zero. Therefore, energy-momentum tensor (EMT) and the the particle current ($N^\mu$) are given by Eqs.\eqref{eq0129} and \eqref{eq0130}. The conservations of EMT and the net baryon number are written as follows
\beqa
\pd_{\mu}T^{\mu \nu}=0~,\,\,\,\,\,\pd_{\mu}N^{\mu}=0 ~.
\label{eq0701}
\eeqa
The evolution equations of energy density and fluid velocity are obtained from  the Eqs.\eqref{eq0701} by taking projection along the directions parallel and perpendicular to $u^{\mu}$ as:
\beqa
\label{eq0707}
D\epsilon=-(\epsilon+P)\theta+(\partial_{\mu} u_{\nu})\Delta T^{\nu\mu}-(\partial_{\mu} q^{\mu})~,\\
\label{eq0708}
(\epsilon+P)D u^{\alpha}=\Delta^{\alpha\mu}\partial_{\mu}-\Delta^{\alpha}_{\nu}\partial_{\mu} \Delta T^{\nu\mu}~,
\eeqa
where, $u^{\mu}u_{\mu}=-1,$ and
\beqa
\theta&=&\partial_{\mu}u^{\mu}=v\gamma^3\frac{\partial v}{\partial t}+\gamma^3\frac{\partial v}{\partial x}~,\\
\partial_{\mu}q^{\mu}&=&q^{x}\frac{\partial v}{\partial t}+v\frac{\partial q^{x}}{\partial t}+\frac{\partial q^{x}}{\partial x}~.
\eeqa
Here, 
\beqa
u_{\mu}q^{\nu}=0,\,\,\, u_{\mu}\pi^{\mu\nu}=0,\,\,\, g_{\mu\nu}\pi^{\mu\nu}=0,\,\,\, \pi^{xy}=\pi^{zx}=\pi^{yz}=0,\,\,\, \pi^{yy}=\pi^{zz}=\pi^{T}~,
\eeqa
which leads to, 
\beqa
\pi^{00}=v^2\pi^{xx},\,\,\, \pi^{xt}=\pi^{tx}=
v^2\pi^{xx},\,\,\, \pi^{T}=\pi^{tx}=\frac{(
v^2-1)}{2}\pi^{xx}~,
\eeqa
and 
\beqa
q^{t}=vq^{x}, \,\,\,q^{y}=q^{z}=
q^{T}~.
\eeqa
Now from the energy-momentum conservation equations {\it{i.e.} } from Eqs.~\eqref{eq0707} and~\eqref{eq0708}, we get,
\beqa
\label{eq0714}
\gamma v\Big(\frac{\partial \epsilon}{\partial t}+\frac{\partial \epsilon}{\partial x}\Big)=&& \Big[ (h+\Pi)v\gamma^3+v \gamma \pi^{xx}+q^{x}\}\frac{\partial v}{\partial t} \Big]\nn\\
&&+\Big[(h+\Pi)\gamma^3+\gamma \pi^{xx}\Big]
\frac{\partial v}{\partial x}+v\frac{\partial q^{x}}{\partial t}+\frac{\partial q^{x}}{\partial x},
\eeqa
\beqa
\label{eq0715}
&&\Big[ (h+\Pi)v\gamma^4+v \gamma^2(1-2v^2)v \pi^{xx}+\gamma^3(1+v^2)v
q^{x}\Big]\frac{\partial v}{\partial t}\nn\\
&+&\Big[ (h+\Pi)v\gamma^4-\gamma^2 v^2\pi^{xx}+\gamma v(1+2\gamma^2)q^{x}\Big]
\frac{\partial v}{\partial x}\nn\\
&=&-v^2\Big[\gamma^2\frac{\partial P_{\zeta}}{\partial t}+\frac{\gamma^2 }{v}\frac{\partial P_{\zeta}}{\partial x}
+\frac{\partial \pi^{xx}}{\partial t}+ \frac{1}{v}\frac{\partial \pi^{xx}}{\partial x}
+\frac{ \gamma}{v}\frac{\partial q^{x}}{\partial t}- \gamma\frac{\partial q^{x}}{\partial x}\Big]
\eeqa
We define $P+\Pi=P_{\zeta}$. The equation governing the net baryon number conservation reads,
\beqa
\label{eq0716}
\frac{\partial n}{\partial t}+ v\frac{\partial n}{\partial x}
= -n \gamma^2\Big[v\frac{\partial v}{\partial t}+\frac{\partial v}{\partial x}\Big]~,
\eeqa
The other three equations originating from dissipative fluxes are given by the Eqs.\eqref{eq0142}, \eqref{eq0143}, and \eqref{eq0144}, which 
can be written as:
\beqa
\label{eq0717}
\Pi+\zeta \beta_{0}\Big[\gamma \frac{\partial \Pi}{\partial t}+v\frac{\partial \Pi}{\partial x}\Big]&=&-\zeta\Big[(v\gamma^3 -\tilde{\alpha_{0}} q^{x})\frac{\partial v}{\partial t}+\gamma^3
\frac{\partial v}{\partial x}-\tilde{\alpha_{0}}(v\frac{\partial q^{x}}{\partial t}-\frac{\partial q^{x}}{\partial x})\Big]~,
\eeqa
\beqa
\label{eq0718}
 &&q^{x}+\kappa T \tilde{\beta_{1}}\Big[\gamma \frac{\partial q^{x}}{\partial t}+v\frac{\partial q^{x}}{\partial x}\Big]= \kappa T\Big[
(-\gamma^4+ \tilde{\beta_{1}}\gamma^3 vq^{x}+\tilde{\alpha_{1}}(2-\gamma^2) \pi^{xx}
 )\frac{\partial v}{\partial t}\nn\\
 &&+(-v\gamma^4+ \tilde{\beta_{1}}\gamma^3 v^2 q^{x}- \tilde{\alpha_{1}}\gamma^2 v \pi^{xx})
\frac{\partial v}{\partial x}
+\tilde{\alpha_{1}}v \frac{\partial \pi^{xx}}{\partial t}+\tilde{\alpha_{1}} \frac{\partial \pi^{xx}}{\partial x}\nn\\
&&-\tilde{\alpha_{0}}v\gamma^2 \frac{\partial \Pi}{\partial t}-\tilde{\alpha_{0}}\gamma^2 \frac{\partial \Pi}{\partial x}
+\frac{1}{T}v\gamma^2 \frac{\partial T}{\partial t}+\frac{1}{T}\gamma^2 \frac{\partial T}{\partial x}\Big]~,\\ \nn\\
\label{eq0719}
&&\pi^{xx}+\frac{4}{3}\eta \beta_{2}\Big[\gamma \frac{\partial\pi^{xx}}{\partial t}+v\frac{\partial \pi^{xx}}{\partial x}\Big]= \frac{4}{3}\eta\Big[
(\gamma^5+\tilde{\alpha_{1}}\gamma^4 v q^{x}+ 2\beta_{2}\gamma^3 \pi^{xx})\frac{\partial v}{\partial t}\nn\\
 &&+(\gamma^5+ \tilde{\alpha_{1}}\gamma^4 v q^{x}+ 2\beta_{2}\gamma^3v^2 \pi^{xx})
\frac{\partial v}{\partial x}-\tilde{\alpha_{1}}\gamma^2 v\frac{\partial q^{x}}{\partial t}-\tilde{\alpha_{1}}\gamma^2 \frac{\partial q^{x}}{\partial x}\Big]~.
\eeqa

\subsection{Derivation of nonlinear wave equations}
The above hydrodynamic equations from MIS theory is used to derive the equations governing the motion of the nonlinear perturbations in the fluid. We have adopted Reductive Perturbative Method (RPM) technique~\cite{Washimi1966,Davidson1974,Leblond2008,Kraenkel1995}, where we need to define `stretched co-ordinates' as,
\beqa
X=\frac{\sigma^{1/2}}{L}(x-c_s t), \text{and} \hspace{0.4cm} Y=\frac{\sigma^{3/2}}{L}(c_s t)~,
\label{eq0720}
\eeqa
where, $L$ is some characteristic length and $c_{s}$ is the speed of sound. Therefore, from the above equations we get
\beqa
\label{eq0721}
\frac{\partial}{\partial x}=\frac{\sigma^{1/2}}{L}\frac{\partial}{\partial X}, \text{and} \hspace{0.4cm} 
\frac{\partial}{\partial t}=-c_{s}\frac{\sigma^{1/2}}{L}\frac{\partial}{\partial X}+c_{s}\frac{\sigma^{3/2}}{L}\frac{\partial}{\partial Y}~,
\eeqa
where, $\sigma$ is the expansion parameter. The coordinate $X$ is measured from the frame of propagating sound waves, whereas the $Y$ represents fast moving coordinate. The RPM technique is designed to preserve the structural form of the parent equation. Now we perform the simultaneous series expansion of hydrodynamic quantities in  powers of
$\sigma$ to get,
\beqa
\hat{\epsilon}&=&\frac{\epsilon}{\epsilon_0}=1+\sigma \epsilon_1+\sigma^2 \epsilon_2+\sigma^3 \epsilon_3+...,\,
\hat{p}=\frac{P}{P_0}=1+\sigma p_1+\sigma^2 p_2+\sigma^3 p_3+...,\nn\\
\hat{v}&=&\frac{v}{c_s}=\sigma v_1+\sigma^2 v_2+\sigma^3 v_3+...,\,
\,\,\,\,\,\hat{\Pi}=\frac{\Pi}{P_0}=\sigma \Pi_1+\sigma^2 \Pi_2+\sigma^3 \Pi_3+...,\nn\\
\hat{q}^x&=&\frac{q^x}{\epsilon_0}=\sigma q^x_1+\sigma^2 q^x_2+\sigma^3 q^x_3+...,\,
\hat{\pi}^{xx}=\frac{\pi^{xx}}{P_0}=\sigma \pi^{xx}_1+\sigma^2 \pi^{xx}_2+\sigma^3 \pi^{xx}_3+...,
\label{eq0722}
\eeqa
where $\epsilon_0$ and $P_0$ are the background equilibrium energy density and pressure respectively on which the perturbations propagate. By collecting terms with different order of $\sigma$ from the above series, different types of equation like Breaking wave equation, Burgers equation or KdV (Korteweg-De Vries) like equation, etc can be
obtained~\cite{Fogaca1,Lick1990,Bhattacharyya:2020sua}. 

We rewrite the equations of motion {\textit{i.e.}} Eq.~\eqref{eq0714}-Eq.\eqref{eq0719}, using Eqs.~\eqref{eq0721} and~\eqref{eq0722} and collect terms with different orders in $\sigma$ 
and then finally revert from $(X,Y) \to (t,x)$ to obtain the evolution equations of different order of perturbations. It is found in Ref.~\cite{Fogaca:2014gwa} that for the conformal background, the evolution equation of the first-order perturbation does not include any relaxation and coupling coefficients arising from the MIS theory. Therefore, to get the effect of second-order, it is required to go to the second-order in perturbation ($\hat{\epsilon_{2}}$). To obtain the time evolution of second order perturbation, we need to collect terms up to third order in $\sigma$ in the expansion, because the equations in $n$-th order in $\sigma$ contains time derivatives of $\hat{\epsilon}_{n-1}$. Therefore, to get equations containing time 
derivative of $\hat{\epsilon}_{2}$ we  go up to 3rd order in $\sigma$. This leads to the following equations for the perturbation in the energy density ($\hat{\epsilon}$) as:
\beqa
\frac{\pd \hat{\epsilon_{1}}}{\pd t}
+\left[1+(1-c_{s}^2) \frac{\epsilon_{0}}{\epsilon_{0}+P_{0}}\hat{\epsilon_{1}}\right]c_{s}\frac{\pd \hat{\epsilon_{1}}}{\pd x}- \left[\frac{1}{2(\epsilon_{0}+P_{0})}(\zeta +\frac{4}{3}\eta)\right]
\frac{\pd^{2} \hat{\epsilon_{1}}}{\pd x^{2}}=0,
\label{eq0723}
\eeqa
{\textrm and}
\beqa
\frac{\pd \hat{\epsilon_{2}}}{\pd t}+\mathcal{S}_{1}\frac{\pd \hat{\epsilon_{2}}}{\pd x}+\mathcal{S}_{2}\frac{\pd \hat{\epsilon_{1}}}{\pd x}+\mathcal{S}_{3} \frac{\pd^{2} \hat{\epsilon_{1}}}{\pd x^{2}}+\mathcal{S}_{4}\frac{\pd^{3} \hat{\epsilon_{1}}}{\pd x^{3}}+\mathcal{S}_{5}\frac{\pd^{2} \hat{\epsilon_{2}}}{\pd x^{2}}=0~.
\label{eq0724}
\eeqa
where $\hat{\epsilon_{i}}= \sigma^i \epsilon_i$ for $i=1,2,.....$.
The coefficients $\mathcal{S}_i$'s for $i=1$ to $5$ are:
\beqa
\mathcal{S}_{1}&=&\frac{1}{\epsilon_{0}+P_{0}}  \Big[c_s \{\epsilon _0 \left(1- \hat{\epsilon }_1\left(c_s^2-1\right)\right)+P_0\}\Big],\nn\\
 \mathcal{S}_{2}&=& \frac{1}{\epsilon_{0}+P_{0}}  \Big[\epsilon _0 c_s \{c_s^2-1\} \{\epsilon _0 \left(\left(2 c_s^2+1\right) \hat{\epsilon }_1{}^2-\hat{\epsilon }_2\right)-P_0 \hat{\epsilon }_2\}\Big],\nn\\
\mathcal{S}_{3}&=& \frac{1}{12(\epsilon_{0}+P_{0})^{2}}  \Big[\epsilon _0 \hat{\epsilon }_1 \{3 c_s^2 (7 \zeta +8 \eta )+3 \zeta +4 \eta \}\Big],\nn\\
 \mathcal{S}_{4}&=& \frac{1}{72 c_s c_V (\epsilon _0+P_{0})^2}
\Big[4 c_s^2 \{3 \kappa  T \epsilon _0(3 \alpha _0 \zeta +4 \alpha _1 \eta )\nn\\
&&+3 \zeta +4 \eta +P_0 (3 \alpha _0 \zeta +4 \alpha _1 \eta )\}+(\epsilon _0+P_{0}) (9 \beta _0 \zeta ^2+16 \beta _2 \eta ^2)-c_{V}(3 \zeta +4 \eta )^2\nn\\
&&+12 \kappa  \{\epsilon _0+P_{0}\} \{\epsilon _0 (3 \alpha _0 \zeta +4 \alpha _1 \eta )\}+3 \zeta +4 \eta +P_0 (3 \alpha _0 \zeta +4 \alpha _1 \eta )\Big],\nn\\
  \mathcal{S}_{5}  &=&-\frac{3 \zeta +4 \eta }{6 \left(\epsilon _0+P_{0}\right)}\,.
  \label{eq0725}
\eeqa
Eqs.~\eqref{eq0723} and \eqref{eq0724} have been solved to analyse the fate of the nonlinear waves propagating through a relativistic viscous fluid. The relevant EoS ({\textit{i.e.}}, relation between $\epsilon_0$ and $P_0$) and the initial conditions required to solve these equations.

We observe that the governing equation Eq.\eqref{eq0723} does not have any effect of second order theory. This equation can be derived from the NS theory as well. However, the second equation contains the second order effects via the coupling and relaxation coefficients of the MIS theory. Here, we have considered all the relevant transport coefficients to provide a general equations for the propagation of nonlinear waves. The dispersive term in Eq.\eqref{eq0724} (third order space derivative term) indicate that the combined effects of shear and bulk viscosities in the diffusive term act against the effect of thermal conductivity. This might weaken the diffusion of nonlinear waves in a dissimilar way.
\section{Results and discussion}
We present here the results on the fate of the nonlinear perturbations in the QGP when it passes near the region of the CEP. The equations we obtained look like KdV equation (ideal background) whose general analytic solution is a sec-hyperbolic function with a shifted argument. To better understand the deviation of propagation from that of the ideal background, the initial profile of perturbations of both the orders are taken as of the same form as the `{\it{sech}}' function. We solve the equations describing the evolution of $\hat{\epsilon_i}$ 
for the following initial profile of the perturbations:
\beqa
\hat{\epsilon}_1&=&A_{1}\big[\mathrm{sech} \frac{(x-x_{0})}{B_{1}}\big]^{2} \nn\\
\hat{\epsilon}_2&=&A_{2}\big[\mathrm{sech} \frac{(x-x_{0})}{B_{2}}\big]^{2},
\label{eq0726}
\eeqa
where, $x_0$ is the initial position of the peak of the perturbations. Since, here we are dealing with 1D propagation in $+x$ direction in an extended and static background, any choice of value of $x_0$ should provide the same salient features of the propagation. Here we have taken $x_0=10$ fm to demonstrate the propagation of the perturbation however, one may choose any other value of $x_0$ as well. The solutions of the Eqs.\eqref{eq0723} and \eqref{eq0724} are solitonic in nature (with {\it{sech}}$^2$ dependence), which is similar to the solutions of KdV equations for small dissipations. This motivates us to choose initial profile given by Eq.~\eqref{eq0726}. Also to note that for small spatial gradient the solitonic behaviour of the solution has also been demonstrated in Ref.~\cite{Fogaca:2014gwa} for Gaussian initial profiles in a conformally invariant hydrodynamic background. In the initial profile, the  $A_{i}$ and $B_{i}$ respectively are determining the height and width of the initial profile. The  effects of the CEP has been taken into consideration through the EoS (Ch.\ref{chapter4}) and the scaling behaviour of  transport coefficients and thermodynamic response functions in Eq.\eqref{eq0619}. 
\begin{figure*}
\includegraphics[width=4.5cm]{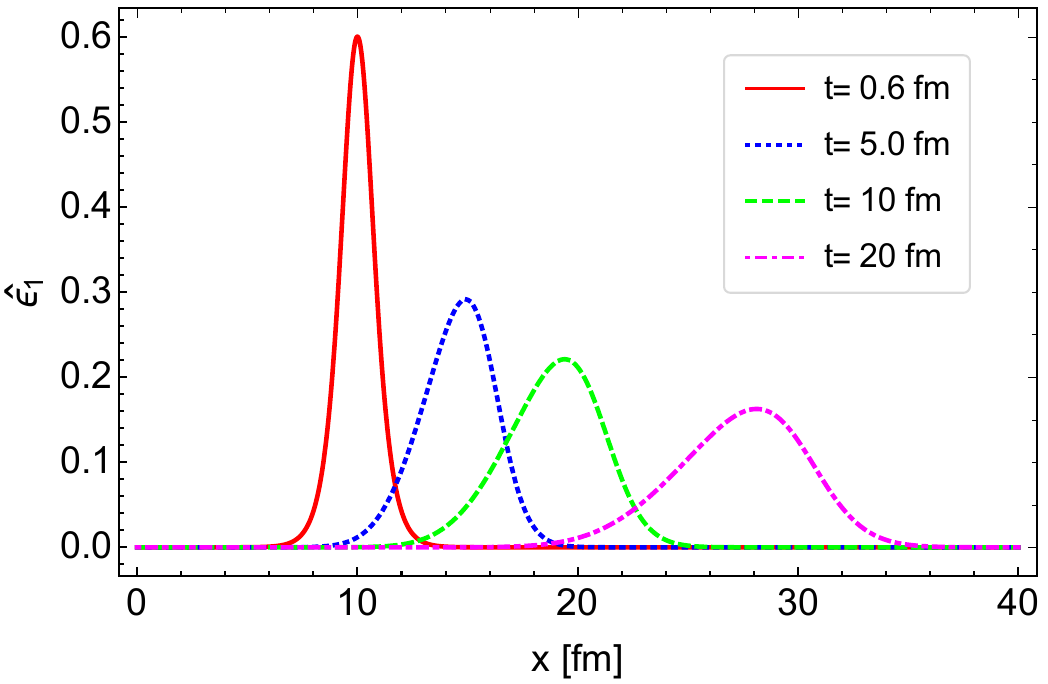}
\includegraphics[width=4.5cm]{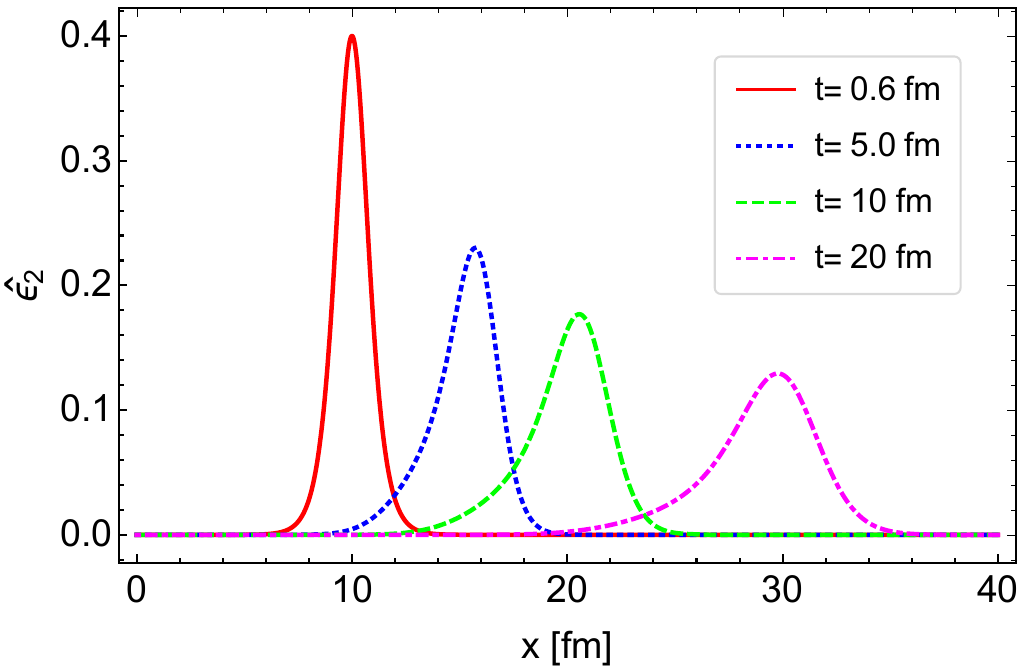}
\includegraphics[width=4.5cm]{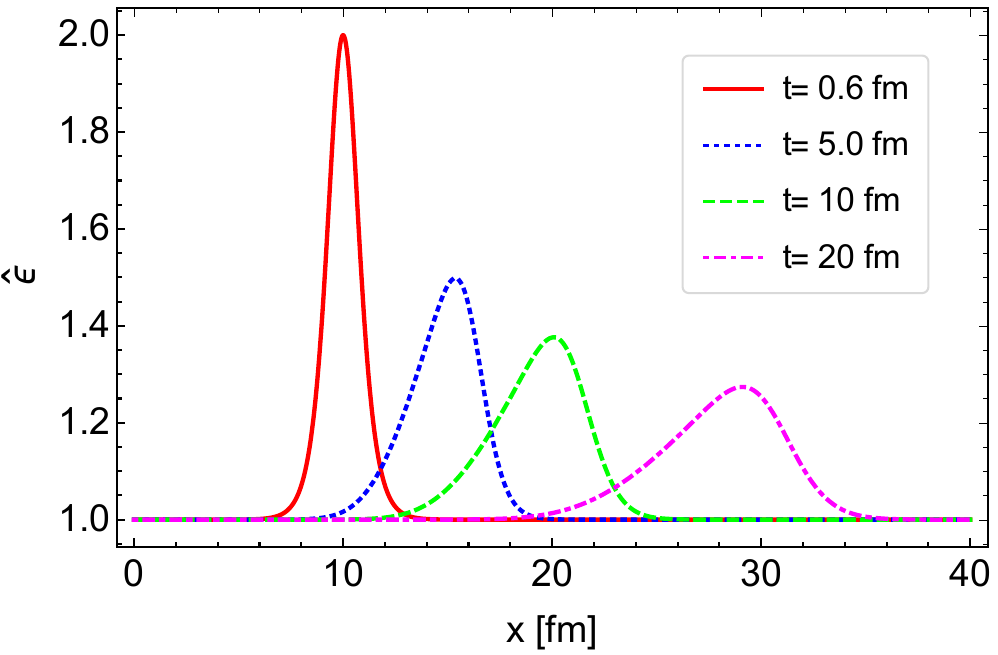}
\caption{ The spatial variation of perturbations
in energy density when the system is away 
from CEP at $\mu=400$ MeV and $T=185$ MeV.
The left panel shows $x$ dependence of $\hat{\epsilon_1}=
\sigma\epsilon_1$.
The middle and right panels indicate the variation 
of $\hat{\epsilon_2}=\sigma^2\epsilon_2$ and $[1+\hat{\epsilon_1}+
\hat{\epsilon_2}]$ (see Eq.~\eqref{eq0722}) respectively with $x$.
It is clear that when the system is 
away from CEP, the perturbations survive despite the dissipative 
effects. 
The results are obtained here for 
$A_{1}=0.6, B_{1}=1, A_{2}=0.4$ and $B_{2}=1$.
}
\label{fig0701}
\end{figure*}

Fig.\ref{fig0701} depicts the propagation of a nonlinear wave in the medium when the system is passing away from the CEP. It is observed that, though the amplitude of the wave is reduced, the nonlinear wave survives the dissipation despite the presence of the non-zero values of transport coefficients of the medium. However, the nonlinear wave is substantially dissipated (almost died) near the CEP as evident from the results shown in Fig.~\ref{fig0702}. It has been observed that the dissipative feature remains unaltered with the variation of the location of the CEP along the transition line in the QCD phase diagram. This clearly indicates that the nonlinear perturbations will exhibit unique detectable effects of the CEP. It also shows that the speed of the attenuated wave is non-vanishing in contrast to linear waves~\cite{Hasan1}. This is due to the  amplitude dependent propagation speed of the nonlinear waves, shown in Fig.\ref{fig0703}. The attenuation of the perturbation is in comparison smaller for the second-order correction ($\hat{\epsilon_2}$). Furthermore, the second-order perturbation travels a bit faster than the first-order one. This is clear from the results displayed in Figs.\ref{fig0701} and Figs.\ref{fig0702}, which can also be understood from Eq.\eqref{eq0724}, where the second-order correction contains the third-order derivatives of the first-order perturbation. This is equivalent to the dispersive term in the KdV equation, which is responsible for height preserving solitonic behaviour~\cite{Lick1990}. Therefore, this dispersive terms compete with the diffusive terms,  results in weakening the damping effect. The relaxation terms in the dissipative fluxes incorporated in MIS theory (Eq.\eqref{eq0724}) makes the dissipation lesser in comparison to the NS theory (Eq.\eqref{eq0723}).
\begin{figure*}
\includegraphics[width=4.5cm]{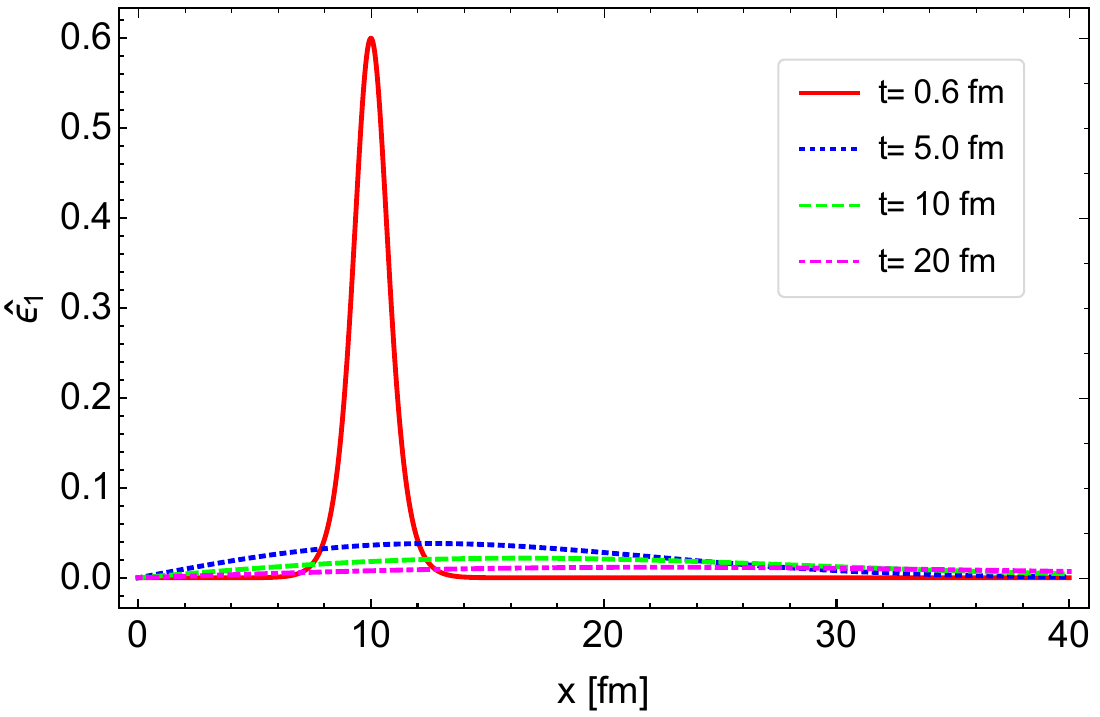}
\includegraphics[width=4.5cm]{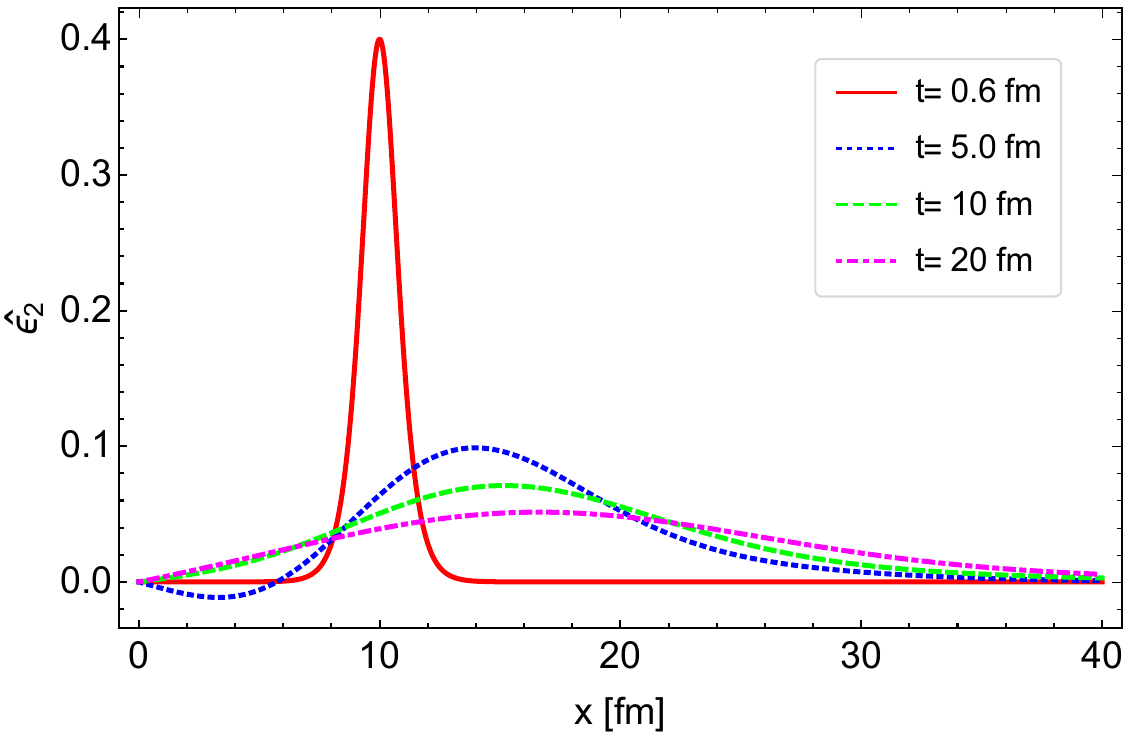}
\includegraphics[width=4.5cm]{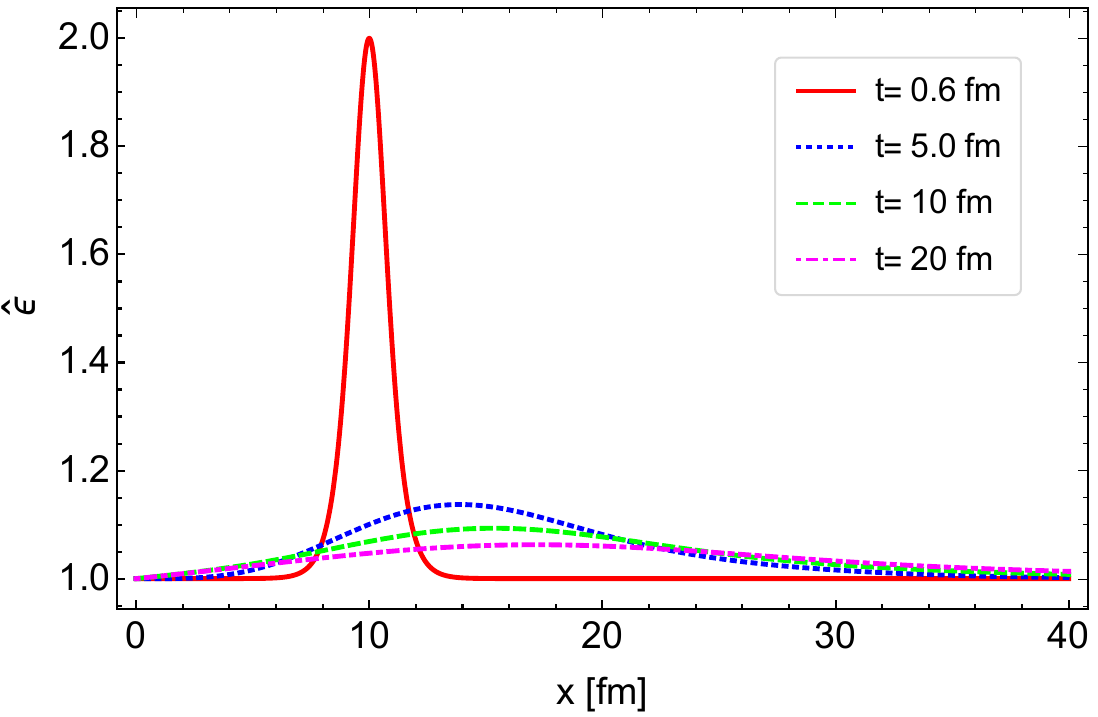}
\caption{ The spatial variation of perturbations
in energy density when the system is close to
CEP at $\mu=367$ MeV and $T=158$ MeV.
The left panel shows $x$ dependence of $\hat{\epsilon_1}=
\sigma\epsilon_1$.
The middle and right panels indicate the variation 
of $\hat{\epsilon_2}=\sigma^2\epsilon_2$ and $[1+\hat{\epsilon_1}+
\hat{\epsilon_2}]$ (see Eq.~\eqref{eq0722}) respectively with $x$.
It is clear that when the system is 
near the CEP, the nonlinear perturbations is strongly dissipated. 
The results are obtained here for 
$A_{1}=0.6, B_{1}=1, A_{2}=0.4$ and $B_{2}=1$.
}
\label{fig0702}
\end{figure*}

Therefore, near the CEP, the disturbances with larger magnitudes, which are created in the medium due to energy deposition by energetic particles or jets, will be highly dissipated. In earlier studies~\cite{Hasan1}, it was shown that linear waves will be completely stopped and dissipated near the CEP. Similar suppression for nonlinear waves too are found in the presence of the CEP. Although the speed of propagation of the nonlinear waves are amplitude dependent, in presence of the CEP, irrespective of their amplitude all the waves are highly suppressed. This may have important consequences in detecting the CEP from the analysis of the hadron spectra.

In has been speculated earlier~\cite{Minami,Hasan1,Hasan2} that the formation of Mach cone is forbidden near the CEP for the propagation of linear perturbations. The remaining question to be reported if the Mach cone formation will  be there for nonlinear waves or not. This is important because nonlinear waves are found to maintain the solitonic nature in comparison to linear perturbations. Nevertheless, we find that even the nonlinear perturbations will not be able to preserve the Mach cone effects if it hits the CEP. So the propagation of both types of the perturbations (linear and nonlinear) are stopped due to the presence of the CEP. These results can be used to detect the CEP by looking into the suppression of the Mach cone effect in two particle correlation~\cite{BetzPRL,Betzthesis}. Due to the high dissipation of the nonlinear waves, the broadening effect of localized waves will also vanish ~\cite{Fogaca1} along with the Mach cone effect.
\begin{figure}
\centering
\includegraphics[width=8cm]{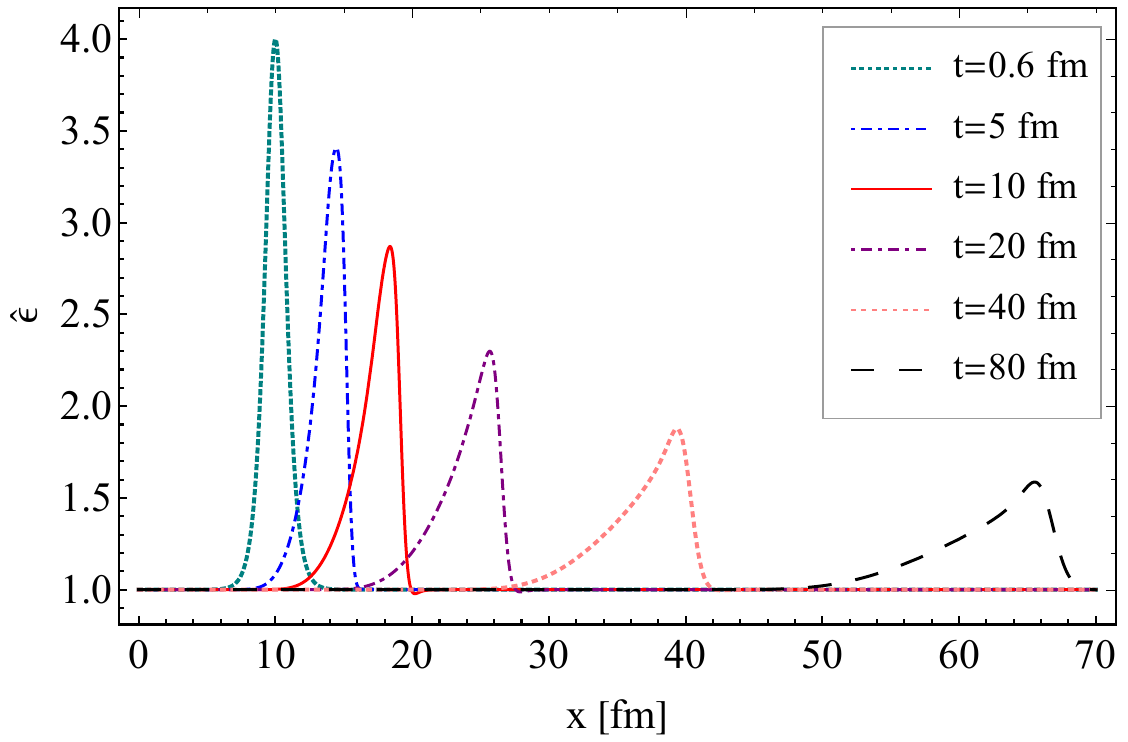}
\caption{ The spatial variation of perturbations
in energy density when the system is away from the CEP. Same plot as Fig.\ref{fig0701}(c), but with larger amplitude.}
\label{fig0703}
\end{figure}
The flow harmonics play important role in characterizing the medium formed in RHIC-E. It was shown in Refs. ~\cite{Stocker:2007pd,Stoecker1976} that some of the flow harmonics will collapse at the CEP (along phase boundary of the QCD phase diagram). Based on linear analysis it has been argued in~\cite{Minami,Hasan1,Hasan2}  that $v_{2}$ or higher harmonics will be reduced near the CEP. The similar conclusion can be drawn for the nonlinear perturbations too near the CEP. If the initial spatial shape of the system formed in HICs is highly distorted azimuthally, the fluid dynamical response to the initial eccentricity can be nonlinear. The suppression of the nonlinear waves also suggests that $v_2$ and higher order flow harmonics will be suppressed near the CEP due to the absorption of sound wave.

In a recent study~\cite{Dore}, it is argued that the path to the critical point is greatly influenced by far-from equilibrium initial conditions, where viscous effects may lead to dramatically different $(T,\mu)$ trajectories. The sound wave propagating through the system will be ceased near the CEP. Therefore, the pressure gradient produced (due to EoS) due to initial spatial anisotropy will not be converted effectively to momentum anisotropy through hydrodynamic expansion in an event of isentropic trajectory passing through the CEP. In such events the flow harmonics get suppressed. In another event, due to the different initial conditions, the system will follow a trajectory away from the CEP (see Ref.\cite{Nonaka}). In such events, sound will propagate accordingly and thus no suppression in flow harmonics should be observed. This difference in flow harmonics (between two events one nearby and the 
other away from the CEP) will lead to large event-by-event fluctuations in flow harmonics. Therefore,  large event-by-event fluctuation in flow harmonics could be a signature of the CEP.

So far it is evident that the nonlinear waves survive dissipation even in the presence of $\eta, \zeta$ and $\kappa$ without the critical effect, and the suppression of nonlinear wave only happens almost exclusively because of the presence of the CEP. This hints that the formation of Mach cone will be forbidden in the presence of the CEP. Therefore, the present theoretical investigation indicates that the Mach cones may disappear in the presence of the CEP. The Mach cone effect manifests as a double hump in the two-particle correlation in the low momentum domain of associated particles, and the CEP plays a unique role in suppressing the double hump in the two-particle correlation contrary to the other mechanisms: {\textit{e.g.}} (i) deflection of away side jets, (ii) Cherenkov radiation and (iii) radiation of gluons which produce the double hump. These mechanisms have the ability to obscure the suppression due to the CEP by creating double hump.

The dip in the azimuthal distribution of two particle-correlation at $\Delta \phi(=\phi-\phi_{trig}=\pi)$ accompanied by two local maxima on the both side of $\Delta \phi=\pi$ for the transverse momentum range $0.15<p_{assoc}^{T}<4$ GeV~\cite{Wang2006,Adams2005}, is attributed to the processes pointed above. Therefore, we contrast the effect of the CEP to the following mechanism. 

(i) Deflection of the away side jet by asymmetric flow and third flow harmonics (triangular flow) due to the initial state fluctuations~\cite{Betzthesis,Wang2013,Cao2020} lead to peaks on the away side jet on the both side of $\Delta \phi=\pi$. However, if the system passes near the CEP, the flow harmonics will get highly suppressed, and hence, the deflection also will be strongly diminished. 

(ii) Cherenkov radiation~\cite{Koch2006,STAR2003} is characterized by strong momentum dependence of the cone angle ($\sim 1/\mu_{r}$), where $\mu_{r}$ is the refractive index of the medium). This process is not likely to be responsible for the double hump due to the lack of momentum dependence of the location of the double maxima of associated particles.

(iii) The radiation of gluons by the
away side jet will deviate it from propagating at an angle $\pi$
with respect to the near side (trigger) jet. However, the quantitative prediction of the Mach cone positions studied through three particle correlation~\cite{Abelev2009}  and the momentum independence of
the location of the double hump indicate that the double hump
may originate from Mach cone effects. The vanishing of the Mach cone-like structure in particle correlation will therefore, indicate
the existence of the CEP.

In this study, the propagation of perturbation is studied in a static background. In presence of expansion, the system will cool 
and move toward the phase transition line with changing transport coefficients. However, during this cooling, if the trajectory goes away from the CEP, 
the response will allow the survival of perturbations. 
But if it passes near the CEP, the 
perturbative effects will be mostly washed out and no effect will survive at the latter stage. Therefore, it is expected that the results obtained in static situation may not
vary with the inclusion of expansion if the system passes near the CEP. We investigate the hydrodynamic propagation of perturbations in the system without considering the 
fluctuations~\cite{Stephanov:2017ghc,An2020} originating from the CEP itself, as they do not create any angular pattern as jets produce in correlations. Though the non-equilibrium fluctuations due to the CEP is not taken into account here for obtaining the non-linear wave equations, the enhancement of thermodynamic fluctuations near the critical point, which affects the hydrodynamic response, is inherently taken into account through the EoS and other thermodynamic quantities with the critical exponents.

\chapter[Summary and outlook]{Summary and outlook}
\label{chapter9}
The study of the critical point in the QCD phase diagram through collision of heavy ions at relativistic energies is highly contemporary. The location of the CEP is still not clear. Even if the first-principle lattice calculations are not applicable to the finite $\mu_{B}$ region, where the CEP is supposed to exist, there are several studies based on effective models to find the location (or at least the region of its existence) of the CEP. In this dissertation, we assume the existence of the CEP at some location ($T_{c},\,\mu_{c}$) and study its consequences on the dispersion relation, on the propagation of perturbation (both linear and nonlinear) and on the structure factor. The effects of the CEP goes into the calculation through an EoS (of 3D Ising universality class) and scaling behaviour of several thermodynamic coefficients. The EoS used here is found to agree with the available lattice results. 

To observe the hydrodynamic response of the system, a space-time-dependent perturbation is placed into the medium. The response of the system when passing through the CEP (nearby to the CEP) or away from the CEP have been investigated to find the difference in the response of the fluid. The dispersion relations are calculated with the ambit of the linearized hydrodynamic equations. The real part and the imaginary part of the dispersion relations are separated out. The imaginary part of the solution provides us the information about the dissipation of the plane wave in the medium, whereas the real part of the solution is for the survival of the wave. Therefore, the interplay between the magnitude of the real and the imaginary part will decide the fate of the perturbation (sound wave). This motivates us to calculate a threshold wave number $k_{th}$, and consequently, a threshold wavelength for the wave, symbolized as $\lambda_{th}$. It is defined in such a way such that any wave with a wavelength smaller than $\lambda_{th}$ will eventually get dissipated in the medium for given values of transport coefficients and other thermodynamic quantities. Interestingly, near the CEP, we have found $\lambda_{th}$ to diverge implying that no wave is allowed to propagate if the system hits the CEP {\it i.e.} waves with all wavelength get dissipated at the CEP irrespective to the values of transport coefficients. The fluidity defined in Eq.\eqref{eq0530} of the system seem to diverge at the CEP.

In conventional condensed matter system, it has been observed experimentally ~\cite{Schneider1951,Botch1965,Kadanoff1968,Kawasaki1970,Kemensky1973,Dengler1987} that the absorption of sound is maximum due to diffraction of sound wave from the critical region, which is similar to the scattering of light at the critical point resulting in the critical opalescence.  Near the CEP, the correlation length $(\xi)$ diverges, therefore, the hydrodynamic limit ($\xi<<\lambda_{s}$, where, $\lambda_{s}$ is the wavelength of the sound mode) is violated. As a result, the development of sound wave is prevented. The forbiddance of the sound wave will eventually lead to the vanishing of Mach cone  (Mach angle, $\alpha=${\it{sin}}$^{-1}(c_s/v)$, where, $c_{s}$ is the speed of sound and $v$ is the fluid velocity). Therefore, the vanishing of Mach cone may indicate the presence of critical point.

The presence of CEP makes the viscous horizon  scale, $R_v\sim 1/k_{th}$ to diverge. This scale is related to various flow harmonics of the azimuthal distribution of produced particles in relativistic heavy-ion collider experiments (RHIC-E), which are useful quantities to characterize the matter of its initial state of the formation. For example, the triangular flow helps to understand the initial fluctuations and elliptic flow can be used to comprehend  the EoS of the system. Since the highest order of surviving flow harmonic is defined as, $n_v\sim 2\pi R/R_v$, thus ideally the vanishing harmonics will indicate the presence of the CEP. However, in experiments, the measurable quantities are superpositions of different densities and temperature from the formation to the freeze-out stage, therefore, even if the system hits the critical point in the $T-\mu$ plane, the harmonics may not vanish, but the critical point may suppress them.

We also derived the expression for the dynamical spectral structure of the density fluctuation to study its behaviour near the QCD critical point using linear response theory and the MIS hydrodynamics, where we have kept all the relevant transport coefficients in the calculations. The change in the spectral structure of the system as it approaches the critical point has been studied. We have found that the Rayleigh and Brillouin peaks are distinctly visible when the system is away from the critical point but the peaks tend to merge near the critical point. The sensitivity of the structure of the spectral function on wave vector $(k)$ of the sound wave has been demonstrated. It has been shown that the Brillouin peaks get merged with the Rayleigh peak because of the absorption of sound waves in the vicinity of the critical point. As consequences, we have argued that Mach cone formation will be prevented and the flow harmonics are suppressed when the system passes near the CEP. We have also the shown mode dependent propagation of the perturbations which helps to find the speed of sound by looking at the position of the Brillouin peaks.

Critical slowing down is an important phenomena, cause the system to behave differently, should be included in the study of critical dynamics. The effects of critical slowing down is put into the calculations by considering a scalar non-hydrodynamic mode in the system. Therefore, we have derived the relevant equation of motions for the slow modes when 
the extensivity condition of thermodynamics is unaltered. The role of extra out-of-equilibrium $\phi$ mode on the dynamic structure factor near the QCD critical point has been investigated. The $\Snn$ in presence of the $\phi$ modes shows four peaks of Lorentzian type, which are asymmetrically situated in frequency axis with uneven magnitudes. While the $\Snn$ without $\phi$ modes admits three Lorentzian peaks, symmetrically situated on frequency axis with even magnitudes. We find that the asymmetry in the peaks originates due to  coupling of the $\phi$ modes with the hydrodynamic modes. The presence of the out-of-equilibrium modes can generate the extra peak in $\Snn$ within the scope causal theory of hydrodynamics when the transverse modes (along with the longitudinal modes) are taken into account. This study may help in the experimental investigation of role of the out-of-equilibrium modes near any of the $\mathcal{O}(4)$ critical points. The out-of-equilibrium mode plays a crucial role in reducing the width of the Lorentzians representing the thermal fluctuation which indicate the enhancement of the relaxation time of the system.

Along with the linear perturbations, we have also studied the nonlinear perturbations in the fluid. We have derived the equations for the propagation of nonlinear perturbations by the Reductive Perturbative method in MIS theory with considering all the relevant transport coefficients. We have derived relevant equations governing the propagation of nonlinear wave within the purview of second order causal hydrodynamics by taking into account the non-zero values of $\eta$, $\zeta$ and $\kappa$ in contrast to earlier works with nonlinear wave equations, where the effects of $\zeta$ and $\kappa$ were ignored~\cite{Fogaca1,Fogaca:2014gwa}. In the presence of the CEP, $\zeta$ and $\kappa$ play significant roles as they tend to diverge near the CEP and hence can not be ignored. We observe, similar to the linear perturbation, the nonlinear perturbations too get suppressed near the CEP. The diverging nature of $\kappa$ near the CEP plays dominant role in the suppression of the nonlinear wave. One expects the vanishing of Mach cone effects (or away side double-peak structure) and broadening of the two and three particle correlation as a consequence of the presence of the CEP. The suppression or collapse of elliptic flow will also indicated the existence of CEP. This may lead to the large event-by-event fluctuation of flow harmonics. Therefore, vanishing Mach cone effect (or away side double-peak structure) on away side jet and the enhancement of fluctuation of flow harmonics in event-by-event analysis accompanied by suppressed flow harmonics could be considered as signals of the CEP.

The possibility of existence and detection of the CEP has been studied in~\cite{KapustaChi}.  The mode coupling theory has been used to calculate the thermal conductivity at the location near and away from the QCD critical point in the $T-\mu$ plane and shown that the $\kappa$ diverges at the critical point.  Authors in Ref.~\cite{KapustaChi} have also demonstrated that the sharp change in $\kappa$ at the critical point is eventually reflected in the two particle-correlation of fluctuations in rapidity space. Therefore, the study pave the way to confirm the existence of the CEP. The suppression of fluctuations in baryonic chemical potential ($\Delta \mu$) and temperature ($\Delta T$) due to the enhanced thermodynamic response functions at the CEP can signal the presence of the CEP~\cite{Stephanov:1998dy,Stephanov:1999zu}. The  suppression in flow harmonics and suppression in fluctuations in $\Delta \mu$ and $\Delta T$ will be reflected through the spectra of hadrons and proton to pion ratio respectively.

In a real scenario, the possibility of the trajectories passing through the CEP may be infrequent, which restricts the magnitude of the fluctuations near the critical point. These fluctuations will stay out of equilibrium because of the expansion of the system and critical slowing down~\cite{Berdnikov:1999ph,Stephanov:2017ghc}. This problem has been analysed~\cite{Akamatsu2019} in the evolution of hydrodynamic fluctuations in the system formed in RHIC-E. The  emergence of the Kibble-Zurek length scale~\cite{Kibble,Zurek} and its relation with short range spatial correlations has been also discussed~\cite{Akamatsu2019}. To note that the non-flow correlations get enhanced in presence of the CEP and such correlations may be measured as a function of $n/s$ (ratio of particle density and entropy density) for detecting the CEP~\cite{Akamatsu2019}.

The fluid dynamical description works in the region where the condition, $k<<q$ is maintained (where  $q$ is the inverse of correlation length, expressed as $q=q_o\,r^{\nu}$, $q_o$ is a constant and $\nu$ is the critical index with a numerical value $\nu=0.73\pm 0.02$~\cite{Rajagopal:1992qz}). The hydrodynamics becomes invalid when $\xi$ diverges at the CEP. However, there could be a region of validity in the neighbourhood of the CEP  where the predictions of fluid dynamical approach may be effective. Following procedure similar to the condensed matter physics~\cite{Stanley} we can write, $k<<q_o r^{\nu}$ for the region of validity of the hydrodynamics, which implies,
\begin{equation}
T>T_c\left[1+\left(\frac{k}{q_o}\right)^{1/\nu}\right]
\end{equation}
As hydrodynamics is an effective theory for soft physics (small wave vector or large wavelength), the fluid dynamical description can be applied in the neighbourhood of the CEP but becomes inappropriate at the CEP due to the divergence of the correlation length. The response of the trajectories in the neighbourhood of the CEP in the $T- \mu$ plane to the initial conditions away from equilibrium has been investigated in Ref.~\cite{Dore}.

In a realistic scenario, the matter formed in RHIC-E 
evolves in space and time from the initial QGP phase to the final hadronic 
freeze-out state through an intermediary phase transition. 
The space time evolution of the locally equilibrated system is described
by relativistic viscous hydrodynamics. The experimentally  detected signals
is the superposition of the yields for all the possible values 
of temperatures and densities realised during the evolution of the system
from its initial to the freeze-out state.
The detection of CEP will require the disentanglement of contributions from the  
neighbourhood ($\mu_c,T_c$) from all other possible values of ($\mu,T$) which the system 
confronts during its evolution history.  
In the present work the expansion dynamics has not been taken into consideration, therefore,
the results obtained here can not be contrasted with experiments directly. The effects of the CEP with
(3+1) dimensional expansion within the scope of second order viscous hydrodynamics will be 
next step in this direction.

The electromagnetic (EM) probes of QGP (see
\cite{ALAM1996243} for review) 
{\it{i.e.}} real photons and lepton pairs can be used to study the evolution of the
system from the pristine partonic stage to the
final hadronic stage through an intermediary
phase transition or crossover.
The photons and lepton can bring the information of the thermodynamic 
state of their production points~\cite{ALAM1996243,ALAM2000159} efficiently as
their  mean free paths are larger than the size of the fireball created
in RHIC-E. This fact should be contrasted with hadrons which  are subjected to rescattering 
in the medium and consequently loose information of their production point. 
Therefore, in principle, the photons can bring the information
of the CEP very efficiently. In Ref.~\cite{Paquet2016} the photon spectra 
has been evaluated  
by using second-order dissipative hydrodynamics 
and the most updated rate of photon productions from QGP and 
hadronic phases. However, the effects of the CEP on EoS, various transport
coefficients and response functions are required to be included 
to get the imprints of CEP on the photon spectra which is beyond the
scope of the present work, although worth exploring and may be taken up in future endeavour.

\appendix
\chapter[Relaxation and coupling coefficients]{Relativistic Hydrodynamics}
\label{appendix01_A}
\section{Relaxation and coupling coefficients}
In this appendix we provide the expressions for the relaxation and coupling coefficients~\cite{Israel:1979wp}
required to solve the Israel-Stewart hydrodynamical equations. 
\beqa
\alpha_{0}&=&(D_{41}D_{20}-D_{31}D_{30})\Lambda \phi \Omega J_{21}J_{31}~,\nn\\
\alpha_{1}&=& (J_{41}J_{42}-J_{31}J_{52}) \Lambda \phi J_{21}J_{31}~, \nn\\
\beta_{0}&=& \frac{3\beta}{\phi^{2}\Omega^{2}}[5J_{52}-\frac{3}{D_{20}}\{J_{31}(J_{31}J_{30}-J_{41}J_{20})+J_{41}(J_{41}J_{10}-J_{31}J_{20})\}]~, \nn\\
\beta_{1}&=& \frac{D_{41}}{\Lambda^{2}nmJ_{21}J_{31}}~, \nn\\
\beta_{2}&=& \frac{\beta J_{52}}{2\phi^{2}}~,
\eeqa
where, 
\beqa
D_{rs}&=&J_{r+1,s}J_{r-1,s}-(J_{rs})^{2},\,\,\,\, \phi=T(\epsilon+P),\,\,\,\, 
\psi=\frac{\epsilon +p}{n_im_i}, \nn\\
\Lambda&=& 1+5(\frac{\phi}{nm})-\psi^{2},\,\,\,\, 
\Omega= 3\Big(\frac{\pd ln \phi}{\pd ln n}\Big)-5~,
\eeqa
where $r$, $s$ are integers, $m_i$ and $n_i$ are respectively the mass and density of $i$-th type particle. 
The quantities, $\phi, \psi$ and  $\Lambda$ are calculated by using their relations with  $\epsilon$, $p$ and $n_i$. 
We take the current quark mass of up and down flavors as 10 MeV. 
$J_{rs}$ is defined as,
\beqa
J_{rs}=\frac{A_{0}}{(2s+1)!!}\int_{0}^{\infty} N \Delta Sinh^{2(s+1)}\mathcal{R} Cosh^{r-2s}\mathcal{R} d\mathcal{R}~,
\eeqa
and
\beqa
N=\frac{1}{exp(\beta Cosh \mathcal{R} -\alpha)-\epsilon}~,
\eeqa
where, 
\beqa
\Delta=1+\epsilon N,\,\,\,\, A_{0}=4\pi m_i^3~.
\label{eq14}
\eeqa

\chapter[Thermodynamic derivatives]{Propagation of perturbation near the QCD critical point}
\label{appendix05_A}
\section{Thermodynamic derivatives}
The expressions for $\omega_{\Re}(k)$ and $\omega_{\Im}(k)$ contain derivatives of several thermodynamics variables. In this appendix
we recast these derivatives
in terms of response functions like: isothermal and adiabatic compressibilities  ($\kappa_T$ and $\kappa_s$), volume expansivity $\alpha_{p}$ 
specific heats ($c_p$ and $c_v$), baryon number susceptibility ($\chi_{B}$), velocity of sound ($c_s$), etc. The 
baryon number density ($n$) and the entropy density ($s$)  are given by
\beqa
n=\Big(\frac{\partial p}{\partial \mu_B}\Big)_{T};\,\,\,\,\, s=\Big(\frac{\partial p}{\partial T}\Big)_{\mu_B}
\eeqa
Baryon number susceptibility ($\chi_{B}$), isothermal compressibility ($\kappa_{T}$), adiabatic compressibility ($\kappa_{s}$) and volume expansivity ($\alpha_{p}$) are given by,
\beqa
\chi_{B}=\Big(\frac{\partial n}{\partial \mu_B}\Big)_{T}; \kappa_{T}=\frac{1}{n}\Big(\frac{\partial n}{\partial p}\Big)_{T}; \kappa_{s}=\frac{1}{n}\Big(\frac{\partial n}{\partial p}\Big)_{s}; \alpha_{p}=\frac{1}{V}\Big(\frac{\partial V}{\partial T}\Big)_{p}=-\frac{1}{n}\Big(\frac{\partial n}{\partial T}\Big)_{p}
\eeqa
Specific heats are given by
\beqa
C_{P}=T\Big(\frac{\partial s}{\partial T}\Big)_{p}; C_{V}=T\Big(\frac{\partial s}{\partial T}\Big)_{V}=T\Big(\frac{\partial s}{\partial T}\Big)_{n}=\Big(\frac{\partial \epsilon}{\partial T}\Big)_{V}=\Big(\frac{\partial \epsilon}{\partial T}\Big)_{n}
\eeqa
We  have to express six quantities such as: $(\frac{\partial P}{\partial T})_{n}, 
(\frac{\partial P}{\partial n})_{T}, (\frac{\partial \epsilon}{\partial T})_{n}, \Big(\frac{\partial \epsilon}{\partial n}\Big)_{T}$ and $(\frac{\partial n}{\partial T})_{\mu}, (\frac{\partial n}{\partial \mu})_{T}$ in terms of various thermodynamic variables.

i) To evaluate: $(\frac{\partial P}{\partial T})_{n}$ we start with
\beqa
\Big(\frac{\partial P}{\partial T}\Big)_{n}&=&\frac{\pd (P, n)}{\pd (T,n)}
=\frac{\pd (P, n)}{\pd (T, P)}\frac{\pd (T, P)}{\pd (s, P)}\frac{\pd (s, P)}{\pd (s, \epsilon)}\frac{\pd (s, \epsilon)}{\pd (s, n)}\frac{\pd (s, n)}{\pd (T, n)} \nn\\
&=& \Big[-\Big(\frac{\partial n}{\partial T}\Big)_{P}\Big]\Big(\frac{\partial T}{\partial s}\Big)_{P}\Big(\frac{\partial P}{\partial \epsilon}\Big)_{s}\Big(\frac{\partial \epsilon}{\partial n}\Big)_{s}\Big(\frac{\partial s}{\partial T}\Big)_{n} \nn\\
&=& n\alpha_{P}\frac{T}{C_{P}}c^{2}_{s}\Big(\frac{\partial \epsilon}{\partial n}\Big)_{s}\frac{C_{V}}{T}
=nc^{2}_{s}\alpha_{P} \frac{C_{V}}{C_{P}}\Big(\frac{\partial \epsilon}{\partial n}\Big)_{s}
\eeqa
Now if we use the relation
\beqa
d\epsilon=Tds+\mu_B dn \,\,\,\,\, \text{and } \mu_B=\Big(\frac{\partial \epsilon}{\partial n}\Big)_{s}
\eeqa
then we can write  \vspace{-.85cm}
\beqa
{\Big(\frac{\partial P}{\partial T}\Big)_{n}=\mu_B nc^{2}_{s}\alpha_{P} \frac{C_{V}}{C_{p}}}
\label{eqb6}
\eeqa 

ii)  Now consider $\Big(\frac{\partial P}{\partial n}\Big)_{T}$: \vspace{-1cm}
\beqa
{\Big(\frac{\partial P}{\partial n}\Big)_{T}= \frac{1}{n\kappa_{T}}}
\label{eqb7}
\eeqa
\beqa
\Big(\frac{\partial P}{\partial n}\Big)_{T}&=&\frac{\pd (P, T)}{\pd (n, T)}
=\frac{\pd (P, T)}{\pd (P, s)}\frac{\pd (P, s)}{\pd (\epsilon, s)}\frac{\pd (\epsilon, s)}{\pd (n,s)}\frac{\pd (n, s)}{\pd (n, T)}\nn\\
&=&\Big(\frac{\partial T}{\partial s}\Big)_{P}\Big(\frac{\partial P}{\partial \epsilon}\Big)_{s}\Big(\frac{\partial \epsilon}{\partial n}\Big)_{s}\Big(\frac{\partial s}{\partial T}\Big)_{n}\nn\\
&=& \frac{T}{C_{P}}c^{2}_{s}\Big(\frac{\partial \epsilon}{\partial n}\Big)_{s}\frac{C_{V}}{T}
\\&=&\mu_B c^{2}_{s} \frac{C_{V}}{C_{P}}
\label{eqb9}
\eeqa
 
iii) The factor, $\Big(\frac{\partial \epsilon}{\partial T}\Big)_{n}$ can be estimated as follows:
\beqa
\Big(\frac{\partial \epsilon}{\partial T}\Big)_{n}=c_n
\eeqa
For fixed net baryon number,  $c_n$ can be written as $c_{n}=c_{V}$.
Therefore, 
\beqa
{\Big(\frac{\partial \epsilon}{\partial T}\Big)_{n}=c_{V}}
\label{eqb11}
\eeqa
iv) $\Big(\frac{\partial \epsilon}{\partial n}\Big)_{T}$ can be estimated as:
\beqa
\Big(\frac{\partial \epsilon}{\partial n}\Big)_{T}&=&\frac{\pd (\epsilon, T)}{\pd (n,T)}
=\frac{\pd (\epsilon, T)}{\pd (\epsilon,s)}\frac{\pd (\epsilon, s)}{\pd (n,s)}\frac{\pd (n, s)}{\pd (n,T)}\nn\\
&=& \Big(\frac{\partial T}{\partial s}\Big)_{\epsilon}\Big(\frac{\partial \epsilon}{\partial s}\Big)_{s}\Big(\frac{\partial s}{\partial T}\Big)_{n} \nn\\
&=& \Big(\frac{\partial T}{\partial s}\Big)_{\epsilon}\Big[\Big(\frac{\partial \epsilon}{\partial p}\Big)_{s}\Big(\frac{\partial p}{\partial n}\Big)_{s}\Big] \frac{c_{V}}{T} \nn\\
&=& \Big(\frac{\partial T}{\partial s}\Big)_{\epsilon}\Big[\frac{1}{c^{2}_{s}}\frac{1}{n\kappa_{s}}\Big]\frac{C_{V}}{T}
= \frac{T}{c_{\epsilon}}\Big[\frac{1}{c^{2}_{s}}\frac{1}{n\kappa_{s}}\Big]\frac{C_{V}}{T}
\\&=&\frac{C_{V}}{c_{\epsilon}}\Big[\frac{1}{c^{2}_{s}}\frac{1}{n\kappa_{s}}\Big]
\label{eqb13}
\eeqa
v) The quantity $\Big(\frac{\partial n}{\partial \mu}\Big)_{T}$ can be calculated as
\beqa
\Big(\frac{\partial n}{\partial \mu}\Big)_{T} =\chi_{B}
\eeqa
vi) For $\Big(\frac{\partial n}{\partial T}\Big)_{\mu}$:
\beqa
\Big(\frac{\partial n}{\partial T}\Big)_{\mu}=-\Big(\frac{\partial n}{\partial \mu}\Big)_{T}\Big(\frac{\partial \mu}{\partial T}\Big)_{n}=-\chi_{B}\Big(\frac{\partial \mu}{\partial T}\Big)_{n}
\eeqa
Now
\beqa
sdT&=&dP-nd\mu \to dP=sdT+nd\mu \\
Tds&=&d\epsilon-\mu dn \to d\epsilon =Tds+\mu dn\\
\Big(\frac{\partial P}{\partial \epsilon}\Big)&=&\frac{sdT+nd\mu }{Tds+\mu dn}\\
&=& \frac{s+n\Big(\frac{\partial \mu}{\partial T}\Big)}{T\Big(\frac{\partial s}{\partial T}\Big)+\mu\Big(\frac{\partial n}{\partial T}\Big)}\\
n\Big(\frac{\partial \mu}{\partial T}\Big)&=&\Big(\frac{\partial P}{\partial \epsilon}\Big)\Big[T\Big(\frac{\partial s}{\partial T}+\mu \Big(\frac{\partial n}{\partial T}\Big)\Big]-s
\eeqa
Thus
\beqa
\Big(\frac{\partial \mu}{\partial T}\Big)_{n}=\frac{c^{2}_{n}C_{V}-s}{n} \hspace*{1cm} \text{where}, c^{2}_{n}=\Big(\frac{\partial P}{\partial \epsilon}\Big)_{n}
\eeqa
Finally 
\beqa
\Big(\frac{\partial n}{\partial T}\Big)_{\mu}=\frac{\chi_{B}}{n}(s-c^{2}_{n}C_{V})
\eeqa

\chapter[Dynamic structure factor near the QCD critical point]{Dynamic structure factor near the QCD critical point}
\label{appendix06_A}
\section{Expression for structure factor and the thermodynamic derivatives}
In this appendix the expression for the dynamical spectral structure, $\Snn$  derived by considering contributions 
up to second order in transport coefficients (i.e $\eta^{2},\zeta^{2},\kappa^{2},\eta \zeta, \eta \kappa,\zeta\kappa$)
has been provided.  The coupling and relaxation coefficients ($\tilde{\alpha_{0}},\tilde{\alpha_{1}},\beta_{0},\tilde{\beta_{1}},\beta_{2}$) 
have been taken non-zero in obtaining  the results displayed in the text, but have been taken 
as zero in the following to avoid  a more lengthy and complex expressions. Here we have used
\begin{equation}
\mathcal{S}_{nn}(\vec{k},\omega)=\frac{\mathcal{S^\prime}_{nn}(\vec{k},\omega)}
   {\Big< \delta n(\vec{k},0)\delta n(\vec{k},0)\Big>}
\end{equation}
\beqa
&&\mathcal{S^\prime}_{nn}(\vec{k},\omega)=k^2 n_0 \Bigg[ \omega ^2 n_0 T^{2}\Big(\frac{\text{$\pd $s}}
{\text{$\pd $T}}\Big)_n \Big\{h_0 \kappa T \omega+
\Big(\frac{\text{$\pd $P}}{\text{$\pd $T}}\Big)_n  k^2 \Big( 
\zeta+\frac{4}{3} \eta  \Big)-k^{2}T^{2} \kappa  \omega ^2\nn\\
&&+h_{0}k T\kappa\Big(\frac{\text{$\pd $s}}{\text{$\pd $n}}\Big)_T  +k^2T \kappa  ( \zeta +\frac{4}{3}\eta )- T^{2} \Big(\frac{\text{$\pd $P}}{\text{$\pd $T}}\Big)_n^2\Big\}+k^2 T \kappa h_{0}  \Big\{- h_0T \Big(\frac{\text{$\pd
   $P}}{\text{$\pd $T}}\Big)_n\nn\\
   &&+k^2T \kappa  (\zeta+\frac{4}{3} \eta  )+T^{2} \Big(\frac{\text{$\pd $P}}{\text{$\pd
   $T}}\Big)_n^2\Big\}+n_0^{2}T^{2} \Big(\frac{\text{$\pd $s}}{\text{$\pd $T}}\Big)_n \Big(\frac{\text{$\pd $s}}{\text{$\pd $n}}\Big)_T \times \nn\\
   &&\Big\{\Big(\frac{\text{$\pd $P}}{\text{$\pd $T}}\Big)_n k ( \zeta +\frac{4}{3} \eta )+T^{3} \kappa  \omega ^2- h_0T \kappa \Big\}+k^4 n_{0} h_{0} (\zeta +\frac{4}{3} \eta )^{2}h_{0}\Big(\frac{\text{$\pd $P}}{\text{$\pd $n}}\Big)_T \times \nn\\
   &&\Big\{n_0 T^2 \omega^{2}
    \Big(k^2 (\zeta +\frac{4}{3} \eta )+ \kappa T \omega ^2- k^2 T\kappa
  \Big) \Big\}\nn\\
  &&+ h_{0}^{2} \kappa T^{2} k^{2}  \Big(\frac{\text{$\pd $P}}{\text{$\pd $T}}\Big)_n- n_0h_{0} \kappa T^{3} k^{2}\omega ^2 \Big(\frac{\text{$\pd
   $s}}{\text{$\pd $n}}\Big)_T\Bigg] \Bigg /\nn\\
   && \Bigg[ h_0^2 \Big\{ k^4 T^{2}\kappa^2 \omega ^4+n_0^2 T^2 \omega ^4\Big(\frac{\text{$\pd $s}}{\text{$\pd $T}}\Big)_n\Big\}+h_0 k^2
   \omega ^2  \kappa^2T^{4}\Big(\frac{\text{$\pd $s}}{\text{$\pd $T}}\Big)_n \Big\{  \omega ^4 -2n_{0} \Big(\frac{\text{$\pd
   $P}}{\text{$\pd $n}}\Big)_T\Big\}\nn\\
   &&-2 k^2 n_0^2 T^{4} \kappa^2 \omega ^2\Big( \frac{\text{$\pd $s}}{\text{$\pd $T}}\Big)_n^{2}-2n_{0}h_{0}k^2 T^{4} \kappa^2
   \omega ^2 \Big(\frac{\text{$\pd $P}}{\text{$\pd $T}}\Big)_n-2n_0^3 T^2k^{4} \omega ^2 \Big(\frac{\text{$\pd $s}}{\text{$\pd $T}}\Big)_n\times\nn\\
   &&  \Big\{\Big(\frac{\text{$\pd $P}}{\text{$\pd $T}}\Big)_n \Big(\frac{\text{$\pd $s}}{\text{$\pd $n}}\Big)_T+ \Big(\frac{\text{$\pd $s}}{\text{$\pd
   $T}}\Big)_n \Big(\frac{\text{$\pd $P}}{\text{$\pd $n}}\Big)_T\Big\}+k^4 n_0^4 T^2 \omega ^2 \Big(\frac{\text{$\pd
   $s}}{\text{$\pd $n}}\Big)_T^2  \Big(\frac{\text{$\pd $P}}{\text{$\pd $n}}\Big)_T^2\nn\\
   &&+2k^2
   n_0 T^{2} \kappa  \omega ^2 \Big(\frac{\text{$\pd $P}}{\text{$\pd $T}}\Big)_n \Big\{ k^4 T\kappa  \Big(\frac{\text{$\pd $P}}{\text{$\pd $n}}\Big)_T+T \omega ^2 \Big(\frac{\text{$\pd $s}}{\text{$\pd $n}}\Big)_T k^2(  \zeta+\frac{4}{3} \eta  )-T^{4} \kappa  \omega ^2\Big\}\nn\\
   &&+n_0^2 T^{2}\Big\{ h_{0}k^{2} \kappa^2 \Big(\frac{\text{$\pd $P}}{\text{$\pd $n}}\Big)_T^2+k^{2}T^4 \kappa^2 \omega ^4-2 k^4 T^{2} \kappa \omega ^4
    \Big(\frac{\text{$\pd $P}}{\text{$\pd $n}}\Big)_T^{2}+2k^2 h_{0}T
   \kappa  \omega ^2 ( \zeta +\frac{4}{3} \eta )\nn\\
   &&+\Big(\frac{\text{$\pd $s}}{\text{$\pd $T}}\Big)_n^2 k^4 ( \zeta+\frac{4}{3}\eta )^{2}+2k^4T^{2} \kappa  \Big(\frac{\text{$\pd $P}}{\text{$\pd $T}}\Big)_n
   \Big(\frac{\text{$\pd $s}}{\text{$\pd $n}}\Big)_T  (\zeta +\frac{4}{3} \eta )\Big\}\Bigg]
   \Big< \delta n(\vec{k},0)\delta n(\vec{k},0)\Big>\nn\\
   ,
   \label{eqc2}
\eeqa

The expression for $\mathcal{S}_{nn}(\vec{k},\omega)$ contain derivatives of several thermodynamics quantities. In this appendix
we recast these derivatives
in terms of response functions like: isothermal and adiabatic compressibilities  ($\kappa_T$ and $\kappa_s$), 
specific heats ($C_p$ and $C_v$), baryon number susceptibility ($\chi_B$) and velocity of sound ($c_s$), etc. The 
baryon number density ($n$) and the entropy density ($s$) can be written as: 
\beqa
n=\Big(\frac{\partial P}{\partial \mu}\Big)_{T};\,\,\,\,\, s=\Big(\frac{\partial P}{\partial T}\Big)_{\mu}
\eeqa
Baryon number susceptibility, isothermal compressibility and adiabatic compressibility are given by,
\beqa
\chi_{B}=\Big(\frac{\partial n}{\partial \mu}\Big)_{T}; \kappa_{T}=\frac{1}{n_{0}}\Big(\frac{\partial n}{\partial P}\Big)_{T}; \kappa_{s}=\frac{1}{n_{0}}\Big(\frac{\partial n}{\partial P}\Big)_{s}
\eeqa
Specific heats  can be expressed as:
\beqa
C_{P}=T\Big(\frac{\partial s}{\partial T}\Big)_{P}; C_{V}&=&T\Big(\frac{\partial s}{\partial T}\Big)_{V}=T\Big(\frac{\partial s}{\partial T}\Big)_{n}\nn\\
&=&\Big(\frac{\partial \epsilon}{\partial T}\Big)_{V}=\Big(\frac{\partial \epsilon}{\partial T}\Big)_{n}
\eeqa
The expression for partial derivatives, $(\frac{\partial P}{\partial T})_{n}, (\frac{\partial P}{\partial n})_{T}, (\frac{\partial \epsilon}{\partial T})_{n}$ and $(\frac{\partial \epsilon}{\partial n})_{T}$ are given in Eq.\eqref{eqb6}, Eq.\eqref{eqb7}, Eq.\eqref{eqb9}, Eq.\eqref{eqb11} and Eq.\eqref{eqb13}

Now we have to evaluate other quantities, such as $\Big(\frac{\partial s}{\partial n}\Big)_{T}$ and $(\frac{\partial s}{\partial T})_{n}$ appear in $\Snn$ in Eq.\eqref{eqc2}. 

i) $\Big(\frac{\partial s}{\partial T}\Big)_{n}$ is evaluated as:

\beqa
 \Big(\frac{\partial s}{\partial T}\Big)_{n}=\frac{1}{T}\Big(\frac{T_{0}\partial s}{\partial T}\Big)_{n}=\frac{C_{V}}{T}
\eeqa
For fixed net baryon number,  $c_n$ can be written as $c_{n}=C_{V}$.
Therefore, 
\beqa
{\Big(\frac{\partial \epsilon}{\partial T}\Big)_{n}=C_{V}}
\eeqa
ii) We  evaluate the derivative   $\Big(\frac{\partial s}{\partial n}\Big)_{T}$  as
\beqa
\Big(\frac{\partial s}{\partial n}\Big)_{T}&=&-\Big(\frac{\partial s}{\partial T}\Big)_{n}\Big(\frac{\partial T}{\partial n}\Big)_{s}=-\frac{1}{T}\Big(\frac{T\partial s}{\partial T}\Big)_{n}\Big(\frac{\partial T}{\partial n}\Big)_{s}\nn\\
&=& -\frac{C_{V}}{T}\frac{1}{n_{0}}\Big[n_{0}\Big(\frac{\partial T}{\partial n}\Big)_{s}\Big]= \frac{C_{V}}{n_{0}T\alpha_{s}}
\eeqa
The speed of sound is can be evaluated as:
\beqa
c_s^2=\Big(\frac{\pd P}{\pd \epsilon}\Big)_{s/n}&=& \frac{nd\mu+sdT}{\mu dn+Tds}\nn\\
&=& \frac{nF dT+sdT}{\mu (\frac{\partial n}{\partial T})_\mu  dT+ \mu(\frac{\partial n}{\partial \mu})_T d\mu+T(\frac{\partial s}{\partial T})_\mu  dT+ T(\frac{\partial s}{\partial \mu})_T d\mu}\nn\\
&=&  \frac{nF dT+sdT}{\mu (\frac{\partial n}{\partial T})_\mu  dT+ \mu F(\frac{\partial n}{\partial \mu})_T dT+T(\frac{\partial s}{\partial T})_\mu  dT+ TF(\frac{\partial s}{\partial \mu})_T dT} \nn\\
&=&  \frac{nF+s}{\mu (\frac{\partial n}{\partial T})_\mu + \mu F(\frac{\partial n}{\partial \mu})_T+T(\frac{\partial s}{\partial T})_\mu+ TF(\frac{\partial s}{\partial \mu})_T}
\label{cs2}
\eeqa
where,
\beqa
F&=& \Big[(\frac{\partial s}{\partial T})_\mu- 
\frac{s}{n}(\frac{\pd n}{\pd T})_{\mu}\Big] \Big/ \Big[{\frac{s}{n}(\frac{\pd n}{\pd \mu})_{T}
-(\frac{\partial s}{\partial \mu})_T}\Big]
\eeqa
\chapter[Understanding the parametric slow mode through the dynamic structure factor]{Understanding the parametric slow mode through the dynamic structure factor}
\label{appendix08_A}

\section{Extensivity condition}
\label{extc}
The infinitesimal change in the total entropy for reversible process of a system of volume $V$ and total conserved number $N$ and total slow mode $\Phi$ at temperature $T$ and the chemical potential $\mu$ is
\begin{eqnarray}
\text{dS}&=&(E+\text{PdV}-\mu  N)/T-\pi  d \Phi \, .
\end{eqnarray}
This will give 
\begin{eqnarray}
\frac{\partial S}{\partial E}\Bigg{|}_{N,V,\Phi}=\frac{1}{T},\,\, \frac{\partial S}{\partial V}\Bigg{|}_{N,E,\Phi}=\frac{P}{T},\,\, \frac{\partial S}{\partial N}\Bigg{|}_{E,V,\Phi}=-\frac{\mu}{T},\,\, \frac{\partial S}{\partial \Phi }\Bigg{|}_{E,V,N}=-\frac{\pi}{T}\,.
\end{eqnarray}
The total entropy $S$ satisfies the following relation:
\begin{eqnarray}
S(\lambda E, \lambda V,\lambda N,\lambda \Phi)&=&\lambda S( E, V, N,\Phi)\,.
\end{eqnarray}
Differentiation with respect to $\lambda$ yield:
\begin{eqnarray}
S&=& E \frac{\partial S}{\partial \lambda E}\Bigg{|}_{\lambda  N, \lambda V,\lambda \Phi}+ N \frac{\partial S}{\partial \lambda N}\Bigg{|}_{\lambda  E, \lambda V,\lambda \Phi}+ V\frac{\partial S}{\partial \lambda V}\Bigg{|}_{\lambda  N, \lambda E,\lambda \Phi}+\Phi \frac{\partial S}{\partial \lambda \Phi}\Bigg{|}_{\lambda  N, \lambda V,\lambda E}\,.
\end{eqnarray}
Putting $\lambda=1$ and dividing by $V$ we get,
\begin{eqnarray}
\label{exentrop}
s&=&\beta(\epsilon+P-\mu n)-\pi \phi\,,
\end{eqnarray}
$s=S/V$, $\epsilon=E/V$, $n=N/V$, $\phi=\Phi/V$ are 
the entropy density, energy density, number (baryon) denisty,
and the density of the scalar field, $\phi$. 
Eq.~\eqref{exentrop} and  Eq.~\eqref{eq0805} give,
\beqa
dP&=&sdT + n d\mu + \phi d (T \pi) \,.
\eeqa
\chapter[Understanding the parametric slow mode through the dynamic structure factor]{Understanding the parametric slow mode through the dynamic structure factor}
\label{appendix08_A}

\section{The set linear equations derived for the perturbations in $\omega-k$ space}
\label{appendixA}
The linearized system of equations obtained 
by using the Fourier-Laplace transformation
of the perturbative quantities (generically denoted 
by $\delta Q$) as:
\beq
\delta {Q}(\vec{k},\omega)=\int^{\infty}_{-\infty}\, d^3 r\int_{0}^{\infty}\, dt e^{-i(\vec{k}\cdot r+ \omega t)} \delta \tilde{Q}(\vec{r},t).
\eeq
and
\beq
\delta {Q}(\vec{k},\,0)=\delta {Q}(\vec{k},t=0)=\int^{\infty}_{-\infty}\, d^3r 
e^{-i(\vec{k}\cdot r)} \delta \tilde{Q}(\vec{r},t=0).
\eeq

The system of equations for the perturbations in hydrodynamic and 
non-hydrodynamic ($\phi$) field in $\omega-k$ space are given  
below.

\begin{eqnarray}
i\omega \delta \epsilon(\vec{k}, \omega) + i k \delta q_{||}(\vec{k}, \omega)+i k(\epsilon_0+P_0)\delta u_{||}(\vec{k},\omega)= 
-\delta {\epsilon}(\vec{k},0),
\label{eq23}
\eeqa

 \beqa
  i\omega (\epsilon_0+P_0)\delta u_{||}(\vec{k},\omega)-i k \delta P(\vec{k},\omega)&-&i k \delta\Pi(\vec{k},\omega)+i \omega \delta q_{||}(\vec{k},\omega)- i k \delta \pi_{||\,||}(\vec{k},\omega)  \nn\\
   &=&(\epsilon_0+P_0){\delta u}_{||}(\vec{k},0)-\delta {q}_{||}(\vec{k},0),
   \eeqa
   \beqa
  -i \omega &&(\epsilon_0+P_0)\delta u_{\perp}(\vec{k},\omega)-i\omega \delta q_{\perp}(\vec{k},\omega)-i k \delta \pi_{\perp \,||}(\vec{k}, \omega)\nn\\
   &=&(\epsilon_0+P_0)\delta {u}_{\perp}(\vec{k},0)-{\delta q}_{\perp}(\vec{k}, 0),
   \eeqa
   \beqa
   i\omega \delta n(\vec{k},\omega)+i k n_0 \delta u_{||}(\vec{k},\omega)=-{\delta n}(\vec{k}, 0),
  \eeqa
  \beqa
   \big(1+i\omega\frac{1}{3}\zeta \beta_0\big)\delta \Pi(\vec{k},\omega)+i k\frac{1}{3}\zeta \big(\delta u_{||}(\vec{k},\omega)- {\alpha}_0 \delta q_{||}(\vec{k},\omega)\big)=-\frac{1}{3}\zeta \beta_0 {\delta\Pi}(\vec{k}, 0),
   \eeqa
  \beqa
   &&(1+i \omega \kappa T_0 \beta_1)\delta q_{||}(\vec{k},\omega)+i k \kappa T_0 \delta u_{||}(\vec{k},\omega)+i k\big[\kappa-C_{T \pi} K_{q\pi} T_0^2\big]\delta T(\vec{k},\omega)\nn\\
   &&-i k T_0^2 K_{q\pi}C_{\phi \pi}\delta\phi(\vec{k},\omega)-i k\kappa T_0{\alpha}_0\delta \Pi (\vec{k},\omega)-i k T_0^2K_{q\pi}C_{n \pi}\delta n(\vec{k},\omega)\nn\\
   &&-i k \kappa T_0 {\alpha}_1 \delta \pi_{||\,||}(\vec{k},\omega)=-\kappa T_0 {\delta u}_{||}(\vec{k}, 0)-\beta_1 \kappa T_0 {\delta q}_{||}(\vec{k}, 0),
  \eeqa
  \beqa
 &&(1+i\omega \beta_1 k T_0)\delta q_{\perp}(\vec{k},\omega)+i\kappa T_0(\omega \delta u_{\perp}(\vec{k},\omega) -k  {\alpha}_1\delta \pi_{\perp ||}(\vec{k},\omega)\big)\nn\\
 &=&-\kappa T_0\big[\delta {u}_{\perp}(\vec{k}, 0)-
\beta_1{\delta q}_{\perp}(\vec{k}, 0)\big]
  \eeqa
  \beqa
  &&(1+2i\omega\eta \beta_2)\delta \pi_{||\,||}(\vec{k},\omega)+i k\frac{4}{3}\eta \delta u_{||}(\vec{k},\omega)-\frac{4}{3}\delta q_{||}(\vec{k},\omega)\nn\\
  &&=-2\eta\beta_2{\delta\pi}_{||\,||}(\vec{k}, 0),
  \eeqa
  \beqa
  && (1+2 i\omega \eta \beta_2)\delta \pi_{\perp\, \perp}
  (\vec{k},\omega)-i k \frac{4}{3} \eta \delta u_{||}(\vec{k},\omega)+i k\frac{4}{3}\eta {\alpha}_1 \delta q_{||}(\vec{k},\omega)\nn\\
  &&=-2\eta \beta_2 {\delta\pi}_{\perp\, \perp}(\vec{k}, 0),
  \eeqa
  \beqa
  &&(1+2 i\omega \eta \beta_2)\delta \pi_{\perp\,\perp}(\vec{k},\omega)+i k \eta \delta u_{\perp}(\vec{k},\omega)-i k \delta q_{\perp}(\vec{k},\omega)\nn\\
  &&=-2\eta \beta_2 {\delta\pi}_{\perp\,||}(\vec{k}, 0),
  \eeqa
  \beqa
  && (i\omega +C_{\phi \pi}(\gamma-T_0^2\frac{K_{q\pi}}{\kappa}k^2))\delta \phi(\vec{k},\omega)+\{\gamma-K_{q\pi}(T_0^2\frac{K_{q\pi}}{\kappa}-\frac{1}{C_{T \pi}})k^2\}C_{T \pi}\delta T(\vec{k},\omega)\nn\\
  &&+(\gamma-T_0^2\frac{K_{q\pi}^2}{\kappa}k^2)C_{n \pi}\delta n(\vec{k},\omega)-i k({\phi}-i\omega T_0K_{q\pi})\delta u_{||}(\vec{k},\omega)\nn\\
  &&=-(1+i k \frac{K_{q\pi}^2}{\kappa}T_0 C_{\phi \pi}){\delta\phi}(\vec{k}, 0)-
i k T_0 K_{q\pi}{\delta u}_{||}(\vec{k}, 0)\,,
  \label{eq33}
\end{eqnarray}
where, subscripts "$||$" and "$\perp$" stand for projection along and perpendicular to $\vec{k}$ respectively. After expressing $\delta \epsilon$ and $\delta p$ as:
\beqa
 \delta \epsilon&=&\big(\frac{\partial \epsilon}{\partial n}\big)\delta n+\big(\frac{\partial \epsilon}{\partial T}\big)\delta T+\big(\frac{\partial \epsilon}{\partial \phi}\big)\delta \phi,\\
  \delta P&=&\big(\frac{\partial P}{\partial n}\big)\delta n+\big(\frac{\partial P}{\partial T}\big)\delta T+\big(\frac{\partial P}{\partial \phi}\big)\delta \phi\,.
 \eeqa
 The equations can be arranged to have the matrix from with matrix:
\beq
\mathcal{M}\, \mathcal{\delta  Q}=\mathcal{A}
\eeq
where, $\mathcal{M}$, is an $11 \times 11$ matrix representing the
coefficients of the column vector comprinsing the
quantities, $\delta n,\delta T,\delta u_{||}, \delta u_{\perp},\delta \Pi, 
\delta q_{||}$, $\delta q_{\perp},\delta \pi_{||\,||},\delta \pi_{||\,\perp},
\delta \pi_{\perp\,\perp}$, and $\delta \phi$.
\beq
{\delta  \mathcal{Q}}=\left( 
\begin{array}{c}
\delta n\\
\delta T\\
\delta \phi\\
\delta u_{||}\\
\delta q_{||}\\
\delta \pi\\
\delta \pi_{|| ||}\\
\delta \pi_{\perp \perp}\\
\delta u_{\perp}\\
\delta q_{\perp}\\
\delta \pi _{|| \perp}
\end{array}
\right),\,\,\,\,
   \mathcal{A}=\left(
\begin{array}{c}
-{\delta {n}}(\vec{k}, 0) \\ 
-\epsilon_n {\delta {n}}(\vec{k}, 0)-\epsilon_{\phi } \delta {\phi}(\vec{k}, 0)-e_T {\delta {T}}(\vec{k}, 0) \\
-\delta {\phi} (\vec{k}, 0)-i k T_0 {\delta {u}}_{||}(\vec{k}, 0) \kappa _{{q\pi }} \\ 
(\epsilon_0+P_0){\delta {u}}_{||}(\vec{k}, 0) -{\delta {q}}_{||} (\vec{k}, 0)\\
-\chi \beta _1 T_0 {\delta {q}}_{||}(\vec{k}, 0)-T_0 \chi  {\delta {u}}_{||}(\vec{k}, 0) \\
-\frac{1}{3} \beta_0 \zeta  \delta {\pi} (\vec{k}, 0)\\
-2 \beta _2 \eta  \delta {\pi} _{||||}(\vec{k}, 0) \\
-2 \beta _2 \eta  \delta {\pi}_{\perp \perp}(\vec{k}, 0) \\
(\epsilon_0+P_0){\delta {u}}_{\perp}(\vec{k}, 0)-{\delta {q}}_{\perp}(\vec{k}, 0) \\
-\chi \beta _1 T_0  {\delta {q}}_{\perp }(\vec{k}, 0)-T_0 \chi  {\delta {u}}_{\perp}(\vec{k}, 0) \\
-2 \beta _2 \eta  \delta {\pi} _{||}\perp(\vec{k}, 0)
\end{array}
\right)
\eeq
\chapter[Dynamic structure factor near the QCD critical point]{Dynamic structure factor near the QCD critical point}
\label{appendix08_A}
\section{Speed of sound and the roots of $\omega(k)$}
In this appendix we provide the expressions for the speed of sound 
and the roots of $\omega(k)$ with the inclusion of 
out-of-equilibrium mode $\phi$.
The speed of sound ($c_s$) is given by,
\beqa
c_s^2&=&\Big(\frac{\pd P}{\pd \epsilon}\Big)_{s/n}\nn\\
&=& \frac{sdT+nd\mu+\phi d\pi}{Tds+\mu dn+\pi d\phi}\nn\\
&=&\mathcal{A}/\mathcal{B}\,,
\eeqa
where, $\mathcal{A}$ and $\mathcal{B}$ are given by 
\beqa
\mathcal{A}= {s+n\mathcal{F} +\phi \mathcal{G} }\,,
\eeqa
\beqa
\mathcal{B}&=&T(\frac{\partial s}{\partial T})_{\mu,\pi}  + T\mathcal{F}(\frac{\partial s}{\partial \mu})_{T,\pi} +T\mathcal{G}(\frac{\partial s}{\partial \pi})_{T,\mu}+\mu (\frac{\partial n}{\partial T})_{\mu,\pi}  + \mu \mathcal{F}(\frac{\partial n}{\partial \mu})_{T,\pi}\nn\\
&+&\mu \mathcal{G}(\frac{\partial n}{\partial \pi})_{T,\mu} +T(\frac{\partial s}{\partial T})_{\mu,\pi}  + T\mathcal{F}(\frac{\partial s}{\partial \mu})_{T,\pi}+T\mathcal{G}(\frac{\partial s}{\partial \pi})_{T,\mu} \,.
\eeqa
where,
\beqa
\mathcal{F}&=& \frac{\Big[(\frac{\partial s}{\partial T})- 
\frac{s}{n}(\frac{\pd n}{\pd T})\Big]+(\frac{\partial \pi}{\partial T})\Big[(\frac{\partial s}{\partial \pi})- 
\frac{s}{n}(\frac{\pd n}{\pd \pi})\Big]}{\frac{s}{n}(\frac{\pd n}{\pd \mu})_{T}
-(\frac{\partial s}{\partial \mu})_T}\,,\\
\mathcal{G}&=& \frac{\Big[(\frac{\partial s}{\partial T})- 
\frac{s}{n}(\frac{\pd n}{\pd T})\Big]+(\frac{\partial \mu}{\partial T})\Big[(\frac{\partial s}{\partial \mu})- 
\frac{s}{n}(\frac{\pd n}{\pd \mu})\Big]}{\frac{s}{n}(\frac{\pd n}{\pd \mu})_{T}
-(\frac{\partial s}{\partial \mu})_T}\,.
\eeqa
The four roots of $\omega$ indicating the location of peaks 
in the structure factor with scalar field ($\phi$), obtained from the dispersion relation are
given below.
\beqa
\omega_{1} =\frac{-\sqrt{-(\epsilon_0+P_0)^2}+i \epsilon_0+i P_0}{2 T_0 \chi }+\frac{\eta  k^2 \sqrt{-(\epsilon_0+P_0){}^2}}{(\epsilon_0+P_0){}^2}+\mathcal{O}(k^3)\,,
\eeqa
\beqa
\omega_{2} =\frac{\sqrt{-(\epsilon_0+P_0){}^2}+i \epsilon_0+i P_0}{2 T_0 \chi }-\frac{\eta  k^2 \sqrt{-(\epsilon_0+P_0){}^2}}{(\epsilon_0+P_0){}^2}+\mathcal{O}(k^3)\,,
\eeqa
\beqa
\omega_{3} &=&\frac{i \gamma  (C_{\phi \pi } \epsilon_T-C_{{T\pi }} \epsilon_{\phi })}{3 \epsilon_T}+\frac{ik^2}{9 \epsilon_T (\epsilon_0+P_0)} \Big[3 T_0^2 \bar{\phi } C_{\phi \pi } \epsilon_T \kappa _{{q\pi }}-3 T_0^2 \bar{\phi } C_{{T\pi }} \epsilon_{\phi } \kappa _{{q\pi }}+3 \chi  \bar{\phi } \epsilon_{\phi }\nn\\
&-&3 T_0^2 C_{{n\pi }} \epsilon_T n_0 \kappa _{{q\pi }}+3 T_0^2 C_{{T\pi }} \epsilon_n n_0 \kappa _{{q\pi }}-3 \chi  \epsilon_n n_0-3 \epsilon_{\phi } P_0 \kappa _{{q\pi }}-3 \epsilon_0 \epsilon_{\phi } \kappa _{{q\pi }}\nn\\
&-&3 T_0 \epsilon_{\phi } P_T \kappa _{{q\pi }}+3 T_0 \epsilon_T P_{\phi } \kappa _{{q\pi }}+\zeta  \epsilon_T+4 \eta  \epsilon_T+3 T_0 \chi  P_T\Big]+\mathcal{O}(k^3)\,,
\label{A8}
\eeqa
\beqa
\omega_{4} &=&\frac{i \gamma  (C_{\phi \pi } \epsilon_T-C_{{T\pi }} e_{\phi })}{3 \epsilon_T}+k \sqrt{\frac{\bar{\phi } \epsilon_{\phi } P_T-\bar{\phi } \epsilon_T P_{\phi }+\epsilon_T n_0 P_n-\epsilon_n n_0 P_T+\epsilon_0 P_T+P_0 P_T}{\epsilon_T (\epsilon_0+P_0)}}\nn\\
&+&\frac{ik^{2}}{{9 \epsilon_T (\epsilon_0+P_0)}}\Big[(3 T_0^2 \bar{\phi } C_{\phi \pi } \epsilon_T \kappa _{{q\pi }}-3 T_0^2 \bar{\phi } C_{{T\pi }} \epsilon_{\phi } \kappa _{{q\pi }}+3 \chi  \bar{\phi } \epsilon_{\phi }-3 T_0^2 C_{{n\pi }} \epsilon_T n_0 \kappa _{{q\pi }}\nn\\
&+&3 T_0^2 C_{{T\pi }} \epsilon_n n_0 \kappa _{{q\pi }}-3 \chi  \epsilon_n n_0-3 \epsilon_{\phi } P_0 \kappa _{{q\pi }}-3 \epsilon_0 \epsilon_{\phi } \kappa _{{q\pi }}-3 T_0 \epsilon_{\phi } P_T \kappa _{{q\pi }}+3 T_0 \epsilon_T P_{\phi } \kappa _{{q\pi }}\nn\\
&+&\zeta  \epsilon_T+4 \eta  \epsilon_T+3 T_0 \chi  P_T\Big]+\mathcal{O}(k^3)\,.
\label{A9}
\eeqa
The width of the Brillouin peaks can be identified as
\beqa
\Gamma _B&=&-{9 \epsilon_T (\epsilon_0+P_0)}\Big[3 T_0^2 \bar{\phi } C_{\phi \pi } \epsilon_T \kappa _{{q\pi }}-3 T_0^2 \bar{\phi } C_{{T\pi }} \epsilon_{\phi } \kappa _{{q\pi }}+3 \chi  \bar{\phi } \epsilon_{\phi }-3 T_0^2 C_{{n\pi }} \epsilon_T n_0 \kappa _{{q\pi }}\nn\\
&+&3 T_0^2 C_{{T\pi }} \epsilon_n n_0 \kappa _{{q\pi }}-3 \chi  \epsilon_n n_0-3 \epsilon_{\phi } P_0 \kappa _{{q\pi }}-3 \epsilon_0 \epsilon_{\phi } \kappa _{{q\pi }}-3 T_0 \epsilon_{\phi } P_T \kappa _{{q\pi }}+3 T_0 \epsilon_T P_{\phi } \kappa _{{q\pi }}\nn\\
&+&\zeta  \epsilon_T+4 \eta  \epsilon_T+3 T_0 \chi  P_T\Big]\,.
\label{A10}
\eeqa
The speed of sound obtained from the dispersion relation is expressed as:
\beqa
c^{2}_s=\frac{\epsilon_T n_0 P_n-\epsilon_n n_0 P_T+\epsilon_0 P_T+P_0 P_T+\bar{\phi } \epsilon_{\phi } P_T-\bar{\phi } \epsilon_T P_{\phi }}{\epsilon_T (\epsilon_0+P_0)}\,.
\label{A11}
\eeqa
where, in Eqs.\eqref{A8}-\eqref{A11}, we have used the notation as
\beqa
X_Y=\Big(\frac{\partial X}{\partial Y}\Big)\,.
\eeqa
\cleardoublepage
 \bibliography{Hasan_thesis}
  \bibliographystyle{elsarticle-num} 
\end{document}